\documentclass[aps,a4paper,superscriptaddress,showpacs,preprintnumbers,amsmath,amssymb]{revtex4}

\usepackage{ulem}

\usepackage{psfrag} \usepackage{graphicx} \usepackage{dcolumn}
\usepackage{color} \usepackage{latexsym,amsfonts} \usepackage{bm}
\usepackage{amssymb}
\begin{document}

\title{Neutron $\beta^-$--Decay as Laboratory for Test of Standard
  Model}

\author{A. N. Ivanov} \affiliation{Atominstitut, Technische
  Universit\"at Wien, Stadionalle 2, A-1020 Wien, Austria},
\author{M. Pitschmann}\affiliation{University of Wisconsin--Madison,
  Department of Physics, 1150 University Avenue, Madison, WI 53706,
  USA}\affiliation{Physics Division, Argonne National Laboratory,
  Argonne, Illinois 60439, USA} \author{
  N. I. Troitskaya}\affiliation{ State Polytechnic University of
  St. Petersburg, Polytechnicheskaya 29, 195251, Russian
  Federation}\email{ivanov@kph.tuwien.ac.at}

\date{\today}

\begin{abstract}
We analyse the sensitivity of all experimentally observable
asymmetries and energy distributions for the neutron $\beta^-$--decay
with a polarized neutron and unpolarised decay proton and electron and
the lifetime of the neutron to contributions of order $10^{-4}$ of
interactions beyond the Standard model (SM). Since the asymmetries and
energy distributions are expressed in terms of the correlation
coefficients of the neutron $\beta^-$--decay, in order to obtain a
theoretical background for the analysis of contributions beyond the SM
we revise the calculation of the correlation coefficients within the
SM. We take into account a complete set of contributions, induced to
next--to--leading order in the large proton mass expansion by the
``weak magnetism'' and the proton recoil, and the radiative
corrections of order $(\alpha/\pi)$, calculated to leading order in
the large proton mass expansion.  We confirm the results, obtained in
literature. The contributions of interactions beyond the SM we analyse
in the linear approximation with respect to the Herczeg
phenomenological coupling constants, introduced at the hadronic
level. Such an approximation is good enough for the analysis of
contributions of order $10^{-4}$ of interactions beyond the SM.  We
show that in such an approximation the correlation coefficients depend
only on the axial coupling constant, which absorbs the contributions
of the Herczeg left--left and left--right lepton--nucleon
current--current interactions (vector and axial--vector interactions
beyond the SM), and the Herczeg scalar and tensor coupling
constants. In the lifetime of the neutron in addition to the axial
coupling constant the contributions of the Herczeg left--left and
left--right lepton--nucleon current--current interactions (vector and
axial--vector interactions beyond the SM) are absorbed by the
Cabibbo--Kobayashi--Maskawa (CKM) matrix element.
\end{abstract}

\maketitle

\section{Introduction}
\label{sec:intro}

In this paper we propose a consistent analysis of the sensitivity of
all observable asymmetries and energy distributions of the neutron
$\beta^-$--decay $n \to p + e^- + \bar{\nu}_e$ with a polarized
neutron and unpolarised decay proton and electron and the lifetime of
the neutron \cite{Abele1,Nico1} to contributions of order $10^{-4}$ of
interactions beyond the Standard model (SM), described at the
phenomenological level. Such an order of corrections beyond the SM has
been pointed out by Ramsey--Musolf and Su \cite{SUSY} within the
Minimal Supersymmetric extension of the SM (MSSM). As has been shown
in \cite{SPT1}--\cite{SPT4}, at the phenomenological level the neutron
$\beta^-$--decay may be described by eight complex coupling constants
determining the strength of interactions beyond the SM. As a result
possible deviations from the SM, causing the contributions of order
$10^{-4}$ to the correlation coefficients of the neutron
$\beta^-$--decay and the lifetime of the neutron, may be determined in
terms of vector, axial--vector, scalar and tensor lepton--baryon weak
coupling constants \cite{SPT1a}.

It is well--known \cite{Abele1,Nico1} that in the non--relativistic
approximation to leading order in the large proton mass expansion,
which is equivalent to the leading order of the heavy--baryon
approximation, and in the rest frame of the neutron the SM with weak
$V - A$ interactions \cite{PDG12} describes the $\beta^-$--decay of
the neutron in terms of two coupling constants $G_V$ and $G_A$
\cite{Abele1,Nico1} (see also \cite{SPT5}).  The coupling constant
$G_V$ is defined by the product $G_V = G_F V_{ud}$ of the Fermi
coupling constant $G_F = g^2_W/8 M^2_W$ \cite{PDG12}, where $g_W$ and
$M_W$ are the electroweak coupling constant and the $W$--boson mass,
and $V_{ud}$ is the matrix element of the Cabibbo--Kobayashi-Maskawa
(CKM) quark mixing matrix \cite{PDG12}.  The coupling constant $G_A$
is equal to $G_A = \lambda G_V$, where $\lambda = G_A/G_V$ is the
axial coupling constant, induced by renormalisation of the axial
hadronic current by strong low--energy interactions \cite{CHP73}.  For
the weak interactions, invariant under time reversal, the coupling
constant $\lambda$ is real.

The observables of the neutron $\beta^-$--decay with unpolarised
particles are the lifetime of the neutron $\tau_n$ and the correlation
coefficient $a_0$, describing correlations between 3--momenta of the
electron and antineutrino to leading order in the large proton mass
expansion in the rest frame of the neutron. Experimentally the
correlation coefficient $a_0$ can be determined, for example, by
measuring either the electron--proton energy distribution $a(E_e,
T_p)$, where $E_e$ and $T_p$ are the total and kinetic energies of the
electron and proton, respectively, or the proton--energy spectrum
$a(T_p)$. As a function of the electron energy $E_e$ the correlation
coefficient of correlations between the electron and antineutrino
3--momenta we denote as $a(E_e)$.

The neutron $\beta^-$--decay with a polarized neutron and unpolarised
decay proton and electron is characterized also by two additional
observable correlation coefficients $A_0$ and $B_0$, describing to
leading order in the large proton mass expansion correlations between
the neutron spin  and 3--momenta of the electron and antineutrino,
respectively. As functions of the electron energy $E_e$ the
correlation coefficients of correlations between the neutron spin  and
3--momenta of the electron and antineutrino we denote as $A(E_e)$ and
$B(E_e)$, respectively. The lifetime of the neutron and the
correlation coefficients under consideration are of order $\tau_n \sim
880\,{\rm s}$, $a_0 \sim a(E_e) \sim A_0 \sim A(E_e) \sim - 0.1$ and
$B_0 \sim B(E_e) \sim 1$ \cite{PDG12}.  The coupling constants $G_V$
and $\lambda$ define the main contributions to the lifetime of the
neutron and the correlation coefficients \cite{Abele1,Nico1} (see also
\cite{SPT5}).

However for the description of the neutron $\beta^-$--decay at the
modern level of experimental accuracies when the experimental data on
the axial coupling constant $\lambda = - 1.2750(9)$ \cite{Abele1} and
the lifetime of the neutron $\tau^{(\exp)}_n = 878.5(8)\,{\rm s}$
\cite{Serebrov1} have been obtained with accuracies $0.07\,\%$ and
$0.09\,\%$, respectively, two coupling constants $G_V$ and $\lambda$
are not enough for the correct description of the properties of the
neutron $\beta^-$--decay. Indeed, it is well--known that the radiative
corrections to the lifetime of the neutron \cite{RC3}--\cite{RC18},
calculated to leading order in the large proton mass expansion within
the SM with the $V - A$ weak interactions and Quantum Electrodynamics
(QED), give the contribution of about $3.9\,\%$ \cite{RC1} (see also
\cite{Abele1}).  This allows to describe correctly the lifetime of the
neutron \cite{SPT5}. The radiative corrections are very important also
for the correct determination of the Fermi coupling constant $G_F$. It
may be extracted from the experimental data on the weak coupling
constant $G_{\mu}$ of the $\mu^-$--decay $\mu^- \to e^- + \nu_{\mu} +
\bar{\nu}_e$ with the account for the radiative corrections
\cite{RC3}--\cite{RC6}.  The contributions of the radiative
corrections to the ratios of the correlation coefficients $a(E_e)/a_0$
and $A(E_e)/A_0$ are by order of magnitude smaller compared with the
contribution to the lifetime of the neutron, whereas the correlation
coefficient $B(E_e)$ has no radiative corrections to order
$\alpha/\pi$ \cite{Gudkov1,Gudkov2}.

Another type of corrections, which may be calculated within the SM and
should be taken into account for the description of the neutron
$\beta^-$--decay on the same footing as the radiative corrections, are
the contributions of the ``weak magnetism'' and the proton recoil
\cite{Holstein, Wilkinson} (see also \cite{Gudkov2}), calculated to
next--to--leading order in the large proton mass expansion.

The radiative corrections and the corrections from the ``weak
magnetism'' and the proton recoil, calculated to leading order and to
next--to--leading order in the large proton mass expansion,
respectively, define a complete set of corrections to the observables
of the neutron $\beta^-$--decay, which should be taken into account
within the SM as a background for the analysis of contributions of
order $10^{-4}$ of interactions beyond the SM.

Since the observable asymmetries and energy distributions are defined
in terms of the correlation coefficients \cite{Abele1}, for the aim of
the paper to analyse the sensitivity of the observables of the neutron
$\beta^-$--decay to contributions of order $10^{-4}$ of interactions
beyond the SM we revise the calculation of the correlation
coefficients of the neutron $\beta^-$--decay with a polarized neutron and
unpolarised decay proton and electron and the lifetime of the neutron
within the SM with $V - A$ weak interactions.  In addition to the
non--relativistic terms $a_0$, $A_0$ and $B_0$, calculated to leading
order in the large proton mass expansion, we take into account 1) the
contributions of the ``weak magnetism'' and the proton recoil to
next--to--leading order in the large $M$ expansion, which provide a
complete set of corrections of order $1/M$, where $M$ is an average
mass of the neutron and proton $M = (m_n + m_p)/2$, and 2) the
radiative corrections of order $\alpha/\pi$, calculated to leading
order in the large $M$ expansion, where $\alpha = 1/137.036$ is the
fine--structure constant \cite{PDG12}. The small parameter of the
large $M$ expansion is $E_0/M \sim 10^{-3}$, where $E_0 \sim 1\,{\rm
  MeV}$ is the end--point energy of the electron--energy spectrum. The
parameter $E_0/M \sim 10^{-3}$ is commensurable with the parameter of
the radiative corrections $\alpha/\pi \sim 10^{-3}$. As we show below
due to the strong dependence on the axial coupling constant $\lambda$
the numerical values of the $1/M$ corrections vary from $10^{-5}$ to
$10^{-1}$. In turn the contributions of the radiative corrections do
not depend on the axial coupling constant $\lambda$. Following Sirlin
\cite{RC8} we show that the contributions of the radiative
corrections, depending on the axial coupling constant $\lambda$, may
be absorbed by renormalisation of the Fermi $G_F$ and axial $\lambda$
coupling constants.  The contributions of radiative corrections to the
ratios $a(E_e)/a_0$ and $A(E_e)/A_0$ are of order $10^{-3}$. The
correlation coefficients, calculated within the SM to order $1/M$ and
$\alpha/\pi$, determine the theoretical background for the analysis of
contributions of order $10^{-4}$ beyond the SM.

A phenomenological analysis of contributions of interactions beyond
the SM model shows that these interactions induce only 1) the energy
independent contributions and 2) the contributions proportional to
$m_e/E_e$, where $m_e$ is the electron mass. We show below that the
contributions of the ``weak magnetism'' and the proton recoil into the
terms proportional to $m_e/E_e$ of the correlation coefficients
$a(E_e)$, $A(E_e)$ and $B(E_e)$ are of order $10^{-3} - 10^{-4}$. Thus, the
obtained theoretical expressions for the correlation coefficients
should be taken into account for a correct experimental determination
of contributions of order $10^{-4}$ beyond the SM.

The radiative corrections to the correlation coefficients $a(E_e)/a_0$
and $A(E_e)/A_0$ determine the most important and complicated part of
the corrections of order $10^{-3}$.  In the neutron $\beta^-$--decay
the radiative corrections are defined by 1) the contributions to the
continuum-state $\beta^-$--decay mode from one--virtual photon
exchanges, $W$--boson and $Z$--boson exchanges and QCD corrections
\cite{RC14,RC16}--\cite{RC18} and 2) the contribution of the radiative
$\beta^-$--decay mode $n \to p + e^- + \bar{\nu}_e + \gamma$ with
emission of a real photon \cite{RC3}--\cite{RC18} (see also
\cite{Gudkov1,Gudkov2}). The sum of the electron-energy and angular
distributions of these two decay modes does not suffer from infrared
divergences of virtual and real photons.

The radiative corrections to the $\beta^-$--decay of the neutron we
calculate within the standard finite-photon mass (FPM) regularization
of infrared divergences \cite{RC3}--\cite{RC18}. As has been shown by
Marciano and Sirlin \cite{RC12}, the FPM regularization is equivalent
to the dimensional regularization \cite{IZ80}.

The radiative corrections to the lifetime of the neutron, calculated
within the FPM regularization, obey the Kinoshita--Lee--Nauenberg
(KLN) theorem \cite{KLN}. According to the KLN theorem, the radiative
corrections to the lifetime of the neutron, integrated over the phase
volume of the final state of the neutron $\beta^-$--decay in the limit
of the massless electron $m_e \to 0$, should not depend on the
electron mass $m_e$. For the first time such an independence of the
electron mass in the limit $m_e \to 0$ has been demonstrated by
Kinoshita and Sirlin \cite{RC5}.

We reproduce fully the radiative corrections to the lifetime of the
neutron and the correlation coefficients $a(E_e)$ and $A(E_e)$,
calculated in \cite{RC6}--\cite{RC18} and \cite{Gudkov1,Gudkov2},
respectively (see Eq.(\ref{label7}) and Appendix D).  We show also
that the correlation coefficient $B(E_e)$ has no radiative corrections
to order $\alpha/\pi$, calculated to leading order in the large $M$
expansion. This agrees well with the results, obtained by Gudkov {\it
  et al.} \cite{Gudkov2}.  The corrections of the ``weak magnetism''
and the proton recoil we calculate in complete agreement with the
results, obtained in \cite{Gudkov2} (see also \cite{Wilkinson}). Above
the background of these corrections the contributions of interactions
beyond the SM are calculated to leading order in the large $M$
expansion and in the linear approximation with respect to the Herczeg
phenomenological coupling constants, introduced at the hadronic level
in terms of the lepton--nucleon current--current interactions (see
Appendix G). Such an approximation is good enough for the analysis of
the sensitivity of the observables of the neutron $\beta^-$--decay to
contributions of order $10^{-4}$ beyond the SM.

The obtained results are applied to the theoretical analysis of 1) the
asymmetries $A_{\exp}(E_e)$, $B_{\exp}(E_e)$, the electron--proton
energy distribution $a(E_e, T_p)$, the proton--energy spectrum
$a(T_p)$ and the proton recoil asymmetry $C_{\exp}$, used for the
experimental determination of the axial coupling constant $\lambda$
and the correlation coefficients $A_0$, $B_0$, $a_0$ and $C_0 = -
x_C(A_0 + B_0)$, where $x_C = 0.27591$ is a theoretical numerical
factor (see section~\ref{sec:Cexp}), respectively, 2) the lifetime of
the neutron and 3) the sensitivity of the electron--proton energy
distribution $a(E_e, T_p)$, the proton--energy spectrum $a(T_p)$, the
asymmetries $A_{\exp}(E_e)$, $B_{\exp}(E_e)$ and $C_{\exp}$ and the
lifetime of the neutron $\tau_n$ to contributions of order $10^{-4}$
of interactions beyond the SM. The experimental analysis of the
asymmetries $A_{\exp}(E_e)$, $B_{\exp}(E_e)$, the proton--energy
spectrum $a(T_p)$ and the proton recoil asymmetry $C_{\exp}$ has been
carried out in \cite{Abele1,Mund}, \cite{Abele2,Schumann},
\cite{Byrne} (see also \cite{Stratowa}) and \cite{Schumann08}),
respectively.

\section{Organization of paper}
\label{sec:content}

The paper is organized as follows. In section~\ref{sec:spectrum} we
calculate the electron--energy and angular distribution of the neutron
$\beta^-$--decay with a polarized neutron and unpolarised decay proton
and electron. We take into account a complete set of corrections of
order $1/M$, caused by the ``weak magnetism'' and the proton recoil,
and the radiative corrections of order $\alpha/\pi$.  Then we analyse
1) in section~\ref{sec:Aexp} the electron asymmetry $A_{\exp}(E_e)$,
which has been used in \cite{Abele1,Mund} for the experimental
determination of the axial coupling constant $\lambda = - 1.2750(9)$
and the correlation coefficient $A^{(\exp)}_0 = - 0.11933(34)$, 2) in
section~\ref{sec:Bexp} the antineutrino asymmetry $B_{\exp}(E_e)$,
which has been used for the experimental determination of the
correlation coefficient $B^{(\exp)}_0 = 0.9802(50)$ in
\cite{Abele2,Schumann}, 3) in section~\ref{sec:a0exp} the
electron--proton energy distribution $a(E_e,T_p)$ and the
proton--energy spectrum $a(T_p)$, which may be used for the
experimental determination of the axial coupling constant $\lambda$
and the correlation coefficient $a_0$, 4) in section~\ref{sec:Cexp}
the proton recoil asymmetry $C_{\exp}$, which has been used for the
experimental determination of the correlation coefficient
$C^{(\exp)}_0 = - x_C (A_0 + B_0) = - 0.2377(26)$ \cite{Schumann08},
and 5) in section~\ref{sec:lifetime} the lifetime of the neutron. In
section~\ref{sec:sensitivity} we propose the theoretical analysis of
the sensitivity of the electron--proton energy distribution
$a(E_e,T_p)$, the proton--energy spectrum $a(T_p)$, the asymmetries
$A_{\exp}(E_e)$, $B_{\exp}(E_e)$ and $C_{\exp}$ and the lifetime of
the neutron to contributions of order $10^{-4}$ of interactions beyond
the SM. We show that to linear approximation with respect to the
Herczeg phenomenological coupling constants the Herczeg coupling
constants $a^h_{LL}$ and $a^h_{LR}$ of the left--left and left--right
lepton--nucleon current--current interactions (vector and
axial--vector interactions beyond the SM) can be fully absorbed by the
axial coupling constants and cannot be determined from the
experimental data on the electron--proton energy distribution $a(E_e,
T_p)$, the proton--energy spectrum $a(T_p)$ and the asymmetries
$A_{\exp}(E_e)$, $B_{\exp}(E_e)$ and $C_{\exp}(E_e)$. This agrees well
with the results obtained in \cite{UMW1}--\cite{UMW3}. In the lifetime
of the neutron the contributions of the Herczeg left--left and
left--right lepton--nucleon current--current interactions (vector and
axial--vector interactions beyond the SM) become unobservable after
the renormalisation of the CKM matrix element $V_{ud} \to
(V_{ud})_{\rm eff} = V_{ud} (1 + a^h_{LL} + a^h_{LR})$ (see
section~\ref{sec:sensitivity}) in agreement with
\cite{UMW1}--\cite{UMW3}.  In the conclusion~\ref{sec:conclusion} we
summarized the obtained results. In Appendix A we give a detailed
calculation of the amplitude, the electron-energy and angular
distribution of the continuum-state $\beta^-$--decay of the neutron to
next--to--leading order in the large $M$ expansion. In Appendix B we
calculate the rate, the electron--energy and angular distribution of
the radiative $\beta^-$--decay of the neutron with a polarized neutron
and unpolarised decay particles and their contributions to the rate,
the electron--energy and angular distribution of the neutron
$\beta^-$--decay.  We compare our results for the branching ratios of
the radiative $\beta^-$--decay of the neutron with the experimental
data, obtained by Nico {\it et al.}  \cite{Nico2} and Cooper {\it et
  al.}  \cite{Cooper} (see also \cite{RC08}), and the theoretical
results, obtained in \cite{RBD1,RBD2}.  In Appendix C within the FPM
regularization \cite{RC3}--\cite{RC18} we analyse the infrared
divergent contributions of one--virtual photon exchanges to the
continuum-state $\beta^-$--decay mode of the neutron. In Appendix D we
give a detailed calculation of the radiative corrections, caused by
one--virtual photon exchanges in the continuum-state $\beta$--decay of
the neutron and the radiative $\beta^-$--decay of the neutron.  We
define the radiative corrections to the lifetime of the neutron and
the correlation coefficients by the functions $g_n(E_e)$ and
$f_n(E_e)$, respectively. We show also that in the rest frame of the
neutron and in the non--relativistic approximation for the proton
one--virtual photon exchanges induce effective scalar and tensor weak
lepton--baryon couplings of order $\alpha G_F/\pi$ and $\alpha \lambda
G_F/\pi$, respectively, depending on the electron energy $E_e$.  In
Appendix E we show that the radiative corrections to the lifetime of
the neutron, defined by the function $g_n(E_e)$, satisfy the KLN
theorem. This agrees with the analysis of the radiative corrections,
carried out in \cite{RC5}. In Appendix F we calculate the radiative
corrections to the lifetime of the neutron, using the procedure
proposed by Sirlin \cite{RC8}. We confirm that unambiguity of
observable radiative corrections to the lifetime of the neutron is
caused by gauge invariance of the amplitude of one--virtual photon
exchanges in the continuum-state $\beta^-$--decay of the neutron
\cite{RC8}.  In Appendix G we calculate the contributions to the
correlation coefficients and the lifetime of the neutron from
phenomenological vector, axial--vector, scalar and tensor interactions
beyond the SM \cite{SPT1}--\cite{SPT4}. In Appendix H we calculate the
proton recoil corrections, caused by the electron--proton Coulomb
interaction in the final state of the neutron $\beta^-$--decay. We
show that in the experimentally used electron energy region $250\,{\rm
  keV} \le T_e = E_e - m_e \le 455\,{\rm keV}$ \cite{Abele1} the
contributions of these corrections to the correlation coefficients are
of order $10^{-6} - 10^{-5}$ and may be neglected for the analysis of
contributions of order $10^{-4}$. In Appendix I we give a detailed
calculation of the electron--proton energy--momentum and angular
distribution of the neutron $\beta^-$--decay with a polarized neutron
and unpolarised decay proton and electron. We calculate the
electron--proton energy distribution $a(E_e,T_p)$, the proton--energy
spectrum $a(T_p)$ and the correlation coefficient $C$ of the proton
recoil asymmetry by taking into account the $1/M$ corrections from the
``weak magnetism'' and the proton recoil and the radiative corrections
of order $\alpha/\pi$, calculated to leading order in the large $M$
expansion.

\section{Electron--energy and angular distribution of
neutron $\beta^-$--decay in Standard model}
\label{sec:spectrum}

For the analysis of the electron--energy  and angular
distribution of the continuum-state $\beta^-$--decay of the neutron
we use the Hamiltonian of $V-A$ interactions with a real axial
coupling constant $\lambda$ and the contribution of the ``weak
magnetism'' \cite{IZ80,BH92}
\begin{eqnarray}\label{label1}
\hspace{-0.3in}{\cal H}_W(x) =
\frac{G_F}{\sqrt{2}}\,V_{ud}\,\Big\{[\bar{\psi}_p(x)\gamma_{\mu}(1 +
  \lambda \gamma^5)\psi_n(x)] + \frac{\kappa}{2 M}
\partial^{\nu}[\bar{\psi}_p(x)\sigma_{\mu\nu}\psi_n(x)]\Big\}
        [\bar{\psi}_e(x)\gamma^{\mu}(1 - \gamma^5)\psi_{\nu}(x)]
\end{eqnarray}
invariant under time reversal, where $\psi_p(x)$, $\psi_n(x)$,
$\psi_e(x)$ and $\psi_{\nu}(x)$ are the field operators of the proton,
neutron, electron and antineutrino, respectively, $\gamma^{\mu}$,
$\gamma^5$ and $\sigma^{\mu\nu} = \frac{i}{2}(\gamma^{\mu}\gamma^{\nu}
- \gamma^{\nu}\gamma^{\mu})$ are the Dirac matrices \cite{IZ80} and
$\kappa = \kappa_p - \kappa_n = 3.7058$ is the isovector anomalous
magnetic moment of the nucleon, defined by the anomalous magnetic
moments of the proton $\kappa_p = 1.7928$ and the neutron $\kappa_n =
- 1.9130$ and measured in nuclear magneton \cite{PDG12}.

For numerical calculations we use $G_F = 1.1664\times 10^{-11}\,{\rm
  MeV}^{-2}$ and $|V_{ud}| = 0.97427(15)$ \cite{PDG12}. The value of
the CKM matrix element $|V_{ud}| = 0.97427(15)$ agrees well with
$|V_{ud}| = 0.97425(22)$, measured from the superallowed $0^+ \to 0^+$
nuclear $\beta^-$--decays \cite{Vud}. It satisfies also well the
unitarity condition $|V_{ud}|^2 + |V_{us}|^2 + |V_{ub}|^2 =
0.99999(41)$ for the CKM matrix elements \cite{PDG12}. The error
$\Delta_{\rm U} = \pm 0.00041$ of the unitarity condition is
determined by the errors of the CKM matrix elements $|V_{ud}|=
0,97427\pm 0.00015$, $|V_{us}| = 0.22534\pm 0.00065$ and $|V_{ub}| =
0.00351^{+0.00015}_{-0.00014}$ (see Eq.(11.27) in p.\,162 of
Ref.\cite{PDG12}). As a result the error of the unitarity condition is
equal to
\begin{eqnarray*}
\Delta_{\rm U} = \sqrt{\sum_q|2V_{uq}\Delta V_{uq}|^2} = 0.00041,
\end{eqnarray*}
where $q = d, s, b$ and $\Delta V_{ud} = 0.00015$, $\Delta V_{us} =
0.00065$ and $\Delta V_{ub} = 0.00015$, respectively.

The amplitude of the continuum-state $\beta^-$--decay of the neutron,
calculated in the rest frame of the neutron and to next--to--leading
order in the large $M$ expansion taking into account the contributions
of the ``weak magnetism'' and the proton recoil, is (see Appendix A)
\begin{eqnarray}\label{label2}
\hspace{-0.3in}&&M(n \to p\,e^- \,\bar{\nu}_e) = -
2 m_n\,\frac{G_F}{\sqrt{2}}\,V_{ud}\,\Big\{
[\varphi^{\dagger}_p\varphi_n]\,[\bar{u}_e\,\gamma^0 (1 -
\gamma^5)v_{\bar{\nu}}] - \tilde{\lambda}\, [\varphi^{\dagger}_p
\vec{\sigma}\,\varphi_n]\cdot [\bar{u}_e\,\vec{\gamma}\,(1 -
\gamma^5)v_{\bar{\nu}}]\nonumber\\
\hspace{-0.3in}&&- \frac{m_e}{2
  M}\,[\varphi^{\dagger}_p\varphi_n][\bar{u}_e\,(1 -
  \gamma^5)v_{\bar{\nu}}] + \frac{\tilde{\lambda}}{2
  M}[\varphi^{\dagger}_p(\vec{\sigma}\cdot \vec{k}_p) \varphi_n
]\,[\bar{u}_e\,\gamma^0 (1 - \gamma^5)v_{\bar{\nu}}] - i\,
\frac{\kappa + 1}{2 M} [\varphi^{\dagger}_p (\vec{\sigma}\times
  \vec{k}_p) \varphi_n]\cdot [\bar{u}_e\,\vec{\gamma}\,(1 -
  \gamma^5)v_{\bar{\nu}}] \Big\},
\end{eqnarray}
where $\varphi_p$ and $\varphi_n$ are the Pauli spinorial wave
functions of the proton and neutron and $u_e$ and $v_{\bar{\nu}}$ are
the Dirac bispinor wave functions of the electron and antineutrino,
respectively. Then, $\vec{k}_p$ is a 3--momentum of the proton related
to 3--momenta of the electron $\vec{k}_e$ and antineutrino $\vec{k}$
as $\vec{k}_p= - \vec{k}_e - \vec{k}$, $\tilde{\lambda} = \lambda (1 -
E_0/2M)$, where $E_0 = (m^2_n - m^2_p + m^2_e)/2 m_n = 1.2927\,{\rm
  MeV}$ is the end--point energy of the electron--energy spectrum,
calculated for $m_n = 939.5654\,{\rm MeV}$, $m_p = 938.2720\,{\rm
  MeV}$ and $m_e = 0.5110\,{\rm MeV}$ \cite{PDG12}. From
Eq.(\ref{label2}) one may see that the parameter of the large $M$
expansion or the $1/M$ corrections to the amplitude of the
$\beta^-$--decay of the neutron is $k_p/M \sim E_0/M \sim
10^{-3}$. The detailed calculation of the amplitude Eq.(\ref{label2})
is given in Appendix A.

The electron--energy and angular distribution of the neutron
$\beta^-$--decay takes the form \cite{Abele1,Gudkov2}
\begin{eqnarray}\label{label3}
\hspace{-0.3in}&&\frac{d^5 \lambda_n(E_e,\vec{k}_e,\vec{k},\vec
  {\xi}_n)}{dE_e d\Omega_e d\Omega} = (1 + 3
\lambda^2)\,\frac{G^2_F|V_{ud}|^2}{32\pi^5}\,(E_0 -
E_e)^2\,\sqrt{E^2_e - m^2_e}\, E_eF(E_e, Z =
1)\,\Phi_{\beta^-_c}(\vec{k}_e,\vec{k}\,)\,\tilde{\zeta}(E_e)\nonumber\\
\hspace{-0.3in}&&\times\,\Big(1 + \tilde{a}(E_e)\,\frac{\vec{k}_e\cdot
  \vec{k}}{E_e E} + \tilde{A}(E_e)\,\frac{\vec{\xi}_n\cdot
  \vec{k}_e}{E_e} + \tilde{B}(E_e)\, \frac{\vec{\xi}_n\cdot
  \vec{k}}{E} + \tilde{K}_n(E_e)\,\frac{(\vec{\xi}_n\cdot
  \vec{k}_e)(\vec{k}_e\cdot \vec{k})}{E^2_e E}+
\tilde{Q}_n(E_e)\,\frac{(\vec{\xi}_n\cdot \vec{k})(\vec{k}_e\cdot
  \vec{k})}{E_e E^2}\nonumber\\
\hspace{-0.3in}&&+ \tilde{D}(E_e)\,\frac{\vec{\xi}_n\cdot
  (\vec{k}_e\times \vec{k}\,)}{E_e E}\Big),
\end{eqnarray}
where $E = E_0 - E_e$ is the antineutrino energy and $d\Omega_e$ and
$d\Omega$ are the infinitesimal elements of the solid angles of the
electron and antineutrino 3--momenta relative to the neutron spin,
respectively.

The function $\Phi_{\beta^-_c}(\vec{k}_e,\vec{k}\,)$ is defined by the
contribution of the proton recoil. To next--to--leading order in the
large $M$ expansion it is equal to (see Eqs.(\ref{labelA.19}) -
(\ref{labelA.23}) in Appendix A)
\begin{eqnarray}\label{label4}
\Phi_{\beta^-_c}(\vec{k}_e,\vec{k}\,) = 1+ \frac{3}{M}\Big(E_e -
\frac{\vec{k}_e\cdot \vec{k}}{E}\Big).
\end{eqnarray}
The function $F(E_e, Z = 1)$ is the relativistic Fermi function
\cite{Jackson1958,EK66} (see also  \cite{Wilkinson})
\begin{eqnarray}\label{label5}
\hspace{-0.3in}F(E_e, Z = 1 ) =  \Big(1 +
\frac{1}{2}\gamma\Big)\,\frac{4(2 r_pm_e\beta)^{2\gamma}}{\Gamma^2(3 +
  2\gamma)}\,\frac{\displaystyle e^{\,\pi
 \alpha/\beta}}{(1 - \beta^2)^{\gamma}}\,\Big|\Gamma\Big(1 + \gamma +
 i\,\frac{\alpha }{\beta}\Big)\Big|^2,
\end{eqnarray}
where $\beta = k_e/E_e = \sqrt{E^2_e - m^2_e}/E_e$ is the electron
velocity, $\gamma = \sqrt{1 - \alpha^2} - 1$, $r_p$ is the electric
radius of the proton and $\alpha = 1/137.036$ is the fine--structure
constant.  In numerical calculations we will use $r_p = 0.841\,{\rm
  fm}$ \cite{LEP}.

Following \cite{Gudkov2} we transcribe the r.h.s. of Eq.(\ref{label3})
into the form
\begin{eqnarray}\label{label6}
\hspace{-0.3in}&&\frac{d^5 \lambda_n(E_e, \vec{k}_e, \vec{k},
  \vec{\xi}_n)}{dE_e d\Omega_e d\Omega} = (1 + 3
\lambda^2)\,\frac{G^2_F|V_{ud}|^2}{32\pi^5}\,(E_0 - E_e)^2
\,\sqrt{E^2_e - m^2_e}\, E_e\,F(E_e, Z = 1)\,\zeta(E_e)\nonumber\\
\hspace{-0.3in}&&\times\,\Big\{1 + a(E_e)\,\frac{\vec{k}_e\cdot
  \vec{k}}{E_e E} + A(E_e)\,\frac{\vec{\xi}_n\cdot \vec{k}_e}{E_e} +
B(E_e)\, \frac{\vec{\xi}_n\cdot \vec{k}}{E} +
K_n(E_e)\,\frac{(\vec{\xi}_n\cdot \vec{k}_e)(\vec{k}_e\cdot
  \vec{k})}{E^2_e E}+ Q_n(E_e)\,\frac{(\vec{\xi}_n\cdot
  \vec{k})(\vec{k}_e\cdot \vec{k})}{ E_e E^2} \nonumber\\
\hspace{-0.3in}&&+ D(E_e)\,\frac{\vec{\xi}_n\cdot (\vec{k}_e\times
  \vec{k}\,)}{E_e E} - 3\,\frac{E_e}{M}\,\frac{1 - \lambda^2}{1 + 3
  \lambda^2}\,\Big(\frac{(\vec{k}_e\cdot \vec{k}\,)^2}{E^2_e E^2} -
\frac{1}{3}\,\frac{k^2_e}{E^2_e}\,\Big) \Big\}.
\end{eqnarray}
The correlation coefficients are given by (see Appendix A, B, C and D)
\begin{eqnarray}\label{label7}
\hspace{-0.3in}&&\zeta(E_e) = \Big(1 +
\frac{\alpha}{\pi}\,g_n(E_e)\Big) + \frac{1}{M}\,\frac{1}{1 + 3
  \lambda^2}\, \Big[- 2\,\lambda\Big(\lambda - (\kappa + 1)\Big)\,E_0
  + \Big(10 \lambda^2 - 4(\kappa + 1)\, \lambda +
  2\Big)\,E_e\nonumber\\
\hspace{-0.3in}&&- 2 \lambda\,\Big(\lambda - (\kappa +
1)\Big)\,\frac{m^2_e}{E_e}\Big],\nonumber\\
\hspace{-0.3in}&&\zeta(E_e)\,a(E_e) = a_0\,\Big(1 +
\frac{\alpha}{\pi}\,g_n(E_e) + \frac{\alpha}{\pi}\,f_n(E_e)\Big) +
\frac{1}{M}\,\frac{1}{1 + 3 \lambda^2}\,\Big[2 \lambda\,\Big(\lambda -
  (\kappa + 1)\Big)\,E_0 - 4 \lambda\Big(3 \lambda - (\kappa +
  1)\Big)\,E_e\Big],\nonumber\\
\hspace{-0.3in}&&\zeta(E_e)\,A(E_e) = A_0\,\Big(1 +
\frac{\alpha}{\pi}\,g_n(E_e) + \frac{\alpha}{\pi}\,f_n(E_e)\Big) +
\frac{1}{M}\,\frac{1}{1 + 3 \lambda^2}\,\Big[\Big(\lambda^2 -
  \kappa\,\lambda - (\kappa + 1)\Big)\,E_0 \nonumber\\
\hspace{-0.3in}&& - \Big(5 \lambda^2 - (3\kappa - 4)\,\lambda -
(\kappa + 1)\Big)\,E_e\Big],\nonumber\\
\hspace{-0.3in}&&\zeta(E_e)\,B(E_e) = B_0\,\Big(1 +
\frac{\alpha}{\pi}\,g_n(E_e)\Big) + \frac{1}{M}\,\frac{1}{1 + 3
  \lambda^2}\,\Big[- 2\,\lambda\,\Big(\lambda - (\kappa +
  1)\Big)E_0\nonumber\\
\hspace{-0.3in}&& + \Big(7 \lambda^2 - (3 \kappa + 8)\, \lambda +
(\kappa + 1)\Big)\,E_e - \Big(\lambda^2 - (\kappa + 2)\, \lambda +
(\kappa + 1)\Big)\,\frac{m^2_e}{E_e}\Big],\nonumber\\
\hspace{-0.3in}&& \zeta(E_e)\,K_n(E_e) = \frac{1}{M}\,\frac{1}{1 + 3
  \lambda^2}\,\Big(5 \lambda^2 - (\kappa - 4)\, \lambda - (\kappa +
1)\Big) E_e,\nonumber\\
\hspace{-0.3in}&&\zeta(E_e)\,Q_n(E_e) = \frac{1}{M}\,\frac{1}{1 + 3
  \lambda^2}\,\Big[ \Big(\lambda^2 - (\kappa + 2) \lambda + (\kappa +
  1)\Big) E_0 - \Big(7 \lambda^2 - (\kappa + 8)\,\lambda + (\kappa +
  1)\Big) E_e\Big],\nonumber\\
\hspace{-0.3in}&& \zeta(E_e)\,D(E_e) = 0,
\end{eqnarray}
where the correlation coefficients $a_0$, $A_0$ and $B_0$ are
determined by \cite{Abele1} (see also \cite{SPT5})
\begin{eqnarray}\label{label8}
a_0 = \frac{1 - \lambda^2}{1 + 3 \lambda^2}\quad,\quad A_0 = -
\,2\,\frac{\lambda(1 + \lambda)}{1 + 3 \lambda^2}\quad,\quad B_0 = -
2\,\frac{ \lambda(1 - \lambda) }{1 + 3 \lambda^2}.
\end{eqnarray}
The radiative corrections are determined by the functions $g_n(E_e)$
and $f_n(E_e)$, which are given in Eq.(\ref{labelD.57}) of Appendix D.

The functions $(\alpha/\pi)\,g_n(E_e)$ and $(\alpha/\pi)\,f_n(E_e)$
describe the radiative corrections to the lifetime of the neutron and
the correlation coefficients $a(E_e)$ and $A(E_e)$, respectively. They
are equal to the radiative corrections, calculated in
\cite{RC8}--\cite{Gudkov2}.  We show in Appendix D that the
correlation coefficient $B(E_e)$ has no radiative corrections to order
$\alpha/\pi$. This agrees also well with the results, obtained in
\cite{Gudkov2}. The densities of the radiative corrections
$(\alpha/\pi)\,g_n(E_e)\rho_{\beta^-_c}(E_e)$ and
$(\alpha/\pi)f_n(E_e)\rho_{\beta^-_c}(E_e)$, where
$\rho_{\beta^-_c}(E_e)$ is the electron--energy spectrum density
Eq.(\ref{labelD.58}), are plotted in Fig.\,1.

\begin{figure}
\centering
\includegraphics[height=0.15\textheight]{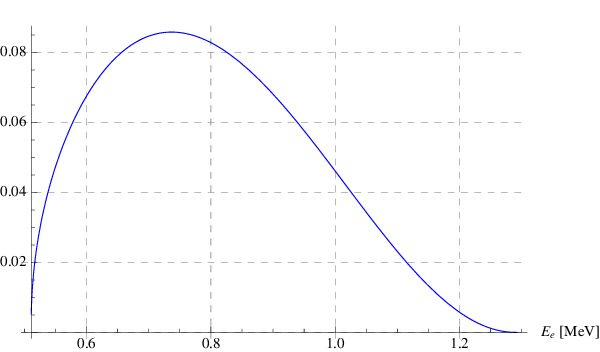}
\includegraphics[height=0.15\textheight]{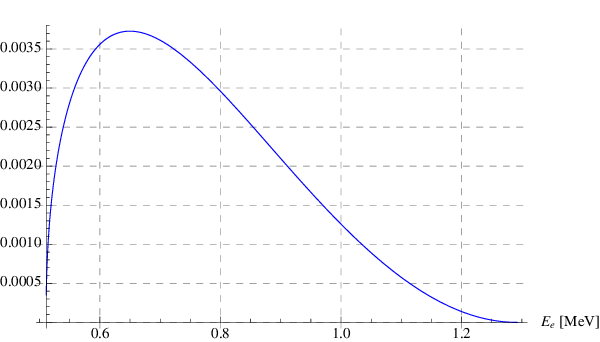}
\caption{The densities $(\alpha/\pi)\,g_n(E_e)\rho_{\beta^-_c}(E_e)$
  (left) and $(\alpha/\pi)\,f_n(E_e)\rho_{\beta^-_c}(E_e)$ (right),
  measured in ${\rm MeV^{-1}}$, of the radiative corrections to the
  lifetime of the neutron and the correlation coefficients $a(E_e)$
  and $A(E_e)$, where $\rho_{\beta^-_c}(E_e)$ is the electron--energy
  spectrum density Eq.(\ref{labelD.58}).}
 \end{figure}

The coefficients $K_n(E_e)$ and $Q_n(E_e)$ have been introduced in
\cite{Gudkov1,Gudkov2} and calculated within the effective quantum
field theory, based on the heavy--baryon Chiral Perturbation Theory
(HBChPT).  Our results of the calculation of the correlation
coefficients, carried out to next--to--leading order in the large $M$
expansion (see Appendix A), agree fully with the expressions,
calculated in \cite{Gudkov2}.

The correlation coefficient $D(E_e)$ relates to a violation of time
reversal invariance.  In the SM a non--vanishing correlation
coefficient $D(E_e)$ may appear due to long--range
\cite{LRM1}--\cite{LRM4} (see also \cite{Holstein}) and short--range
\cite{SRM} mechanisms of time reversal violation. In the long--range
mechanism of time reversal violation the correlation coefficient
$D(E_e)$ is induced by the electron--proton interaction in the
final--state of the decay due to the distortion of the electron wave
function in the Coulomb field of the proton \cite{Jackson1958,EK66},
the ``weak magnetism'' and the proton recoil. In the short--range
mechanism of time reversal violation the correlation coefficient
$D(E_e)$ takes a contribution from the CP--violating phase $\delta$ of
the CKM quark mixing matrix \cite{PDG12}. According to \cite{SRM}, the
contribution of the long--range mechanism of time reversal violation
dominates by many orders of magnitude in comparison with the
contribution of the short--range one. As has been shown in
\cite{LRM1}--\cite{LRM4} the correlation coefficient $D(E_e)$ is a
function of the electron energy $E_e$. Using the results obtained in
\cite{LRM1}, for the electron kinetic energies $250\,{\rm keV} \le T_e
\le 455\,{\rm keV}$ \cite{Abele1,Mund,Schumann} and the axial coupling
constant $\lambda = - 1.2750$ we obtain that $D(E_e) \sim
10^{-5}$. Hence the contribution of the long--range mechanism of time
reversal violation to the correlation coefficient $D(E_e)$ is smaller
compared with contributions of order $10^{-4}$, which may be induced
by interactions beyond the SM. Recently the correlation coefficient
$D(E_e)$ has been calculated within heavy--baryon effective field
theory by Ando {\it et al.} \cite{AndoD09}. The authors have
reproduced the result, obtained by Callan and Treiman \cite{LRM1}, and
have found a correction, which is smaller compared with $10^{-7}$ in
the experimental region of electron kinetic energies $250\,{\rm keV}
\le T_e \le 455\,{\rm keV}$ \cite{Abele1,Mund,Schumann}.

From Eq.(\ref{label7}) the correlation coefficients under
consideration, taking into account the contributions of order $1/M$
and $\alpha/\pi$, we define as follows
\begin{eqnarray}\label{label9}
\hspace{-0.3in}a(E_e) &=& a_0\,\Big(1 +
\frac{\alpha}{\pi}\,f_n(E_e)\Big) + \frac{1}{M}\,\Big[
  \frac{2\,\lambda(\lambda - (\kappa + 1))}{1 + 3 \lambda^2}\,E_0 -
  \frac{4 \lambda (3 \lambda - (\kappa + 1))}{1 + 3 \lambda^2}
  E_e\Big] - a_0\,\delta \zeta(E_e),\nonumber\\
\hspace{-0.3in}A(E_e) &=& A_0\,\Big(1 +
\frac{\alpha}{\pi}\,f_n(E_e)\Big) + \frac{1}{M}\Big[\frac{\lambda^2 -
    \kappa\,\lambda - (\kappa + 1)}{1 + 3 \lambda^2}\,E_0 - \frac{5
    \lambda^2 - (3\kappa - 4)\,\lambda - (\kappa + 1)}{1 + 3
    \lambda^2}\,E_e\Big] - A_0\,\delta \zeta(E_e),\nonumber\\
\hspace{-0.3in}B(E_e) &=& B_0 + \frac{1}{M}\Big[-
  \frac{2\,\lambda\,(\lambda - (\kappa + 1))}{1 + 3 \lambda^2}\,E_0 +
  \frac{7 \lambda^2 - (3 \kappa + 8)\, \lambda + (\kappa + 1)}{1 + 3
    \lambda^2}\,E_e - \frac{\lambda^2 - (\kappa + 2)\,\lambda + (\kappa
  + 1)}{1 + 3 \lambda^2}\,\frac{m^2_e}{E_e}\Big] \nonumber\\
\hspace{-0.3in}&-& B_0\,\delta \zeta(E_e),\nonumber\\
\hspace{-0.3in}K_n(E_e) &=& \frac{1}{M}\,\frac{5 \lambda^2 - (\kappa
  - 4)\, \lambda - (\kappa + 1)}{1 + 3 \lambda^2}\,E_e,\nonumber\\
\hspace{-0.3in}Q_n(E_e) &=& \frac{1}{M}\,\Big[\frac{\lambda^2 -
    (\kappa + 2)\, \lambda + (\kappa + 1)}{1 + 3 \lambda^2}\,E_0 -
  \frac{7 \lambda^2 - (\kappa + 8)\,\lambda + (\kappa + 1)}{1 + 3
    \lambda^2}\,E_e\Big].
\end{eqnarray}
Using the following expansion
\begin{eqnarray}\label{label10}
\hspace{-0.3in}\frac{1}{\zeta(E_e)} &=& \Big(1 -
\frac{\alpha}{\pi}\,g_n(E_e)\Big) - \delta \zeta(E_e),\nonumber\\
\hspace{-0.3in}\delta\zeta(E_e) &=& \frac{1}{M}\, \frac{1}{1 + 3
  \lambda^2}\,\Big(\zeta_1E_0 + \zeta_2 E_e + \zeta_3 \frac{m^2_e}{E_e}\Big),\nonumber\\
\hspace{-0.3in}\zeta_1 &=& - 2\,\lambda\,(\lambda - (\kappa +
1)),\nonumber\\
\hspace{-0.3in}\zeta_2 &=& 10 \lambda^2 - 4(\kappa + 1)\,\lambda +
2,\nonumber\\
 \hspace{-0.3in}\zeta_3 &=& - 2 \lambda\,(\lambda - (\kappa +
1))
\end{eqnarray}
we transcribe the correlation coefficients $a(E_e)$, $A(E_e)$ and
$B(E_e)$ into the form, which is similar to that proposed by Wilkinson
for the correlation coefficient $A^{(W)}(E_e)$ (see
Eq.(\ref{label17})). We get
\begin{eqnarray}\label{label11}
\hspace{-0.3in}&&a(E_e) = a_0\Big\{1 + \frac{1}{M}\,\frac{1}{(1 -
  \lambda^2)(1 + 3 \lambda^2)}\,\Big(a_1 E_0 + a_2 E_e +
a_3\frac{m^2_e}{E_e}\Big)\Big\}\,\Big(1 +
\frac{\alpha}{\pi}\,f_n(E_e)\Big),\nonumber\\
\hspace{-0.3in}&&a_1 = 4 \lambda (\lambda^2 + 1)(\lambda -(\kappa +
1)),\nonumber\\
\hspace{-0.3in}&&a_2 = - 26 \lambda^4 + 8(\kappa + 1)\,\lambda^3 - 20
\lambda^2 + 8(\kappa + 1)\,\lambda - 2,\nonumber\\
\hspace{-0.3in}&&a_3 = - 2 \lambda (\lambda^2 - 1)(\lambda - (\kappa +
1))
\end{eqnarray}
and
\begin{eqnarray}\label{label12}
\hspace{-0.3in}&&A(E_e) = A_0\,\Big\{1 -
\frac{1}{M}\,\frac{1}{2\lambda(1 + \lambda)(1 + 3 \lambda^2)}\, \Big(A
_1 E_0 + A_2 E_e + A_3\frac{m^2_e}{E_e}\Big)\Big\}\,\Big(1 +
\frac{\alpha}{\pi}\,f_n(E_e)\Big),\nonumber\\
\hspace{-0.3in}&&A_1 = - \lambda^4 + \kappa \,\lambda^3 + (\kappa +
2)\,\lambda^2 - \kappa\, \lambda - (\kappa + 1),\nonumber\\
\hspace{-0.3in}&&A_2 = 5 \lambda^4 + \kappa\, \lambda^3 - (5 \kappa +
6)\,\lambda^2 + 3 \kappa\,\lambda + (\kappa + 1),\nonumber\\
\hspace{-0.3in}&&A_3 = -\,4\lambda^2(\lambda + 1)\,(\lambda - (\kappa +
1))
\end{eqnarray}
 and
\begin{eqnarray}\label{label13}
\hspace{-0.3in}&&B(E_e) = B_0\,\Big\{1 -
\frac{1}{M}\,\frac{1}{2\lambda(1 - \lambda)(1 + 3 \lambda^2)}\,\Big(B_1
E_0 + B_2 E_e + B_3\frac{m^2_e}{E_e}\Big)\Big\},\nonumber\\
\hspace{-0.3in}&&B_1 = - 2\,\lambda (\lambda + 1)^2 (\lambda - (\kappa
+ 1)),\nonumber\\
\hspace{-0.3in}&&B_2 = \lambda^4 - (\kappa - 4)\,\lambda^3 - (5 \kappa
+ 2)\,\lambda^2 - (3 \kappa + 4)\,\lambda + (\kappa + 1),\nonumber\\
\hspace{-0.3in}&&B_3 = (\lambda^2 - 1)(\lambda - 1)(\lambda - (\kappa
+ 1)).
\end{eqnarray}
For the derivation of Eq.(\ref{label11}), Eq.(\ref{label12}) and
Eq.(\ref{label13}) we have neglected the terms of order
$(\alpha/\pi)(E_0/M) \sim 10^{-6}$, which are smaller compared with
contributions of order $10^{-4}$ beyond the SM.

In order to estimate the values of the obtained $1/M$ corrections we
calculate them at $\lambda = - 1.2750$.  This gives
\begin{eqnarray}\label{label13a}
\hspace{-0.3in}&&\delta \zeta(E_e) = \frac{1}{M}\frac{1}{1 + 3
  \lambda^2}\Big(\zeta_1 E_0 + \zeta_2 E_e + \zeta_3
\frac{m^2_e}{E_e}\Big) = - 3.57\times 10^{-3} + 9.90\times
10^{-3}\,\frac{E_e}{E_0} - 1.41\times
10^{-3}\,\frac{m_e}{E_e},\nonumber\\
\hspace{-0.3in}&&\frac{\delta a(E_e)}{a_0} = \frac{1}{M}\,\frac{1}{(1
  - \lambda^2)(1 + 3 \lambda^2)} \Big(a_1 E_0 + a_2 E_e +
a_3\frac{m^2_e}{E_e}\Big) = - 3.00\times 10^{-2} + 8.59\times
10^{-2}\frac{E_e}{E_0} + 1.41\times
10^{-3}\,\frac{m_e}{E_e},\nonumber\\
\hspace{-0.3in}&& \frac{\delta A(E_e)}{A_0} = - \frac{1}{M}\,\frac{1}{2\lambda(1 +
  \lambda)(1 + 3 \lambda^2)}\Big(A_1 E_0 + A_2 E_e +
A_3\frac{m^2_e}{E_e}\Big)= 3.44 \times 10^{-4} + 1.46\times
10^{-2}\frac{E_e}{E_0} + 1.41\times
10^{-3}\,\frac{m_e}{E_e},\nonumber\\
\hspace{-0.3in}&&\frac{\delta B(E_e)}{B_0} = -\frac{1}{M}\,\frac{1}{2
  \lambda(1 - \lambda)(1 + 3 \lambda^2)}\Big(B_1 E_0 + B_2 E_e +
B_3\frac{m^2_e}{E_e}\Big)= - 4.66 \times 10^{-5} - 2.97\times
10^{-4}\frac{E_e}{E_0} + 1.36\times
10^{-4}\,\frac{m_e}{E_e},\nonumber\\
\hspace{-0.3in}&&K_n(E_e) = 7.14\times 10^{-4}\frac{E_e}{E_0},\nonumber\\
\hspace{-0.3in}&&Q_n(E_e) = 3.19\times 10^{-3} - 7.27\times 10^{-3}\frac{E_e}{E_0}.
\end{eqnarray}
Due to strong dependence on the axial coupling constant $\lambda$ the
numerical values of the contributions of the ``weak magnetism'' and
the proton recoil, calculated at $\lambda = - 1.2750$, vary from
$10^{-5}$ to $10^{-1}$ for energy independent and energy dependent
terms.

\noindent{\bf Summary}. The correlation coefficients of the
electron--energy and angular distribution of the neutron
$\beta^-$--decay with a polarized neutron and unpolarised decay proton
and electron are calculated by taking into account a complete set of
the $1/M$ corrections, caused by the ``weak magnetism'' and the proton
recoil, and the radiative corrections of order $\alpha/\pi$,
calculated to leading order in the large $M$ expansion. The obtained
expressions for the correlation coefficients should be used as a
theoretical background for contributions of order $10^{-4}$ of
interactions beyond the SM, calculated in Appendix G (see
Eq.(\ref{labelG.6} and Eq.(\ref{labelG.6a})).  Contributions of order
$10^{-4}$ of interactions beyond the SM may be determined by measuring
the asymmetries $A_{\exp}(E_e)$, $B_{\exp}(E_e)$ and $C_{\exp}$
between the neutron spin and the 3--momenta of the decay particles,
the electron--proton energy distribution $a(E_e, T_p)$ and the
proton--energy spectrum $a(T_p)$, related to correlations between the
3--momenta of the electron and proton, and the lifetime of the neutron
$\tau_n$ (see section~\ref{sec:sensitivity}).

\section{Standard model analysis of Experimental determination of 
correlation coefficient $A_0$. Electron asymmetry $A_{\exp}(E_e)$}
\label{sec:Aexp}

For the experimental determination of the correlation coefficient
$A_0$, defining correlations between the neutron spin  and the electron
3--momentum in the SM to leading order in the large $M$ expansion
\cite{Abele1}, the directions of the emission of the antineutrino are not
fixed and one has to integrate over the antineutrino 3--momentum
$\vec{k}$.  As a result we arrive at the following electron--energy
and angular distribution \cite{Abele1,Mund}
\begin{eqnarray}\label{label14}
\hspace{-0.3in}\frac{d^2 \lambda_n(E_e, \vec{k}_e,
  \vec{\xi}_n)}{dE_e d\Omega_e } &=& (1 + 3
\lambda^2)\,\frac{G^2_F|V_{ud}|^2}{8\pi^4} \,(E_0 - E_e)^2\,\sqrt{E^2_e - m^2_e}\, E_e
\,F(E_e, Z = 1)\,\zeta(E_e)\nonumber\\
\hspace{-0.3in}&&\times\,\Big(1 + A^{(W)}(E_e)\,\Big(1 +
\frac{\alpha}{\pi}\,f_n(E_e)\Big)\,\vec{\xi}_n\cdot
\vec{\beta}\,\Big),
\end{eqnarray}
where we have denoted 
\begin{eqnarray}\label{label15}
\hspace{-0.3in}A(E_e) + \frac{1}{3}\,Q_n(E_e) = A^{(W)}(E_e)\,\Big(1
+ \frac{\alpha}{\pi}\,f_n(E_e)\Big),
\end{eqnarray}
$d\Omega_e = 2\pi \sin\theta_e d\theta_e$ is an infinitesimal solid
angle of the electron 3--momentum with respect to the neutron spin and
$\vec{\xi}_n\cdot \vec{\beta} = P \beta \cos\theta_e$ with the neutron
polarization $P = |\vec{\xi}_n| \le 1$. The correlation coefficients
$A(E_e)$ and $Q_n(E_e)$ are given in Eq.(\ref{label12}) and
Eq.(\ref{label9}), respectively. The contribution, proportional to
$Q_n(E_e)$, with structure $\vec{\xi}_n\cdot \vec{k}_e/E_e$ we obtain
having integrated the term with the structure
$(\vec{\xi}_n\cdot\vec{k}\,)(\vec{k}_e\cdot \vec{k}\,)/E_eE^2$ in
Eq.(\ref{label6}) over directions of the antineutrino 3--momentum
$\vec{k}$.

The asymmetry, which may be used for the experimental determination of
the axial coupling constant $\lambda$ and the correlation coefficient
$A_0$, takes the form
\begin{eqnarray}\label{label16}
\hspace{-0.3in}A_{\exp}(E_e) = \frac{N^{+}(E_e) -
  N^{-}(E_e)}{N^{+}(E_e) + N^{-}(E_e)} = \frac{1}{2}\,A^{(W)}(E_e)
\,\Big(1 + \frac{\alpha}{\pi}\,f_n(E_e)\Big)\, P
\beta\,(\cos\theta_1 + \cos\theta_2),
\end{eqnarray}
where $N^{\pm}(E_e)$ are the numbers of events of the emission of the
electron forward $(+)$ and backward $(-)$ with respect to the neutron
spin into the solid angle $\Delta \Omega_{12} = 2\pi (\cos\theta_1 -
\cos\theta_2)$ with $0 \le \varphi \le 2\pi$ and $\theta_1 \le
\theta_e \le \theta_2$. They are determined by \cite{Dubbers}
\begin{eqnarray}\label{label16a}
\hspace{-0.3in}&&N^{+}(E_e) = 2\pi N(E_e)
\int^{\theta_2}_{\theta_1} \Big(1 + A^{(W)}(E_e)\,\Big(1 +
\frac{\alpha}{\pi}\,f_n(E_e)\Big)\,P\beta\,\cos\theta_e\Big)\sin\theta_e\,
d\theta_e
=\nonumber\\
\hspace{-0.3in}&&= 2\pi N(E_e)\Big(1 + \frac{1}{2}\,A^{(W)}(E_e)\,\Big(1
+ \frac{\alpha}{\pi}\,f_n(E_e)\Big)\,P\beta\,(\cos\theta_1 +
\cos\theta_2)\Big)\,(\cos\theta_1 - \cos\theta_2),\nonumber\\
\hspace{-0.3in}&&N^{-}(E_e) = 2\pi N(E_e)\int^{\pi - \theta_2}_{\pi
  - \theta_1} \Big(1 + A^{(W)}(E_e)\,\Big(1 +
\frac{\alpha}{\pi}\,f_n(E_e)\Big)\,P\beta\,\cos\theta_e\Big)\sin\theta_e\,
d\theta_e
= \nonumber\\
\hspace{-0.3in}&&= 2\pi N(E_e)\Big(1 -
\frac{1}{2}\,A^{(W)}(E_e)\,\Big(1 +
\frac{\alpha}{\pi}\,f_n(E_e)\Big)\,P\beta\,(\cos\theta_1 +
\cos\theta_2)\Big)\,(\cos\theta_1 - \cos\theta_2),
\end{eqnarray}
where $N(E_e)$ is the normalization factor equal to
\begin{eqnarray}\label{label21}
\hspace{-0.3in}N(E_e) = (1 + 3
\lambda^2)\,\frac{G^2_F|V_{ud}|^2}{8\pi^4} (E_0 - E_e)^2\,\sqrt{E^2_e
  - m^2_e}\,E_e\,F(E_e, Z = 1)\,\zeta(E_e).
\end{eqnarray}
The correlation coefficient $A^{(W)}(E_e)$ is
\begin{eqnarray}\label{label17}
\hspace{-0.3in}&&A^{(W)}(E_e) = A_0\Big\{1
-\frac{1}{M}\,\frac{1}{2\lambda (1 + \lambda)(1 + 3 \lambda^2)}\,
\Big(A^{(W)}_1 E_0 + A^{(W)}_2 E_e +
A^{(W)}_3\frac{m^2_e}{E_e}\Big)\Big\},\nonumber\\
\hspace{-0.3in}&&A^{(W)}_1 = \frac{2}{3}\,\Big(- 3 \lambda^3 + (3
\kappa + 5)\,\lambda^2 - (2 \kappa + 1)\,\lambda - (\kappa + 1)\Big) =
- 2\Big(\lambda - (\kappa + 1)\Big)\Big(\lambda^2 -
\frac{2}{3}\,\lambda - \frac{1}{3}\Big),\nonumber\\
\hspace{-0.3in}&&A^{(W)}_2 = \frac{2}{3}\,\Big(- 3 \lambda^4 + (3
\kappa + 12)\, \lambda^3 - (9 \kappa + 14)\,\lambda^2 + (5\kappa +
4)\,\lambda + (\kappa + 1)\Big) = - 2\Big(\lambda - (\kappa +
1)\Big)\, \Big(\lambda^3 - 3 \lambda^2 + \frac{5}{3}\,\lambda +
\frac{1}{3}\Big),\nonumber\\
\hspace{-0.3in}&&A^{(W)}_3 = - 4\,\lambda^2(\lambda + 1)\,\Big(\lambda
- (\kappa + 1)\Big).
\end{eqnarray}
It agrees well with the result, obtained by Wilkinson
\cite{Wilkinson,Wilkinson1}. We note that the correlation coefficient
$A(E_e) + \frac{1}{3}Q_n(E_e)$ differs from the Wilkinson correlation
coefficient $A^{(W)}(E_e)$ by the contribution of the radiative
corrections, described by the function $(\alpha/\pi)\,f_n(E_e)$.  In
the replacement $A(E_e) + \frac{1}{3}Q_n(E_e) \to A^{(W)}(E_e)(1 +
(\alpha/\pi)f_n(E_e))$ we have neglected the contributions of order
$(\alpha/\pi)(E_0/M) \sim 10^{-6}$, which are smaller compared with
contributions of order $10^{-4}$ of our interest. The contribution to
the correlation coefficient $A^{(W)}(E_e)$ of the ``weak magnetism''
and the proton recoil, calculated at $\lambda = - 1.2750$, is equal to
\begin{eqnarray}\label{label18a}
\hspace{-0.3in}\frac{\delta A^{(W)}(E_e)}{A_0} &=& -
\frac{1}{M}\,\frac{1}{2\lambda (1 + \lambda)(1 + 3 \lambda^2)}
\Big(A^{(W)}_1 E_0 + A^{(W)}_2 E_e + A^{( W)}_3\frac{m^2_e}{E_e}\Big)
= \nonumber\\
\hspace{-0.3in}&=&- 8.56\times 10^{-3} + 3.49\times
10^{-2}\,\frac{E_e}{E_0} + 1.41\times 10^{-3}\,\frac{m_e}{E_e}.
\end{eqnarray}
In the experimentally used region of electron kinetic energies
$250\,{\rm keV} \le T_e \le 455\,{\rm keV}$
\cite{Abele1,Mund,Schumann} the radiative corrections
$(\alpha/\pi)\,f_n(E_e)$ vary over the region $1.53\times 10^{-3} \ge
(\alpha/\pi)\,f_n(E_e) \ge 1.04\times 10^{-3}$ and increase the
absolute value of the correlation coefficient $A^{(W)}(E_e)$.

\noindent{\bf Summary}. The Wilkinson expression for the correlation
coefficient $A^{(\rm W)}(E_e)$ (see Eq.(\ref{label17})), taking into
account a complete set of the $1/M$ corrections from the ``weak
magnetism'' and the proton recoil, is improved by the account for the
radiative corrections, described by the function $2.81\times 10^{-3}
\ge (\alpha/\pi)\,f_n(E_e) \ge 0.62\times 10^{-3} $ for the electron
energies $m_e \le E_e \le E_0$ (see Eq.(\ref{labelD.57})). The
contribution of the radiative corrections increases the absolute value
of the correlation coefficient $A^{(\rm W)}(E_e)$.  The results,
obtained in this section, should be used as a theoretical background
for the experimental determination of contributions of order $10^{-4}$
of interactions beyond the SM from the experimental data on the
electron asymmetry $A_{\exp}(E_e)$ (see
section~\ref{sec:sensitivity}).

\section{Standard model analysis of experimental determination  of 
correlation coefficient $B_0$. Antineutrino asymmetry $B_{\exp}(E_e)$}
\label{sec:Bexp}

In the SM to leading order in the large $M$ expansion the correlations
between the neutron spin  and the antineutrino 3--momentum are defined
by the correlation coefficient $B_0$ \cite{Abele1}, which may be
determined from the experimental data on the asymmetry
$B_{\exp}(E_e)$, defined by \cite{Abele2}
\begin{eqnarray}\label{label19}
  B_{\exp}(E_e) = \frac{N^{--}(E_e) - N^{++}(E_e)}{N^{--}(E_e) +
  N^{++}(E_e)}.
\end{eqnarray}
It defines the asymmetry of the emission of the antineutrinos into the
forward and backward hemisphere with respect to the neutron spin,
where $N^{\mp \mp}(E_e)$ is the number of events of the emission of
the electron--proton pairs as functions of the electron energy
$E_e$. The signs $(++)$ and $(--)$ show that the electron--proton
pairs were emitted parallel $(++)$ and antiparallel $(--)$ to a
direction of the neutron spin. This means that antineutrinos were
emitted antiparallel $(++)$ and parallel $(--)$ to a direction of the
neutron spin.  The number of events $N^{--}(E_e)$ and $N^{++}(E_e)$
are defined by the electron--energy and angular distribution of the
neutron $\beta^-$--decay, integrated over the forward and backward
hemisphere relative to the neutron spin, respectively.

The integration region for the electron--proton pairs, emitted
parallel to a direction $(++)$ of the neutron spin, is defined by the
constraints \cite{FG95}: $\vec{\xi}_n\cdot \vec{k}_e = P
k_e\,\cos\theta_e > 0$ and $\vec{\xi}_n\cdot \vec{k}_p =
\vec{\xi}_n\cdot (-\vec{k}_e - \vec{k}\,) = P E(-r \cos\theta_e -
\cos\theta) > 0$ or $-r \cos\theta_e > \cos\theta$, where $r = k_e/E =
\sqrt{E^2_e - m^2_e}/(E_0 - E_e)$ and $P = |\vec{\xi}_n| \le 1$ is the
neutron polarization.  For $N^{++}(E_e)$ and $r < 1$ we obtain the
following expression
\begin{eqnarray}\label{label20}
\hspace{-0.3in}&& N^{++}(E_e) = 2\pi\, N(E_e)\Big\{\Big(1 -
\frac{1}{4}\,a\beta  + \frac{1}{2}P\beta\Big(A - \frac{1}{3}\,K_n\,
\beta\,\Big) - \frac{1}{2}P\Big(B - \frac{1}{3}\,Q_n\,\beta\,\Big)\Big)\nonumber\\
\hspace{-0.3in}&& -\frac{1}{2} r\Big(1 + \frac{2}{3}P A \beta\Big) +
  \frac{1}{8}r^2\Big(a\beta + \frac{4}{3}\,P B + \frac{4}{5}\,P\, K_n\,\beta^2 \Big)
-\frac{1}{15}\,r^3 \,P \,Q_n\, \beta - \frac{1}{8}\,a_0\,\beta^2\,r(1
- r^2)\,\frac{E_e}{M}\Big\},
\end{eqnarray}
where $N(E_e)$ is the normalization factor Eq.(\ref{label21}).  For $r
> 1$ the upper limit of the integration over $\cos\theta_e$ is
restricted by $\cos\theta_e \le 1/r$. The result of the interaction is
\begin{eqnarray}\label{label22}
\hspace{-0.3in}&& N^{++}(E_e) = 2\pi\,
N(E_e)\Big\{\frac{1}{2}\frac{1}{r}\Big(1 -\frac{2}{3}P B\Big) -
\frac{1}{8}\frac{1}{r^2}\Big(a\beta - \frac{4}{3}P A \beta -
\frac{4}{5}P Q_n \beta\, \Big)- \frac{1}{15}\frac{1}{r^3}P K_n \beta^2
+ \frac{1}{8}\,a_0\,\beta^2\,\frac{1}{r}\Big(1 -
\frac{1}{r^2}\Big)\,\frac{E_e}{M}\Big\}.\nonumber\\
\hspace{-0.3in}&& 
\end{eqnarray}
\begin{figure}
\centering 
\includegraphics[height=0.20\textheight]{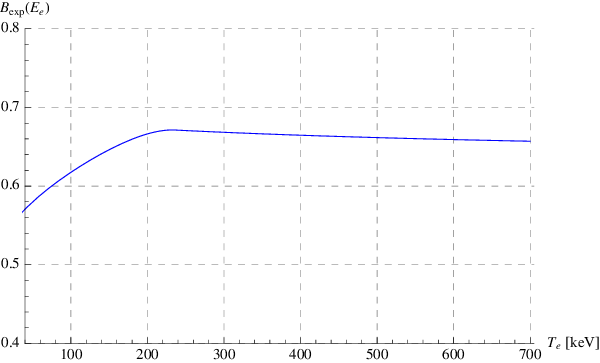}
\caption{The antineutrino asymmetry $B_{\exp}(E_e)$, including a
  complete set of the $1/M$ corrections, caused by the ``weak
  magnetism'' and the proton recoil, and radiative corrections of
  order $\alpha/\pi$, calculated to leading order in the large $M$
  expansion; $T_e = E_e - m_e$ is the electron kinetic energy.}
\end{figure}
For the calculation of $N^{--}(E_e)$ we have to integrate over the
region \cite{FG95}: $\vec{\xi}_n\cdot \vec{k}_e = P k_e\,\cos\theta_e <
0$ and $\vec{\xi}_n\cdot \vec{k}_p = P E(-r \cos\theta_e
- \cos\theta) < 0$ or $ \cos\theta >
-r \cos\theta_e$. The number of events $N^{--}(E_e)$ for $r < 1$ is
given by
\begin{eqnarray}\label{label23}
\hspace{-0.3in}&& N^{--}(E_e) = 2\pi\, N(E_e)\Big\{\Big(1 -
\frac{1}{4}\,a\beta  - \frac{1}{2}P\beta\Big(A - \frac{1}{3}\,K_n
\beta\,\Big) + \frac{1}{2}P\Big(B - \frac{1}{3}Q_n \beta\,\Big)\Big)\nonumber\\
\hspace{-0.3in}&& -\frac{1}{2} r\Big(1 - \frac{2}{3}PA \beta\Big) +
\frac{1}{8}r^2\Big(a\beta - \frac{4}{3}\,P B- \frac{4}{5}\,P K_n
\beta^2 \Big) + \frac{1}{15}\,r^3 P Q_n \beta -
\frac{1}{8}\,a_0\,\beta^2\,r(1 - r^2)\,\frac{E_e}{M}\Big\}.
\end{eqnarray}
For $r > 1$ the lower limit of the integration over $\cos\theta_e$ is
restricted by $\cos\theta_e > -1/r$. The number of events
$N^{--}(E_e)$, calculated for $r > 1$, is equal to
\begin{eqnarray}\label{label24}
\hspace{-0.3in}&& N^{--}(E_e) = 2\pi\,
N(E_e)\Big\{\frac{1}{2}\frac{1}{r}\Big(1 + \frac{2}{3}P B
\Big) - \frac{1}{8}\frac{1}{r^2}\Big(a\beta + \frac{4}{3}P
A \beta + \frac{4}{5}P Q_n\beta\,\Big) + \frac{1}{15}\frac{1}{r^3}P K_n
\beta^2 + \frac{1}{8}\,a_0\,\beta^2\,\frac{1}{r}\Big(1 -
\frac{1}{r^2}\Big)\,\frac{E_e}{M}\Big\}.\nonumber\\
\hspace{-0.3in}&& 
\end{eqnarray}
Using our formulas for the numbers of events we calculate the
asymmetry $B_{\exp}(E_e)$. For $r \le 1$ or $0 \le T_e \le (E_0 -
m_e)^2/2 E_0 = 236\,{\rm keV}$ and for $r \ge 1$ or $(E_0 - m_e)^2/2
E_0 = 236\,{\rm keV}\le T_e \le E_0 - m_e$ the asymmetry
$B_{\exp}(E_e)$ is equal to
\begin{eqnarray}\label{label25}
\hspace{-0.3in}&&B^{(r < 1)}_{\exp}(E_e) = \frac{2
  P}{3}\,\frac{\displaystyle (3 - r^2) B - (3 - 2 r)A\beta + \Big(1 -
  \frac{3}{5} r^2\Big)\,K_n\beta^2 - \Big(1 - \frac{2}{5} r^3\Big)
  Q_n\beta}{\displaystyle (4 - 2r) - \Big(1 -
\frac{1}{2}\,r^2\Big) a\beta - \frac{1}{2}\,a_0\beta^2\, r(1 -
r^2)\,\frac{E_e}{M}}
\end{eqnarray}
and
\begin{eqnarray}\label{label26}
\hspace{-0.3in}B^{(r > 1)}_{\exp}(E_e) =
\frac{2P}{3}\,\frac{\displaystyle B - \frac{1}{2}\,\Big(A +
  \frac{3}{5}Q_n\Big)\, \frac{\beta}{r} + \frac{1}{5}
  K_n\,\frac{\beta^2}{r^2}}{\displaystyle 1 - a\,\frac{\beta}{4 r} +
  \frac{1}{4}\,a_0\,\beta^2\,\Big(1 -
  \frac{1}{r^2}\Big)\frac{E_e}{M}},
\end{eqnarray}
respectively. At $r = 1$ or $T_e = (E_0 - m_e)^2/2E_0 = 236\,{\rm
  keV}$ the asymmetry $B_{\exp}(E_e)$ is continuous.  To leading order
in the large $M$ expansion the asymmetry $B_{\exp}(E_e)$ reduces to
the form
\begin{eqnarray}\label{label26a}
\hspace{-0.3in}&&B_{\exp}(E_e) \Big|_{M \to \infty}= \frac{2
  P}{3}\,\left\{\begin{array}{r@{\;,}l}\frac{\displaystyle B
  \Big(\frac{3}{2} - \frac{1}{2} r^2\Big) - \Big(\frac{3}{2} - r\Big)
  A \beta}{\displaystyle (2 - r) - \frac{1}{2}\Big(1 -
  \frac{1}{2}r^2\Big) a \beta}&\; r \le 1\\ \frac{\displaystyle B -
  \frac{1}{2}\,A\,\frac{\beta}{r}}{\displaystyle 1 -
  \frac{1}{4}\,a\,\frac{\beta}{r}} &\; r \ge 1,
\end{array}\right.
\end{eqnarray}
where the correlation coefficients $B$, $A$ and $a$ are equal to
\begin{eqnarray}\label{label26b}
\hspace{-0.3in}B = B_0\quad,\quad A = A_0\Big(1 +
\frac{\alpha}{\pi}\,f_n(E_e)\Big) \quad,\quad a = a_0\Big(1 +
\frac{\alpha}{\pi}\,f_n(E_e)\Big).
\end{eqnarray}
In Fig.\,2 we plot the asymmetry $B_{\exp}(E_e)$, given by
Eq.(\ref{label25}) and Eq.(\ref{label26}) and obtained in the SM with
the account for the contributions of the $1/M$ corrections, caused by
the ``weak magnetism'' and the proton recoil and the radiative
corrections of order $\alpha/\pi$, calculated to leading order in the
large $M$ expansion and described by the function
$(\alpha/\pi)\,f_n(E_e)$. In \cite{Abele2,Schumann} for the
experimental determination of the correlation coefficient $B_0$ the
experimental data on the asymmetry $B_{\exp}(E_e)$ were fitted by
Eq.(\ref{label26a}) with the replacement $B \to B_0$, $ A \to A_0$ and
$a \to a_0$, respectively. Such an asymmetry was calculated in
\cite{FG95}. The ``weak magnetism'', the proton recoil and the
radiative corrections to Eq.(\ref{label26a}) with the replacement $B
\to B_0$, $ A \to A_0$ and $a \to a_0$ have been calculated
numerically in \cite{FG95a}. They are positive and negative for $T_e <
470\,{\rm keV}$ and $T_e > 470\,{\rm keV}$, respectively. The
corrections, calculated in this paper, are positive and of order of
magnitude larger compared with the absolute values of the corrections,
calculated in \cite{FG95a}.

The theoretical value $B_0 = 0.9871(1)$ of the correlation coefficient
$B_0$, calculated for $\lambda = - 1.2750(9)$, agrees within 1.5
standard deviations with the experimental values $B^{(\exp)}_0 =
0.9802(50)$, $B^{(\exp)}_0 = 0.9821(40)$ and $B^{(\exp)}_0 =
0.9894(83)$, obtained in \cite{Abele2,Schumann} (see also
\cite{Abele1}), \cite{SAA1} and \cite{SAA2}, respectively, and within
two standard deviations with the experimental one $B^{(\exp)}_0 =
0.967(12)$, obtained in \cite{SAA3}.

\noindent{\bf Summary}. The antineutrino asymmetry $B_{\exp}(E_e)$,
calculated in \cite{FG95} and used in \cite{Abele2,Schumann} for the
experimental determination of the correlation coefficient
$B^{(\exp)}_0 = 0.9802(50)$, is improved by the contributions of a
complete set of the $1/M$ corrections, caused by the ``weak
magnetism'' and the proton recoil, and the radiative corrections
$(\alpha/\pi)\,f_n(E_e)$, calculated to leading order in the large $M$
expansion. The obtained expression of the antineutrino asymmetry
$B_{\exp}(E_e)$ should be used as a theoretical background for the
experimental determination of contributions of order $10^{-4}$ of
interactions beyond the SM (see section~\ref{sec:sensitivity}).

\section{Standard model analysis of experimental determination of 
correlation coefficient $a_0$. Electron--proton energy distribution
$a(E_e, T_p)$ and proton--energy spectrum $a(T_p)$}
\label{sec:a0exp}

For the experimental determination of the correlation coefficient
$a_0$ the correlation coefficient $a(E_e)$, given by
Eq.(\ref{label11}), can be hardly used, since the antineutrino is hard
to detect.  Thus, for the determination of $a_0$ and the contributions  of
interactions beyond the SM one should measure correlations of the
3--momenta of the decay charged particles. Using the results, obtained
in Appendix I, the electron--proton energy spectrum for the neutron
$\beta^-$--decay with unpolarised particles can be given in the
following form
\begin{eqnarray}\label{label28}
\hspace{-0.3in}\frac{d^2{\rm BR}_{\beta^-_c}(E_e,T_p)}{dE_edT_p} =
(\tau_n)_{\rm SM} M\, (1 + 3 \lambda^2)
\frac{G^2_F|V_{ud}|^2}{4\pi^3}\,a(E_e,T_p)\,\Big(1 +
\frac{\alpha}{\pi}\,g_n(E_e)\Big)\,F(E_e, Z = 1)\,E_e,
\end{eqnarray}
where $(\tau_n)_{\rm SM}$ is the theoretical lifetime of the neutron,
calculated in the SM (see section~\ref{sec:lifetime}) and $T_p$ is the
kinetic energy of the proton varying from zero to its maximal value
$(T_p)_{\rm max} = (m_n - m_p)^2 - m^2_e/2m_n = (E^2_0 - m^2_e)/2 M =
0.751\,{\rm keV}$, i.e. $0 \le T_p \le 0.751\,{\rm keV}$. The limits
of the integration over the electron energy $E_e$, i.e. $(E_e)_{\rm
  min} \le E_e \le (E_e)_{\rm max}$, are the functions of the proton
kinetic energy $T_p$. They are adduced in Appendix I. The
electron--proton energy distribution $a(E_e,T_p)$ is defined by (see
Appendix I)
\begin{eqnarray}\label{label30}
\hspace{-0.3in}a(E_e,T_p) = \zeta_1(E_e,T_p) + a_0\,\Big(1 +
\frac{1}{1 - \lambda^2}\,\frac{E_0}{M}\Big)\,\Big(1 +
\frac{\alpha}{\pi}\,f_n(E_e)\Big)\,\zeta_2(E_e,T_p).
\end{eqnarray}
The functions $\zeta_1(E_e,T_p)$ and $\zeta_2(E_e,T_p)$ are given by
in Appendix I. They are calculated to leading order in the large $M$
and do not contain the radiative corrections. In the electron energy
region $(E_e)_{\rm min} \le E_e \le (E_e)_{\rm max}$ the contribution
of the radiative corrections $(\alpha/\pi)\,f_n(E_e)$ to the
electron--proton energy distribution $a(E_e, T_p)$ relative to the
correlation coefficient $a_0$ is of order $ 10^{-3}$. The account for
the radiative corrections is very important for a correct experimental
determination of contributions of order $10^{-4}$ of interactions
beyond the SM.

Integrating the electron--proton energy spectrum Eq.(\ref{label28})
over the electron energy $(E_e)_{\rm min} \le E_e \le (E_e)_{\rm max}$
we obtain the proton--energy spectrum
\begin{eqnarray}\label{label28a}
\hspace{-0.3in}\frac{d{\rm BR}_{\beta^-_c}(T_p)}{dT_p} = (\tau_n)_{\rm
  SM} M\, (1 + 3 \lambda^2) \frac{G^2_F|V_{ud}|^2}{4\pi^3}\,a(T_p),
\end{eqnarray}
where $a(T_p)$ is defined by
\begin{eqnarray}\label{label28b}
\hspace{-0.3in}a(T_p) = g_1(T_p) + a_0\,\Big(1 + \frac{1}{1 -
  \lambda^2}\,\frac{E_0}{M}\Big)\,g_2(T_p).
\end{eqnarray}
The functions $g_1(T_p)$ and $g_2(T_p)$ are given in Appendix
I. Recently the correlation coefficient $a_0$ and the axial coupling
constant $\lambda$ have been determined from the proton--energy
spectrum by Byrne {\it et al.} \cite{Byrne}. The obtained value
$a^{(\exp)}_0 = - 0.1054(55)$ can be fitted by the axial coupling
constant $\lambda = - 1.271(18)$. In turn, the experimental value
$a^{(\exp)}_0 = - 0.1017(51)$, obtained in \cite{Stratowa}, defines
the axial coupling constant equal to $\lambda = - 1.259(17)$. These
experimental values agree with the theoretical value of the correlation
coefficient $a_0 = -\,0.1065(3)$, calculated at $\lambda = -
1.2750(9)$, and with the axial coupling constant $\lambda = -
1.2750(9)$ within one standard deviation.

From the point of view of the experimental determination of
contributions of order $10^{-4}$ of interactions beyond the SM (see a
discussion in section VII) the experimental accuracy of the
determination of the axial coupling constant $\lambda$ by measuring
the proton--energy spectrum as well as the electron--proton energy
distribution $a(E_e, T_p)$ should be improved by more than two orders
of magnitude in comparison with the accuracy of Byrne's experiment
\cite{Byrne}. There are three major, funded experiments, which are
currently attempting to do this. They are i) aSPECT experiment at
Institute Laue--Langevin (ILL) in Grenoble, invented to perform
precise measurements of the correlation coefficient $a_0$ by measuring
the proton--energy spectrum in the decay of unpolarised neutrons, ii)
aCORN at the National Institute of Standards and Technology (NIST) and
iii) Nab at the new Spallation Neutron Source (SNS) in Oak Ridge at
Tennessee \cite{Konrad}.  The expected experimental accuracy of the
determination of the correlation coefficient $a_0$ in the aCORN and
Nab experiments is better than $1\,\%$ (see also \cite{Wang2012}).

We have discussed the experimental determination of the correlation
coefficient $a_0$ by measuring the electron--proton energy
distribution $a(E_e, T_p)$ and the proton--energy spectrum
$a(T_p)$. However, there are another possibilities to determine
experimentally the correlation coefficient $a_0$. It can be extracted
from the experimental data on the $(T_e,\cos\theta_{e\bar{\nu}})$ and
$(T_e,\cos\theta_{ep})$ distributions \cite{FG93}, where
$\theta_{e\bar{\nu}}$ and $\theta_{ep}$ are angles of the
electron--antineutrino and electron-proton correlations, respectively.

\noindent{\bf Summary}. The electron--proton energy distribution
$a(E_e, T_p)$ is calculated by taking into account a complete set of
the $1/M$ corrections, caused by the ``weak magnetism'' and the proton
recoil, and the radiative corrections of order $\alpha/\pi$,
calculated to leading order in the large $M$ expansion. The
proton--energy spectrum is improved in comparison with that, used in
\cite{Byrne}, by the account for the $1/M$ and radiative corrections.
The obtained expression for the electron--proton energy distribution
$a(E_e, T_p)$ as well as the proton--energy spectrum $a(T_p)$ should
provide a theoretical background for the experimental determination of
contributions of order $10^{-4}$ of interactions beyond the SM (see
section~\ref{sec:sensitivity}). For the analysis of contributions of
interactions beyond the SM we propose to multiply the electron--proton
energy spectrum of the rate of the neutron $\beta^-$--decay by the
lifetime of the neutron $\tau_n$, calculated with the account for the
contributions of interactions beyond the SM (see Appendix G). As a
result the analysis of the contributions of vector and axial--vector
interactions beyond the SM by means of the electron--proton energy
distribution $\tau_n a(E_e, T_p)$ and the proton--energy spectrum
$\tau_n a(T_p)$ reduces to the analysis of these contributions to the
axial coupling constant $\lambda$ only (see
section~\ref{sec:sensitivity}).

\section{Standard model analysis of experimental determination of
 correlation coefficient $C_0$. Proton asymmetry $C_{\exp}$}
\label{sec:Cexp}

The correlations between the neutron spin and the proton 3--momentum
are described by the correlation coefficient $C$
\cite{PRA1}--\cite{PRA5}.  It defines the proton recoil asymmetry $C_0
= - x_C (A_0 + B_0)$, where $x_C$ is the theoretical numerical factor,
calculated within the SM to leading order in the large $M$
expansion. The first measurement of the correlation coefficient
$C^{(\exp)}_0 = - 0.2377(26)$ has been performed by Schumann {\it et
  al.}  \cite{Schumann08}. The angular distribution of the probability
of the neutron $\beta^-$--decay, related to the proton recoil
asymmetry, is given by \cite{PRA1}
\begin{eqnarray}\label{label34a}
\hspace{-0.3in}4\pi \frac{dW(\theta_p)}{d\Omega_p} = 1 + 2 P C
\cos\theta_p,
\end{eqnarray}
where $d\Omega_p = 2\pi\sin\theta_pd\theta_p$ is the infinitesimal
solid angle of the proton 3--momentum with respect to the neutron
polarization $\vec{\xi}_n$, i.e. $\vec{\xi}_n\cdot \vec{k}_p = P k_p
\cos\theta_p$ and $P = |\vec{\xi}_n|$ is the neutron polarization. The
correlation coefficient $C$, calculated at the account for the $1/M$
and $\alpha/\pi$-radiative corrections, is equal
to (see Appendix I Eq.(\ref{labelI.29}))
\begin{eqnarray}\label{label34b}
\hspace{-0.3in} &&C = - \frac{1}{2}\frac{X_8}{X_1}\,(A_0 + B_0) +
\frac{1}{2}\frac{X_9}{X_1}\,A_0 +
\frac{\alpha}{\pi}\,\frac{1}{2}\frac{X_{10}}{X_1}\,A_0 -
\frac{\alpha}{\pi}\,\frac{1}{2}\frac{X_{11}}{X_1}\,B_0 + \frac{1}{M}\,\frac{1}{1 +
  3\lambda^2}\,\Big(\lambda\,\frac{1}{2}\frac{X_{12}}{X_1} - (\kappa +
1)\,\lambda\,\frac{1}{2}\frac{X_{13}}{X_1}\nonumber\\
\hspace{-0.3in}&&- (2\kappa +
1)\,\lambda\,\frac{1}{2}\frac{X_{14}}{X_1} - \lambda(1 +
\lambda)\,\frac{1}{2}\frac{X_{15} + Y_3}{X_1} + \lambda(1 -
\lambda)\,\frac{1}{2}\frac{X_{16} + Y_4}{X_1}\Big) + (A_0 +
B_0)\,\frac{X_8}{X_1}\,\Big\{
\frac{\alpha}{\pi}\,\frac{1}{2}\frac{X_2}{X_1} +
\frac{1}{M}\,\frac{1}{1 + 3\lambda^2}\nonumber\\
\hspace{-0.3in}&&\times\,\Big(\frac{1}{2}\frac{X_3}{X_1} + (1 +
3\lambda^2)\,\frac{1}{2}\frac{X_4 + Y_1}{X_1} - (1 - \lambda^2)\,
\frac{1}{2}\frac{X_5 + Y_2}{X_1}+ (\lambda^2 + 2(\kappa +
1)\lambda)\,\frac{1}{2}\frac{X_6}{X_1} - (\lambda^2 - 2(\kappa +
1)\lambda)\,\frac{1}{2}\frac{X_7}{X_1}\Big)\Big\},
\end{eqnarray}
where the numerical factors $X_j$ ($j = 1, \ldots, 14$) and $Y_j$ ($j
= 1,2,3,4$) are calculated in Appendix I (see Eq.(\ref{labelI.22}) and
Eq.(\ref{labelI.26})). The factor $x_C = X_8/2X_1 = 0.27591$ agrees
well with the factor $x_C = 0.27594$, calculated by Gl\"uck
\cite{PRA2}. As a result the correlation coefficient $C_0$ is equal to
\begin{eqnarray}\label{label34e}
C_0 = - 0.27591\,(A_0 + B_0).
\end{eqnarray}
The numerical value $C_0 = - 0.2386$, calculated at $\lambda = -
1.2750$, agrees well with the experimental one $C^{\exp}_0 = -
0.2377(26)$ \cite{Schumann08}. Defining the proton recoil asymmetry
$C_{\exp}$ as we have defined the asymmetry $A_{\exp}(E_e)$ (see
Eq.(\ref{label16})) we obtain
\begin{eqnarray}\label{label34c}
\hspace{-0.3in}C_{\exp} = P C \,(\cos\vartheta_1 + \cos\vartheta_2),
\end{eqnarray}
where the polar angles $\theta_1$ and $\theta_2$ define the solid
angle $\Delta \Omega_{12} = 2\pi (\cos\theta_1 - \cos\theta_2)$ of the
proton emission to the forward and backward hemisphere with respect to
the neutron spin.

The account for the contributions of the proton--photon correlations
to the proton recoil energy and angular distribution of the radiative
$\beta^-$--decay of the neutron \cite{Gluck1997,Ivanov2013} changes
the proton recoil asymmetry $C$ as follows \cite{Ivanov2013} (see also
Appendix I and Eq.(\ref{labelI.32}))
\begin{eqnarray}\label{label34f}
\hspace{-0.3in} &&C = - \Big(x_C + \frac{\alpha}{\pi}\,x_{\rm
  eff}\Big)\,(A_0 + B_0) + \frac{1}{2}\frac{X_9}{X_1}\,A_0 +
\frac{1}{M}\,\frac{1}{1 +
  3\lambda^2}\,\Big(\lambda\,\frac{1}{2}\frac{X_{12}}{X_1} - (\kappa +
1)\,\lambda\,\frac{1}{2}\frac{X_{13}}{X_1} - (2\kappa +
1)\,\lambda\,\frac{1}{2}\frac{X_{14}}{X_1}\nonumber\\
\hspace{-0.3in}&& -\lambda(1 + \lambda)\,\frac{1}{2}\frac{X_{15} +
  Y_3}{X_1} + \lambda(1 - \lambda)\,\frac{1}{2}\frac{X_{16} +
  Y_4}{X_1}\Big) + (A_0 +
B_0)\,\frac{X_8}{X_1}\,\Big\{\frac{\alpha}{\pi}\,\frac{1}{2}
\frac{X_2}{X_1} + \frac{1}{M}\,\frac{1}{1 +
  3\lambda^2}\,\Big(\frac{1}{2}\frac{X_3}{X_1} + (1 +
3\lambda^2)\nonumber\\
\hspace{-0.3in}&&\times\,\frac{1}{2}\frac{X_4 + Y_1}{X_1} - (1 -
\lambda^2)\, \frac{1}{2}\frac{X_5 + Y_2}{X_1}+ (\lambda^2 + 2(\kappa +
1)\lambda)\,\frac{1}{2}\frac{X_6}{X_1} - (\lambda^2 - 2(\kappa +
1)\lambda)\,\frac{1}{2}\frac{X_7}{X_1}\Big)\Big\},
\end{eqnarray}
where $x_{\rm eff} = X_{\rm eff}/2X_1 = 4.712120$ and $X_{\rm eff} =
2.215111\,{\rm MeV^5}$ \cite{Ivanov2013}. One may see that the
contributions of the proton--photon correlations make the radiative
corrections to the proton recoil asymmetry $C$ symmetric with respect
to a change $A_0 \longleftrightarrow B_0$ as well as the main term
$C_0 = - x_C (A_0 + B_0)$. 

\noindent{\bf Summary}. The correlation coefficient $C$, describing
correlations between the neutron spin and the proton 3--momentum, is
calculated by taking into account a complete set of the $1/M$
corrections, caused by the ``weak magnetism'' and the proton recoil,
and the radiative corrections of order $\alpha/\pi$, calculated to
leading order in the large $M$ expansion. The obtained result should
be used as a theoretical background for the experimental determination
of contributions of order $10^{-4}$ of interactions beyond the SM (see
section~\ref{sec:sensitivity}).

\section{Standard model analysis of Lifetime of neutron}
\label{sec:lifetime}

Having integrated the electron--energy and angular distribution
Eq.(\ref{label6}) over the directions of the electron 3--momentum
$\vec{k}_e$, the antineutrino 3--momentum $\vec{k}$ and the electron
energy $E_e$ within the limits $m_e \le E_e \le E_0$, we obtain the
rate of the neutron $\beta^-$--decay. It is equal to
\cite{SPT5}
\begin{eqnarray}\label{label35}
 ( \lambda_n)_{\rm SM} = (1 + 3 \lambda^2)\, \frac{G^2_F |V_{ud}|^2}{2
    \pi^3}\,f_n(E_0, Z = 1),
\end{eqnarray}
where the Fermi integral $f_n(E_0, Z = 1)$, given  by
\begin{eqnarray}\label{label36}
\hspace{-0.3in}&& f_n(E_0, Z = 1) = \int^{E_0}_{m_e} (E_0 - E_e
)^2\,\sqrt{E^2_e - m^2_e}\,E_e \, F(E_e, Z = 1)\,\Big(1 +
\frac{\alpha}{\pi}\,g_n(E_e)\Big)\nonumber\\
\hspace{-0.3in}&&\times\,\Big\{1 + \frac{1}{M}\,\frac{1}{1 + 3
  \lambda^2}\Big[\Big(10 \lambda^2 - 4(\kappa + 1)\,\lambda +
  2\Big)\,E_e - 2 \lambda\,\Big(\lambda - (\kappa + 1)\Big)\,\Big(E_0
  + \frac{m^2_e}{E_e}\Big)\Big]\Big\}\, dE_e,
\end{eqnarray}
contains the contributions of the ``weak magnetism'', the proton
recoil and the radiative corrections, described by the function
$g_n(E_e)$.  The calculation of the lifetime of the neutron with the
Fermi function, determined by Eq.(\ref{label5}), the axial coupling
constant $\lambda = - 1.2750(9)$ and the CKM matrix element $V_{ud} =
0.97427(15)$, gives $(\tau_n)_{\rm SM} = 879.6(1.1)\,{\rm s}$. The
error bars $\pm 1.1\,{\rm s}$ are defined by the error bars of the
experimental value of the coupling constant and the CKM matrix
element. The theoretical value of the lifetime of the neutron
$(\tau_n)_{\rm SM} = 879.6(1.1)\,{\rm s}$ agrees well with the
experimental values $\tau^{(\exp)}_n = 878.5(8)\,{\rm s}$ and
$\tau^{(\exp)}_n = 880.7(1.8)\,{\rm s}$ and $\tau^{(\exp)}_n =
881.6(2.1)\,{\rm s}$, measured by Serebrov {\it et al.}
\cite{Serebrov1}, Pichlmaier {\it et al.}  \cite{Pichlmaier} and
Arzumanov {\it et al.} \cite{Arzumanov}, respectively. It agrees also
well with the new world average values (w.a.v.) of the neutron
lifetime $\tau^{(\rm w.a.v.)}_n = 880.1(1.1)\,{\rm s}$ \cite{PDG12},
$\tau^{(\rm w.a.v.)}_n = 880.0(9)\,{\rm s}$ \cite{Serebrov2} and
$\tau^{(\rm w.a.v.)}_n = 881.9(1.3)\,{\rm s}$ \cite{DD2011},
respectively.

\noindent{\bf Summary}. The lifetime of the neutron is calculated by
taking into account the radiative corrections by Sirlin {\it et al.},
calculated to leading order in the large $M$ expansion, and a complete
set of the $1/M$ corrections, caused by the ``weak magnetism'' and the
proton recoil. The obtained result should be used as a theoretical
background for the experimental determination of contributions of
order $10^{-4}$ of interactions beyond the SM (see
section~\ref{sec:sensitivity}). The numerical value of the lifetime of
the neutron $(\tau_n)_{\rm SM} = 879.6(1.1)\,{\rm s}$, calculated in
this section, is left unchanged after the absorption of the Herczeg
phenomenological coupling constants $a^h_{LL}$ and $a^h_{LR}$ of the
left--left and left--right lepton--nucleon current--current
interactions (vector and axial--vector interactions beyond the SM) by
the axial coupling constant $\lambda \to \lambda_{\rm eff} = (\lambda
- a^h_{LL} + a^h_{LR})/(1 + a^h_{LL} + a^h_{LR}) = \lambda - a^h_{LL}
+ a^h_{LR} - \lambda(a^h_{LL} + a^h_{LR})$ and the CKM matrix element
$V_{ud} \to (V_{ud})_{\rm eff} = V_{ud}(1 + a^h_{LL} + a^h_{LR})$ (see
section~\ref{sec:sensitivity}).

\section{Sensitivity of electron--proton energy distribution, 
asymmetries $A_{\exp}(E_e)$, $B_{\exp}(E_e)$ and $C_{\exp}$ and
lifetime of neutron $\tau_n$ to contributions of order $10^{-4}$ of
interactions beyond Standard model}
\label{sec:sensitivity}

In this section we propose a theoretical analysis of the sensitivity
of the electron--proton energy distribution $a(E_e,T_p)$, the
proton--energy spectrum $a(T_p)$, the asymmetries $A_{\exp}(E_e)$,
$B_{\exp}(E_e)$ and $C_{\exp}$ and the lifetime of the neutron
$\tau_n$ to contributions of order $10^{-4}$ of interactions beyond
the SM.

The electron--proton energy distribution $a(E_e, T_p)$ and the
correlation coefficients $a(E_e)$, $A(E_e)$ and $B(E_e)$, including
the contributions of interactions beyond the SM, can be written in the
form
\begin{eqnarray}\label{label37}
a(E_e, T_p) &=&\zeta_1(E_e, T_p)_{\rm eff} + \bar{a}_{\rm eff}(E_e)\Big(1 +
\frac{\alpha}{\pi}\,f_n(E_e)\Big)\,
\zeta_2(E_e,T_p),\nonumber\\ a(E_e) &=& a_{\rm eff}(E_e) \Big(1 +
\frac{\alpha}{\pi}\,f_n(E_e)\Big),\nonumber\\ A(E_e) &=& A_{\rm
  eff}(E_e) \Big(1 +
\frac{\alpha}{\pi}\,f_n(E_e)\Big),\nonumber\\ B(E_e) &=& B_{\rm
  eff}(E_e),
\end{eqnarray}
where $\zeta_1(E_e, T_p)_{\rm eff}$, $\bar{a}_{\rm eff}(E_e)$, $a_{\rm
  eff}(E_e)$, $A_{\rm eff}(E_e)$ and $B_{\rm eff}(E_e)$ are the
electron--proton energy distribution and the correlation coefficients,
defined by the contributions of the ``weak magnetism'', the proton
recoil and interactions beyond the SM only. They are given by
\begin{eqnarray}\label{label38}
\hspace{-0.3in}&&\zeta_1(E_e, T_p)_{\rm eff} = \Big(1
- \frac{a_4}{a_0}\Big(\frac{m_e}{E_e} - \Big\langle
\frac{m_e}{E_e}\Big\rangle_{\rm SM}\Big)\Big)\,\zeta_1(E_e, T_p),\nonumber\\
\hspace{-0.3in}&&\bar{a}_{\rm eff}(E_e)= (a_0)_{\rm eff} +
\frac{1}{M}\frac{1}{1 + 3\lambda^2}\,E_0 + a_4\Big\langle\frac{m_e}{E_e}\Big\rangle_{\rm SM},\nonumber\\
\hspace{-0.3in}&&a_{\rm eff}(E_e) = (a_0)_{\rm eff} +
\frac{1}{M}\,\frac{1}{(1 + 3 \lambda^2)^2}\,\Big(a_1 E_0 + a_2 E_e +
a_3\frac{m^2_e}{E_e}\Big) + a_4\frac{m_e}{E_e},\nonumber\\
\hspace{-0.3in}&&A_{\rm eff}(E_e) = (A_0)_{\rm eff} +
\frac{1}{M}\,\frac{1}{(1 + 3 \lambda^2)^2}\,\Big(A_1 E_0 + A_2 E_e +
A_3\frac{m^2_e}{E_e}\Big) + A_4\frac{m_e}{E_e},\nonumber\\
\hspace{-0.3in}&&B_{\rm eff}(E_e) = (B_0)_{\rm eff} +
\frac{1}{M}\,\frac{1}{(1 + 3 \lambda^2)^2}\, \Big(B_1 E_0 + B_2 E_e +
B_3\frac{m^2_e}{E_e}\Big) + B_4\frac{m_e}{E_e},
\end{eqnarray}
where $\langle m_e/E_e\rangle_{\rm SM}$ is the average value,
calculated in the SM with the electron--energy spectrum density
Eq.(\ref{labelD.58}).  The terms, proportional to $\langle
m_e/E_e\rangle_{\rm SM}$, come from the lifetime of the neutron,
calculated at the account for the contributions  of interactions beyond the
SM (see Appendix G). The coefficients $a_j$, $A_j$ and $B_j$ for $j =
1,2,3$ are given in Eq.(\ref{label11}), Eq.(\ref{label12}) and
Eq.(\ref{label13}), respectively. They are calculated to
next--to--leading order in the large $M$ expansion and include the
contributions of the ``weak magnetism'' and the proton recoil
only. The numerical values of these corrections are estimated in
Eq.(\ref{label13a}) at $\lambda = - 1.2750$.  The terms $(a_0)_{\rm
  eff}$, $(A_0)_{\rm eff}$ and $(B_0)_{\rm eff}$ are the sums of the
correlation coefficients $a_0$, $A_0$ and $B_0$ and energy independent
contributions of interactions beyond the SM (see Appendix G). They
read
\begin{eqnarray}\label{label39}
\hspace{-0.3in}&&(a_0)_{\rm eff} = a_0 + \frac{1}{(1 + 3
  \lambda^2)^2}\,\Big(4 \lambda^2 \,{\rm Re}(\delta C_V - \delta
\bar{C}_V ) + 4 \lambda\, {\rm Re}(\delta C_A -
\delta\bar{C}_A)\Big),\nonumber\\
\hspace{-0.3in}&&(A_0)_{\rm eff} = A_0 + \frac{1}{(1 + 3
  \lambda^2)^2}\,\Big(- \lambda(3 \lambda^2 - 2 \lambda -
1)\, {\rm Re}(\delta C_V - \delta \bar{C}_V) - (3
\lambda^2 - 2 \lambda - 1){\rm Re}( \delta C_A - \delta\bar{C}_A)\Big)
,\nonumber\\
\hspace{-0.3in}&&(B_0)_{\rm eff} = B_0 + \frac{1}{(1 + 3
  \lambda^2)^2}\Big(- \lambda (3 \lambda^2 + 2 \lambda - 1)\, {\rm
  Re}(\delta C_V - \delta \bar{C}_V) - (3\lambda^2 + 2\lambda - 1)\,
       {\rm Re}(\delta C_A - \delta \bar{C}_A)\Big).
\end{eqnarray}
The coefficients $a_4$, $A_4$ and $B_4$ are induced by interactions
beyond the SM only. They are equal to
\begin{eqnarray}\label{label40}
\hspace{-0.3in}&&a_4 = -\frac{1 - \lambda^2}{(1 + 3 \lambda^2)^2}
\Big({\rm Re}(C_S - \bar{C}_S) + 3\lambda {\rm Re}(C_T -
\bar{C}_T)\Big),\nonumber\\
\hspace{-0.3in}&&A_4 =  2\,\frac{\lambda(1 + \lambda)}{(1 + 3
  \lambda^2)^2}\Big({\rm Re}(C_S - \bar{C}_S) + 3 \lambda {\rm
  Re}(C_T - \bar{C}_T)\Big) ,\nonumber\\
\hspace{-0.3in}&&B_4 = \frac{(1 + \lambda)(1 - 3\lambda)}{(1 + 3
  \lambda^2)^2}\Big(\lambda {\rm Re}(C_S - \bar{C}_S) - {\rm Re}(C_T -
\bar{C}_T)\Big).
\end{eqnarray}
For the derivation of Eqs.(\ref{label39}) and (\ref{label40}) we have
used Eq.(\ref{labelG.6}) and Eq.(\ref{labelG.6a} in Appendix G with
the coupling constants $C_V = 1 + \delta C_V$, $\bar{C}_V = - 1 +
\delta \bar{C}_V$, $C_A = - \lambda + \delta C_A$ and $\bar{C}_A =
\lambda + \delta \bar{C}_A$ and have kept only the linear
contributions of the deviations from the coupling constants of the SM.
By assumption the axial coupling constant $\lambda$ in
Eqs.(\ref{label39}) and (\ref{label40}) is determined by all
interactions within the SM only.

The differences of the phenomenological coupling constants $\delta C_V
- \delta \bar{C}_V$, $\delta C_A - \delta \bar{C}_A$, $C_S -
\bar{C}_S$ and $C_T - \bar{C}_T$, calculated in terms of the Herczeg
phenomenological lepton--nucleon coupling constants \cite{SPT3} (see
Appendix G), take the form
\begin{eqnarray}\label{label40a}
\hspace{-0.3in}\delta C_V - \delta \bar{C}_V &=& 2\,(a^h_{LL} +
a^h_{LR}),\nonumber\\
\hspace{-0.3in}\delta C_A - \delta \bar{C}_A &=&  2\,(a^h_{LL} -
a^h_{LR}),\nonumber\\
\hspace{-0.3in}C_S - \bar{C}_S &=& 2\,(A^h_{LL} + A^h_{LR}) = 4\,a_S,\nonumber\\
\hspace{-0.3in}C_T - \bar{C}_T &=& 4\,\alpha^h_{LL} = 4\, a_T.
\end{eqnarray}
This means that the phenomenological coupling constants $\delta C_V -
\delta \bar{C}_V$, $\delta C_A - \delta \bar{C}_A$, $C_S - \bar{C}_S$
and $C_T - \bar{C}_T$ describe interactions beyond the SM of
left--handed leptonic currents and left(right)--handed hadronic
currents, i.e. $L\otimes L$ and $L\otimes R$, respectively.

The experimental determination of the correlation coefficients $a_0$,
$A_0$ and $B_0$ runs as follows. Using the theoretical expressions for
the electron--proton energy distribution $a(E_e, T_p)$ and the
asymmetries $A_{\exp}(E_e)$ and $B_{\exp}(E_e)$ the experimental data
on them are being fitted by a tuning of the axial coupling constant
\cite{Abele1}. Such a procedure gives the axial-coupling constants
$\lambda_a$, $\lambda_A$ and $\lambda_B$, obtained from the fit of the
experimental data on $a(E_e,T_p)$, $A_{\exp}(E_e)$ and
$B_{\exp}(E_e)$, respectively. In terms of these axial coupling
constants one may define the correlation coefficients
$(a_0)^{(\exp)}_{\rm eff}$, $(A_0)^{(\exp)}_{\rm eff}$ and
$(B_0)^{(\exp)}_{\rm eff}$, respectively, as follows
\begin{eqnarray}\label{label41}
\hspace{-0.3in}(a_0)^{(\exp)}_{\rm eff} = \frac{1 - \lambda^2_a}{1 + 3
  \lambda^2_a} \quad ,\quad (A_0)^{(\exp)}_{\rm eff} = -
2\,\frac{\lambda_A(1 + \lambda_A)}{1 + 3 \lambda^2_A}\quad ,\quad
(B_0)^{(\exp)}_{\rm eff} =- 2\,\frac{\lambda_B(1 - \lambda_B)}{1 + 3
  \lambda^2_B}
\end{eqnarray}
The theoretical expressions for the axial coupling constants
$\lambda_a$, $\lambda_A$ and $\lambda_B$ in terms of the axial
coupling constant $\lambda$, defined by interactions within the SM
only, and the contributions of interactions beyond the SM are
\begin{eqnarray}\label{label42}
\lambda_a &=& \lambda - \frac{1}{2} \Big(\lambda{\rm Re}(\delta C_V -
\delta \bar{C}_V ) + {\rm Re}(\delta C_A -
\delta\bar{C}_A)\Big),\nonumber\\ \lambda_A &=& \lambda -
\frac{1}{2}\Big(\lambda {\rm Re}(\delta C_V - \delta \bar{C}_V) + {\rm
  Re}( \delta C_A - \delta\bar{C}_A)\Big),\nonumber\\ \lambda_B &=&
\lambda - \frac{1}{2} \Big(\lambda{\rm Re}(\delta C_V -
\delta \bar{C}_V ) + {\rm Re}(\delta C_A -
\delta\bar{C}_A)\Big).
\end{eqnarray}
Thus, in the linear approximation with respect to the deviations of
the phenomenological coupling constants of interactions beyond the SM
from the coupling constants of the SM we obtain that $\lambda_a =
\lambda_A = \lambda_B$. This agrees well with the results obtained in
\cite{UMW1}--\cite{UMW3}.  Replacing $\lambda_a$, $\lambda_A$ and
$\lambda_B$ by $\lambda_{\rm eff}$, which includes the contributions
of vector and axial--vector interactions beyond the SM in addition to
the contributions of interactions within the SM, and denoting
\begin{eqnarray}\label{label42a}
\hspace{-0.3in}b_F = \frac{1}{1 + 3\lambda^2_{\rm eff}}\Big({\rm Re}(C_S -
\bar{C}_S) + 3\lambda_{\rm eff} {\rm Re}(C_T - \bar{C}_T)\Big) = \frac{4}{1 +
  3\lambda^2_{\rm eff}}\Big({\rm Re}(a_S) + 3\lambda_{\rm eff} {\rm Re}(a_T)\Big)
\end{eqnarray}
and 
\begin{eqnarray}\label{label44a}
c_{ST} = \frac{1}{1 + 3\lambda^2_{\rm eff}}\Big(\lambda_{\rm
  eff}\,{\rm Re}(C_S - \bar{C}_S) - {\rm Re}(C_T - \bar{C}_T)\Big) =
\frac{4}{1 + 3\lambda^2_{\rm eff}}\Big(\lambda_{\rm eff} {\rm Re}(a_S)
- {\rm Re}(a_T)\Big),
\end{eqnarray}
where $b_F$ is the Fierz term (see Appendix G), we obtain the
electron--proton energy distribution $a(E_e, T_p)$ and the correlation
coefficients $a(E_e)$, $A(E_e)$ and $B(E_e)$ in the following form
\begin{eqnarray}\label{label42b}
\hspace{-0.3in}a(E_e,T_p) &=& \Big(1 - b_F \Big\langle
\frac{m_e}{E_e}\Big\rangle_{\rm SM}\Big)\Big\{\Big(1 + b_F
\frac{m_e}{E_e}\Big)\,\zeta_1(E_e, T_p) + a_0\,\Big(1 + \frac{1}{1 -
  \lambda^2_{\rm eff}}\,\frac{E_0}{M}\Big) \,\Big(1 +
\frac{\alpha}{\pi}\,f_n(E_e)\Big)\,\zeta_2(E_e, T_p)\Big\},\nonumber\\
\hspace{-0.3in}a(E_e) &=& a_0\,\Big(1 - b_F\,
\frac{m_e}{E_e}\Big)\,\Big(1 +
\frac{\alpha}{\pi}\,f_n(E_e)\Big)\Big\{1 + \frac{1}{M}\,\frac{1}{(1 -
  \lambda^2_{\rm eff})(1 + 3 \lambda^2_{\rm eff})}\, \Big(a_1 E_0 + a_2 E_e +
a_3\frac{m^2_e}{E_e}\Big)\Big\},\nonumber\\
\hspace{-0.3in}A(E_e) &=& A_0\Big(1- b_F\,
\frac{m_e}{E_e}\Big)\,\Big(1 +
\frac{\alpha}{\pi}\,f_n(E_e)\Big)\Big\{1 - \frac{1}{M}\,\frac{1}{2
  \lambda_{\rm eff}(1 + \lambda_{\rm eff})(1 + 3 \lambda^2_{\rm
    eff})}\, \Big(A_1 E_0 + A_2 E_e +
A_3\frac{m^2_e}{E_e}\Big)\Big\},\nonumber\\
\hspace{-0.3in}B(E_e) &=& B_0\,\Big(1 - \frac{(1 + \lambda_{\rm
    eff})(1 - 3\lambda_{\rm eff})}{2 \lambda_{\rm eff}(1 -
  \lambda_{\rm eff})}\,c_{ST}\frac{m_e}{E_e}\Big)\Big\{1 - \frac{1}{M}\,\frac{1}{2\lambda_{\rm eff} (1 -
  \lambda_{\rm eff})(1 + 3 \lambda^2_{\rm eff})} \Big(B_1 E_0 + B_2 E_e +
B_3\frac{m^2_e}{E_e}\Big)\Big\},
\end{eqnarray}
where the coefficients $a_j$, $A_j$ and $B_j$ for $j = 1,2,3$ are
defined in Eq.(\ref{label11}), Eq.(\ref{label12}) and
Eq.(\ref{label13}), respectively, with the replacement $\lambda \to
\lambda_{\rm eff}$. The same replacement defines the correlation
coefficients $a_0$, $A_0$ and $B_0$ in terms of $\lambda_{\rm eff}$
(see Eq.(\ref{label8})).

The correlation coefficients $a(E_e)$, $A(E_e)$ and $B(E_e)$, extended
by the contributions of interactions beyond the SM, together with the
correlation coefficients $K_n(E_e)$ and $Q_n(E_e)$, caused by the
contributions of the $1/M$ corrections from the ``weak magnetism'' and
the proton recoil only, determine the asymmetries $A_{\exp}(E_e)$ and
$B_{\exp}(E_e)$. For the calculation of the electron asymmetry
$A_{\exp}(E_e)$, taking into account the contributions of interactions
beyond the SM, we have to replace the correlation coefficient $A^{(\rm
  W)}(E_e)$ in Eq.(\ref{label16}) by the expression
\begin{eqnarray}\label{label44c}
\hspace{-0.3in}&&A^{(\rm W)}(E_e) = A_0\Big(1 -
b_F\frac{m_e}{E_e}\Big)\Big\{1 - \frac{1}{M}\,\frac{1}{2 \lambda_{\rm
    eff} (1 + \lambda_{\rm eff})(1 + 3 \lambda^2_{\rm eff})}\, \Big(A^{(\rm W)}_1 E_0 +
A^{(\rm W)}_2 E_e + A^{(\rm W)}_3\frac{m^2_e}{E_e}\Big)\Big\},
\end{eqnarray}
where the coefficients $A^{(\rm W)}_j$ are given in
Eq.(\ref{label17}) with the replacement $\lambda \to \lambda_{\rm eff}$.

The antineutrino asymmetry $B_{\exp}(E_e)$ is defined by
Eq.(\ref{label25}) and Eq.(\ref{label26}) for $r \le 1$ and $r \ge 1$,
respectively. For the account for the contributions of interactions
beyond the SM the correlation coefficients $a(E_e)$, $A(E_e)$ and
$B(E_e)$ should be taken in the form, given by Eq.(\ref{label42b}).

The proton recoil asymmetry $C_{\exp}$ completes the set of
asymmetries, which can be measured in the neutron $\beta^-$--decay
with a polarized neutron and unpolarised proton and electron. The
correlation coefficient $C_{\rm eff}$ (see Appendix I), defining the
asymmetry $C_{\exp}$ and extended by the contributions of interactions
beyond the SM, takes the form
\begin{eqnarray}\label{label34d}
\hspace{-0.3in}C_{\rm eff} = - x_C \Big(A_0 + B_0\Big) +
\frac{\lambda_{\rm eff} (1 + \lambda_{\rm eff})}{1 + 3\lambda^2_{\rm
    eff}}\,b_F\,\frac{X_{17}}{X_1} - \frac{1}{2}\,\frac{(1 +
  \lambda_{\rm eff})(1 - 3 \lambda_{\rm eff})}{1 + 3\lambda^2_{\rm
    eff}}\,c_{ST}\,\frac{X_{18}}{X_1} + C_{\rm SM},
\end{eqnarray}
where $C_{\rm SM} = C + x_C (A_0 + B_0)$ and $x_C = 0.27591$ (see
Eq.(\ref{label34f}) and Eq.(\ref{label34e})). The numerical factors
$X_{17}/X_1 = - 0.90187$ and $X_{18}/X_1 = 0.39806$ are calculated in
Appendix I (see Eq.(\ref{labelI.22})).

Using the results, obtained in Appendix G, the theoretical expression
for the rate of the neutron $\beta^-$--decay, including the
contributions of interactions beyond the SM, is 
\begin{eqnarray}\label{label49}
\hspace{-0.3in}&&\lambda_n = (\lambda_n)_{\rm SM} \Big(1 + \frac{1}{1
  + 3\lambda^2}\Big({\rm Re}(\delta C_V - \delta \bar{C}_V) -
3\lambda\,{\rm Re}(\delta C_A - \delta \bar{C}_A)\Big) + b_F
\Big\langle\frac{m_e}{E_e}\Big\rangle_{\rm SM}\Big) =
\nonumber\\ \hspace{-0.3in}&&=
\frac{G^2_F|V_{ud}|^2}{2\pi^3}\,f_n(E_0, Z = 1)\,(1 + 3
\lambda^2)\Big(1 + \frac{1}{1 + 3\lambda^2}\Big({\rm Re}(\delta C_V -
\delta \bar{C}_V) - 3\lambda\,{\rm Re}(\delta C_A - \delta
\bar{C}_A)\Big) + b_F \Big\langle\frac{m_e}{E_e}\Big\rangle_{\rm
  SM}\Big),
\end{eqnarray}
where $(\lambda_n)_{\rm SM}$ is the lifetime of the neutron,
calculated within the SM (see Eq.(\ref{label35}) and
Eq.(\ref{label36})), and $\langle m_e/E_e\rangle_{\rm SM}$ is the
average value, calculated with the electron--energy spectrum density
Eq.(\ref{labelD.58}). Now we have to define the rate of the neutron
$\beta^-$--decay Eq.(\ref{label49}) in terms of the axial coupling
constant $\lambda_{\rm eff}$, which is related to the axial coupling
constant $\lambda$ as
\begin{eqnarray}\label{label49a}
\lambda = \lambda_{\rm eff} + \frac{1}{2}(\lambda_{\rm eff}{\rm
  Re}(\delta C_V - \delta \bar{C}_V) + {\rm Re}(\delta C_A - \delta
\bar{C}_A)).
\end{eqnarray}
Substituting Eq.(\ref{label49a}) into Eq.(\ref{label49}) and keeping
only the linear terms in power of ${\rm Re}(\delta C_V - \delta
\bar{C}_V)$ and ${\rm Re}(\delta C_A - \delta \bar{C}_A)$ one may show
that
\begin{eqnarray}\label{label49b}
\hspace{-0.3in}&&(1 + 3\lambda^2)\,\Big(1 + \frac{1}{1 +
  3\lambda^2}\Big({\rm Re}(\delta C_V - \delta \bar{C}_V) -
3\lambda\,{\rm Re}(\delta C_A - \delta \bar{C}_A)\Big) + b_F
\Big\langle\frac{m_e}{E_e}\Big\rangle_{\rm SM}\Big) = \nonumber\\
\hspace{-0.3in}&&= \Big(1 +
{\rm Re}(\delta C_V - \delta \bar{C}_V)\Big)(1 +
3\lambda^2_{\rm eff})\Big(1 + b_F
\Big\langle\frac{m_e}{E_e}\Big\rangle_{\rm SM}\Big).
\end{eqnarray}
This gives the rate of the neutron $\beta^-$--decay equal to
\begin{eqnarray}\label{label49c}
\hspace{-0.3in}&&\lambda_n = \frac{G^2_F|V_{ud}|^2}{2\pi^3}\,f_n(E_0,
Z = 1)\,\Big(1 + {\rm Re}(\delta C_V - \delta
\bar{C}_V)\Big)(1 + 3\lambda^2_{\rm eff})\Big(1 + b_F
\Big\langle\frac{m_e}{E_e}\Big\rangle_{\rm SM}\Big),
\end{eqnarray}
where the Fermi integral is given by Eq.(\ref{label36}) with the
replacement $\lambda \to \lambda_{\rm eff}$. 

If we introduce again $(\lambda_n)_{\rm SM}$, defined by
Eq.(\ref{label35}) and Eq.(\ref{label36}) with the replacement
$\lambda \to \lambda_{\rm eff}$, we get the following expression for
the rate of the neutron $\beta^-$--decay, corrected by the
contributions of interactions beyond the SM taken to linear
approximation with respect to the Herczeg phenomenological coupling
constants
\begin{eqnarray}\label{label49d}
\hspace{-0.3in}&&\lambda_n = (\lambda_n)_{\rm SM}\,\Big(1 + 2\,{\rm
  Re}(a^h_{LL} + a^h_{LR})\Big)\,\Big(1 + b_F
\Big\langle\frac{m_e}{E_e}\Big\rangle_{\rm SM}\Big),
\end{eqnarray}
where we have set ${\rm Re}(\delta C_V - \delta \bar{C}_V) = 2 {\rm
  Re}(a^h_{LL} + a^h_{LR})$.  Thus, at first glimpse a deviation of
the experimental values $\tau_n = 1/\lambda_n$ of the lifetime of the
neutron considered relative to the theoretical value of the lifetime
of the neutron $(\tau_n)_{\rm SM} = 1/(\lambda_n)_{\rm SM}$,
calculated in the SM at zero Herczeg coupling constants, may give an
information about the contribution of the Herczeg left--left and
left--right lepton--nucleon current--current interactions (vector and
axial--vector interactions beyond the SM) by using the experimental
value of the Fierz term $b_F$, determined from the experimental data
on the electron--proton energy distribution $a(E_e, T_p)$, the
proton--energy spectrum $a(T_p)$ and the asymmetries $A_{\exp}(E_e)$,
$B_{\exp}(E_e)$ and $C_{\exp}$. However, the problem is that the
Herczeg phenomenological interactions beyond the SM, if they exist,
should exist always together with the interactions of the SM, and a
separation of these interactions is rather artificial.

Hence, using the Hamilton Eq.(\ref{labelG.1}) and Eq.(\ref{labelG.2})
we may redefine the axial coupling constant and the CKM matrix element
as follows $\lambda_{\rm eff} = (\lambda - a^h_{LL} + a^h_{LR})/(1 +
a^h _{LL} + a^h_{LR})$ and $(V_{ud})_{\rm eff} = V_{ud} (1 + a^h_{LL} +
a^h_{LR})$ as it has been proposed in \cite{UMW1}--\cite{UMW3}. After
such a change one may show that the rate of the neutron
$\beta^-$--decay may contain only the contribution of the Fierz term
\begin{eqnarray}\label{label49f}
\lambda_n = (\lambda_n)_{\rm SM}\,\Big(1 + b_F
\Big\langle\frac{m_e}{E_e}\Big\rangle_{\rm SM}\Big),
\end{eqnarray}
where $(\lambda_n)_{\rm SM}$ is given by Eq.(\ref{label35}) and
Eq.(\ref{label36}) with the replacement $\lambda \to \lambda_{\rm
  eff}$ and $V_{ud} \to (V_{ud})_{\rm eff}$.  

The definition of the effective coupling constant $\lambda_{\rm eff} =
(\lambda - a^h_{LL} + a^h_{LR})/(1 + a^h_{LL} + a^h_{LR})$ and the CKM
matrix element $(V_{ud})_{\rm eff} = V_{ud} (1 + a^h_{LL} + a^h_{LR})$
at the Hamiltonian level introduces the imaginary parts to the axial
coupling constant and the CKM matrix element
\begin{eqnarray}\label{label49g}
\hspace{-0.3in}\lambda_{\rm eff} &=& {\rm Re}\lambda_{\rm eff} +
i\,{\rm Im}\lambda_{\rm eff},\nonumber\\ 
\hspace{-0.3in}(V_{ud})_{\rm eff} &=& V_{ud}(1 + {\rm Re}(a^h_{LL} +
a^h_{LR}))(1 + i\,{\rm Im}(a^h_{LL} + a^h_{LR}))
\end{eqnarray}
where ${\rm Im}\lambda_{\rm eff}$ is equal to
\begin{eqnarray}\label{label49h}
\hspace{-0.3in}{\rm Im}\lambda_{\rm eff} = - (1 + {\rm Re}\lambda_{\rm
  eff}){\rm Im}(a^h_{LL}) + (1 - {\rm Re}\lambda_{\rm eff}){\rm
  Im}(a^h_{LR}).
\end{eqnarray}
An information about the imaginary part of the axial coupling constant
$\lambda_{\rm eff}$ one may obtain by measuring the correlation
coefficient $D(E_e)$, describing a violation of time reversal
invariance. From Eq.(\ref{labelG.6}), Eq.(\ref{labelG.6a}) and
Eq.(\ref{label49h}) we obtain
\begin{eqnarray}\label{label45}
\hspace{-0.3in}&&D(E_e) = D_{\rm SM}(E_e) - \frac{2}{1 + 3
  \lambda^2_{\rm eff}}\Big(\lambda_{\rm eff}{\rm Im}(a^h_{LL} + a^h_{LR})
+ {\rm Im}(a^h_{LL} - a^h_{LR})\Big) = \nonumber\\ &&= D_{\rm SM}(E_e) -
\frac{2 }{1 + 3 \lambda^2_{\rm eff}}\Big((1 + \lambda_{\rm eff}){\rm
  Im}(a^h_{LL}) - (1 - \lambda_{\rm eff}){\rm Im}(a^h_{LR})\Big) = D_{\rm
  SM}(E_e) + \frac{2 {\rm Im}\lambda_{\rm eff}}{1 + 3 \lambda^2_{\rm
    eff}},
\end{eqnarray}
where we have replaced ${\rm Re}\lambda_{\rm eff}$ by $\lambda_{\rm
  eff}$, having neglected the contribution of the imaginary part, and
$D_{\rm SM}(E_e)$ is the contribution to the correlation coefficient
$D(E_e)$, calculated within the SM \cite{LRM1}--\cite{AndoD09}. As we
have estimated in section~\ref{sec:spectrum} such a contribution,
caused by the electron--proton interaction in the final state
\cite{LRM1}--\cite{LRM4,AndoD09}, is of order $ 10^{-5}$ for the
electron kinetic energies $250\,{\rm keV} \le T_e \le 455\,{\rm
  keV}$. Thus, the correlation coefficient $D(E_e)$, defined in the
same energy region, should be sensitive to the contributions beyond
the SM of $10^{-4}$.  Recently the experimental value $D_{\exp}(E_e) =
(- 4\pm 6)\times 10^{-4}$ of the correlation coefficient $D(E_e)$
\cite{Abele1,PDG12} has been substantially improved with a result
$D_{\exp}(E_e) = ( - 0.96\pm 1.89_{\rm stat} \pm 1.01_{\rm
  syst})\times 10^{-4}$ \cite{Nico3}. However, the new experimental
value as well as the old one still implies that to order $10^{-4}$ the
correlation coefficient $D(E_e)$ is commensurable with zero.  Hence,
to order $10^{-4}$ the imaginary part of the axial coupling constant
$\lambda_{\rm eff}$ is also commensurable with zero, i.e.  ${\rm
  Im}\lambda_{\rm eff} = 0$. Nevertheless, setting $\lambda_{\rm eff}
= {\rm Re}\lambda _{\rm eff} = - 1.2750$ and using the relation ${\rm
  Im}\lambda_{\rm eff} = 0$ we may obtain that ${\rm Im}(a^h_{LR}) = -
0.12\,{\rm Im}(a^h_{LL})$. This adds an additional phase shift
$e^{\,i\,0.88\,{\rm Im}(a^h_{LL})}$ to the CKM matrix element
$(V_{ud})_{\rm eff}$.

The axial coupling constant $\lambda_a$ and the correlation
coefficient $a_0$ may be also determined by measuring the
proton--energy spectrum Eq.(\ref{label28a}) (see \cite{Byrne}). The
proton--energy spectrum, taking into account the contributions of
interactions beyond the SM, is
\begin{eqnarray}\label{label50e}
\hspace{-0.3in}&&a_{\rm eff}(T_p) = \Big(1 - b_F\Big\langle
\frac{m_e}{E_e}\Big\rangle_{\rm SM}\Big)\Big\{g_1(T_p)_{\rm eff} + a_0 \Big(1 +
\frac{1}{1 - \lambda^2}\,\frac{E_0}{M}\Big)\,g_2(T_p)\Big\},
\end{eqnarray}
where the functions $g_1(T_p)_{\rm eff}$ and $g_2(T_p)$ are defined by the
integrals
\begin{eqnarray}\label{label50d}
\hspace{-0.3in}&&g_1(T_p)_{\rm eff} = \int^{(E_e)_{\rm
    max}}_{(E_e)_{\rm min}} \Big(1 + b_F\,\frac{m_e}{E_e}\Big)\,\Big(1
+ \frac{\alpha}{\pi}\,g_n(E_e)\Big)\,\zeta_1(E_e,T_p) \,F(E_e, Z =
1)\,E_e\,dE_e,\nonumber\\
\hspace{-0.3in}&&g_2(T_p) = \int^{(E_e)_{\rm max}}_{(E_e)_{\rm min}}
\Big(1 + \frac{\alpha}{\pi}\,g_n(E_e) +
\frac{\alpha}{\pi}\,f_n(E_e)\Big)\,\zeta_2(E_e,T_p) \,F(E_e, Z =
1)\,E_e\,dE_e,
\end{eqnarray}
where the limits of integration $(E_e)_{\rm max/min}$ are given in
Appendix I.

\noindent{\bf Summary}. We have analysed the sensitivity of the
electron--proton energy distribution $a(E_e, T_p)$, the proton--energy
spectrum $a(T_p)$, the asymmetries $A_{\exp}(E_e)$, $B_{\exp}(E_e)$,
$C_{\exp}$ and the lifetime of the neutron $\tau_n$ to contributions
of order $10^{-4}$ of interactions beyond the SM, taken to linear
approximation with respect to the Herczeg phenomenological coupling
constants of weak lepton--nucleon current--current interactions.  We
have shown that in such an approximation the axial coupling constant
$\lambda_{\rm eff}$ and the CKM matrix element $(V_{ud})_{\rm eff}$
absorb the contributions of the Herczeg left--left and left--right
lepton--nucleon current--current interactions with the coupling
constants $a^h_{LL}$ and $a^h_{LR}$.  In this approximation the axial
coupling constant does not acquire an imaginary part to order
$10^{-4}$, but the CKM matrix element becomes an additional phase
$e^{\,i\,{\rm Im}(a^h_{LL} + a^h_{LR})}$. Thus, after the measurements
of the electron--proton energy spectrum $a(E_e,T_p)$, the
proton--energy spectrum $a(T_p)$, the asymmetries $A_{\exp}(E_e)$,
$B_{\exp}(E_e)$ and $C_{\exp}$ and the lifetime of the neutron
$\tau_n$ one may determine the axial coupling constant $\lambda_{\rm
  eff}$ and the real parts of the scalar and tensor coupling constants
$a_S$ and $a_T$, defined in Eq.(\ref{label40a}),
\begin{eqnarray}\label{label50c}
\hspace{-0.3in}{\rm Re}(a_S) + 3\lambda_{\rm eff}\,{\rm Re}(a_T) &=& \frac{1 +
  3\lambda^2_{\rm eff}}{4}\,b_F\nonumber\\
\hspace{-0.3in}\lambda_{\rm eff}\,{\rm Re}(a_S) - {\rm Re}(a_T) &=&
\frac{1 + 3\lambda^2_{\rm eff}}{4}\,c_{ST},
\end{eqnarray}
where in the r.h.s. of the algebraical equations $\lambda_{\rm eff}$,
$b_F$ and $c_{ST}$ are the experimental values with their experimental
errors. 

\section{Conclusion}
\label{sec:conclusion}

We have analysed the sensitivity of the electron--proton energy
distribution $a(E_e, T_p)$, the proton--energy spectrum $a(T_p)$, and
the asymmetries $A_{\exp}(E_e)$, $B_{\exp}(E_e)$ and $C_{\exp}$ of the
correlations between the neutron spin and 3--momenta of the decay
electron, antineutrino and proton, respectively, for the neutron
$\beta^-$--decay with a polarized neutron and unpolarised proton and
electron to contributions of order $10^{-4}$ of interactions beyond
the SM. For the analysis of contributions of order $10^{-4}$ we have
used the linear approximation for the correlation coefficient with
respect to the Herczeg phenomenological coupling constants of weak
lepton--nucleon current--current interactions.  We have shown that in
such an approximation the Herczeg right--left and right--right
lepton--nucleon current--current interactions with the coupling
constants $a^h_{RL}$ and $a^h_{RR}$ give no contributions to the
correlation coefficients of the neutron $\beta^-$--decay and the
lifetime of the neutron. Then, the contributions of the Herczeg
left--left and left--right lepton--nucleon current--current
interactions with the coupling constants $a^h_{LL}$ and $a^h_{LR}$ may
be absorbed by the axial coupling constant, which we denote as
$\lambda_{\rm eff} = \lambda - {\rm Re}(a^h_{LL} - a^h_{LR}) +
\lambda\, {\rm Re}(a^h_{LL} + a^h_{LR})$ and the CKM matrix element
$(V_{ud})_{\rm eff} = V_{ud}(1 + {\rm Re}(a^h_{LL} + a^h_{LR}))$. We
have shown that the Herczeg coupling constants $a^h_{LL}$ and
$a^h_{LR}$ in the effective Hamiltonian of weak interactions may be
absorbed by the axial coupling constant $\lambda_{\rm eff} = (\lambda
- a^h_{LL} + a^h_{LR})/(1 + a^h_{LL} + a^h_{LR})$ and the CKM matrix
element $(V_{ud})_{\rm eff} =V_{ud}(1 + a^h_{LL} + a^h_{LR})$. Using
the experimental data on the correlation coefficient $D(E_e)$ we have
shown that at the level of $10^{-4}$ the imaginary part of the axial
coupling constant $\lambda_{\rm eff}$ is equal to zero, whereas the
CKM matrix element acquires an additional phase $e^{\, i\,{\rm
    Im}(a^h_{LL} + a^h_{LR})}$. This shows that in addition to the
background, calculated in the SM, the correlation coefficients of the
neutron $\beta^-$--decay and the lifetime of the neutron, calculated
to linear approximation with respect to the Herczeg coupling constants
of lepton--nucleon current--current interactions, depend on the
contributions of the scalar and tensor interactions only. This agrees
with recent results, obtained in \cite{UMW1}--\cite{UMW3}.

We have shown that the contributions of the scalar and tensor
interactions beyond the SM are described by the Fierz term $b_F$ and
the coupling constant $c_{ST}$. The Fierz term may be determined from
the experimental data on the asymmetry $A_{\exp}(E_e)$ and the
electron--proton energy distribution $a(E_e, T_p)$ (or the
proton--energy spectrum $a(T_p)$). The coupling constant $c_{ST}$ may
be determined from the experimental data on the asymmetry
$B_{\exp}(E_e)$ and the proton recoil asymmetry $C_{\exp}$. This
allows to determine the scalar ${\rm Re}(a_S)$ and tensor ${\rm
  Re}(a_T)$ coupling constant by solving the system of algebraical
equations (see Eq.(\ref{label50c}))
\begin{eqnarray*}
\hspace{-0.3in}{\rm Re}(a_S) + 3\lambda_{\rm eff}\,{\rm Re}(a_T) &=& \frac{1 +
  3\lambda^2_{\rm eff}}{4}\,b_F\nonumber\\
\hspace{-0.3in}\lambda_{\rm eff}\,{\rm Re}(a_S) - {\rm Re}(a_T) &=& \frac{1 +
  3\lambda^2_{\rm eff}}{4}\,c_{ST},
\end{eqnarray*}
The lifetime of the neutron is defined by the background, calculated
in the SM, and the Fierz term (see Eq.(\ref{label49f}))
\begin{eqnarray*}
\hspace{-0.3in}\tau_n = (\tau_n)_{\rm SM}\Big(1 - b_F \,\Big\langle
\frac{m_e}{E_e}\Big\rangle_{\rm SM}\Big),
\end{eqnarray*}
where $(\tau_n)_{\rm SM}$ and $\langle m_e/E_e\rangle_{\rm SM}$ are
calculated in the SM. 

We would like to note that the experimental analysis of the neutron
$\beta^-$--decay with a polarized neutron and unpolarised proton and
electron may be carried out in terms the electron--proton energy
distribution $a(E_e, T_p)$, the proton--energy spectrum $a(T_p)$, the
asymmetries $A_{\exp}(E_e)$, $B_{\exp}(E_e)$, $C_{\exp}$ and the
lifetime of the neutron $\tau_n$. From the fit of the experimental
data on these energy distributions, asymmetries and the lifetime we
determine three parameters, i.e. the axial coupling constant
$\lambda_{\rm eff}$, the Fierz term $b_F$ and the coupling constant
$c_{ST}$. In order to determine these parameters without correlations
between them it suffices to use experimental data, obtained only in
three out of five independent experiments. This means that experimental
data, obtained from other two independent experiments, should be
described well by the parameters $(\lambda_{\rm eff}, b_F, c_{ST})$,
determined from the first three experiments. The deviations from the
predicted values should be much smaller compared with $10^{-4}$, since
they may be explained only by the contributions of higher powers of
the Herczeg coupling constants.

We have to note that the theoretical analysis of the sensitivity of
the electron--proton energy distribution $a(E_e, T_p)$, the
proton--energy spectrum $a(T_p)$, the asymmetries $A_{\exp}(E_e)$,
$B_{\exp}(E_e)$ and $C_{\exp}$ of the neutron $\beta^-$--decay and the
lifetime of the neutron to contributions of order $10^{-4}$ of
interactions beyond the SM we have carried out above the background,
calculated within the SM. We have taken into account a complete set of
the $1/M$ corrections, caused by the ``weak magnetism'' and the proton
recoil, calculated to next--to--leading order in the large $M$ or the
large proton mass expansion, and the radiative corrections of order
$\alpha/\pi$, calculated to leading order in the large $M$ or the
large proton mass expansion.

The corrections, caused by the ``weak magnetism'' and the proton
recoil are calculated in analytical agreement with the results,
obtained by Wilkinson \cite{Wilkinson} and Gudkov {\it et al.}
\cite{Gudkov2}. The radiative corrections to the lifetime of the
neutron and the correlation coefficients of the neutron
$\beta^-$--decay with a polarized neutron and unpolarised decay proton
and electron, calculated in this paper, we have given in terms of two
functions $(\alpha/\pi)\,g_n(E_e)$ and $(\alpha/\pi)\,f_n(E_e)$. The
analytical expressions of these functions are given in
Eq.(\ref{labelD.57}) of Appendix D.  They are in analytical agreement
with the radiative corrections, calculated in \cite{RC8}--\cite{RC18}
and in \cite{Gudkov1,Gudkov2}, respectively. We have confirmed
Sirlin's assertion that an unambiguous definition of the observable
radiative corrections to the lifetime of the neutron is fully caused
by the requirement of gauge invariance of the amplitude of
one--virtual photon exchanges of the continuum-state $\beta^-$--decay
of the neutron \cite{RC8}.

We have improved the theoretical expressions for the asymmetries
$A_{\exp}(E_e)$, $B_{\exp}(E_e)$ and $C_{\exp}$ with respect to the
expressions, used in \cite{Abele1,Mund}, \cite{Abele2,Schumann} and
\cite{Schumann08} for the experimental determination of the axial
coupling constant $\lambda = -1.2750(9)$ and the correlation
coefficients $A^{(\exp)}_0 = -0.11933(34)$, $B^{(\exp)}_0 =
0.9802(50)$ and $C^{(\exp})_0 = - 0.2377(26)$, respectively. We have
added the radiative corrections to the electron asymmetry
$A_{\exp}(E_e)$ and the $1/M$ and radiative corrections to the
antineutrino $B_{\exp}(E_e)$ and proton $C_{\exp}$ asymmetries. In
connection with the experimental analysis of contributions of order
$10^{-4}$ of interactions beyond the SM to the neutron
$\beta^-$--decay we have calculated the electron--proton energy
distribution $a(E_e, T_p)$ and the proton--energy spectrum $a(T_p)$ by
taking into account the complete set of the $1/M$ corrections, caused
by the ``weak magnetism'' and the proton recoil, and the radiative
corrections of order $\alpha/\pi$.

As has been pointed out by Gl\"uck \cite{Gluck1997}, the contributions
of the radiative $\beta^-$--decay of the neutron to the proton--energy
spectrum and angular distribution demand a detailed analysis of the
proton--photon correlations, which appear in the proton recoil energy
and angular distribution of the radiative $\beta^-$--decay of the
neutron.  For the aim of a consistent calculation of the contributions
of the nucleus--photon and hadron--photon correlations in the
radiative nuclear and hadronic $\beta$--decays Gl\"uck has used the
Monte Carlo simulation method. Recently the calculation of the
proton--photon correlations in the radiative $\beta^-$--decay of the
neutron has been performed in \cite{Ivanov2013}. There it has been
shown that the contributions of the proton--photon correlations to the
lifetime of the neutron $\tau_n$, the proton--energy spectrum $a(T_p)$
and the electron--proton energy distribution $a(E_e, T_p)$ are smaller
compared with the contributions of the radiative corrections,
described by the functions $g_n(E_e)$ and $f_n(E_e)$. At the level of
$10^{-5}$ accuracy the contributions of the proton--photon
correlations to part of the proton recoil angular distribution,
independent of $\cos\theta_p$, can be neglected. In turn, the
contributions of the proton--photon correlations to the proton recoil
asymmetry $C$, i.e. in part of the proton recoil angular distribution
proportional to $\cos\theta_p$, are of order $10^{-4}$. The account
for these contributions makes the radiative corrections to the proton
recoil angular distribution and the proton recoil asymmetry $C$
symmetric with respect to a change $A_0 \longleftrightarrow B_0$ as
well as the main term $C_0 = - x_C (A_0 + B_0)$. A detailed analysis
of the proton--energy spectrum $a(T_p)$ has been recently carried out
in \cite{Ivanov2013}. In addition to the proton--energy spectrum
$a(T_p)$, calculated in this paper, the authors of the paper
\cite{Ivanov2013} have included the contributions of the
proton--photon correlations and analysed the proton energy regions,
convenient for measurements of the Fierz term $b_F$, caused by scalar
and tensor interactions beyond the SM.

We have also shown that at the present level of the experimental
accuracy the lifetime of the neutron is described well by the SM with
the account for a complete set of the $1/M$ corrections, caused by the
``weak magnetism'' and the proton recoil, calculated to
next--to--leading order in the large proton mass expansion, and the
radiative corrections of order $\alpha/\pi$, calculated to leading
order in the large proton mass expansion. The theoretical value of the
lifetime of the neutron $(\tau_n)_{\rm SM} = 879.6(1.1)\,{\rm s}$,
where the error bars are defined by the error bars of the axial
coupling constant $\lambda = - 1.2750(9)$ and the CKM matrix element
$V_{ud} = 0.97427(15)$, agrees well with the experimental values
$\tau^{(\exp)}_n = 878.5(8)\,{\rm s}$, $\tau^{(\exp)}_n =
880.7(1.8)\,{\rm s}$ and $\tau^{(\exp)}_n = 881.6(2.1)\,{\rm s}$,
measured by Serebrov {\it et al.}  \cite{Serebrov1}, Pichlmaier {\it
  et al.}  \cite{Pichlmaier} and Arzumanov {\it et al.}
\cite{Arzumanov}, respectively, and the world average values of the
neutron lifetime $\tau^{(\rm w.a.v.)}_n = 880.1(1.1)\,{\rm s}$,
$\tau^{(\rm w.a.v.)}_n = 880.0(9)\,{\rm s}$ and $\tau^{(\rm w.a.v.)}_n
= 881.9(1.3)\,{\rm s}$, obtained in \cite{PDG12,Serebrov2,DD2011},
respectively.

In order to demonstrate the sensitivity of the lifetime of the
neutron, calculated in the SM, to the contributions of the radiative
and $1/M$ corrections we propose to rewrite the Fermi integral
$f_n(E_0, Z = 1)$, given by Eq.(\ref{label36}), in the following form
\begin{eqnarray}\label{label52}
\hspace{-0.3in}&& f_n(E_0, Z = 1) = \int^{E_0}_{m_e} (E_0 - E_e
)^2\,\sqrt{E^2_e - m^2_e}\,E_e\, F(E_e, Z = 1)\,\Big\{\Big(1 +
k_1\frac{\alpha}{\pi}\,g_n(E_e)\Big)\nonumber\\
\hspace{-0.3in}&&+ k_2\,\frac{1}{M}\,\frac{1}{1 + 3
  \lambda^2}\Big[\Big(10 \lambda^2 - 4(\kappa + 1)\,\lambda +
  2\Big)\,E_e - 2 \lambda\,\Big(\lambda - (\kappa + 1)\Big)\,\Big(E_0
  + \frac{m^2_e}{E_e}\Big)\Big]\Big\}\, dE_e,
\end{eqnarray}
where the coefficients $k_j$ for $j = 1,2$ are equal to $k_j = 0$ or
$k_j = 1$ that means that without $k_j = 0$ and with $k_j = 1$
corresponding corrections. The numerical values of the lifetime of the
neutron for different $k_j$ are adduced in Table I. It is seen that
the most important contributions come from the radiative corrections.

\begin{table}[h]
\begin{tabular}{|c|c|c|}
\hline $\tau_n$& $k_1$ & $k_2$\\ \hline $915.3\,{\rm s}$&0& 0\\\hline
$913.7\,{\rm s} $ & 0 & 1\\ \hline $881.0\,{\rm s} $ & 1 & 0\\\hline
$879.6\,{\rm s}$ & 1 & 1\\\hline
\end{tabular} 
\caption{The neutron lifetime, calculated for $\lambda = - 1.2750$,
  the radiative corrections $k_1 = 0,1$ and the $1/M$ corrections $k_2
  = 0,1$, caused by the ``weak magnetism'' and the proton recoil.}
\end{table}
For the comparison of the theoretical lifetime of the neutron, defined
by Eq.(\ref{label35}), with the expression, which is usually used for
the measurement of the CKM matrix element $V_{ud}$ \cite{RC1} (see
also \cite{Mund12}), we transcribe
Eq.(\ref{label35}) into the form \cite{RC1,Mund12}
\begin{eqnarray}\label{label53}
\hspace{-0.3in}\frac{1}{\tau_n} = C_n |V_{ud}|^2(1 +
3\lambda^2) f(1 + {\rm RC}),
\end{eqnarray}
where we have denoted $C_n = G^2_F m^5_e/2\pi^3 = 1.1614\times
10^{-4}\,{\rm s^{-1}}$ and ${\rm RC} = \langle
(\alpha/\pi)g_n(E_e)\rangle = 0.03886$, defining the radiative
corrections \cite{Abele1,RC1} integrated over the phase volume with
the account for the proton--electron final--state Coulomb
interaction. Then, the phase--space factor $f$, including the $1/M$
corrections from the ``weak magnetism'' and the proton recoil, is
determined by
\begin{eqnarray}\label{label54}
\hspace{-0.3in}&& f = \frac{1}{m^5_e}\int^{E_0}_{m_e} (E_0 - E_e
)^2\,\sqrt{E^2_e - m^2_e}\,E_e \, F(E_e, Z = 1)\,\Big\{1 +
\frac{1}{M}\,\frac{1}{1 + 3 \lambda^2}\Big[\Big(10 \lambda^2 -
  4(\kappa + 1)\,\lambda + 2\Big)\,E_e\nonumber\\
\hspace{-0.3in}&&- 2 \lambda\,\Big(\lambda - (\kappa +
1)\Big)\,\Big(E_0 + \frac{m^2_e}{E_e}\Big)\Big]\Big\}\, dE_e = 1.6894.
\end{eqnarray}
The numerical value agrees well with the value $f = 1.6887$,
calculated in \cite{RC1} (see also \cite{Mund12}). The factor $1 +
{\rm RC}$ we may represent in the following form $1 + {\rm RC} = (1 +
\delta_R)(1 + \Delta_R)$, where $\delta_R = \langle
(\alpha/\pi)\,(g_n(E_e) - C_{WZ})\rangle = 0.01505$ is defined by
one--photon exchanges and emission only \cite{RC1,RC17} and $\Delta_R
= (\alpha/\pi)\,C_{WZ} = 0.02381$ is the part of the radiative
corrections, induced by $W$--boson and $Z$--boson exchanges and QCD
corrections \cite{RC1,RC17} (see also \cite{Gudkov2}). The
phase--space factor $f_R$, including the contributions of the
radiative corrections, caused by one--photon exchanges and emission
only, is equal to $f_R = f(1 + \delta_R) = 1.71483$. It does not
contradict the value $f_R = 1.71385(34)$, used in \cite{Mund12} (see
also \cite{Gertrud}).

Currently the lifetime of the neutron is proposed to measuring in TU
M\"unchen within the project PENeLOPE, using a superconducting
magneto--gravitational trap of ultracold neutrons (UCN) for a precise
neutron lifetime measurement \cite{Paul}. In this experiment the UCN
are trapped in a multipole field of a flux density up to $2\,{\rm T}$
and bound by a gravitational force at the top. This makes the
extraction and detection of the protons possible and allows a direct
measurement of neutron decay.  A planing accuracy of $0.1\,{\rm s}$
and better demands high storage times and good knowledge of systematic
errors, which could result from neutron spin flip and high energetic
UCN which leave the storage volume only slowly. Therefore, the neutron
spectrum is cleaned by an absorber. The big storage volume of
$800\,{\rm dm^3}$ and the expected high neutron flux of FRMII give
more than $10^7$ neutrons per filling of the storage volume and meet
statistical demands. Of course, the experimental data on the lifetime
of the neutron, which should be obtained within this project with a
planning accuracy better than $0.1\,{\rm s}$, should place new
constraints on contributions of interactions beyond the SM.

For the completeness of our analysis we have calculated (see Appendix
H) the contributions of the proton recoil corrections of order
$\alpha/M$, caused by the electron--proton Coulomb interaction in the
final state of the neutron $\beta^-$--decay. We have shown that these
corrections to the lifetime of the neutron and the correlation
coefficients are of order $10^{-6} - 10^{-5}$. This allows to neglect
them for the analysis of contributions of order $10^{-4}$ of
interactions beyond the SM.

We would like to note that we have used the experimental value of the
axial coupling constant $\lambda = - 1.2750(9)$, determined from the
experimental data on the electron asymmetry $A_{\exp}(E_e)$
\cite{Abele1,Mund}. Such an experimental value of the axial coupling
constant has been obtained with an unprecedented accuracy of about
$0.07\,\%$. The axial coupling constant $\lambda = - 1.2750(9)$ agrees
well with the axial coupling constants $\lambda = -
1.2761^{+14}_{-17}$, $\lambda = - 1.2759^{+40.9}_{-44.5}$ and $\lambda
= - 1.2756(30)$, obtained recently by the PERKEO (PERKEO II)
Collaboration \cite{Mund12} and the UCNA (Ultra-cold Neutron
Asymmetry) Collaboration \cite{AX4,AX5}, respectively, the accuracies
of which are large compared with the accuracy of the axial coupling
constant $\lambda = - 1.2750(9)$. The lifetimes of the neutron $\tau_n
= 879\,(2)\,{\rm s}$, $\tau_n = 879\,(6)\,{\rm s}$ and $\tau_n =
879\,(4)\,{\rm s}$, calculated for the axial coupling constants
$\lambda = - 1.2761^{+14}_{-17}$, $\lambda = - 1.2759^{+40.9}_{-44.5}$
and $\lambda = - 1.2756(30)$, respectively, agree with the
experimental data $\tau^{(\exp)}_n = 878.5(8)\,{\rm s}$,
$\tau^{(\exp)}_n = 880.7(1.8)\,{\rm s}$ and $\tau^{(\exp)}_n =
881.6(2.1)\,{\rm s}$, measured by Serebrov {\it et al.}
\cite{Serebrov1}, Pichlmaier {\it et al.}  \cite{Pichlmaier} and
Arzumanov {\it et al.}  \cite{Arzumanov}, respectively, and the world
average values of the neutron lifetime $\tau^{(\rm w.a.v.)}_n =
880.1(1.1)\,{\rm s}$, $\tau^{(\rm w.a.v.)}_n = 880.0(9)\,{\rm s}$ and
$\tau^{(\rm w.a.v.)}_n = 881.9(1.3)\,{\rm s}$, obtained in
\cite{PDG12,Serebrov2,DD2011}, respectively.

The other experimental values of the axial coupling
constant $\lambda = - 1.266(4)$, $\lambda = - 1.2594(38)$ and $\lambda
= - 1.262(5)$, obtained in \cite{AX1,AX2} and \cite{AX3},
respectively, and cited by \cite{PDG12}, lead to the lifetimes of the
neutron $\tau_n = 890(5)\,{\rm s}$, $\tau_n = 898(5)\,{\rm s}$ and
$\tau_n = 895(7)\,{\rm s}$, which do not agree with the world average 
values of the neutron lifetime $\tau^{(\rm w.a.v.)}_n =
880.1(1.1)\,{\rm s}$, $\tau^{(\rm w.a.v.)}_n = 880.0(9)\,{\rm s}$ and
$\tau^{(\rm w.a.v.)}_n = 881.9(1.3)\,{\rm s}$, obtained in
\cite{PDG12,Serebrov2,DD2011}, respectively.  Moreover, the
experimental methods, used in \cite{AX1,AX2} and \cite{AX3} for the
measurements of the electron asymmetry $A_{\exp}(E_e)$, has been
recently criticized in \cite{Mund12}. As has been pointed out by Mund
{\it et al.}  \cite{Mund12}, in the experiments \cite{AX1,AX2} and
\cite{AX3} large corrections of about $15\,\% - 30\,\%$ should be
applied to 1) neutron polarization, 2) magnetic mirror effects, 3)
solid angle and 4) background.

For a long time \cite{RC3}--\cite{RC18} (see also
\cite{Gudkov1,Gudkov2}) due to infrared divergences the calculation of
the radiative $\beta^-$--decay of the neutron has been associated with
the calculation of the radiative corrections to the neutron
$\beta^-$--decay.  As has been shown already in \cite{RC3}, the sum of
the rates as well as the electron--energy and angular distributions of
the continuum-state and radiative $\beta^-$--decay modes of the
neutron does not suffer from infrared divergences, caused by
one--virtual photon exchanges in the continuum-state $\beta^-$--decay
mode and by the emission of real photons in radiative $\beta^-$--decay
mode of the neutron.

Nevertheless, the radiative $\beta^-$--decay of the neutron $n \to p +
e^- + \bar{\nu}_e + \gamma$ may be treated as a physical process,
which may be observed separately from the neutron $\beta^-$--decay $n
\to p + e^- + \bar{\nu}_e$. For the first time, the theoretical
analysis of the radiative $\beta^-$--decay of the neutron $n \to p +
e^- + \bar{\nu}_e + \gamma$ as a physical observable process has been
carried out in \cite{RBD1,RBD2}.  First reliable experimental data on
the branching ratio of the radiative $\beta^-$--decay of the neutron
${\rm BR}^{(\exp)}_{\beta^-_c\gamma} = 3.13(35)\times 10^{-3}$,
measured for the photon energy region $\omega_{\rm min} = 15\,{\rm
  keV} \le \omega \le \omega_{\rm max}= 340\,{\rm keV}$, have been
reported by Nico {\it et al.}  \cite{Nico2}. Then this result has been
updated by Cooper {\it et al.} \cite{Cooper,RC08}, who have obtained
${\rm BR}^{(\exp)}_{\beta^-_c\gamma} = 3.09(32)\times 10^{-3}$. These
experimental values agree well with the theoretical value ${\rm
  BR}_{\beta^-_c\gamma} = 2.85\times 10^{-3}$, calculated by Gardner
within HB$\chi$PT for the same photon energy region
\cite{RBD2,Nico2,Cooper}.  In Appendix B we have carried out the
calculation of the rate, the electron--photon energy and
photon--energy spectra and angular distributions of the radiative
$\beta^-$--decay of the neutron with a polarized neutron and
unpolarised decay particles.  Our results for the branching ratios
${\rm BR}_{\beta^-_c\gamma} = 2.87\times 10^{-3}$ and ${\rm
  BR}_{\beta^-_c\gamma} = 4.45 \times 10^{-3}$, calculated for the
photon energy regions $\omega_{\rm min} = 15\,{\rm keV} \le \omega \le
350\,{\rm keV}$ and $\omega_{\rm min} = 5\,{\rm keV} \le \omega \le
E_0 - m_e$, respectively, agree well with the results ${\rm
  BR}_{\beta^-_c\gamma} = 2.85\times 10^{-3}$ and ${\rm
  BR}_{\beta^-_c\gamma} = 4.41\times 10^{-3}$, obtained by Gardner
\cite{Nico2,Cooper} and Bernard {\it et al.} \cite{RBD2},
respectively. Within one standard deviation the branching ratio ${\rm
  BR}_{\beta^-_c\gamma} = 2.87\times 10^{-3}$ agrees also with the
experimental data \cite{Nico2,Cooper}.

The rate of the radiative $\beta^-$--decay of the neutron, depending
on a photon polarization, has been calculated in \cite{RBD2}. We argue
that the more precise theoretical and experimental analysis of the
energy spectra and angular distributions of the radiative
$\beta^-$--decay of the neutron, depending on the polarisations of the
neutron and photon, should be of great deal of importance for a test of
the SM. We are planning to perform such a theoretical analysis in our
forthcoming publication.

\subsection{Universality of radiative corrections to order $\alpha/\pi$}

The radiative corrections of $\alpha/\pi \sim 10^{-3}$ to the
electron--energy spectrum of the neutron $\beta^-$--decay, described
by the function $g_n(E_e)$, are universal for the electron (positron)
energy spectra of nuclear and neutron $\beta$--decays
\cite{Sirlin2004,Sirlin1987}. A universality of the radiative
corrections to order $\alpha/\pi \sim 10^{-3}$ to neutrino
(antineutrino) reactions, induced by weak charged currents, has been
pointed out by Kurylov, Ramsey--Musolf and Vogel \cite{RamseyM2003} by
example of the neutrino (antineutrino) disintegration of the deuteron
with the electron (positron) in the final state. Such a universality
has been confirmed in \cite{Ivanov2013a} for the cross section for the
inverse $\beta$--decay. As has been shown in \cite{Ivanov2013a} the
radiative corrections, calculated in \cite{RamseyM2003}, can be
described by the function $f_A(E_{\bar{\nu}})$ of the antineutrino
energy $E_{\bar{\nu}}$, calculated by Vogel \cite{IBD1}, Fayans
\cite{IBD2}, Fukugita and Kubota \cite{IBD3}, and Raha, Myhrer and
Kudobera \cite{IBD4} (see also \cite{Ivanov2013a}) and caused by
one--virtual photon exchanges and the radiative inverse
$\beta$--decay, and the constant part, caused by the electroweak boson
exchanges. In turn, as has been shown by Sirlin \cite{Sirlin2011}, the
radiative corrections, caused by one--virtual photon exchanges and the
bremsstrahlung, to neutrino (antineutrino) energy spectra of the
$\beta$--decays are also described by the function
$f_A(E_{\bar{\nu}})$ (see also \cite{Ivanov2013a}). A nice review of
the radiative corrections in precision electroweak physics has been
recently written by Sirlin and Ferroglia \cite{Sirlin2013}.

\section{Acknowledgements}

We are very grateful to H. Abele and A. P. Serebrov for numerous
discussions of the results, obtained in this paper, and advices on the
content and structure of the paper. We thank G. Konrad for discussions
of the measurements of the correlation coefficient $a_0$ by means of
the measurements of the electron-proton energy distribution and the
proton-energy spectrum of the neutron $\beta^-$--decay. We acknowledge
fruitful discussions with M. Ramsey--Musolf and M. Gonz\'alez-Alonso.
We are grateful to W. Marciano for the discussions of the radiative
corrections to the $\beta^-$--decay of the neutron. We thank
S. Gardner and V. Gudkov for the discussions of their results on the
analysis of the $\beta^-$--decay modes of the neutron, which were put
in the ground of our paper as well as the results, obtained by Sirlin
{\it et al.}.

The theoretical analysis of the sensitivity of 1) the electron--proton
energy distribution and the proton--energy spectrum, 2) the electron,
antineutrino and proton asymmetries of correlations between the
neutron spin and the 3--momenta of the decay particles and 3) the
lifetime of the neutron is carried out according to the experimental
program on the contract I534-N20 PERC, the theoretical program on the
contract I689-N16, supported both by the Austrian ``Fonds zur
F\"orderung der Wissenschaftlichen Forschung'' (FWF), and the
experimental program of the experimental group of the Petersburg
Nuclear Physics Institute (PNPI), headed by A. P. Serebrov, on the
contract No. 11-02-91000 -ANF$_-$a, supported by the Russian
Foundation for Basic Research.

This work was supported by the Austrian ``Fonds zur F\"orderung der
Wissenschaftlichen Forschung'' (FWF) under the contracts I689-N16,
I534-N20 PERC and I862-N20 and by the Russian Foundation for Basic
Research under the contract No. 11-02-91000 -ANF$_-$a and in part by
the U.S. Department of Energy contract No. DE-FG02-08ER41531,
No. DE-AC02-06CH11357 and the Wisconsin Alumni Research Foundation.

\section*{Appendix A: Amplitude of continuum-state $\beta^-$--decay 
of neutron with ``weak magnetism'' and proton recoil corrections to
order $1/M$} \renewcommand{\theequation}{A-\arabic{equation}}
\setcounter{equation}{0}

The amplitude of the continuum-state $\beta^-$--decay of the neutron
we rewrite as follows
\begin{eqnarray}\label{labelA.1} 
M(n \to pe^-\bar{\nu}_e) = -\frac{G_F}{\sqrt{2}}\,V_{ud}\,{\cal
  M}_{\beta^-_c},
\end{eqnarray}
where ${\cal M}_{\beta^-_c}= [\bar{u}_p O_{\mu} u_n][\bar{u}_e\gamma^{\mu}(1 -
\gamma^5)v_{\bar{\nu}}] $ and the matrix $O_{\mu}$ takes the form
\begin{eqnarray}\label{labelA.2} 
O_{\mu} = \gamma_{\mu}(1 + \lambda \gamma^5) + i\,\frac{\kappa}{2
  M}\,\sigma_{\mu\nu}(k_p - k_n)^{\nu}.
\end{eqnarray}
In terms of the time and space components of the matrix $O_{\mu} =
(O^0, -\vec{O}\,)$ the amplitude ${\cal M}_{\beta^-_c}$ is defined by
\begin{eqnarray}\label{labelA.3} 
{\cal M}_{\beta^-_c} = [\bar{u}_p O^0 u_n][\bar{u}_e\gamma^0(1 -
\gamma^5)v_{\bar{\nu}}] - [\bar{u}_p \vec{O}
u_n]\cdot [\bar{u}_e \vec{\gamma}\,(1 - \gamma^5)v_{\bar{\nu}}].
\end{eqnarray}
The time $O^0$ and spacial $\vec{O}$ components of the matrix
$O_{\mu}$ we determine to first in the large $M$ expansion. They read
\begin{eqnarray}\label{labelA.4}
 \hspace{-0.3in} O^0 =\left(\begin{array}{ccc} 1 & {\displaystyle
     \lambda + \frac{\kappa}{2 M}\,(\vec{\sigma}\cdot
     \vec{k}_p)}\\ {\displaystyle - \lambda + \frac{\kappa}{2
       M}\,(\vec{\sigma}\cdot \vec{k}_p) } & - 1\\
    \end{array}\right)
\end{eqnarray}
and 
\begin{eqnarray}\label{labelA.5}
 \hspace{-0.3in}&& \vec{O} =\left(\begin{array}{ccc} {\displaystyle 
     \lambda \vec{\sigma} + i\,\frac{\kappa}{2 M}\,(\vec{\sigma}\times
     \vec{k}_p)} & {\displaystyle \vec{\sigma}\,\Big(1 -
     \frac{\kappa}{2 M}\,E_0\Big)}\\ {\displaystyle
     -\,\vec{\sigma}\,\Big(1 + \frac{\kappa}{2 M}\,E_0\Big)} &
   {\displaystyle - \lambda \vec{\sigma}+ i \frac{\kappa}{2
       M}\,(\vec{\sigma}\times \vec{k}_p)} \\
    \end{array}\right),\nonumber\\
 \hspace{-0.3in}&&
\end{eqnarray}
where we have kept the terms of order $1/M$ only. For the calculation
of the amplitude of the $\beta^-$--decay of the neutron we use the
Dirac bispinorial wave functions of the neutron and the proton
\begin{eqnarray}\label{labelA.6}
 \hspace{-0.3in} u_n(\vec{0},\sigma_n) = \sqrt{2
   m_n}\Big(\begin{array}{c}\varphi_n \\ 0
 \end{array}\Big) \quad,\quad u_p(\vec{k}_p,\sigma_p) = \sqrt{E_p +
 m_p}\left(\begin{array}{c}\varphi_p \\ {\displaystyle
 \frac{\vec{\sigma}\cdot \vec{k}_p}{E_p + m_p}\,\varphi_p }
 \end{array}\right),
\end{eqnarray}
where the Pauli spinorial wave functions $\varphi_n$ and $\varphi_p$
depend on the polarisations $\sigma_n$ and $\sigma_p$,
respectively. The matrix elements $[\bar{u}_p O^0 u_n]$ and
$[\bar{u}_p \vec{O} u_n]$ are equal to
\begin{eqnarray}\label{labelA.7}
 \hspace{-0.3in}&&[\bar{u}_p O^0 u_n] = \sqrt{2 m_n(E_p +
   m_p)}\,\Big\{[\varphi^{\dagger}_p\varphi_n] + \frac{\lambda}{2
   M}\,[\varphi^{\dagger}_p(\vec{\sigma}\cdot
   \vec{k}_p)\varphi_n]\Big\}
\end{eqnarray}
and 
\begin{eqnarray}\label{labelA.8}
 \hspace{-0.3in}&&[\bar{u}_p \vec{O} u_n] = \sqrt{2 m_n(E_p +
   m_p)}\,\Big\{\lambda [\varphi^{\dagger}_p\vec{\sigma}\,\varphi_n] +
 i\,\frac{\kappa }{2 M}\,[\varphi^{\dagger}_p(\vec{\sigma}\times
   \vec{k}_p)\varphi_n] + \frac{1}{2
   M}\,[\varphi^{\dagger}_p(\vec{\sigma}\cdot
   \vec{k}_p)\vec{\sigma}\,\varphi_n]\Big\},
\end{eqnarray}
where in curly brackets we have kept the contributions of the terms of
order $1/M$ only. Using the relation $(\vec{\sigma}\cdot
\vec{k}_p)\vec{\sigma} = \vec{k}_p + i\,(\vec{\sigma}\times
\vec{k}_p)$ we rewrite the r.h.s. of Eq.(\ref{labelA.8}) as follows
\begin{eqnarray}\label{labelA.9}
 \hspace{-0.3in}[\bar{u}_p \vec{O} u_n] = \sqrt{2 m_n(E_p +
   m_p)}\,\Big\{\lambda [\varphi^{\dagger}_p\vec{\sigma}\,\varphi_n] +
 i\,\frac{\kappa + 1}{2 M}\,[\varphi^{\dagger}_p(\vec{\sigma}\times
   \vec{k}_p)\varphi_n] + \frac{\vec{k}_p}{2
   M}\,[\varphi^{\dagger}_p\varphi_n]\Big\}.
\end{eqnarray}
Thus, the amplitude ${\cal M}_{\beta^-_c}$ is given by
\begin{eqnarray}\label{labelA.10}
\hspace{-0.3in}&&{\cal M}_{\beta^-_c} = \sqrt{2 m_n(E_p +
 m_p)}\Big\{[\varphi^{\dagger}_p\varphi_n][\bar{u}_e\gamma^0(1 -
 \gamma^5)v_{\bar{\nu}}] - \lambda
        [\varphi^{\dagger}_p\vec{\sigma}\,\varphi_n]\cdot
        [\bar{u}_e\vec{\gamma}\,(1 -
          \gamma^5)v_{\bar{\nu}}] + \frac{\lambda}{2
   M}\,[\varphi^{\dagger}_p(\vec{\sigma}\cdot
   \vec{k}_p)\varphi_n]\nonumber\\
 \hspace{-0.3in}&&\times\,[\bar{u}_e\gamma^0(1 -
   \gamma^5)v_{\bar{\nu}}] - i\,\frac{\kappa + 1}{2
   M}\,[\varphi^{\dagger}_p(\vec{\sigma}\times
   \vec{k}_p)\varphi_n]\cdot [\bar{u}_e\vec{\gamma}\,(1 -
   \gamma^5)v_{\bar{\nu}}] - \frac{\vec{k}_p}{2
   M}\,[\varphi^{\dagger}_p\varphi_n]\cdot [\bar{u}_e\vec{\gamma}\,(1
   - \gamma^5)v_{\bar{\nu}}]\Big\}.
\end{eqnarray}
For the transformation of the last term we use the following identity
\begin{eqnarray}\label{labelA.11}
\hspace{-0.3in}- \frac{\vec{k}_p}{2
  M}\,[\varphi^{\dagger}_p\varphi_n]\cdot [\bar{u}_e\vec{\gamma}\,(1 -
  \gamma^5)v_{\bar{\nu}}] = \frac{E_0}{2
  M}[\varphi^{\dagger}_p\varphi_n][\bar{u}_e \gamma^0 (1 -
  \gamma^5)v_{\bar{\nu}}] - \frac{m_e}{2
  M}[\varphi^{\dagger}_p\varphi_n][\bar{u}_e (1 -
  \gamma^5)v_{\bar{\nu}}],
\end{eqnarray}
based on the Dirac equation for the electron and
antineutrino. Substituting Eq.(\ref{labelA.11}) into
Eq.(\ref{labelA.10}) we obtain
\begin{eqnarray}\label{labelA.12}
\hspace{-0.3in}&&{\cal M}_{\beta^-_c} = \sqrt{2 m_n(E_p +
  m_p)}\,\Big\{\Big(1 + \frac{E_0}{2
  M}\Big)\,[\varphi^{\dagger}_p\varphi_n][\bar{u}_e\gamma^0(1 -
  \gamma^5)v_{\bar{\nu}}] - \lambda
 [\varphi^{\dagger}_p\vec{\sigma}\,\varphi_n]\cdot
 [\bar{u}_e\vec{\gamma}\,(1 - \gamma^5)v_{\bar{\nu}}]+ \frac{\lambda}{2
   M}\,[\varphi^{\dagger}_p(\vec{\sigma}\cdot
   \vec{k}_p)\varphi_n]\nonumber\\
 \hspace{-0.3in}&&\times\,[\bar{u}_e\gamma^0(1 -
   \gamma^5)v_{\bar{\nu}}] - i\,\frac{\kappa + 1}{2
   M}\,[\varphi^{\dagger}_p(\vec{\sigma}\times
   \vec{k}_p)\varphi_n]\cdot [\bar{u}_e\vec{\gamma}\,(1 -
   \gamma^5)v_{\bar{\nu}}] - \frac{m_e}{2
  M}[\varphi^{\dagger}_p\varphi_n][\bar{u}_e (1 -
  \gamma^5)v_{\bar{\nu}}]\Big\}.
\end{eqnarray}
The next step of the calculation is to expand the normalization factor
$\sqrt{E_p + m_p}$ of the bispinorial wave function of the
proton. This gives
\begin{eqnarray}\label{labelA.13}
\sqrt{2 m_n(E_p + m_p)} = 2m_n\,\Big(1 - \frac{E_0}{2 M}\Big),
\end{eqnarray}
where we have kept the next--to--leading terms in the large $M$
expansion.  

Thus the amplitude ${\cal M}_{\beta^-_c}$, calculated to
next--to--leading order order in the large $M$ expansion, is
\begin{eqnarray}\label{labelA.14}
\hspace{-0.3in}&&{\cal M}_{\beta^-_c} = 2
 m_n\,\Big\{[\varphi^{\dagger}_p\varphi_n][\bar{u}_e\gamma^0(1 -
 \gamma^5)v_{\bar{\nu}}] - \tilde{\lambda}
 [\varphi^{\dagger}_p\vec{\sigma}\,\varphi_n]\cdot
 [\bar{u}_e\vec{\gamma}\,(1 - \gamma^5)v_{\bar{\nu}}] + \frac{\tilde{\lambda}}{2
   M}\,[\varphi^{\dagger}_p(\vec{\sigma}\cdot
   \vec{k}_p)\varphi_n]\nonumber\\
 \hspace{-0.3in}&&\times\,[\bar{u}_e\gamma^0(1 -
   \gamma^5)v_{\bar{\nu}}] - i\,\frac{\kappa + 1}{2
   M}\,[\varphi^{\dagger}_p(\vec{\sigma}\times
   \vec{k}_p)\varphi_n]\cdot [\bar{u}_e\vec{\gamma}\,(1 -
   \gamma^5)v_{\bar{\nu}}] - \frac{m_e}{2
   M}[\varphi^{\dagger}_p\varphi_n][\bar{u}_e (1 -
   \gamma^5)v_{\bar{\nu}}]\Big\},
\end{eqnarray}
where we have denoted $\tilde{\lambda} = \lambda (1 - E_0/2M)$. The
hermitian conjugate amplitude takes the form
\begin{eqnarray}\label{labelA.15}
\hspace{-0.3in}&&{\cal M}^{\dagger}_{\beta^-_c} = 2
 m_n\,\Big\{[\varphi^{\dagger}_n\varphi_p][\bar{v}_{\bar{\nu}}\gamma^0(1
 - \gamma^5)u_e] - \tilde{\lambda}^*
 [\varphi^{\dagger}_n\vec{\sigma}\,\varphi_p]\cdot
 [\bar{v}_{\bar{\nu}}\vec{\gamma}\,(1 - \gamma^5)u_e] + \frac{\tilde{\lambda}^*}{2
   M}\,[\varphi^{\dagger}_n(\vec{\sigma}\cdot
   \vec{k}_p)\varphi_p]\nonumber\\
 \hspace{-0.3in}&&\times\,[\bar{v}_{\bar{\nu}}\gamma^0(1 - \gamma^5)
   u_e] + i\,\frac{\kappa + 1}{2
   M}\,[\varphi^{\dagger}_n(\vec{\sigma}\times
   \vec{k}_p)\varphi_p]\cdot [\bar{v}_{\bar{\nu}}\vec{\gamma}\,(1 -
   \gamma^5) u_e] - \frac{m_e}{2
   M}[\varphi^{\dagger}_n\varphi_p][\bar{v}_{\bar{\nu}}(1 + \gamma^5)
   u_e]\Big\}.
\end{eqnarray}
Substituting Eq.(\ref{labelA.14}) into Eq.(\ref{labelA.1}) we arrive
at the amplitude of the continuum-state $\beta^-$--decay of the
neutron, taking into account the contributions of the ``weak
magnetism'' and the proton recoil, calculated to next--to--leading
order in the $1/M$ expansion.

The electron--energy and angular distribution is
proportional to $\frac{1}{2}\sum_{\rm pol}|{\cal M}_{\beta^-_c}|^2$,
where we sum over all polarisations of the interacting
particles. Recall that the antineutrino is polarized in the direction
parallel to its 3--momentum. The quantity $\frac{1}{2}\sum_{\rm
  pol}|{\cal M}_{\beta^-_c}|^2$ is equal to
\begin{eqnarray}\label{labelA.16}
\hspace{-0.3in}&& \sum_{\rm pol}\frac{|{\cal
    M}_{\beta^-_c}|^2}{4m^2_n} = {\rm tr}\{1 +
\vec{\xi}_n\cdot\vec{\sigma}\}{\rm tr}\{(\hat{k}_e + m_e) \gamma^0
\hat{k} \gamma^0 (1 - \gamma^5) \} - \tilde{\lambda} {\rm tr}\{(1 +
\vec{\xi}_n\cdot \vec{\sigma}\,)\vec{\sigma}\}\cdot {\rm tr}\{
(\hat{k}_e + m_e)\vec{\gamma}\,\hat{k}\gamma^0 (1 - \gamma^5)\}\nonumber\\
\hspace{-0.3in}&& - \tilde{\lambda}^*{\rm tr}\{(1 + \vec{\xi}_n\cdot
\vec{\sigma}\,)\vec{\sigma}\}\cdot {\rm tr}\{(\hat{k}_e + m_e)
\gamma^0 \hat{k}\vec{\gamma}\, (1 - \gamma^5)\} +
|\tilde{\lambda}|^2{\rm tr}\{(1 + \vec{\xi}_n\cdot
\vec{\sigma}\,)\sigma^i\sigma^j\}{\rm tr}\{ (\hat{k}_e + m_e)
\gamma^j\hat{k}\gamma^i (1 - \gamma^5)\}\nonumber\\
\hspace{-0.3in}&& - \frac{m_e}{2M}\,{\rm tr}\{(1 +
\vec{\xi}_n\cdot\vec{\sigma}\,)\}{\rm tr}\{(\hat{k}_e + m_e)\gamma^0
\hat{k} (1 + \gamma^5)\} - \frac{m_e}{2M}\,{\rm tr}\{(1 +
\vec{\xi}_n\cdot\vec{\sigma}\,)\}{\rm tr}\{(\hat{k}_e + m_e) \hat{k}
\gamma^0(1 - \gamma^5)\}\nonumber\\
\hspace{-0.3in}&& + \tilde{\lambda}\frac{m_e}{2 M}{\rm tr}\{(1 +
  \vec{\xi}_n\cdot \vec{\sigma}\,)\vec{\sigma}\}\cdot {\rm
  tr}\{(\hat{k}_e + m_e)\vec{\gamma}\,\hat{k} (1 +
  \gamma^5)\} + \tilde{\lambda}^* \frac{m_e}{2 M}{\rm tr}\{(1 +
\vec{\xi}_n\cdot \vec{\sigma}\,)\vec{\sigma}\}\cdot {\rm
  tr}\{(\hat{k}_e + m_e) \hat{k} \vec{\gamma}\,(1 -
\gamma^5)\}\nonumber\\
\hspace{-0.3in}&& +\frac{\tilde{\lambda}}{2 M}\,{\rm tr}\{(1 +
\vec{\xi}_n\cdot \vec{\sigma}\,)(\vec{\sigma}\cdot \vec{k}_p)\}\, {\rm
  tr}\{(\hat{k}_e + m_e) \gamma^0 \hat{k} \gamma^0 (1 - \gamma^5)\}\nonumber\\
\hspace{-0.3in}&& + \frac{\tilde{\lambda}^*}{2 M}\,{\rm tr}\{(1 +
\vec{\xi}_n\cdot \vec{\sigma}\,)(\vec{\sigma}\cdot \vec{k}_p)\}\, {\rm
  tr}\{(\hat{k}_e + m_e) \gamma^0 \hat{k} \gamma^0 (1 - \gamma^5)\}
\nonumber\\
\hspace{-0.3in}&&- \frac{|\tilde{\lambda}|^2}{2M}{\rm tr}\{(1 +
\vec{\xi}_n\cdot \vec{\sigma}\,)(\vec{\sigma}\cdot
\vec{k}_p)\vec{\sigma}\,\}\cdot {\rm tr}\{(\hat{k}_e +
m_e)\vec{\gamma}\, \hat{k} \gamma^0 (1 - \gamma^5)\}\nonumber\\
\hspace{-0.3in}&& - \frac{|\tilde{\lambda}|^2}{2M} {\rm tr}\{(1 +
\vec{\xi}_n\cdot \vec{\sigma}\,)\vec{\sigma}(\vec{\sigma}\cdot
\vec{k}_p)\}\cdot {\rm tr}\{(\hat{k}_e + m_e)\gamma^0 \hat{k}
\vec{\gamma}(1 - \gamma^5)\} \nonumber\\
\hspace{-0.3in}&&- i \,\frac{\kappa + 1}{2 M}{\rm tr}\{(1 +
\vec{\xi}_n\cdot\vec{\sigma})(\vec{\sigma}\times \vec{k}_p)\} \cdot
    {\rm tr}\{(\hat{k}_e + m_e) \vec{\gamma}\,\hat{k} \gamma^0 (1 -
    \gamma^5)\}\nonumber\\
\hspace{-0.3in}&& + i\,\frac{\kappa + 1}{2 M}{\rm tr}\{(1 +
\vec{\xi}_n\cdot\vec{\sigma})(\vec{\sigma}\times \vec{k}_p)\} \cdot
    {\rm tr}\{(\hat{k}_e + m_e) \gamma^0\hat{k} \vec{\gamma}\, (1 -
    \gamma^5)\}\nonumber\\
\hspace{-0.3in}&&+ i\,\tilde{\lambda}^* \frac{\kappa + 1}{2 M}{\rm tr}\{(1 +
 \vec{\xi}_n\cdot\vec{\sigma})\sigma^j(\vec{\sigma}\times
 \vec{k}_p)^{\ell}\}\,{\rm tr}\{(\hat{k}_e + m_e)
 \gamma^{\ell}\,\hat{k}\gamma^j (1 - \gamma^5)\}\nonumber\\
\hspace{-0.3in}&&- i\, \tilde{\lambda} \frac{\kappa + 1}{2 M}{\rm
  tr}\{(1 + \vec{\xi}_n\cdot\vec{\sigma})(\vec{\sigma}\times
\vec{k}_p)^{\ell}\sigma^j\}\, {\rm tr}\{(\hat{k}_e + m_e)
 \gamma^j\,\hat{k} \gamma^{\ell}(1 - \gamma^5)\}.
\end{eqnarray}
Since in our analysis the axial coupling constant $\lambda$ is real,
below we set $\lambda^* = \lambda$.  Calculating the traces with the
real axial coupling constant we obtain the following result
\begin{eqnarray*}
\hspace{-0.3in}&& \sum_{\rm pol}\frac{|{\cal
    M}_{\beta^-_c}|^2}{32m^2_n} = (1 + 3 \tilde{\lambda}^2)\,E_e E +
(1 - \tilde{\lambda}^2)\,(\vec{k}_e\cdot \vec{k}\,) - 2
\tilde{\lambda} (1 + \tilde{\lambda})\,(\vec{\xi}_n\cdot \vec{k}_e)E -
2 \tilde{\lambda} (1 - \tilde{\lambda})\,(\vec{\xi}_n\cdot
\vec{k}\,)E_e \nonumber\\
\hspace{-0.3in}&& + \frac{1}{M}\,\Big\{- m^2_e E + \tilde{\lambda}\,
m^2_e (\vec{\xi}_n\cdot \vec{k}\,) -
\tilde{\lambda}\,(\vec{\xi}_n\cdot \vec{k}_e + \vec{\xi}_n\cdot
\vec{k}\,)\,(E_e E + \vec{k}_e\cdot \vec{k}\,) + \tilde{\lambda}^2
\Big[( E_e - \vec{\xi}_n\cdot \vec{k}_e) (\vec{k}_e\cdot \vec{k} +
  E^2)\nonumber\\
\hspace{-0.3in}&& + (E + \vec{\xi}_n\cdot \vec{k}\,)(k^2_e +
\vec{k}_e\cdot \vec{k}\,)\Big] - (\kappa + 1)\Big[(\vec{\xi}_n\cdot
  \vec{k}_e)(\vec{k}_e\cdot \vec{k} + E^2) - (\vec{\xi}_n\cdot
  \vec{k}\,)(k^2_e + \vec{k}_e\cdot \vec{k}\,)\Big]\nonumber\\
\hspace{-0.3in}&& - 2\,(\kappa + 1)\,\tilde{\lambda} \Big[E(k^2_e +
  \vec{k}_e\cdot \vec{k}\,) - E_e (\vec{k}_e\cdot \vec{k} + E^2)\Big]+
2\, (\kappa + 1)\,\tilde{\lambda}\, (\vec{\xi}_n\cdot \vec{k}_e +
\vec{\xi}_n\cdot \vec{k}\,)\,E_e E
\end{eqnarray*}
\begin{eqnarray}\label{labelA.17}
\hspace{-0.3in}&&  - (\kappa +
1)\,\tilde{\lambda}\Big[(\vec{\xi}_n\cdot \vec{k}_e)(\vec{k}_e\cdot
  \vec{k} + E^2) + (\vec{\xi}_n\cdot \vec{k}\,)(k^2_e + \vec{k}_e\cdot
  \vec{k}\,)\Big]\Big\}.
\end{eqnarray}
Taking into account the contribution of the phase volume
Eq.(\ref{label4}) (see also Eq.(\ref{labelA.23})) and keeping only the
terms of order $1/M$ we have
\begin{eqnarray}\label{labelA.18}
\hspace{-0.3in}&&\Phi_{\beta^-_c}(\vec{k}_e,\vec{k}\,) \sum_{\rm
 pol}\frac{|{\cal M}_{\beta^-_c}|^2}{32m^2_n} = (1 + 3 \lambda^2)\,E_e E +
 (1 - \lambda^2)\,(\vec{k}_e\cdot \vec{k}\,) - 2 \lambda (1 +
\lambda)\,(\vec{\xi}_n\cdot \vec{k}_e)E - 2 \lambda (1 -
\lambda)\,(\vec{\xi}_n\cdot \vec{k}\,)E_e \nonumber\\
\hspace{-0.3in}&& + \frac{1}{M}\,\Big\{\Big[- 3 \lambda^2 E_e E +
  \lambda^2 (\vec{k}_e\cdot \vec{k}\,)+ \lambda(2 \lambda +
  1)\,(\vec{\xi}_n\cdot \vec{k}_e)E - \lambda(2 \lambda -
  1)\,(\vec{\xi}_n\cdot \vec{k}\,)E_e\Big]E_0 + \Big[3 (1 + 3
  \lambda^2) E_e E\nonumber\\
\hspace{-0.3in}&& + 3 (1 - \lambda^2)\, (\vec{k}_e\cdot \vec{k}\,) - 6 \lambda (1 +
\lambda)(\vec{\xi}_n\cdot \vec{k}_e)E - 6 \lambda (1 -
\lambda)\,(\vec{\xi}_n\cdot \vec{k}\,)E_e\Big] E_e +
\Big[- 3 (1 + 3 \lambda^2)(\vec{k}_e\cdot \vec{k}\,) E_e - 3 (1 -
  \lambda^2)\nonumber\\
\hspace{-0.3in}&&\times\, \frac{(\vec{k}_e\cdot \vec{k}\,)^2}{E} + 6
\lambda (1 + \lambda)(\vec{\xi}_n\cdot \vec{k}_e)(\vec{k}_e\cdot
\vec{k}\,) + 6 \lambda (1 - \lambda)\,(\vec{\xi}_n\cdot \vec{k}\,)(\vec{k}_e\cdot
\vec{k}\,)\frac{E_e}{E}\Big] - m^2_e E + \lambda\, m^2_e
(\vec{\xi}_n\cdot \vec{k}\,)\nonumber\\
\hspace{-0.3in}&& - \lambda\,(\vec{\xi}_n\cdot \vec{k}_e +
\vec{\xi}_n\cdot \vec{k}\,)\,(E_e E + \vec{k}_e\cdot \vec{k}\,) +
\lambda^2 \Big[( E_e - \vec{\xi}_n\cdot \vec{k}_e)\, (\vec{k}_e\cdot \vec{k} + E^2)+ (E +
\vec{\xi}_n\cdot \vec{k}\,)(k^2_e + \vec{k}_e\cdot \vec{k}\,)\Big]\nonumber\\
\hspace{-0.3in}&& - (\kappa + 1)\,\Big[(\vec{\xi}_n\cdot
  \vec{k}_e)(\vec{k}_e\cdot \vec{k} + E^2) - (\vec{\xi}_n\cdot
  \vec{k}\,)(k^2_e + \vec{k}_e\cdot \vec{k}\,)\Big] - 2\,(\kappa + 1)
\lambda \Big[E(k^2_e + \vec{k}_e\cdot \vec{k}\,) - E_e (\vec{k}_e\cdot
  \vec{k} + E^2)\Big]\nonumber\\
\hspace{-0.3in}&& + 2(\kappa + 1) \lambda (\vec{\xi}_n\cdot \vec{k}_e
+ \vec{\xi}_n\cdot \vec{k}\,)\,E_e E - (\kappa +
1)\,\lambda\,\Big[(\vec{\xi}_n\cdot
\vec{k}_e)(\vec{k}_e\cdot \vec{k} + E^2)+ (\vec{\xi}_n\cdot
\vec{k}\,)(k^2_e + \vec{k}_e\cdot \vec{k}\,)\Big]\Big\}.
\end{eqnarray}
The function $\Phi_{\beta^-_c}(\vec{k}_e,\vec{k}\,)$ is defined by the
integral over the antineutrino energy $E$
\begin{eqnarray}\label{labelA.19}
\hspace{-0.3in}\Phi_{\beta^-_c}(\vec{k}_e,\vec{k}\,) =
\int^{\infty}_0\delta(f(E))\,\frac{m_n}{E_p}\,\frac{E^2\,dE}{(E_0 -
  E_e)^2},
\end{eqnarray}
where the function $f(E)$ is $f(E) = m_n - E_p - E_e - E$ and $E_p =
\sqrt{m^2_p + (\vec{k}_e + \vec{k}\,)^2}$ is the proton energy after
the integration over the 3--momentum of the proton, giving $\vec{k}_p
= - \vec{k}_e - \vec{k}$. Using the properties of the
$\delta$--function the result of the integration over $E$ is equal to
\begin{eqnarray}\label{labelA.20}
\hspace{-0.3in}\Phi_{\beta^-_c}(\vec{k}_e,\vec{k}\,) =
\frac{m_n}{E_p}\,\frac{E^2}{(E_0 - E_e)^2}\,\frac{1}{\displaystyle
  \Big|\frac{df(E)}{dE}\Big|}\Bigg|_{E = E_r},
\end{eqnarray}
where $E_r$ is the root of the equation $f(E_r) = 0$. To
next--to--leading order in the large $M$ expansion the root $E_r$ is
equal to
\begin{eqnarray}\label{labelA.21}
\hspace{-0.3in}E_r = (E_0 - E_e)\Big(1 + \frac{1}{M}(E_e -
k_e\cos\vartheta_{e\bar{\nu}})\Big),
\end{eqnarray}
where $\cos\vartheta_{e\bar{\nu}} = \vec{k}_e\cdot \vec{k}/k_eE$. Since
\begin{eqnarray}\label{labelA.22}
\hspace{-0.3in}&&\frac{m_n}{E_p}\,\frac{1}{\displaystyle
  \Big|\frac{df(E)}{dE}\Big|}\Bigg|_{E = E_r} = \frac{m_n}{m_n - E_e +
  k_e\cos\vartheta_{e\bar{\nu}}} = 1 + \frac{1}{M}(E_e -
k_e\cos\vartheta_{e\bar{\nu}}),
\end{eqnarray}
where we have kept the terms of order $1/M$ only, using
Eq.(\ref{labelA.21}) and Eq.(\ref{labelA.22}) for the function
$\Phi_{\beta^-_c}(\vec{k}_e,\vec{k}\,)$ we obtain the following
expression
\begin{eqnarray}\label{labelA.23}
\hspace{-0.3in}\Phi_{\beta^-_c}(\vec{k}_e,\vec{k}\,) = 1 +
\frac{3}{M}(E_e - k_e \cos\vartheta_{e\bar{\nu}}) = 1 +
\frac{3}{M}\Big(E_e - \frac{\vec{k}_e\cdot \vec{k}}{E}\Big).
\end{eqnarray}
We have used this expression for the calculation of
Eq.(\ref{labelA.18}), which defines the correlation coefficients of
the continuum-state $\beta^-$--decay of the neutron, calculated to
next--to--leading order in the large $M$ expansion (see
Eq.(\ref{label6})).

\section*{Appendix B: Radiative $\beta^-$--decay of neutron}
\renewcommand{\theequation}{B-\arabic{equation}}
\setcounter{equation}{0}

In this Appendix we calculate the amplitude, the rate, the
photon--electron and photon energy and angular distributions of the
radiative $\beta^-$--decay $n \to p + e^- + \bar{\nu}_e + \gamma$ of
the neutron. The Hamilton operator of weak interactions is defined by
Eq.(\ref{label1}), whereas the Hamilton operator of electromagnetic
interactions is
\begin{eqnarray}\label{labelB.1}
\hspace{-0.3in}{\cal H}_{\rm em}(x) =
e\,[\bar{\psi}_p(x)\gamma^{\mu}\psi_p(x) -
\bar{\psi}_e(x)\gamma^{\mu}\psi_e(x)]\,A_{\mu}(x),
\end{eqnarray}
where $e$ is the electric charge of the proton and $A_{\mu}(x) =
(0,-\vec{A}(x))$ is the electromagnetic vector potential, taken in the
Coulomb gauge ${\rm div}\vec{A}(x) = 0$ \cite{IZ80}.

For the calculation of the amplitude of the radiative $\beta^-$--decay
$n \to p + e^- + \bar{\nu}_e + \gamma$ we take into account the
contributions of the intermediate proton and electron states and drop
the ``weak magnetism'' and proton recoil corrections.  This gives
\begin{eqnarray}\label{labelB.2}
\hspace{-0.3in}M(n\to p\,e^- \bar{\nu}_e\gamma) &=&
e\,\frac{G_F}{\sqrt{2}}\,V_{ud}\,[\bar{u}_p \gamma^{\mu}(1 + \lambda
  \gamma^5) u_n]\,\frac{1}{2 q \cdot k_e}\, [\bar{u}_e (2
  \varepsilon^* \cdot k_e + \hat{\varepsilon}^*\hat{q})\,\gamma_{\mu}
  (1 - \gamma^5) v_{\bar{\nu}}]\nonumber\\
\hspace{-0.3in}&-&
e\,\frac{G_F}{\sqrt{2}}\,V_{ud}\,\frac{1}{2 q \cdot k_p}\,[\bar{u}_p
  (2 \varepsilon^* \cdot k_p +
  \hat{\varepsilon}^*\hat{q})\gamma^{\mu}(1 + \lambda \gamma^5)
  u_n]\, [\bar{u}_e\gamma_{\mu} (1 - \gamma^5)
  v_{\bar{\nu}}],  
\end{eqnarray}
where $\varepsilon = (0,\vec{\varepsilon}\,)$ and $q =
 (\omega,\vec{q}\,)$ are the polarization vector and 4--momentum of
 the photon, obeying the constraint $\varepsilon \cdot q = 0$.

We calculate the amplitude of the radiative $\beta^-$--decay of the
neutron to leading order in the large proton mass (or large $M$)
expansion. This agrees well with 1) the neglect of the contributions
of order $(\alpha/\pi)\,(E_0/M) \sim 10^{-6}$ (see the discussion
below Eq.(\ref{label17})) and 2) the assertion \cite{Cooper,RC08}, that
the electron--photon energy spectrum, the photon polarization
observables and the rate of the radiative $\beta^-$--decay of the
neutron are dominated by the electron emission of photons. In such an
approximation the amplitude of the radiative $\beta^-$--decay takes
the form
\begin{eqnarray}\label{labelB.3}
M(n\to p\,e^- \bar{\nu}_e\gamma)= e\,
\frac{G_F}{\sqrt{2}}\,V_{ud}\,\frac{m_n}{\omega}\,\frac{{\cal
M}_{\beta^-_c\gamma}}{E_e - \vec{n}\cdot \vec{k}_e},
\end{eqnarray}
where $\vec{n} = \vec{q}/\omega$.  The amplitude ${\cal
M}_{\beta^-_c\gamma}$ and its hermitian conjugate are determined by
\begin{eqnarray}\label{labelB.4}
\hspace{-0.3in}{\cal M}_{\beta^-_c\gamma} =
  [\varphi^{\dagger}_p \varphi_n] [\bar{u}_e Q \gamma^0 (1 -
  \gamma^5)v_{\bar{\nu}}] - \lambda
  [\varphi^{\dagger}_p\vec{\sigma}\,\varphi_n]\cdot [\bar{u}_e Q
  \vec{\gamma}\,(1 - \gamma^5)v_{\bar{\nu}}],
\end{eqnarray}
and 
\begin{eqnarray}\label{labelB.5}
\hspace{-0.3in}{\cal M}^{\dagger}_{\beta^-_c\gamma} =
       [\varphi^{\dagger}_n\varphi_p] [\bar{v}_{\bar{\nu}} \gamma^0
         \bar{Q}\, (1 - \gamma^5) u_e] -  \lambda
       [\varphi^{\dagger}_n\vec{\sigma}\,\varphi_p]\cdot
       [\bar{v}_{\bar{\nu}} \vec{\gamma}\, \bar{Q}\, (1 - \gamma^5)
         u_e],
\end{eqnarray}
where $Q = 2(\varepsilon^*\cdot k_e) + \hat{\varepsilon}^*
\hat{q}$ and  $\bar{Q} = \gamma^0 Q^{\dagger} \gamma^0 =
2(\varepsilon\cdot k_e) + \hat{q}\hat{\varepsilon}$.

The squared absolute value of the amplitude Eq.(\ref{labelB.4}),
summed up over the polarisations of the proton and the electron in the
final state accounting for the polarization of the neutron is given by
\begin{eqnarray}\label{labelB.6}
\hspace{-0.3in}&&\sum_{\rm pol.}|{\cal M}_{\beta^-_c\gamma}|^2 = {\rm
  tr}\{(1 + \vec{\xi}_n\cdot \vec{\sigma}\,)\}\,{\rm tr}\{\hat{k}_e Q
\gamma^0 \hat{k}\gamma^0 \bar{Q} (1 - \gamma^5)\} - \lambda {\rm
  tr}\{(1 + \vec{\xi}_n\cdot \vec{\sigma}\,) \vec{\sigma}\,\}\cdot
      {\rm tr}\{\hat{k}_e Q \gamma^0 \hat{k}\vec{\gamma}\, \bar{Q} (1
      - \gamma^5)\}\nonumber\\
\hspace{-0.3in}&&- \lambda {\rm tr}\{(1 + \vec{\xi}_n\cdot
\vec{\sigma}\,) \vec{\sigma}\,\}\cdot {\rm tr}\{\hat{k}_e Q
\vec{\gamma}\, \hat{k}\gamma^0 \bar{Q} (1 - \gamma^5)\} + \lambda^2
    {\rm tr}\{(1 + \vec{\xi}_n\cdot \vec{\sigma}\,) \sigma^m\sigma^n
    \} {\rm tr}\{ \hat{k}_e Q \gamma^n \hat{k}\gamma^m\, \bar{Q} (1 -
    \gamma^5)\}.
\end{eqnarray}
Calculating the traces over the nucleon degrees of freedom and using
the properties of the Dirac matrices 
\begin{eqnarray}\label{labelB.7}
\hspace{-0.3in}\gamma^{\alpha}\gamma^{\nu}\gamma^{\mu} = \gamma^{\alpha}g^{\nu\mu} -
\gamma^{\nu}g^{\mu\alpha} + \gamma^{\mu}g^{\alpha\nu} +
i\,\varepsilon^{\alpha\nu\mu\beta}\,\gamma_{\beta}\gamma^5,
\end{eqnarray}
where $\varepsilon^{\alpha\nu\mu\beta}$ is the Levi--Civita tensor
defined by $\varepsilon^{0123} = 1$ and
$\varepsilon_{\alpha\nu\mu\beta}= - \varepsilon^{\alpha\nu\mu\beta}$
\cite{IZ80}, we transcribe the r.h.s. of Eq.(\ref{labelB.6}) into the
form
\begin{eqnarray}\label{labelB.8}
\hspace{-0.3in}\sum_{\rm pol.}|{\cal M}_{\beta^-_c\gamma}|^2 &=& 2
E\,\Big[\Big((1 + 3 \lambda^2) - 2 \lambda(1 -
  \lambda)\,\frac{\vec{\xi}_n\cdot \vec{k}}{E}\Big)\,{\rm
    tr}\{\hat{k}_e Q \gamma^0 \bar{Q}(1 - \gamma^5)\}\nonumber\\
\hspace{-0.3in}&+& \Big((1 - \lambda^2)\,\frac{\vec{k}}{E} - 2
\lambda (1 + \lambda)\,\vec{\xi}_n\,\Big)\cdot {\rm tr}\{\hat{k}_e Q
\vec{\gamma}\,\bar{Q}(1 - \gamma^5)\}\Big].
\end{eqnarray}
The traces in Eq.(\ref{labelB.8}) are equal to
\begin{eqnarray}\label{labelB.9}
\hspace{-0.3in}&&\frac{1}{16}\,{\rm tr}\{\hat{k}_e Q \gamma^{\mu}
\bar{Q}(1 - \gamma^5)\} = (\varepsilon^*\cdot k_e)(\varepsilon\cdot
k_e)\,(k_e + q)^{\mu} - \frac{1}{2}\,(\varepsilon^*\cdot
\varepsilon)\,\Big((k_e\cdot q)\,q^{\mu} -
\frac{1}{2}\,q^2\,k^{\mu}_e\Big)\nonumber\\
\hspace{-0.3in}&& - \frac{1}{2}\,\Big(\varepsilon^*\cdot
k_e)\,\varepsilon^{\mu} + (\varepsilon \cdot
k_e)\,\varepsilon^{*\mu}\Big)\,\Big(k_e\cdot q + \frac{1}{2}\,q^2\Big)
+
\frac{1}{2}\,i\,\varepsilon^{\mu\alpha\beta\nu}\Big((\varepsilon^*\cdot
k_e)\,\varepsilon_{\alpha} - (\varepsilon \cdot
k_e)\,\varepsilon^*_{\alpha}\Big)\,q_{\beta} k_{e \nu}\nonumber\\
\hspace{-0.3in}&& + \frac{1}{2}\,i\,\Big(q^{\mu}\,q_{\rho} -
\frac{1}{2}\, q^2\,g^{\mu}_{\rho}\Big)\, \varepsilon^{\rho
\alpha\beta\nu}\,\varepsilon^*_{\alpha}\, \varepsilon_{\beta}\,
k_{e\nu},
\end{eqnarray}
where $q^2 = 0$ for a real transverse photon. Summing up over the
photon polarisations we obtain the following photon--electron energy
and angular distribution of the radiative $\beta^-$--decay of the
neutron
\begin{eqnarray}\label{labelB.10}
\hspace{-0.3in}&&\frac{d^8\lambda_{\beta^-_c\gamma}(E_e,\omega,\vec{k}_e,
  \vec{k},\vec{q},\vec{\xi}_n)}{d\omega d E_e
  d\Omega_{\vec{k}_e}d\Omega_{\vec{k}}d\Omega_{\vec{n}}} =
\frac{\alpha}{2\pi}\,(1 + 3
\lambda^2)\,\frac{G^2_F|V_{ud}|^2}{(2\pi)^6}\,\sqrt{E^2_e -
  m^2_e}\,E_e\,F(E_e, Z = 1)\,(E_0 - E_e -
\omega)^2\,\frac{1}{\omega}\nonumber\\
\hspace{-0.3in}&&\times\,\Bigg\{ \Big(1 + B_0\,\frac{\vec
{\xi}_n\cdot \vec{k}}{E}\Big)\Bigg(\frac{k^2_e - (\vec{n}\cdot
\vec{k}_e)^2}{(E_e - \vec{n}\cdot \vec{k}_e)^2}\Big(1 +
\frac{\omega}{E_e}\Big) + \frac{1}{E_e - \vec{n}\cdot
  \vec{k}_e}\,\frac{\omega^2}{E_e}\Bigg)\nonumber\\
\hspace{-0.3in}&& + \Big(a_0\,\frac{\vec{k}}{E} +
A_0\,\vec{\xi}_n\,\Big)\cdot \Bigg[\Bigg(\frac{k^2_e - (\vec{n}\cdot
    \vec{k}_e)^2}{(E_e - \vec{n}\cdot \vec{k}_e)^2} +
  \frac{\omega}{E_e - \vec{n}\cdot
    \vec{k}_e}\Bigg)\,\frac{\vec{k}_e}{E_e} + \Bigg(-
  \frac{m^2_e}{(E_e - \vec{n}\cdot \vec{k}_e)^2} + \frac{E_e +
    \omega}{E_e - \vec{n}\cdot
    \vec{k}_e}\Bigg)\,\frac{\vec{q}}{E_e}\Bigg)\Bigg]\Bigg\}.
\end{eqnarray}
For the unpolarised neutron Eq.(\ref{labelB.10}) coincides with the
spectrum, adduced in \cite{Cooper} (see Eq.(\ref{label3}) of
Ref.\cite{Cooper} and a comment in \cite{Cooper1}).

After the integration over the directions of the photon momentum the
photon--electron energy and angular distribution takes the form
\begin{eqnarray}\label{labelB.11}
\hspace{-0.3in}\frac{d^6\lambda_{\beta^-_c\gamma}(E_e,\omega,\vec{k}_e,
  \vec{k},\vec{q}, \vec{\xi}_n)}{d\omega d E_e
  d\Omega_{\vec{k}_e}d\Omega_{\vec{k}}} &=& \frac{\alpha}{\pi}\,(1 + 3
\lambda^2)\,\frac{G^2_F|V_{ud}|^2}{(2\pi)^5}\,\sqrt{E^2_e -
  m^2_e}\,E_e\,F(E_e, Z = 1)\,(E_0 - E_e -
\omega)^2\,\frac{1}{\omega}\nonumber\\
\hspace{-0.3in}&&\times\,\Bigg\{ \Big(1 + B_0\,\frac{\vec
  {\xi}_n\cdot \vec{k}}{E}\Big)\,\Big\{\Big(1 + \frac{\omega}{E_e} +
  \frac{1}{2}\,\frac{\omega^2}{E^2_e}\Big)\,\Big[\frac{1}{\beta}\,{\ell
  n}\Big(\frac{1 + \beta}{1 - \beta}\Big) - 2\Big] +
  \frac{\omega^2}{E^2_e}\Big\}\nonumber\\
\hspace{-0.3in}&& + \Big(a_0\,\frac{\vec{k}\cdot \vec{k}_e}{E E_e} +
A_0\,\frac{\vec{\xi}_n\cdot \vec{k}_e}{E_e}\,\Big)\,\Big[1 +
  \frac{1}{\beta^2}\,\frac{\omega}{E_e}\Big(1 +
  \frac{1}{2}\,\frac{\omega}{E_e}\Big)\Big]\,
  \Big[\frac{1}{\beta}\,{\ell n}\Big(\frac{1 + \beta}{1 - \beta}\Big)
  - 2\Big]\Bigg\}.
\end{eqnarray}
Integrating over the phase volume of the final state we obtain the
rate of the radiative $\beta^-$--decay of the neutron
\begin{eqnarray}\label{labelB.12}
\hspace{-0.3in}&&\lambda_{\beta^-_c\gamma}(\omega_{\rm min}) =
\frac{\alpha}{\pi}\,(1 + 3 \lambda^2)\,\frac{G^2_F|V_{ud}|^2}{2\pi^3}\int^{E_0 - \omega_{\rm min}}_{m_e}dE_e
\,\sqrt{E^2_e - m^2_e}\,E_e\,F(E_e, Z = 1)\nonumber\\
\hspace{-0.3in}&&\times\int^{E_0 - E_e}_{\omega_{\rm
    min}}\frac{d\omega}{\omega}\,(E_0 - E_e - \omega)^2\Big\{\Big(1 +
\frac{\omega}{E_e} +
\frac{1}{2}\frac{\omega^2}{E^2_e}\Big)\,\Big[\frac{1}{\beta}\,{\ell
    n}\Big(\frac{1 + \beta}{1 - \beta}\Big) -2\Big] +
\frac{\omega^2}{E^2_e}\Big\}.
\end{eqnarray}
The lowest photon energy $\omega_{\rm min}$ may be treated as the
photon energy threshold of the detector.

For the photon energy interval $\omega_{\rm min} \le \omega \le
\omega_{\rm max}$ the rate of the radiative $\beta^-$--decay of the neutron reads
\begin{eqnarray}\label{labelB.13}
\hspace{-0.3in}\lambda_{\beta^-_c\gamma}(\omega_{\rm
  max},\omega_{\rm min}) &=& \frac{\alpha}{\pi}\,(1 + 3
\lambda^2)\,\frac{G^2_F|V_{ud}|^2}{2\pi^3}\int^{\omega_{\rm max}}_{\omega_{\rm
min}}\frac{d\omega}{\omega}\int^{E_0 - \omega}_{m_e}dE_e \,\sqrt{E^2_e
- m^2_e}\,E_e\,F(E_e, Z = 1)\,(E_0 - E_e -
\omega)^2\nonumber\\
\hspace{-0.3in}&&\times\Big\{\Big(1 + \frac{\omega}{E_e} +
\frac{1}{2}\frac{\omega^2}{E^2_e}\Big)\,\Big[\frac{1}{\beta}\,{\ell
    n}\Big(\frac{1 + \beta}{1 - \beta}\Big) - 2\Big] +
\frac{\omega^2}{E^2_e}\Big\}.
\end{eqnarray}
For the region of photon energies $15\,{\rm keV} \le \omega \le
340\,{\rm keV}$, used in the experiments by \cite{Nico2,Cooper} , the
branching ratio is equal to ${\rm BR}_{\beta^-_c\gamma} = 2.87\times
10^{-3}$.  This result agrees well with the experimental values ${\rm
  BR}_{\beta^-_c\gamma} = 3.13(35)\times 10^{-3}$ and ${\rm
  BR}_{\beta^-_c\gamma} = 3.09(32)\times 10^{-3}$ and the result ${\rm
  BR}_{\beta^-_c\gamma} = 2.85\times 10^{-3}$, calculated by Gardner
\cite{Nico2,Cooper}.

For the comparison with the analysis of the radiative
$\beta^-$--decay, carried out by Bernard {\it et al.} \cite{RBD2}, we
calculate the branching ratio for the photon energy region $5\,{\rm
  keV} \le \omega \le E_0 - m_e$. The result ${\rm
  BR}_{\beta^-_c\gamma} = 4.45\times 10^{-3}$ is in a agreement with
${\rm BR}_{\beta^-_c\gamma} = 4.41\times 10^{-3}$, calculated in
\cite{RBD2}.

 For the electron--energy and angular distribution of the radiative
 $\beta^-$--decay of the neutron we obtain the following expression
\begin{eqnarray}\label{labelB.14}
\hspace{-0.3in}&&\frac{d^5\lambda_{\beta^-_c\gamma}(E_e,\vec{k}_e,
  \vec{k}, \vec{\xi}_n, \omega_{\rm min})}{d E_e
  d\Omega_{\vec{k}_e}d\Omega_{\vec{k}}} = \frac{\alpha}{\pi}\,(1 + 3
\lambda^2)\,\frac{G^2_F|V_{ud}|^2}{(2\pi)^5}\,(E_0 -
E_e)^2\,\sqrt{E^2_e - m^2_e}\,E_e\,F(E_e, Z = 1)\nonumber\\
\hspace{-0.3in}&&\times\,\Big\{\Big(1 + B_0\,\frac{\vec {\xi}_n\cdot
  \vec{k}}{E}\Big)\,g^{(1)}_{\beta^-_c\gamma}(E_e,\omega_{\rm min}) +
\Big(a_0\,\frac{\vec{k}\cdot \vec{k}_e}{E E_e} +
A_0\,\frac{\vec{\xi}_n\cdot
  \vec{k}_e}{E_e}\,\Big)\,g^{(2)}_{\beta^-_c\gamma}(E_e,\omega_{\rm
  min})\Big\},
\end{eqnarray}
where the functions $g^{(1)}_{\beta^-_c\gamma}(E_e,\omega_{\rm min})$
and $g^{(2)}_{\beta^-_c\gamma}(E_e,\omega_{\rm min})$ are defined by
\begin{eqnarray*}
\hspace{-0.3in}&&g^{(1)}_{\beta^-_c\gamma}(E_e,\omega_{\rm
min}) = \int^{E_0 - E_e}_{\omega_{\rm
    min}}\frac{d\omega}{\omega}\frac{(E_0 - E_e - \omega)^2}{(E_0 -
  E_e)^2}\,\Big\{\Big(1 + \frac{\omega}{E_e} +
\frac{1}{2}\frac{\omega^2}{E^2_e}\Big) \,\Big[\frac{1}{\beta}\,{\ell n}\Big(\frac{1 +
    \beta}{1 - \beta}\Big) - 2\Big] + \frac{\omega^2}{E^2_e}\Big\} =
\end{eqnarray*}
\begin{eqnarray}\label{labelB.15}
\hspace{-0.3in}&&= \Big[2{\ell n}\Big(\frac{E_0 - E_e}{\omega_{\rm
      min}}\Big) - 3 + \frac{2}{3}\,\frac{E_0 - E_e}{E_e}\Big(1 +
  \frac{1}{8}\,\frac{E_0 - E_e}{E_e}
  \Big)\Big]\Big[\frac{1}{2\beta}\,{\ell n}\Big(\frac{1 + \beta}{1 -
    \beta}\Big) - 1\Big] + \frac{1}{12}\,\frac{(E_0 -
  E_e)^2}{E^2_e},\nonumber\\
\hspace{-0.3in}&&g^{(2)}_{\beta^-_c\gamma}(E_e,\omega_{\rm min}) =
\int^{E_0 - E_e}_{\omega_{\rm min}}\frac{d\omega}{\omega}\,\frac{(E_0
- E_e - \omega)^2}{(E_0 - E_e)^2}\,\Big[1 +
\frac{1}{\beta^2}\frac{\omega}{E_e}\Big(1 +
\frac{1}{2}\frac{\omega}{E_e}\Big)\Big]\,\Big[\frac{1}{\beta}\,{\ell
n}\Big(\frac{1 + \beta}{1 - \beta}\Big) - 2\Big] = \nonumber\\
\hspace{-0.3in}&&= \Big[2 {\ell n}\Big(\frac{E_0 - E_e}{\omega_{\rm
min}}\Big) - 3 + \frac{2}{3}\,\frac{E_0 - E_e}{\beta^2 E_e}\Big(1 +
\frac{1}{8}\,\frac{E_0 - E_e}{E_e}\Big) \Big]\,\Big[\frac{1}{2\beta}\,{\ell n}\Big(\frac{1 +
    \beta}{1 - \beta}\Big) - 1\Big].
\end{eqnarray}
We would like to note that as we show below the functions
$g^{(1)}_{\beta^-_c\gamma}(E_e,\omega_{\rm min})$ and
$g^{(2)}_{\beta^-_c\gamma}(E_e,\omega_{\rm min})$, calculated by means
of the infrared cut--off regularization $\omega_{\rm min}$, have
energy dependencies different in comparison with the functions,
calculated in the FPM regularization (see also  \cite{RC2}).

The discrepancy between the electron--energy and angular
distributions, obtained for a real photon emission with the infrared
cut--off and FPM regularization, respectively, we discuss first in
terms of the logarithmically divergent integral, which defines the
infrared divergent contribution to the amplitude of the radiative
$\beta^-$--decay of the neutron. This integral is
\begin{eqnarray}\label{labelB.16}
J(\beta) = \int \frac{d^3q}{4\pi q_0}\,\frac{\beta^2 - (\vec{v}\cdot
\vec{\beta}\,)^2}{(q_0 - \vec{q}\cdot \vec{\beta}\,)^2},
\end{eqnarray}
where $q_0 = \sqrt{\vec{q}^{\;2} + \mu^2}$ is an energy of a photon
with mass $\mu$, $\vec{\beta} = \vec{k}_e/E_e$ and $\vec{v} =
\vec{q}/q_0$ are the velocities of the electron and massive photon,
respectively. The region of the integration in Eq.(\ref{labelB.16}) is
restricted by $q_1 \le q \le q_2$, where $q = |\vec{q}\,|$.

For the first time the integral Eq.(\ref{labelB.16}) has been
discussed by Kinoshita and Sirlin in \cite{RC5}.  It is obvious that
replacing the lower and upper limits of the integral by $ \omega_{\rm
  min}$ and $E_0 - E_e$, respectively, and setting $\mu$ zero in the
integrand the function $J(\beta)$, given by
\begin{eqnarray}\label{labelB.17}
J(\beta) = 2{\ell n}\Big(\frac{E_0 - E_e}{\omega_{\rm
min}}\Big)\,\Big[\frac{1}{2\beta}\,{\ell n}\Big(\frac{1 + \beta}{1 -
\beta}\Big) - 1\Big],
\end{eqnarray}
defines the infrared divergent part of the functions
$g^{(1)}_{\beta^-_c\gamma}(E_e,\omega_{\rm min})$ and
$g^{(2)}_{\beta^-_c\gamma}(E_e,\omega_{\rm min})$ in
Eq.(\ref{labelB.15}).

Now let us calculate the integral Eq.(\ref{labelB.16}) for $\mu \neq
0$. Following Kinoshita and Sirlin \cite{RC5}, we may rewrite the
integral Eq.(\ref{labelB.16}) as follows
\begin{eqnarray}\label{labelB.18}
\hspace{-0.3in}J(\beta; q_2, q_1) = \frac{\beta^2}{2}\int^{+1}_{-1}dx
\int^{q_2}_{q_1}\frac{dq q^2}{q^3_0}\,\frac{1 - v^2 x^2}{(1 - \beta v
x)^2}.
\end{eqnarray}
Making a change of variables $q \to v$, proposed by Kinoshita and
Sirlin \cite{RC5}, we arrive at the integral
\begin{eqnarray}\label{labelB.19}
\hspace{-0.3in}J(\beta; q_2, q_1) = \frac{\beta^2}{2}\int^{+1}_{-1}dx 
\int^{v_2}_{v_1}\frac{dv
v^2}{1 - v^2}\,\frac{1 - v^2x^2}{(1 - \beta v x)^2}.
\end{eqnarray}
Integrating over $x$ 
\begin{eqnarray}\label{labelB.20}
J(\beta; q_2, q_1) = \int^{v_2}_{v_1}\Big[1 + \frac{1}{1 - \beta^2
v^2} - \frac{2}{1 - v^2} + \frac{1}{\beta}\,\frac{v}{1 -
v^2}\,{\ell n}\Big(\frac{1 + \beta v}{1 - \beta v}\Big)\Big]\,dv
\end{eqnarray}
and over $v$ we obtain 
\begin{eqnarray}\label{labelB.21}
\hspace{-0.3in}&&J(\beta; q_2, q_1) = {\ell n}\Big[\frac{(1 - v_1)}{(1 -
v_2)}\frac{(1 + v_2)}{(1 + v_1)}\Big]\Big[\frac{1}{2 \beta}{\ell
n}\Big(\frac{1 + \beta}{1 - \beta}\Big) - 1\Big] + (v_2 - v_1) + \frac{1}{2\beta}\,{\ell
n}\Big[\frac{(1 + \beta v_2)}{(1 - \beta v_2)}\frac{(1 - \beta
v_1)}{(1 + \beta v_1)}\Big]\nonumber\\
\hspace{-0.3in}&& - \frac{1}{2\beta}\,L\Big(\frac{\beta}{1 + \beta}(1
- v_2)\Big) + \frac{1}{2\beta}\,L\Big(\frac{\beta}{1 + \beta}(1 -
v_1)\Big) + \frac{1}{2\beta}\,L\Big(- \frac{\beta}{1 - \beta}(1 -
v_2)\Big) - \frac{1}{2\beta}\,L\Big(- \frac{\beta}{1 - \beta}(1 -
v_1)\Big)\nonumber\\
\hspace{-0.3in}&& + \frac{1}{2\beta}\,L\Big(\frac{\beta}{1 + \beta}(1
+ v_2)\Big) -  \frac{1}{2\beta}\,L\Big(\frac{\beta}{1 +
  \beta}(1 + v_1)\Big) - \frac{1}{2\beta}\,L\Big(- \frac{\beta}{1 -
\beta}(1 + v_2)\Big) + \frac{1}{2\beta}\,L\Big(-
\frac{\beta}{1 - \beta}(1 + v_1)\Big),
\end{eqnarray}
where $L(x)$ is the Spence function, defined by \cite{RC5,HMF72}
\begin{eqnarray}\label{labelB.22}
L(x) = \int^x_0\frac{dt}{t}\,{\ell n}|1 - t|.
\end{eqnarray}
For $q_1 \gg \mu$ the function $J(\beta)$ reduces to the form
\begin{eqnarray}\label{labelB.23}
\hspace{-0.3in}J(\beta; q_2, q_1) = 2{\ell
  n}\Big(\frac{q_2}{q_1}\Big)\Big[\frac{1}{2\beta}{\ell
    n}\Big(\frac{1+ \beta}{1 - \beta}\Big) - 1\Big].
\end{eqnarray}
One may see that non--trivial finite and energy--dependent terms
appear in the function $J(\beta)$, given by Eq.(\ref{labelB.21}), for
$q_1 = 0$ and the integration over the region $0 \le q \le q_2$. To
show this we set $q_1 = 0$ and $q_2 = q_{\rm max}$ in
Eq.(\ref{labelB.18}) and arrive at the expression
\begin{eqnarray}\label{labelB.24}
\hspace{-0.3in}J(\beta; q_{\rm max}, 0) =  2{\ell n}\Big(\frac{2 q_{\rm
max}}{\mu}\Big)\Big[\frac{1}{2\beta}{\ell n}\Big(\frac{1 +
\beta}{1 - \beta}\Big) - 1\Big] + 1 +
\frac{1}{2\beta}\,{\ell n}\Big(\frac{1 + \beta}{1 -
\beta}\Big) + \frac{1}{2\beta}\,\Big[L\Big(\frac{2 \beta}{1 +
\beta} \Big) - L\Big(- \frac{2 \beta}{1 - \beta}\Big)\Big].
\end{eqnarray}
Due to the relation between Spence's functions \cite{RC9,RC11,HMF72}
\begin{eqnarray}\label{labelB.25}
\hspace{-0.3in}L\Big(\frac{2 \beta}{1 + \beta} \Big) - L\Big(-
\frac{2 \beta}{1 - \beta}\Big) = 2 L\Big(\frac{2 \beta}{1 +
\beta} \Big) - \frac{1}{2}\,{\ell n}^2\Big(\frac{1 + \beta}{1 -
\beta}\Big)
\end{eqnarray}
we derive the following expression for $J(\beta)$
\begin{eqnarray}\label{labelB.26}
\hspace{-0.3in}J(\beta; q_2, q_1) = 2{\ell n}\Big(\frac{2 q_{\rm
    max}}{\mu}\Big)\Big[\frac{1}{2\beta}{\ell n}\Big(\frac{1 +
    \beta}{1 - \beta}\Big) - 1\Big] + 1 + \frac{1}{2\beta}\,{\ell
  n}\Big(\frac{1 + \beta}{1 - \beta}\Big) - \frac{1}{4\beta}\,{\ell
  n}^2\Big(\frac{1 + \beta}{1 - \beta}\Big) +
\frac{1}{\beta}\,L\Big(\frac{2 \beta}{1 + \beta} \Big).
\end{eqnarray}
In turn, using the relation \cite{RC9,RC11,HMF72}
\begin{eqnarray}\label{labelB.27}
\hspace{-0.3in} L\Big(\frac{2 \beta}{1 + \beta} \Big) = L(\beta) -
L(-\beta) + \frac{1}{2}\Big[L\Big(\frac{1 - \beta}{2}\Big) -
  L\Big(\frac{1 + \beta}{2}\Big)\Big] + \frac{1}{2}\,{\ell
  n}\Big(\frac{1 + \beta}{2}\Big){\ell n}\Big(\frac{1 + \beta}{1 -
  \beta}\Big)
\end{eqnarray}
we reduce Eq.(\ref{labelB.26}) to the form, obtained by Kinoshita and
Sirlin (see Eq.(\ref{labelC.4}) of \cite{RC5}).

As a result, the functions $g^{(1)}_{\beta^-_c\gamma}(E_e,\mu)$ and
$g^{(2)}_{\beta^-_c\gamma}(E_e,\mu)$, having the form
\begin{eqnarray}\label{labelB.28}
\hspace{-0.3in}&&g^{(1)}_{\beta^-_c\gamma}(E_e,\mu) = \Big[2{\ell
n}\Big(\frac{2(E_0 - E_e)}{\mu}\Big) - 3 + \frac{2}{3}\,\frac{E_0 -
E_e}{E_e}\, \Big(1 + \frac{1}{8} \frac{E_0 - E_e}{E_e}
\Big)\Big]\Big[\frac{1}{2\beta}\,{\ell n}\Big(\frac{1 + \beta}{1 -
    \beta}\Big) - 1\Big] + 1 \nonumber\\
\hspace{-0.3in}&& + \frac{1}{12} \frac{(E_0 - E_e)^2}{E^2_e}+
\frac{1}{2\beta}\,{\ell n}\Big(\frac{1 + \beta}{1 - \beta}\Big) -
\frac{1}{4\beta}\,{\ell n}^2\Big(\frac{1 + \beta}{1 -
  \beta}\Big) + \frac{1}{\beta}\,L\Big(\frac{2 \beta}{1 +
\beta} \Big),\nonumber\\
\hspace{-0.3in}&&g^{(2)}_{\beta^-_c\gamma}(E_e,\mu) = \Big[2 {\ell
    n}\Big(\frac{2(E_0 - E_e)}{\mu}\Big) - 3 + \frac{2}{3}\,\frac{E_0
    - E_e}{\beta^2 E_e}\, \Big(1 + \frac{1}{8}\,\frac{E_0 -
  E_e}{E_e}\Big) \Big]\Big[\frac{1}{2\beta}\,{\ell n}\Big(\frac{1 +
    \beta}{1 - \beta}\Big) - 1\Big] + 1 \nonumber\\
\hspace{-0.3in}&&+ \frac{1}{2\beta}\,{\ell n}\Big(\frac{1 +
\beta}{1 - \beta}\Big) - \frac{1}{4\beta}\,{\ell n}^2\Big(\frac{1 +
\beta}{1 - \beta}\Big) + \frac{1}{\beta}\,L\Big(\frac{2 \beta}{1 +
\beta} \Big),
\end{eqnarray}
agree fully with the functions, obtained in \cite{RC2} (see Eqs.(12)
and (13) of Ref.\cite{RC2}). We note that the same result, given by
Eq.(\ref{labelB.26}), for the logarithmically divergent integral
Eq.(\ref{labelB.16}) may be obtained within the dimensional
regularization \cite{RC11}. We will use the functions
Eq.(\ref{labelB.28}) for the calculation of the radiative corrections
to the rate and the correlation coefficients of the $\beta^-$--decay
of the neutron (see Appendix D).

\section*{Appendix C: Analysis of infrared divergences 
of radiative corrections to continuum-state $\beta^-$--decay of
neutron} \renewcommand{\theequation}{C-\arabic{equation}}
\setcounter{equation}{0}

The radiative corrections, caused by one--virtual photon exchanges,
are described by three irreducible diagrams. They are shown in
Fig.\,3. The diagrams in Figs.\,3a and 3b define the self--energy
corrections to the masses and wave functions of the proton and
electron \cite{RC3}--\cite{RC8}, respectively.
\begin{figure}
\centering
\includegraphics[height=0.30\textheight]{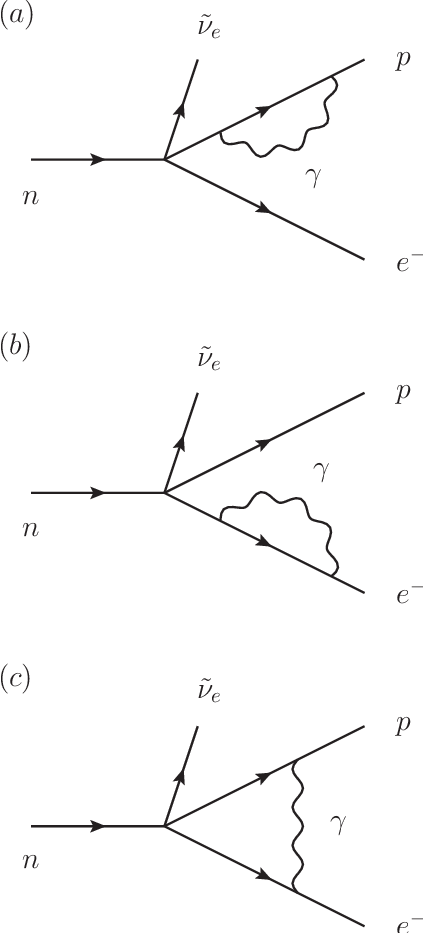}
\caption{Feynman diagrams of radiative corrections to the
  continuum-state $\beta^-$--decay of the neutron, caused by
  one--virtual photon exchanges.}
\end{figure}
As has been shown in \cite{RC3}--\cite{RC8}, the self--energy
corrections to the masses my be removed by the mass renormalisation,
whereas the contributions of the self--energy diagrams to the wave
functions cannot be removed fully by renormalisation of the wave
functions of the proton and electron and give some observable terms.
We calculate the contributions of the self--energy diagrams in
Appendix D. In this Appendix we analyse the contribution of the vertex
diagram in Fig.\,3c.

The contribution of the vertex diagram to the amplitude of the
continuum-state $\beta^-$--decay of the neutron may be written as (see
Appendix D)
\begin{eqnarray}\label{labelC.1}
M^{(\gamma)}(n \to p e^-\bar{\nu}_e) =
e^2\,\frac{G_F}{\sqrt{2}}\,V_{ud}\,{\cal M}^{(\gamma)}.
\end{eqnarray}
The amplitude ${\cal M}^{(\gamma)}$ is defined by
\begin{eqnarray}\label{labelC.2}
\hspace{-0.3in}&&{\cal M}^{(\gamma)} = 4 (k_e\cdot
k_p)\,[\bar{u}_p\gamma^{\mu}(1 + \lambda
  \gamma^5)u_n][\bar{u}_e\gamma_{\mu}(1 - \gamma^5)v_{\bar{\nu}}]\int
\frac{d^4q}{(2\pi)^4i} \frac{1}{[q^2 + i0][q^2 - 2 k_e\cdot q +
    i0][q^2 + 2 k_p\cdot q + i0]} + \ldots,\nonumber\\
\hspace{-0.3in}&&
\end{eqnarray}
where we have kept only the term, which suffers from the infrared
divergences.  The detailed calculation of the vertex diagram in
Fig.\,3c is given in Appendix D.

For the calculation of the integral over the 4--momentum $q$ we follow
the standard procedure, using Feynman's parametrization of the
integrals \cite{RC3,RC8} (see also \cite{Rohrlich}). We perform the
calculation of the 4--momentum integral in Eq.(\ref{labelC.2}) by
using the finite-photon mass regularization \cite{RC3}--\cite{RC8}.

\subsection{finite-photon mass regularization}

Below we calculate the integral in Eq.(\ref{labelC.2}) by using the
finite-photon mass (FPM) regularization of the infrared
divergences. Applying Feynman's parametrization \cite{Rohrlich} we
transform the integral in Eq.(\ref{labelC.2}) into the form
\begin{eqnarray}\label{labelC.3}
\hspace{-0.3in}J(E_e; \mu) = \frac{1}{16\pi^2}\int^1_0 dx\int^1_0
dy\,2 y\int \frac{d^4q}{\pi^2 i}\,\frac{1}{[(q - p(x) y)^2 - p^2(x) y^2 - \mu^2(1 -
    y) + i0]^3},
\end{eqnarray}
where $p(x) = k_ex - k_p(1 - x)$ and $\mu$ is an infinitesimal photon
mass, introduced for the infrared regularization of the 4--momentum
integrals \cite{RC8,Rohrlich}. For the derivation of
Eq.(\ref{labelC.3}) we have used the relation \cite{Rohrlich}
\begin{eqnarray}\label{labelC.4}
\hspace{-0.3in}\frac{1}{A_e A_p} =  \int^1_0\frac{dx}{[A_e x +
A_p(1-x)]^2}\quad,\quad
\frac{1}{A^2 B} = \int^1_0 \frac{2 y dy}{[Ay + B(1 -
y)]^3},
\end{eqnarray}
where $A_e = Q^2 - 2 k_e\cdot Q + i0$, $A_p = Q^2 + 2k_p\cdot Q + i0$,
$A = Q^2 - 2 p(x) \cdot Q + i0$ and $B = Q^2 - \mu^2 + i0$,
respectively.  Making a shift of variables $ q - p(x) y \to q$, a
Wick rotation $q_0 \to iq_4$ \cite{Rohrlich} and integrating over $y$
we arrive at the expression \cite{RC8,RC9}
\begin{eqnarray}\label{labelC.5}
\hspace{-0.3in}J(E_e; \mu) = -
\frac{1}{32\pi^2}\int^1_0\frac{dx}{p^2(x)}\,{\ell
n}\Big[\frac{p^2(x)}{\mu^2}\Big],
\end{eqnarray}
where $p^2(x) = m^2_e x^2 + m^2_p (1-x)^2 - 2 m_e m_p \gamma x(1 - x)$
with $\gamma = 1/\sqrt{1 - \beta^2}$. The r.h.s. of
Eq.(\ref{labelC.5}) may be represented in the form
\begin{eqnarray}\label{labelC.6}
\hspace{-0.3in}J(E_e; \mu) = - \frac{1}{32\pi^2}\frac{1}{m_e m_p
  c}\int^1_0\frac{dx}{(a - x)^2 - b^2}\,{\ell n}\Big[\frac{m_em_p
    c}{\mu^2}\Big((a - x)^2 - b^2\Big)\Big],
\end{eqnarray}
where we have denoted
\begin{eqnarray}\label{labelC.7}
\hspace{-0.3in} a = \frac{\rho + \gamma}{c}\;,\, b =
\frac{\sqrt{\gamma^2 - 1}}{c}\,,\,c = \frac{1}{\rho} + \rho + 2
\gamma
\end{eqnarray}
with $\rho = m_p/m_e$. Following then \cite{RC8,RC9}, we make a
change of variables $a - x = b\,\coth\varphi$. This gives
\begin{eqnarray}\label{labelC.8}
\hspace{-0.3in}&&J(E_e; \mu) = - \frac{1}{32\pi^2}\frac{1}{m_e m_p b
c}\int^{\varphi_2}_{\varphi_1}d\varphi\,{\ell n}\Big[\frac{m_em_p
c}{\mu^2}\,\frac{b^2}{\sinh^2\varphi}\Big] =\nonumber\\
\hspace{-0.3in}&&= - \frac{1}{32\pi^2}\frac{1}{m_e m_p b c}\Big\{{\ell
  n}\Big[\frac{4 m_em_p c}{\mu^2}\,b^2\Big]\,(\varphi_2 - \varphi_1) -
(\varphi^2_2 - \varphi^2_1) -
2\int^{\varphi_2}_{\varphi_1}d\varphi\,{\ell n}\Big(1 - e^{\, -
  2\varphi}\Big)\Big\},
\end{eqnarray}
where $\varphi_2$ and $\varphi_1$ are equal to
\begin{eqnarray}\label{labelC.9}
\hspace{-0.3in}&&\varphi_2 = \frac{1}{2}\,{\ell n}\Big(\frac{1 - a -
  b}{1 - a + b}\Big) = \frac{1}{2}\,{\ell n}\Bigg(\frac{\displaystyle
  \frac{1}{\rho} + \gamma - \sqrt{\gamma^2 - 1}}{\displaystyle
  \frac{1}{\rho} + \gamma + \sqrt{\gamma^2 - 1}}\Bigg),\nonumber\\
\hspace{-0.3in}&&\varphi_1 = \frac{1}{2}\,{\ell n}\Big(\frac{a + b}{a
- b}\Big) = \frac{1}{2}\,{\ell n}\Big(\frac{\rho + \gamma +
\sqrt{\gamma^2 - 1}}{\rho + \gamma - \sqrt{\gamma^2 - 1}}\Big).
\end{eqnarray}
For the calculation of the last integral in Eq.(\ref{labelC.8}) we
make a change of variables $\varphi = - \frac{1}{2}\,{\ell n}t$. This
gives
\begin{eqnarray}\label{labelC.10}
\hspace{-0.3in}J(\beta; \mu) = - \frac{1}{32\pi^2}\frac{1}{m_e m_p
b c}\Big\{{\ell n}\Big[\frac{4 m_em_p c}{\mu^2}\,b^2\Big]\,(\varphi_2
- \varphi_1) - (\varphi^2_2 - \varphi^2_1) + L(e^{\, -2
\varphi_2}) - L(e^{\, -2 \varphi_1})\Big\},
\end{eqnarray}
where the last two terms are the Spence functions, defined by
Eq.(\ref{labelB.22}).

Keeping the leading order contributions in the large $\rho$ expansion
we arrive at the expression
\begin{eqnarray}\label{labelC.11}
\hspace{-0.3in}J(E_e; \mu) &=& - \frac{1}{32\pi^2}\frac{1}{m_e m_p
  \sqrt{\gamma^2 - 1}}\Big\{- {\ell n}\Big[\frac{4
    m^2_e}{\mu^2}\,(\gamma^2 - 1)\Big]\,\frac{1}{2}\,{\ell n}\Big(\frac{\gamma +
\sqrt{\gamma^2 - 1}}{\gamma - \sqrt{\gamma^2 - 1}}\Big) -
\frac{1}{4}\,{\ell n}^2\Big(\frac{\gamma + \sqrt{\gamma^2 - 1}}{\gamma
- \sqrt{\gamma^2 - 1}}\Big)\nonumber\\
\hspace{-0.3in}&+& L\Big(\frac{\gamma + \sqrt{\gamma^2 -
1}}{\gamma - \sqrt{\gamma^2 - 1}}\Big) - L(1)\Big\}.
\end{eqnarray}
In terms of the electron velocity $\beta$ it reads
\begin{eqnarray}\label{labelC.12}
\hspace{-0.3in}J(\beta; \mu) = \frac{1}{32\pi^2}\frac{1}{E_e m_p
\beta}\Big\{ {\ell n}\Big[\frac{4 m^2_e}{\mu^2}\,\frac{\beta^2}{1 -
\beta^2}\Big]\,\frac{1}{2}\,{\ell n}\Big(\frac{1 + \beta}{1 -
\beta}\Big) + \frac{1}{4}\,{\ell n}^2\Big(\frac{1 + \beta}{1 -
  \beta}\Big) - L\Big(\frac{1 + \beta}{1 - \beta}\Big) + L(1)\Big\},
\end{eqnarray}
where $E_e = m_e/\sqrt{1 - \beta^2}$. For further transformation of
the r.h.s. of Eq.(\ref{labelC.12}) we use the following relation for
the Spence functions \cite{HMF72}
\begin{eqnarray}\label{labelC.13}
\hspace{-0.3in}L\Big(\frac{1 + \beta}{1 - \beta}\Big) - L(1) &=&
L\Big(\frac{2\beta}{1 + \beta}\Big) + {\ell n}\Big(\frac{2\beta}{1 -
  \beta}\Big) {\ell n}\Big(\frac{1 + \beta}{1 - \beta}\Big) -
\frac{1}{2}\,{\ell n}^2\Big(\frac{1 + \beta}{1 - \beta}\Big).
\end{eqnarray}
Substituting Eq.(\ref{labelC.13}) into Eq.(\ref{labelC.12}) we obtain
\begin{eqnarray}\label{labelC.14}
\hspace{-0.3in}J(\beta; \mu) = \frac{1}{32\pi^2}\frac{1}{E_e m_p \beta}\Big\{
{\ell n}\Big(\frac{ m_e}{\mu}\Big)\,{\ell n}\Big(\frac{1 + \beta}{1 -
\beta}\Big) + \frac{1}{4}\,{\ell n}^2\Big(\frac{1 + \beta}{1 -
\beta}\Big)- L\Big(\frac{2\beta}{1 + \beta}\Big)\Big\}.
\end{eqnarray}
Thus, the integral under consideration is equal to
\begin{eqnarray}\label{labelC.15}
\hspace{-0.3in}&&J(E_e; \mu) =  \int \frac{d^4q}{(2\pi)^4i}\frac{1}{[q^2 +
i0][q^2 - 2 k_e\cdot q + i0][q^2 + 2 k_p\cdot q + i0]}\nonumber\\
\hspace{-0.3in}&& = \frac{1}{32\pi^2}\frac{1}{E_e m_p \beta}\Big\{
       {\ell n}\Big(\frac{ m_e}{\mu}\Big)\,{\ell n}\Big(\frac{1 +
         \beta}{1 - \beta}\Big) + \frac{1}{4}\,{\ell n}^2\Big(\frac{1
         + \beta}{1 - \beta}\Big) - L\Big(\frac{2\beta}{1 + \beta}\Big)\Big\}.
\end{eqnarray}
This result agrees with the expression, obtained in \cite{RC8} and
\cite{RC9} (see Eq.(B21) of Ref.\cite{RC9}).

\section*{Appendix D: Total contribution of one--virtual photon exchanges 
to continuum-state $\beta^-$--decay of neutron}
\renewcommand{\theequation}{D-\arabic{equation}}
\setcounter{equation}{0}

The contributions of one--virtual photon exchanges to the
continuum-state $\beta^-$--decay of the neutron are shown in
Fig.\,3. The correction to the amplitude of the continuum-state
$\beta^-$--decay of the neutron, caused by one--virtual photon
exchanges, we represent in the following form
\begin{eqnarray}\label{labelD.1}
\hspace{-0.3in}M^{(\gamma)}(n\to p e^-\bar{\nu}_e) =
-\,\frac{G_F}{\sqrt{2}}V_{ud}\Big({\cal M}^{(\gamma)}_{pp} + {\cal
  M}^{(\gamma)}_{ee} + {\cal M}^{(\gamma)}_{pe}\Big),
\end{eqnarray}
where the amplitudes ${\cal M}^{(\gamma)}_{pp}$ and $ {\cal
  M}^{(\gamma)}_{ee}$ define the contributions of the self--energy
diagrams of the proton and electron in Fig.\,3a and Fig.\,3b,
respectively, and ${\cal M}^{(\gamma)}_{pe}$ is defined by the vertex
diagram in Fig.\,3c. Following \cite{RC9} we calculate the diagrams in
Fig.\,3 in the Feynman gauge. As a result they are determined by the
following analytical expressions
\begin{eqnarray}\label{labelD.2}
\hspace{-0.3in}&&{\cal M}^{(\gamma)}_{pp} =
e^2\int\frac{d^4q}{(2\pi)^4i}\,\frac{1}{q^2+i0}\Big[\bar{u}_p\gamma^{\alpha}\frac{1}{m_p
    - \hat{k}_p + \hat{q} -i0}\gamma_{\alpha}\,\frac{1}{m_p - \hat{k}_p - i0}\gamma^{\mu}(1
+ \lambda \gamma^5)u_n\Big][\bar{u}_e\gamma_{\mu}(1 -
  \gamma^5)v_{\bar{\nu}}]\nonumber\\
\hspace{-0.3in}&& + [\bar{u}_p\Big( - \delta m_p + \frac{Z^{(p)}_2 -
    1}{2}(m_p - \hat{k}_p)\Big)\,\frac{1}{m_p - \hat{k}_p -
    i0}\,\gamma^{\mu}(1 + \lambda
  \gamma^5)u_n][\bar{u}_e\gamma_{\mu}(1 -
  \gamma^5)v_{\bar{\nu}}],\nonumber\\
\hspace{-0.3in}&&{\cal M}^{(\gamma)}_{ee} = e^2[\bar{u}_p
  \gamma^{\mu}(1 + \lambda
  \gamma^5)u_n]\int\frac{d^4q}{(2\pi)^4i}\,\frac{1}{q^2+i0}\,
       \Big[\bar{u}_e\gamma^{\alpha}\frac{1}{m_e - \hat{k}_e - \hat{q}
           -i0}\gamma_{\alpha}\frac{1}{m_e - \hat{k}_e -
           i0}\gamma_{\mu}(1 - \gamma^5)v_{\bar{\nu}}\Big]\nonumber\\
\hspace{-0.3in}&& + [\bar{u}_p \gamma^{\mu}(1 + \lambda
  \gamma^5)u_n][\bar{u}_e\Big( - \delta m_e + \frac{Z^{(e)}_2 -
    1}{2}(m_e - \hat{k}_e)\Big)\,\frac{1}{m_e - \hat{k}_e - i0}\gamma_{\mu}(1
- \gamma^5)v_{\bar{\nu}}],\nonumber\\
\hspace{-0.3in}&&{\cal M}^{(\gamma)}_{pe} = -\,
e^2\int\frac{d^4q}{(2\pi)^4i}\,\frac{1}{q^2+i0}\Big[\bar{u}_p
\gamma^{\alpha}\frac{1}{m_p
    - \hat{k}_p + \hat{q} -i0}\,\gamma^{\mu}(1 + \lambda \gamma^5)
u_n\Big]\Big[\bar{u}_e\gamma_{\alpha}\frac{1}{m_e - \hat{k}_e - \hat{q}
    -i0}\gamma_{\mu}(1 - \gamma^5) v_{\bar{\nu}}\Big],
\end{eqnarray}
where $(\delta m_p, Z^{(p)}_2)$ and $(\delta m_e, Z^{(e)}_2)$ are the
renormalisation constants of the masses and wave functions of the
proton and electron, respectively.

The dependence on the electron energy is defined by the amplitude
${\cal M}^{(\gamma)}_{pe}$ only. For the calculation of ${\cal
  M}^{(\gamma)}_{pe}$ we reduce it to the form
\begin{eqnarray}\label{labelD.3}
\hspace{-0.3in}{\cal M}^{(\gamma)}_{pe} &=& -\, \frac{\alpha}{4\pi}
\int\frac{d^4q}{\pi^2i}\,\frac{1}{q^2+i0}\,\frac{1}{q^2 - 2k_p\cdot q
  + i0}\frac{1}{q^2 + 2k_e\cdot q + i0}\nonumber\\
\hspace{-0.3in}&&\times\,[\bar{u}_p\gamma^{\alpha}(m_p + \hat{k}_p -
  \hat{q})\gamma^{\mu}(1 + \lambda \gamma^5)
  u_n]\,[\bar{u}_e\gamma_{\alpha} (m_e + \hat{k}_e +
  \hat{q})\gamma_{\mu}(1 - \gamma^5) v_{\bar{\nu}}],
\end{eqnarray}
where we have set $e^2 = 4\pi \alpha$.  The numerator of the integrand
of ${\cal M}^{(\gamma)}_{pe}$ we transcribe into the form
\begin{eqnarray}\label{labelD.4}
\hspace{-0.3in}&&[\bar{u}_p\gamma^{\alpha}(m_p + \hat{k}_p -
  \hat{q})\gamma^{\mu}(1 + \lambda \gamma^5) u_n]\,[\bar{u}_e\gamma_{\alpha}(m_e + \hat{k}_e +
  \hat{q}) \gamma_{\mu}(1 - \gamma^5) v_{\bar{\nu}}]=\nonumber\\
\hspace{-0.3in}&&= [\bar{u}_p((m_p - \hat{k}_p +
  \hat{q})\gamma^{\alpha} + 2 (k_p - q)^{\alpha}))\gamma^{\mu}(1 + \lambda
  \gamma^5) u_n]\,[\bar{u}_e((m_e - \hat{k}_e -
  \hat{q})\gamma_{\alpha} + 2 (k_e + q)_{\alpha})\gamma_{\mu}(1 -
  \gamma^5) v_{\bar{\nu}}]=\nonumber\\
\hspace{-0.3in}&&= [\bar{u}_p(\hat{q}\gamma^{\alpha} + 2 (k_p -
  q)^{\alpha}))\gamma^{\mu}(1 + \lambda \gamma^5) u_n]\,[\bar{u}_e(-
  \hat{q} \gamma_{\alpha} + 2 (k_e + q)_{\alpha})\gamma_{\mu}(1 -
  \gamma^5) v_{\bar{\nu}}]=\nonumber\\
\hspace{-0.3in}&&= - [\bar{u}_p\hat{q}\gamma^{\alpha} \gamma^{\mu}(1 +
  \lambda \gamma^5) u_n]\,[\bar{u}_e \hat{q} \gamma_{\alpha}\gamma_{\mu}(1
  - \gamma^5) v_{\bar{\nu}}] + [\bar{u}_p 2\hat{q} (\hat{k}_e + \hat{q})
  \gamma_{\mu}(1 + \lambda \gamma^5) u_n][\bar{u}_e\gamma_{\mu}(1 -
  \gamma^5) v_{\bar{\nu}}]\nonumber\\
\hspace{-0.3in}&&- [\bar{u}_p \gamma_{\mu}(1 + \lambda \gamma^5)
  u_n][\bar{u}_e 2\hat{q} (\hat{k}_p - \hat{q})\gamma_{\mu}(1 -
  \gamma^5) v_{\bar{\nu}}] + 4(k_e + q)\cdot (k_p -
q)\,[\bar{u}_p\gamma^{\mu}(1 + \lambda \gamma^5) u_n]\,[\bar{u}_e
  \gamma_{\mu}(1 - \gamma^5) v_{\bar{\nu}}] = \nonumber\\
\hspace{-0.3in}&&= - [\bar{u}_p\hat{q}\gamma^{\alpha} \gamma^{\mu}(1 +
  \lambda \gamma^5) u_n]\,[\bar{u}_e \hat{q} \gamma_{\alpha}\gamma_{\mu}(1
  - \gamma^5) v_{\bar{\nu}}] + 2q^2[\bar{u}_p\gamma^{\mu}(1 + \lambda \gamma^5)
  u_n]\,[\bar{u}_e \gamma_{\mu}(1 - \gamma^5)
  v_{\bar{\nu}}]\nonumber\\
\hspace{-0.3in}&& + (q^2 + 2k_e\cdot q) [\bar{u}_p\gamma^{\mu}(1 + \lambda
  \gamma^5) u_n]\,[\bar{u}_e \gamma_{\mu}(1 - \gamma^5)
  v_{\bar{\nu}}] + (q^2 - 2k_p\cdot q) [\bar{u}_p\gamma^{\mu}(1 + \lambda
  \gamma^5) u_n]\,[\bar{u}_e \gamma_{\mu}(1 - \gamma^5)
  v_{\bar{\nu}}]\nonumber\\
\hspace{-0.3in}&&-
2i[\bar{u}_p\sigma_{\alpha\beta}q^{\alpha}k^{\beta}_e \gamma^{\mu}(1 +
  \lambda \gamma^5) u_n]\,[\bar{u}_e \gamma_{\mu}(1 - \gamma^5)
  v_{\bar{\nu}}] + 2i[\bar{u}_p\gamma^{\mu}(1 + \lambda \gamma^5)
  u_n]\,[\bar{u}_e \sigma_{\alpha\beta}q^{\alpha}k^{\beta}_p
  \gamma_{\mu}(1 - \gamma^5) v_{\bar{\nu}}]\nonumber\\
\hspace{-0.3in}&&- 2(q^2 + 2k_e\cdot q) [\bar{u}_p\gamma^{\mu}(1 + \lambda
  \gamma^5) u_n]\,[\bar{u}_e \gamma_{\mu}(1 - \gamma^5)
  v_{\bar{\nu}}] - 2(q^2 - 2k_p\cdot q) [\bar{u}_p\gamma^{\mu}(1 +
  \lambda \gamma^5) u_n]\,[\bar{u}_e \gamma_{\mu}(1 - \gamma^5)
  v_{\bar{\nu}}]\nonumber\\\hspace{-0.3in}&& + 4(k_e\cdot
k_p)[\bar{u}_p\gamma^{\mu}(1 + \lambda \gamma^5) u_n]\,[\bar{u}_e
  \gamma_{\mu}(1 - \gamma^5) v_{\bar{\nu}}] = - [\bar{u}_p\hat{q}\gamma^{\alpha} \gamma^{\mu}(1 +
  \lambda \gamma^5) u_n]\,[\bar{u}_e \hat{q} \gamma_{\alpha}\gamma_{\mu}(1
  - \gamma^5) v_{\bar{\nu}}]\nonumber\\
\hspace{-0.3in}&& + 2q^2[\bar{u}_p\gamma^{\mu}(1 + \lambda \gamma^5)
  u_n]\,[\bar{u}_e \gamma_{\mu}(1 - \gamma^5)
  v_{\bar{\nu}}] -
2i[\bar{u}_p\sigma_{\alpha\beta}q^{\alpha}k^{\beta}_e \gamma^{\mu}(1 +
  \lambda \gamma^5) u_n]\,[\bar{u}_e \gamma_{\mu}(1 - \gamma^5)
  v_{\bar{\nu}}]\nonumber\\
\hspace{-0.3in}&&+ 2i[\bar{u}_p\gamma^{\mu}(1 + \lambda \gamma^5)
  u_n]\,[\bar{u}_e \sigma_{\alpha\beta}q^{\alpha}k^{\beta}_p
  \gamma_{\mu}(1 - \gamma^5) v_{\bar{\nu}}] - (q^2 + 2k_e\cdot q)
       [\bar{u}_p\gamma^{\mu}(1 + \lambda \gamma^5) u_n]\,[\bar{u}_e
         \gamma_{\mu}(1 - \gamma^5) v_{\bar{\nu}}]\nonumber\\
\hspace{-0.3in}&& - (q^2 - 2k_p\cdot q) [\bar{u}_p\gamma^{\mu}(1 + \lambda
  \gamma^5) u_n]\,[\bar{u}_e \gamma_{\mu}(1 - \gamma^5)
  v_{\bar{\nu}}] + 4(k_e\cdot k_p)[\bar{u}_p\gamma^{\mu}(1 + \lambda
  \gamma^5) u_n]\,[\bar{u}_e \gamma_{\mu}(1 - \gamma^5)
  v_{\bar{\nu}}],
\end{eqnarray}
where we have used the identity $\gamma^{\alpha}\gamma^{\beta} =
g^{\alpha\beta} - i\sigma^{\alpha\beta}$.  For the transformation of
the product \[- [\bar{u}_p\hat{q}\gamma^{\alpha}\gamma^{\mu}(1 +
  \lambda \gamma^5)u_n][\bar{u}_e\hat{q}\gamma_{\alpha}\gamma_{\mu}(1
  - \gamma^5)v_{\bar{\nu}}]\] we propose to use the relations
\begin{eqnarray}\label{labelD.5}
\hspace{-0.3in}&&\hat{q}\gamma^{\alpha}\gamma^{\mu} =
\gamma^{\mu}q^{\alpha} - \gamma^{\alpha}q^{\mu} + \hat{q}\,g^{\alpha\mu}
+ i \varepsilon^{\beta \alpha \mu \lambda}\gamma_{\lambda}\gamma^5
q_{\beta},\nonumber\\
\hspace{-0.3in}&&\varepsilon^{\beta\alpha\mu\lambda}
\varepsilon_{\rho\alpha\mu\varphi} q_{\beta} q^{\rho} = -
2q^2g^{\lambda}_{\varphi} + 2q^{\lambda}q_{\varphi},\nonumber\\
\hspace{-0.3in}&&\lambda + \gamma^5 = - (1 + \lambda \gamma^5) +
(\lambda + 1)(1 + \gamma^5)
\end{eqnarray}
This gives 
\begin{eqnarray}\label{labelD.6}
\hspace{-0.3in}&&- [\bar{u}_p\hat{q}\gamma^{\alpha}\gamma^{\mu}(1 +
  \lambda \gamma^5)u_n][\bar{u}_e\hat{q}\gamma_{\alpha}\gamma_{\mu}(1 -
  \gamma^5)v_{\bar{\nu}}] = - 4q^2[\bar{u}_p\gamma^{\mu}(1 + \lambda
  \gamma^5)u_n][\bar{u}_e\gamma_{\mu}(1 -
  \gamma^5)v_{\bar{\nu}}]\nonumber\\
\hspace{-0.3in}&& - 2(\lambda + 1)[\bar{u}_p\hat{q}(1 +
  \gamma^5)u_n][\bar{u}_e\hat{q}(1 -
  \gamma^5)v_{\bar{\nu}}] + 2(\lambda + 1)q^2 [\bar{u}_p\gamma^{\mu}(1 +
  \gamma^5)u_n][\bar{u}_e\gamma_{\mu}(1 - \gamma^5)v_{\bar{\nu}}].
\end{eqnarray}
Substituting Eq.(\ref{labelD.6}) into Eq.(\ref{labelD.4}) we arrive at
the following expression of the numerator of the integrand of ${\cal
  M}^{(\gamma)}_{pe}$ in Eq.(\ref{labelD.3})
\begin{eqnarray}\label{labelD.7}
\hspace{-0.3in}&&[\bar{u}_p\gamma^{\alpha}(m_p + \hat{k}_p -
  \hat{q})\gamma^{\mu}(1 + \lambda \gamma^5)
  u_n]\,[\bar{u}_e\gamma_{\alpha} (m_e + \hat{k}_e +
  \hat{q})\gamma_{\mu}(1 - \gamma^5) v_{\bar{\nu}}]=\nonumber\\
\hspace{-0.3in}&&=- (q^2 + 2k_e\cdot q) [\bar{u}_p\gamma^{\mu}(1 +
  \lambda \gamma^5) u_n]\,[\bar{u}_e \gamma_{\mu}(1 - \gamma^5)
  v_{\bar{\nu}}]- (q^2 - 2k_p\cdot q) [\bar{u}_p\gamma^{\mu}(1 +
  \lambda \gamma^5) u_n]\,[\bar{u}_e \gamma_{\mu}(1 - \gamma^5)
  v_{\bar{\nu}}]\nonumber\\
\hspace{-0.3in}&& - 2q^2[\bar{u}_p\gamma^{\mu}(1 + \lambda
  \gamma^5)u_n][\bar{u}_e\gamma_{\mu}(1 - \gamma^5)v_{\bar{\nu}}] +
4(k_e\cdot k_p)[\bar{u}_p\gamma^{\mu}(1 + \lambda \gamma^5)
  u_n]\,[\bar{u}_e \gamma_{\mu}(1 - \gamma^5)
  v_{\bar{\nu}}]\nonumber\\
\hspace{-0.3in}&&- 2(\lambda + 1)[\bar{u}_p\hat{q}(1 +
  \gamma^5)u_n][\bar{u}_e\hat{q}(1 - \gamma^5)v_{\bar{\nu}}] +
2(\lambda + 1)q^2 [\bar{u}_p\gamma^{\mu}(1 +
  \gamma^5)u_n][\bar{u}_e\gamma_{\mu}(1 -
  \gamma^5)v_{\bar{\nu}}]\nonumber\\
\hspace{-0.3in}&&-
2i[\bar{u}_p\sigma_{\alpha\beta}q^{\alpha}k^{\beta}_e \gamma^{\mu}(1 +
  \lambda \gamma^5) u_n]\,[\bar{u}_e \gamma_{\mu}(1 - \gamma^5)
  v_{\bar{\nu}}] + 2i[\bar{u}_p\gamma^{\mu}(1 + \lambda \gamma^5)
  u_n]\,[\bar{u}_e \sigma_{\alpha\beta}q^{\alpha}k^{\beta}_p
  \gamma_{\mu}(1 - \gamma^5) v_{\bar{\nu}}].
\end{eqnarray}
For the subsequent calculations it is convenient to represent the
amplitude ${\cal M}^{(\gamma)}_{pe}$ as a sum of two contributions
\begin{eqnarray}\label{labelD.8}
{\cal M}^{(\gamma)}_{pe} = \bar{{\cal M}}^{(\gamma)}_{pe} + \delta {\cal M}^{(\gamma)}_{pe},
\end{eqnarray}
where $\bar{{\cal M}}^{(\gamma)}_{pe}$ and $\delta {\cal
  M}^{(\gamma)}_{pe}$ are given by
\begin{eqnarray}\label{labelD.9}
\hspace{-0.3in}&& \bar{{\cal M}}^{(\gamma)}_{pe} =
\frac{\alpha}{4\pi}\,[\bar{u}_p\gamma^{\mu}(1 + \lambda
  \gamma^5)u_n][\bar{u}_e\gamma_{\mu}(1 -
  \gamma^5)v_{\bar{\nu}}]\Big\{\int \frac{d^4q}{\pi^2i}\frac{1}{q^2 +
  i0}\frac{1}{q^2 - 2k_p\cdot q + i0} + \int \frac{d^4q}{\pi^2i}\frac{1}{q^2 +
  i0}\frac{1}{q^2 + 2k_e\cdot q + i0}\nonumber\\
\hspace{-0.3in}&&+ 2\int \frac{d^4q}{\pi^2i}\frac{1}{q^2 - 2k_p\cdot q
  + i0}\frac{1}{q^2 + 2k_e\cdot q + i0} - 4(k_e\cdot k_p) \int
\frac{d^4q}{\pi^2i}\frac{1}{q^2 + i0}\frac{1}{q^2 - 2k_p\cdot q +
  i0}\frac{1}{q^2 + 2k_e\cdot q + i0}\Big\}.
\end{eqnarray}
and
\begin{eqnarray}\label{labelD.10}
\hspace{-0.3in}&&\delta {\cal M}^{(\gamma)}_{pe} =\delta^{(1)} {\cal
  M}^{(\gamma)}_{pe} + \delta^{(2)} {\cal M}^{(\gamma)}_{pe} +
\delta^{(3)} {\cal M}^{(\gamma)}_{pe} = \nonumber\\
\hspace{-0.3in}&&= \frac{\alpha}{4\pi}\,\Big\{- 2(\lambda + 1)\,
       [\bar{u}_p\gamma^{\mu}(1 +
         \gamma^5)u_n][\bar{u}_e\gamma_{\mu}(1 -
         \gamma^5)v_{\bar{\nu}}] \int \frac{d^4q}{\pi^2i} \frac{1}{q^2 -
  2k_p\cdot q + i0}\frac{1}{q^2 + 2k_e\cdot q + i0}\nonumber\\
\hspace{-0.3in}&& + 2(\lambda + 1)\int \frac{d^4q}{\pi^2i}\frac{1}{q^2
  + i0}\frac{1}{q^2 - 2k_p\cdot q + i0}\frac{1}{q^2 + 2k_e\cdot q +
  i0}\,[\bar{u}_p\hat{q}(1 +
  \gamma^5)u_n][\bar{u}_e\hat{q}(1 - \gamma^5)v_{\bar{\nu}}]
\nonumber\\
\hspace{-0.3in}&& + 2i\int \frac{d^4q}{\pi^2i}\frac{1}{q^2 +
  i0}\frac{1}{q^2 - 2k_p\cdot q + i0}\frac{1}{q^2 + 2k_e\cdot q +
  i0}\Big([\bar{u}_p\sigma_{\alpha\beta}
  q^{\alpha}k^{\beta}_e\gamma^{\mu}(1 + \lambda
  \gamma^5)u_n][\bar{u}_e\gamma_{\mu}(1 -
  \gamma^5)v_{\bar{\nu}}]\nonumber\\
\hspace{-0.3in}&& - [\bar{u}_p\gamma^{\mu}(1 + \lambda
  \gamma^5)u_n][\bar{u}_e\sigma_{\alpha\beta}q^{\alpha}k^{\beta}_p\gamma_{\mu}(1
  - \gamma^5)v_{\bar{\nu}}]\Big)\Big\},
\end{eqnarray}
respectively. Now let us proceed to calculate $\bar{{\cal M}}^{\gamma}_{pe}$, given
by Eq.(\ref{labelD.9}).  Since, the last term in Eq.(\ref{labelD.9})
is calculated in Appendix C, so we should calculate the first three
terms only. Using the Pauli--Villars regularization for the
ultra--violet divergent integrals \cite{IZ80}
\begin{eqnarray}\label{labelD.11}
\hspace{-0.3in}\frac{1}{q^2 + i0} \to \frac{1}{q^2 + i0} -
\frac{1}{q^2 - \Lambda^2 + i0} = - \frac{1}{q^2 +
  i0}\,\frac{\Lambda^2}{q^2 - \Lambda^2 + i0},
\end{eqnarray}
where $\Lambda$ is an ultraviolet cut--off, then Feynman's unification
of the denominators, the shift of the virtual 4--momentum and the Wick
rotation \cite{IZ80} for the first three integrals in
Eq.(\ref{labelD.9}) we obtain the following expressions
\begin{eqnarray*}
\hspace{-0.3in}&&\int \frac{d^4q}{\pi^2i}\frac{1}{q^2 +
  i0}\frac{1}{q^2 - 2k_p\cdot q + i0} \to \int
\frac{d^4q}{\pi^2i}\,\Big[\frac{1}{q^2 + i0}\,\frac{1}{q^2 - 2k_p\cdot
    q + i0} - \frac{1}{q^2 - \Lambda^2 + i0}\,\frac{1}{q^2 - 2k_p\cdot
    q + i0}\Big] = \nonumber\\
\hspace{-0.3in}&&= \int^1_0 dx\int \frac{d^4q}{\pi^2}\,
\Big[\frac{1}{(q^2 + m^2_px^2)^2} - \frac{1}{(q^2 + m^2_px^2 +
    \Lambda^2 (1 - x))^2}\Big] = \int^1_0 dx\,{\ell n}\Big[\frac
  {\Lambda^2 (1 - x) }{ m_p^2 x^2}\Big] = 2{\ell
  n}\Big(\frac{\Lambda}{m_p}\Big) + 1,\nonumber\\
\hspace{-0.3in}&&\int \frac{d^4q}{\pi^2i}\frac{1}{q^2 +
  i0}\frac{1}{q^2 + 2k_e\cdot q + i0} = 2{\ell
  n}\Big(\frac{\Lambda}{m_e}\Big) + 1,\nonumber\\
\end{eqnarray*}
\begin{eqnarray}\label{labelD.12}
\hspace{-0.3in}&&\int \frac{d^4q}{\pi^2i} \frac{1}{q^2 - 2k_p\cdot q +
  i0}\frac{1}{q^2 + 2k_e\cdot q + i0} = \int
\frac{d^4q}{\pi^2i}\,\frac{1}{q^2 + i0}\,\frac{q^2}{q^2 - 2k_p\cdot q
  + i0}\frac{1}{q^2 + 2k_e\cdot q + i0} \to\nonumber\\
\hspace{-0.3in}&&\to \int^1_0dx\int^1_0 dy\,2y \int
\frac{d^4q}{\pi^2i} \Big[ \frac{q^2}{(q - p(x)y)^2 - p^2(x)y^2 - \mu^2
    (1 - y))^3} - \frac{q^2}{(q - p(x)y)^2 - p^2(x)y^2 - \Lambda^2
  (1 - y))^3}\Big] = \nonumber\\
\hspace{-0.3in}&&= \int^1_0dx\int^1_0 dy\,2y \int \frac{d^4q}{\pi^2}
\Big[ \frac{q^2 - p^2(x)y^2}{(q^2 + p^2(x)y^2 + \mu^2 (1 -
    y))^3} - \frac{q^2 - p^2(x) y^2}{(q^2 + p^2(x)y^2 +
  \Lambda^2 (1 - y))^3}\Big] =\nonumber\\
\hspace{-0.3in}&& = \int^1_0dx\int^1_0 dy\,2y\int
\frac{d^4q}{\pi^2}\Big\{\frac{1}{(q^2 + p^2(x)y^2 + \mu^2(1 - y))^2} -
\frac{1}{(q^2 + p^2(x)y^2 + \Lambda^2(1 - y))^2}\nonumber\\
\hspace{-0.3in}&& - \frac{2 p^2(x)y^2}{(q^2 + p^2(x)y^2 + \mu^2(1 -
  y))^3} + \frac{2 p^2(x)y^2 + \Lambda^2(1 - y)}{(q^2 + p^2(x)y^2 +
  \Lambda^2(1 - y))^3}\Big\} = \int^1_0dx\int^1_0 dy\,2y\,\Big\{{\ell
  n}\Big[\frac{\Lambda^2(1 - y)}{p^2(x) y^2}\Big] - \frac{1}{2}\Big\} = \nonumber\\
\hspace{-0.3in}&& = \int^1_0dx\,\Big\{{\ell
  n}\Big[\frac{\Lambda^2}{p^2(x)}\Big] - 1\Big\} = 2\,{\ell
n}\Big(\frac{\Lambda}{m_p}\Big) + 1.
\end{eqnarray}
For the calculation of the third integral we have replaced $p^2(x) =
(k_ex - k_p(1-x))^2$ by $m^2_p(1-x)^2$. For the calculation of the
amplitudes ${\cal M}^{(\gamma)}_{ee}$ and ${\cal M}^{(\gamma)}_{pp}$
we have to perform the following standard transformations
\begin{eqnarray}\label{labelD.13}
\hspace{-0.3in}&&\int \frac{d^4q}{\pi^2 i}\,\bar{u} \frac{1}{q^2 +
  i0}\gamma^{\alpha}\frac{1}{\hat{k} + \hat{q} - m}\gamma_{\alpha}
=\bar{u}\int^1_0 dx\int \frac{d^4q}{\pi^2 i}\, \frac{4m - 2\hat{k} -
  2\hat{q}}{[(q + kx) - m^2 x + k^2 x(1 - x)]^2} =\nonumber\\
\hspace{-0.3in}&&= \int^1_0 dx\,\bar{u} \int \frac{d^4q}{\pi^2 }\,
\frac{4m - 2\hat{k} (1 - x)}{[q^2 + m^2 x - k^2 x(1 - x)]^2} =
\int^1_0 dx \,\bar{u} \int \frac{d^4q}{\pi^2 }\, \frac{2m (1 + x) +
  2(1 - x)(m - \hat{k})}{[q^2 + m^2 x - k^2 x(1 - x)]^2} =\nonumber\\
\hspace{-0.3in}&&= \int^1_0 dx \,\bar{u} \Big[\int \frac{d^4q}{\pi^2
  }\, \frac{2m (1 + x)}{[q^2 + m^2 x - k^2 x(1 - x)]^2}+ \frac{2(1-x)}
  {[q^2 + m^2 x - k^2 x(1 - x)]^2}\,(m - \hat{k} ) - \frac{4m x(1-x^2)(m + \hat{k})}{[q^2 + m^2_e x -
    k^2 x(1 - x)]^3}\nonumber\\
\hspace{-0.3in}&&\times\,(m - \hat{k})\Big]= \bar{u} \Big\{\int^1_0 dx
\int \frac{d^4q}{\pi^2 }\, \frac{2m (1 + x)}{(q^2 + m^2 x^2)^2} + (m -
\hat{k})\int^1_0 dx \int \frac{d^4q}{\pi^2 }\,\Big[\frac{2(1-x)}{(q^2
    + m^2 x^2)^2} - \frac{8m^2 x(1-x^2)}{(q^2 + m^2
    x^2)^3}\Big]\Big\}.
\end{eqnarray}
This results in the relation
\begin{eqnarray}\label{labelD.14}
\hspace{-0.3in}\int \frac{d^4q}{\pi^2 i}\,\bar{u} \frac{1}{q^2 +
  i0}\gamma^{\alpha}\frac{1}{\hat{q} + \hat{k} - m}\gamma_{\alpha}
&=& \bar{u} \Big\{\int^1_0 dx \int \frac{d^4q}{\pi^2
}\, \frac{2m (1 + x)}{(q^2 + m^2 x^2)^2}\nonumber\\
\hspace{-0.3in}&+& (m -
\hat{k}) \int^1_0 dx \int \frac{d^4q}{\pi^2
}\,\Big[\frac{2(1-x)}{(q^2 + m^2 x^2)^2} - \frac{8m^2x(1-x^2)}{(q^2 +
    m^2x^2)^3}\Big]\Big\},
\end{eqnarray}
where $m = m_e$ and $m = m_p$ for the electron and proton self--energy
contributions, respectively.  The term, which is not proportional to
$(m - \hat{k})$, can be removed by the mass renormalisation. So the
amplitudes ${\cal M}^{(\gamma)}_{ee}$ and ${\cal M}^{(\gamma)}_{pp}$
are defined by the contributions of the terms, proportional to $(m_e -
\hat{k}_e)$ and $(m_p - \hat{k}_p)$, respectively. This gives
\begin{eqnarray}\label{labelD.15}
\hspace{-0.3in}{\cal M}^{(\gamma)}_{ee} &=& [\bar{u}_p\gamma^{\mu}(1 +
  \lambda \gamma^5)u_n][\bar{u}_e\gamma_{\mu}(1 -
  \gamma^5)v_{\bar{\nu}}]\,\Big(\frac{\alpha}{4\pi}\,I(m_e) +
\frac{Z^{(e)}_2 - 1}{2}\Big),\nonumber\\
\hspace{-0.3in}{\cal M}^{(\gamma)}_{pp} &=& [\bar{u}_p\gamma^{\mu}(1 +
  \lambda \gamma^5)u_n][\bar{u}_e\gamma_{\mu}(1 -
  \gamma^5)v_{\bar{\nu}}]\,\Big(\frac{\alpha}{4\pi}\,I(m_p) +
\frac{Z^{(p)}_2 - 1}{2}\Big),
\end{eqnarray}
where $I(m)$ is given by
\begin{eqnarray}\label{labelD.16}
\hspace{-0.3in}I(m) = \int^1_0 dx \int \frac{d^4q}{\pi^2}\,\Big[-
  \frac{2(1-x)}{(q^2 + m^2 x^2)^2} +  \frac{8m^2x(1-x^2)}{(q^2 + m^2 x^2 + \mu^2(1 -
  x))^3}\Big]
\end{eqnarray}
for $m = m_e$ and $m_p$, respectively. For the ultra--violet and
infrared regularization we use the following expressions for the
photon Green function
\begin{eqnarray}\label{labelD.17}
\hspace{-0.3in}\frac{1}{q^2 + i0} \to \frac{1}{q^2 - \mu^2 + i0} -
\frac{1}{q^2 - \Lambda^2 + i0}.
\end{eqnarray}
The regularized function $I(m)$ takes the form
\begin{eqnarray}\label{labelD.18}
\hspace{-0.3in}I(m) &=& \int^1_0 dx \int \frac{d^4q}{\pi^2}\,\Big[-
  \frac{2(1-x)}{(q^2 + m^2 x^2)^2} + \frac{2(1-x)}{(q^2 + m^2 x^2 +
    \Lambda^2(1 - x))^2}\nonumber\\
 &+& \frac{8m^2x(1-x^2)}{(q^2 + m^2 x^2 +
    \mu^2(1 - x))^3} - \frac{8m^2x(1-x^2)}{(q^2 + m^2 x^2 +
    \Lambda^2(1 - x))^3}\Big].
\end{eqnarray}
The calculation of the integrals in Eq.(\ref{labelD.18}) runs as
follows. For the first two integrals, which are ultra--violet
divergent, we obtain the expression
\begin{eqnarray}\label{labelD.19}
\hspace{-0.3in}&&\int^1_0 dx \int \frac{d^4q}{\pi^2}\,
\Big[\frac{2(1-x)}{(q^2 + m^2 x^2 + \Lambda^2(1 - x))^2} -
  \frac{2(1-x)}{(q^2 + m^2 x^2)^2}\Big] = - \int^1_0 dx\,2(1 -
x)\,{\ell n}\Big[\frac{(1 - x)\,\Lambda^2}{x^2 m^2}\Big]-\,2\,{\ell n}\Big(\frac{\Lambda}{m}\Big) -
\frac{5}{2}.\nonumber\\
\hspace{-0.3in}&&
\end{eqnarray}
The third and fourth integrals in Eq.(\ref{labelD.18}) are
ultra--violet convergent. Moreover in the limit $\Lambda \to \infty$
the fourth integral vanishes. The third integral is infrared divergent
and its calculation gives
\begin{eqnarray}\label{labelD.20}
\hspace{-0.3in}\int^1_0 dx \int
\frac{d^4q}{\pi^2}\,\frac{8m^2x(1-x^2)}{(q^2 + \mu^2 (1 - x) + m^2
  x^2)^3} = 4m^2\int^1_0 dx \,\frac{x(1-x^2)}{\mu^2 (1 - x) + m^2 x^2}
= - 4\,{\ell n}\Big(\frac{\mu}{m}\Big) - 2.
\end{eqnarray}
Thus, the function $I(m)$ takes the form
\begin{eqnarray}\label{labelD.21}
\hspace{-0.3in}I(m) = - 2{\ell n}\Big(\frac{\Lambda}{m}\Big) - 4 {\ell
  n}\Big(\frac{\mu}{m}\Big) - \frac{9}{2}.
\end{eqnarray}
Summing up the contributions of the amplitudes ${\cal
  M}^{(\gamma)}_{ee}$, ${\cal M}^{(\gamma)}_{pp}$ and $\bar{{\cal
    M}}^{(\gamma)}_{pe}$ we obtain the following expression
\begin{eqnarray}\label{labelD.22}
\hspace{-0.3in}&&{\cal M}^{(\gamma)}_{ee} + {\cal M}^{(\gamma)}_{pp} +
\bar{{\cal M}}^{(\gamma)}_{pe} =[\bar{u}_p\gamma^{\mu}(1 + \lambda
  \gamma^5)u_n]\,[\bar{u}_e\gamma_{\mu}(1 -
  \gamma^5)v_{\bar{\nu}}]\,\Big\{\frac{Z^{(e)}_2 - 1}{2} +
\frac{Z^{(p)}_2 - 1}{2}\nonumber\\
\hspace{-0.3in}&& + \frac{\alpha}{4\pi}\Big[- 2{\ell
    n}\Big(\frac{\Lambda}{m_e}\Big)- 4 {\ell
    n}\Big(\frac{\mu}{m_e}\Big) - \frac{9}{2} - 2{\ell
    n}\Big(\frac{\Lambda}{m_p}\Big) - 4 {\ell
    n}\Big(\frac{\mu}{m_p}\Big) - \frac{9}{2} + 2{\ell
    n}\Big(\frac{\Lambda}{m_p}\Big) + 1 \nonumber\\
\hspace{-0.3in}&& + 2{\ell n}\Big(\frac{\Lambda}{m_e}\Big) + 1 +
4{\ell n}\Big(\frac{\Lambda}{m_p}\Big) + 2 + 2\Big\{{\ell n}\Big(\frac{
  \mu}{m_e}\Big)\,\frac{1}{\beta}\,{\ell n}\Big(\frac{1 + \beta}{1 -
  \beta}\Big) - \frac{1}{4\beta}\,{\ell n}^2\Big(\frac{1 + \beta}{1 -
  \beta}\Big) + \frac{1}{\beta}L\Big(\frac{2\beta}{1 +
  \beta}\Big)\Big\}\Big]\Big\}.
\end{eqnarray}
After the cancellation of some terms we arrive at the expression
\begin{eqnarray}\label{labelD.23}
\hspace{-0.3in}&&{\cal M}^{(\gamma)}_{ee} + {\cal M}^{(\gamma)}_{pp} +
\bar{{\cal M}}^{(\gamma)}_{pe} =[\bar{u}_p\gamma^{\mu}(1 + \lambda
  \gamma^5)u_n]\,[\bar{u}_e\gamma_{\mu}(1 -
  \gamma^5)v_{\bar{\nu}}]\,\Big\{\frac{Z^{(e)}_2 - 1}{2} +
\frac{Z^{(p)}_2 - 1}{2}\nonumber\\
\hspace{-0.3in}&& + \frac{\alpha}{2\pi}\Big[2{\ell
    n}\Big(\frac{\Lambda}{m_p}\Big) - 2 {\ell
    n}\Big(\frac{\mu}{m_e}\Big) - 2{\ell n}\Big(\frac{\mu}{m_p}\Big)
  - \frac{5}{2} + \frac{1}{\beta}{\ell n}\Big(\frac{\mu}{
  m_e}\Big)\,{\ell n}\Big(\frac{1 + \beta}{1 - \beta}\Big) -
\frac{1}{4\beta}\,{\ell n}^2\Big(\frac{1 + \beta}{1 -
  \beta}\Big) + \frac{1}{\beta}L\Big(\frac{2\beta}{1 +
  \beta}\Big)\Big]\Big\}.\nonumber\\
\hspace{-0.3in}&&
\end{eqnarray}
Then, we transcribe the r.h.s. of Eq.(\ref{labelD.23}) into the
following symmetric form
\begin{eqnarray}\label{labelD.24}
\hspace{-0.3in}&&{\cal M}^{(\gamma)}_{ee} + {\cal M}^{(\gamma)}_{pp} +
      \bar{{\cal M}}^{(\gamma)}_{pe} = [\bar{u}_p\gamma^{\mu}(1 + \lambda
  \gamma^5)u_n]\,[\bar{u}_e\gamma_{\mu}(1 -
  \gamma^5)v_{\bar{\nu}}]\nonumber\\
\hspace{-0.3in}&&\times\,\Big\{\Big[\frac{Z^{(e)}_2 - 1}{2} -
  \frac{\alpha}{2\pi}\Big(\frac{1}{2}{\ell
    n}\Big(\frac{\Lambda}{m_e}\Big) + {\ell
    n}\Big(\frac{\mu}{m_e}\Big) + \frac{9}{8}\Big)\Big] +
\Big[\frac{Z^{(p)}_2 - 1}{2} -
  \frac{\alpha}{2\pi}\Big(\frac{1}{2}{\ell
    n}\Big(\frac{\Lambda}{m_p}\Big) + {\ell
    n}\Big(\frac{\mu}{m_p}\Big) + \frac{9}{8}\Big)\Big]\nonumber\\
\hspace{-0.3in}&& + \frac{\alpha}{2\pi}\Big[3{\ell
    n}\Big(\frac{\Lambda}{m_p}\Big) + \frac{3}{2}{\ell
    n}\Big(\frac{m_p}{m_e}\Big) - \frac{1}{4} + 2 {\ell
    n}\Big(\frac{\mu}{ m_e}\Big)\,\Big[\frac{1}{2\beta}\,{\ell
      n}\Big(\frac{1 + \beta}{1 - \beta}\Big) - 1 \Big] -
  \frac{1}{4\beta}\,{\ell n}^2\Big(\frac{1 + \beta}{1 - \beta}\Big) +
  \frac{1}{\beta}L\Big(\frac{2\beta}{1 + \beta}\Big) \Big]\Big\}.
\end{eqnarray}
Defining the constants of the renormalisation of the wave functions of
the electron and proton in the standard form as \cite{IZ80}
\begin{eqnarray}\label{labelD.25}
\hspace{-0.3in}&&\frac{Z^{(e)}_2 - 1}{2} =
\frac{\alpha}{2\pi}\Big[\frac{1}{2}{\ell
  n}\Big(\frac{\Lambda}{m_e}\Big) + {\ell n}\Big(\frac{\mu}{m_e}\Big)
+ \frac{9}{8}\Big],\nonumber\\
\hspace{-0.3in}&&\frac{Z^{(p)}_2 - 1}{2} =
\frac{\alpha}{2\pi}\Big[\frac{1}{2}{\ell
  n}\Big(\frac{\Lambda}{m_p}\Big) + {\ell n}\Big(\frac{\mu}{m_p}\Big)
+ \frac{9}{8}\Big]\nonumber\\
\hspace{-0.3in}&&
\end{eqnarray}
we reduce the r.h.s. of Eq.(\ref{labelD.24}) to the form
\begin{eqnarray}\label{labelD.26}
\hspace{-0.3in}&&{\cal M}^{(\gamma)}_{ee} + {\cal M}^{(\gamma)}_{pp} +
\bar{{\cal M}}^{(\gamma)}_{pe} = [\bar{u}_p\gamma^{\mu}(1 + \lambda
  \gamma^5)u_n]\,[\bar{u}_e\gamma_{\mu}(1 -
  \gamma^5)v_{\bar{\nu}}]\nonumber\\
\hspace{-0.3in}&&\times\, \frac{\alpha}{2\pi}\Big\{3{\ell
  n}\Big(\frac{\Lambda}{m_p}\Big) + \frac{3}{2}{\ell
  n}\Big(\frac{m_p}{m_e}\Big) - \frac{1}{4} + 2 {\ell
  n}\Big(\frac{\mu}{ m_e}\Big)\,\Big[\frac{1}{2\beta}\,{\ell
    n}\Big(\frac{1 + \beta}{1 - \beta}\Big) - 1 \Big] +
\frac{1}{\beta}L\Big(\frac{2\beta}{1 + \beta}\Big) -
\frac{1}{4\beta}\,{\ell n}^2\Big(\frac{1 + \beta}{1 -
  \beta}\Big)\Big\}.
\end{eqnarray}
In the non--relativistic approximation for the proton
Eq.(\ref{labelD.26}) reads
\begin{eqnarray}\label{labelD.27}
\hspace{-0.3in}&& {\cal M}^{(\gamma)}_{ee} + {\cal M}^{(\gamma)}_{pp}
+ \bar{{\cal M}}^{(\gamma)}_{pe} = \frac{\alpha}{2\pi}\Big\{3{\ell
  n}\Big(\frac{\Lambda}{m_p}\Big) + \frac{3}{2}{\ell
  n}\Big(\frac{m_p}{m_e}\Big) - \frac{1}{4} + 2 {\ell
  n}\Big(\frac{\mu}{ m_e}\Big) \,\Big[\frac{1}{2\beta}\,{\ell
    n}\Big(\frac{1 + \beta}{1 - \beta}\Big) - 1 \Big]\nonumber\\
\hspace{-0.3in}&& +
\frac{1}{\beta}L\Big(\frac{2\beta}{1 + \beta}\Big) -
\frac{1}{4\beta}\,{\ell n}^2\Big(\frac{1 + \beta}{1 -
  \beta}\Big)\Big\}\,
\Big\{[\varphi^{\dagger}_p\varphi_n][\bar{u}_e\gamma^0(1 - \gamma^5)
  v_{\bar{\nu}}] - \lambda
      [\varphi^{\dagger}_p\vec{\sigma}\,\varphi_n] \cdot
      [\bar{u}_e\vec{\gamma}\,(1 - \gamma^5) v_{\bar{\nu}}]\Big\}.
\end{eqnarray}
Now let us take into account the contribution of the first two terms
in $\delta {\cal M}^{(\gamma)}_{pe}$, given by
Eq.(\ref{labelD.10}). The first term in Eq.(\ref{labelD.10}) is equal
to
\begin{eqnarray}\label{labelD.28}
\hspace{-0.3in}&&\delta^{(1)} {\cal M}^{(\gamma)}_{pe} = - (\lambda +
1)\, \frac{\alpha}{2\pi}\, [\bar{u}_p\gamma^{\mu}(1 +
  \gamma^5)u_n][\bar{u}_e\gamma_{\mu}(1 - \gamma^5)v_{\bar{\nu}}] \int
\frac{d^4q}{\pi^2i} \frac{1}{q^2 - 2k_p\cdot q + i0}\frac{1}{q^2 +
  2k_e\cdot q + i0}=\nonumber\\
\hspace{-0.3in}&&= - (\lambda + 1)\,\frac{\alpha}{2\pi}\,\Big[2{\ell
    n}\Big(\frac{\Lambda}{m_p}\Big) + 1\Big]\,
       [\bar{u}_p\gamma^{\mu}(1 +
         \gamma^5)u_n][\bar{u}_e\gamma_{\mu}(1 -
         \gamma^5)v_{\bar{\nu}}],
\end{eqnarray}
where we have used Eq.(\ref{labelD.12}). In the non---relativistic
approximation for the proton it reads
\begin{eqnarray}\label{labelD.29}
\hspace{-0.3in}&&\delta^{(1)} {\cal M}^{(\gamma)}_{pe} = - 2 m_n
\,(\lambda + 1)\,\frac{\alpha}{2\pi}\,\Big[2\,{\ell
    n}\Big(\frac{\Lambda}{m_p}\Big) + 1\Big]\Big\{
  [\varphi^{\dagger}_p\varphi_n]\, [\bar{u}_e\gamma^0 (1 -
    \gamma^5)v_{\bar{\nu}}] -
  [\varphi^{\dagger}_p\vec{\sigma}\,\varphi_n]\cdot
  [\bar{u}_e\vec{\gamma}\,(1 -
    \gamma^5)v_{\bar{\nu}}]\Big\}.
\end{eqnarray}
The calculation of the second term in Eq.(\ref{labelD.10}) runs as
follows
\begin{eqnarray}\label{labelD.30}
\hspace{-0.3in}&&\delta^{(2)} {\cal M}^{(\gamma)}_{pe} = (\lambda +
1)\,\frac{\alpha}{2\pi}\,[\bar{u}_p\gamma^{\alpha}(1 +
  \gamma^5)u_n][\bar{u}_e\gamma^{\beta}(1 - \gamma^5)v_{\bar{\nu}}]
\int \frac{d^4q}{\pi^2i}\frac{ q_{\alpha}q_{\beta}}{q^2 +
  i0}\frac{1}{q^2 - 2k_p\cdot q + i0}\frac{1}{q^2 + 2k_e\cdot q + i0}
\to \nonumber\\
\hspace{-0.3in}&&\to (\lambda +
1)\,\frac{\alpha}{2\pi}\,[\bar{u}_p\gamma^{\alpha}(1 +
  \gamma^5)u_n][\bar{u}_e\gamma^{\beta}(1 -
  \gamma^5)v_{\bar{\nu}}]\int \frac{d^4q}{\pi^2i}\Big[\frac{
    q_{\alpha}q_{\beta}}{q^2 + i0} - \frac{ q_{\alpha}q_{\beta}}{q^2 -
    \Lambda^2 + i0}\Big]\,\frac{1}{q^2 - 2k_p\cdot q + i0}\frac{1}{q^2
  + 2k_e\cdot q + i0} = \nonumber\\
\hspace{-0.3in}&&= (\lambda +
1)\,\frac{\alpha}{2\pi}\,[\bar{u}_p\gamma^{\alpha}(1 +
  \gamma^5)u_n][\bar{u}_e\gamma^{\beta}(1 -
  \gamma^5)v_{\bar{\nu}}]\int^1_0dx \int^1_0 dy 2y\int
\frac{d^4q}{\pi^2i}\Big[\frac{ q_{\alpha}q_{\beta}}{(q + p(x)y)^2 -
  p^2(x)y^2)^3}\nonumber\\ 
\hspace{-0.3in}&& - \frac{ q_{\alpha}q_{\beta}}{(q + p(x)y)^2 -
  p^2(x)y^2 - \Lambda^2(1 - y))^3}\Big].
\end{eqnarray}
After the shift of variables, the integration over the 4--dimensional
solid angle and the Wick rotation we arrive at the expression
\begin{eqnarray}\label{labelD.31}
\hspace{-0.3in}&&\delta^{(2)} {\cal M}^{(\gamma)}_{pe} = (\lambda +
1)\frac{\alpha}{8\pi}\,[\bar{u}_p\gamma^{\alpha}(1 +
  \gamma^5)u_n][\bar{u}_e\gamma^{\beta}(1 -
  \gamma^5)v_{\bar{\nu}}]\int^1_0dx \int^1_0 dy\, 2y\int
\frac{d^4q}{\pi^2}\Big[\frac{ q^2g_{\alpha\beta} - 4y^2
  p_{\alpha}(x)p_{\beta}(x)}{(q^2 + p^2(x)y^2)^3}\nonumber\\
\hspace{-0.3in}&& - \frac{ q^2g_{\alpha\beta} - 4y^2
  p_{\alpha}(x)p_{\beta}(x)}{(q^2 + p^2(x)y^2 + \Lambda^2(1 -
  y))^3}\Big]=  (\lambda +
1)\frac{\alpha}{8\pi}\,[\bar{u}_p\gamma^{\alpha}(1 +
  \gamma^5)u_n][\bar{u}_e\gamma^{\beta}(1 -
  \gamma^5)v_{\bar{\nu}}]\Big\{\int^1_0dx \int^1_0 dy\, 2y\int
  \frac{d^4q}{\pi^2}\frac{g_{\alpha\beta}}{(q^2 + p^2(x)y^2)^2}\nonumber\\ 
\hspace{-0.3in}&& - \int^1_0dx\int^1_0 dy 2y^3\int
\frac{d^4q}{\pi^2}\frac{ p^2(x)g_{\alpha\beta} +
  4p_{\alpha}(x)p_{\beta}(x)}{(q^2 + p^2(x)y^2)^3} - \int^1_0dx \int^1_0 dy\, 2y\int
\frac{d^4q}{\pi^2}\frac{g_{\alpha\beta}}{(q^2 + p^2(x)y^2 + \Lambda^2
  (1 - y))^2}\nonumber\\
\hspace{-0.3in}&& + \int^1_0dx\int^1_0 dy 2y\int
\frac{d^4q}{\pi^2}\frac{ (p^2(x)y^2 + \Lambda^2(1 -
  y))\,g_{\alpha\beta}}{(q^2 +
  p^2(x)y^2 + \Lambda^2( 1- y))^3} + \int^1_0dx\int^1_0 dy 2y\int
\frac{d^4q}{\pi^2}\frac{4 y^2 p_{\alpha}(x)p_{\beta}(x)}{(q^2 +
  p^2(x)y^2 + \Lambda^2( 1- y))^3}\Big\}.
\end{eqnarray}
After the integration over $q$ the term $\delta^{(2)} {\cal
  M}^{(\gamma)}_{pe}$ is 
\begin{eqnarray}\label{labelD.32}
\hspace{-0.3in}&&\delta^{(2)} {\cal M}^{(\gamma)}_{pe} = (\lambda +
1)\frac{\alpha}{8\pi}\,[\bar{u}_p\gamma^{\alpha}(1 +
  \gamma^5)u_n][\bar{u}_e\gamma^{\beta}(1 -
  \gamma^5)v_{\bar{\nu}}] \Big\{g_{\alpha\beta}\int^1_0dx\int^1_0dy\,2y\,{\ell
  n}\Big[\frac{\Lambda^2 (1 - y)}{p^2(x) y^2}\Big] - 2\int^1_0
dx\,\frac{p_{\alpha}(x)p_{\beta}(x)}{p^2(x)}\Big\}.\nonumber\\
\hspace{-0.3in}&&
\end{eqnarray}
Keeping the leading terms in the large $m_p$ expansion, i.e. replacing
$p^2(x) = (k_e x - k_p(1-x))^2$ by $p^2(x) \to m^2_p (1- x)^2$, and
integrating over $y$ and $x$ we obtain
\begin{eqnarray}\label{labelD.33}
\hspace{-0.3in}\delta^{(2)} {\cal M}^{(\gamma)}_{pe} = (\lambda +
1)\frac{\alpha}{4\pi}\,[\bar{u}_p\gamma^{\alpha}(1 +
  \gamma^5)u_n][\bar{u}_e\gamma^{\beta}(1 - \gamma^5)v_{\bar{\nu}}]
\Big\{g_{\alpha\beta}\Big[ {\ell n}\Big(\frac{\Lambda}{m_p}\Big) +
  \frac{3}{4}\Big] - g_{o\alpha}g_{0\beta}\Big\}.
\end{eqnarray}
In the non--relativistic approximation for the proton 
Eq.(\ref{labelD.33}) takes the form
\begin{eqnarray}\label{labelD.34}
\hspace{-0.3in}&&\delta^{(2)} {\cal M}^{(\gamma)}_{pe} =
2m_n\frac{\alpha}{2\pi}\Big\{(\lambda + 1)\,\Big[\frac{1}{2}\,{\ell
    n}\Big(\frac{ \Lambda}{m_p}\Big) +
  \frac{1}{4}\Big]\Big\}\,[\varphi^{
    \dagger}_p\varphi_n][\bar{u}_e\gamma^0(1 -
  \gamma^5)v_{\bar{\nu}}]\nonumber\\
\hspace{-0.3in}&& + 2m_n\frac{\alpha}{2\pi}\Big\{ (\lambda +
1)\,\Big[-\frac{1}{2}\,{\ell n}\Big(\frac{ \Lambda}{m_p}\Big) -
  \frac{3}{4}\Big]\Big\}\,[\varphi^{
    \dagger}_p\vec{\sigma}\,\varphi_n]\cdot [\bar{u}_e\vec{\gamma}\,(1
  - \gamma^5)v_{\bar{\nu}}].
\end{eqnarray}
The sum of the contributions of $\delta^{(1)} {\cal
  M}^{(\gamma)}_{pe}$ and $\delta^{(2)} {\cal M}^{(\gamma)}_{pe}$ is
\begin{eqnarray}\label{labelD.35}
\hspace{-0.3in}\delta^{(1)} {\cal M}^{(\gamma)}_{pe} + \delta^{(2)}
       {\cal M}^{(\gamma)}_{pe} &=& 2m_n\Big\{\frac{\alpha}{2\pi}(\lambda +
1)\Big[- \frac{3}{2}{\ell n}\Big(\frac{\Lambda}{m_p}\Big) -
    \frac{3}{4}\Big]\Big\}\,[\varphi^{
    \dagger}_p\varphi_n][\bar{u}_e\gamma^0(1 -
  \gamma^5)v_{\bar{\nu}}]\nonumber\\
\hspace{-0.3in}&+& 2m_n\Big\{\frac{\alpha}{2\pi}(\lambda + 1)\Big[+
  \frac{3}{2}\,{\ell n}\Big(\frac{\Lambda}{m_p}\Big) +
  \frac{1}{4}\Big]\Big\}\,[\varphi^{
    \dagger}_p\vec{\sigma}\,\varphi_n]\cdot [\bar{u}_e\vec{\gamma}\,(1
  - \gamma^5)v_{\bar{\nu}}].
\end{eqnarray}
Now we have to calculate the contribution of the last term
$\delta^{(3)} {\cal M}^{(\gamma)}_{pe}$ to the amplitude $\delta {\cal
  M}^{(\gamma)}_{pe}$, given by
\begin{eqnarray}\label{labelD.36}
\hspace{-0.3in}&&\delta^{(3)} {\cal M}^{(\gamma)}_{pe} =  \frac{\alpha}{2\pi}\,\int
\frac{d^4q}{\pi^2i}\frac{1}{q^2 + i0}\frac{1}{q^2 - 2k_p\cdot q +
  i0}\frac{1}{q^2 + 2k_e\cdot q + i0}\nonumber\\
\hspace{-0.3in}&&\times\Big([\bar{u}_pi\sigma_{\alpha\beta}
  q^{\alpha}k^{\beta}_e\gamma^{\mu}(1 + \lambda
  \gamma^5)u_n][\bar{u}_e\gamma_{\mu}(1 -
  \gamma^5)v_{\bar{\nu}}] - [\bar{u}_p\gamma^{\mu}(1 + \lambda
  \gamma^5)u_n][\bar{u}_ei\sigma_{\alpha\beta}q^{\alpha}k^{\beta}_p\gamma_{\mu}(1
  - \gamma^5)v_{\bar{\nu}}]\Big).
\end{eqnarray}
Following the procedure expounded above, we reduce the r.h.s of
Eq.(\ref{labelD.36}) to the form
\begin{eqnarray}\label{labelD.37}
  \hspace{-0.3in}&&\delta^{(3)} {\cal M}^{(\gamma)}_{pe} =
\frac{\alpha}{2\pi}\Big\{[\bar{u}_pi\sigma_{\alpha\beta}
  k^{\alpha}_ek^{\beta}_p\gamma^{\mu}(1 + \lambda
  \gamma^5)u_n][\bar{u}_e\gamma_{\mu}(1 -
  \gamma^5)v_{\bar{\nu}}]\int^1_0\frac{dx\,(1 - x)}{p^2(x)}\nonumber\\
\hspace{-0.3in}&& - [\bar{u}_p\gamma^{\mu}(1 + \lambda
  \gamma^5)u_n][\bar{u}_ei
  \sigma_{\alpha\beta}k^{\alpha}_ek^{\beta}_p\gamma_{\mu}(1 -
  \gamma^5)v_{\bar{\nu}}]\int^1_0\frac{dx\,x}{p^2(x)}\Big\},
\end{eqnarray}
where $p^2(x) =(k_ex - k_p(1 - x))^2 = m^2_e x^2 + m^2_p (1 - x)^2 -
2m_em_p\gamma x(1 - x)$.

The calculation of the integrals over $x$ in Eq.(\ref{labelD.37}) has
been carried out in detail in Appendix C. Using these results we obtain
\begin{eqnarray}\label{labelD.38}
\hspace{-0.3in}&&\int^1_0\frac{dx}{p^2(x)} = -
\frac{1}{m_em_p}\,\frac{\sqrt{1 - \beta^2}}{2 \beta}\,{\ell
  n}\Big(\frac{1 + \beta}{1 - \beta}\Big),\nonumber\\
\hspace{-0.3in}&&\int^1_0\frac{dx\,x}{p^2(x)} = -
\frac{1}{m_em_p}\,\frac{\sqrt{1 - \beta^2}}{2 \beta}\,{\ell
  n}\Big(\frac{1 + \beta}{1 - \beta}\Big)\nonumber\\
\hspace{-0.3in}&& - \frac{1}{m^2_p}\Big[{\ell
    n}\Big(\frac{m_p}{m_e}\Big) - \frac{1}{2\beta}\,{\ell
    n}\Big(\frac{1 + \beta}{1 - \beta}\Big)\Big] + \ldots,\nonumber\\
 \hspace{-0.3in}&&\int^1_0\frac{dx\,(1 - x)}{p^2(x)} =
 \frac{1}{m^2_p}\Big[{\ell n}\Big(\frac{m_p}{m_e}\Big) -
   \frac{1}{2\beta}\,{\ell n}\Big(\frac{1 + \beta}{1 -
     \beta}\Big)\Big] + \ldots,
\end{eqnarray}
where the ellipses denote the contributions of the terms of higher
order in the large $m_p$ expansion.

Keeping the leading terms in the large $m_p$ expansion, the
contribution of  $\delta^{(3)} {\cal M}^{(\gamma)}_{pe}$ is
\begin{eqnarray}\label{labelD.39}
  \hspace{-0.3in}\delta^{(3)} {\cal M}^{(\gamma)}_{pe} =
  \frac{\alpha}{2\pi}\Big[\frac{\sqrt{1 - \beta^2}}{2\beta}\,{\ell
      n}\Big(\frac{1 + \beta}{1 - \beta}\Big)\Big]\,[\bar{u}_p\gamma^{\mu}(1 + \lambda
  \gamma^5)u_n][\bar{u}_e\frac{E_e - m_e\gamma^0}{m_e}\gamma_{\mu}(1 -
  \gamma^5)v_{\bar{\nu}}],\nonumber\\
\hspace{-0.3in}&&
\end{eqnarray}
where we have used the Dirac equation for the free electron
$\bar{u}_e(\vec{k}_e\cdot \vec{\gamma}\,) = \bar{u}_e(E_e\gamma^0 -
m_e)$. Thus $\delta^{(3)} {\cal M}^{(\gamma)}_{pe}$ is given by
\begin{eqnarray}\label{labelD.40}
  \hspace{-0.3in}\delta^{(3)} {\cal M}^{(\gamma)}_{pe} &=&
  \frac{\alpha}{2\pi}\Big[\frac{1}{2\beta}\,{\ell n}\Big(\frac{1 +
      \beta}{1 - \beta}\Big)\Big]\,[\bar{u}_p\gamma^{\mu}(1 + \lambda
  \gamma^5)u_n][\bar{u}_e\gamma_{\mu}(1
  - \gamma^5)v_{\bar{\nu}}]\nonumber\\
 \hspace{-0.3in}&-& \frac{\alpha}{2\pi}\Big[\frac{\sqrt{1 -
       \beta^2}}{2\beta}\,{\ell n}\Big(\frac{1 + \beta}{1 -
     \beta}\Big)\Big]\,[\bar{u}_p\gamma^{\mu}(1 + \lambda
   \gamma^5)u_n][\bar{u}_e\gamma^0\gamma_{\mu}(1 -
   \gamma^5)v_{\bar{\nu}}].
\end{eqnarray}
In the non--relativistic approximation for the proton
Eq.(\ref{labelD.40}) reads
\begin{eqnarray}\label{labelD.41}
  \hspace{-0.3in}&&\delta^{(3)} {\cal M}^{(\gamma)}_{pe} = 2 m_n\,
  \frac{\alpha}{2\pi}\,\frac{1}{2\beta}\,{\ell n}\Big(\frac{1 +
    \beta}{1 -
    \beta}\Big)\,\Big\{[\varphi^{\dagger}_p\varphi_n][\bar{u}_e\gamma^0
    (1 - \gamma^5)v_{\bar{\nu}}] - \lambda
       [\varphi^{\dagger}_p\vec{\sigma}\,\varphi_n]\cdot
       [\bar{u}_e\vec{\gamma}\,(1 -
         \gamma^5)v_{\bar{\nu}}]\Big\}\nonumber\\
\hspace{-0.3in}&& - 2m_n\,\frac{\alpha}{2\pi}\,\frac{\sqrt{1 -
    \beta^2}}{2\beta}\,{\ell n}\Big(\frac{1 + \beta}{1 -
  \beta}\Big)\,\Big\{[\varphi^{\dagger}_p\varphi_n][\bar{u}_e (1 -
  \gamma^5)v_{\bar{\nu}}] - \lambda
       [\varphi^{\dagger}_p\vec{\sigma}\,\varphi_n]\cdot
       [\bar{u}_e\gamma^0 \vec{\gamma}\,(1 -
         \gamma^5)v_{\bar{\nu}}]\Big\}.
\end{eqnarray}
The second term can be identified with the contributions of the scalar
and tensor lepton--nucleon weak interactions.  In order to show this
we use the Hamilton density operator of weak interactions, taken in
the following form \cite{SPT5}
\begin{eqnarray}\label{labelD.42}
\hspace{-0.3in}{\cal H}_W(x) &=&
\frac{G_F}{\sqrt{2}}\,V_{ud}\,\Big\{[\bar{\psi}_p(x)\gamma_{\mu}(1 +
  \lambda\gamma^5)\psi_n(x)]\, [\bar{\psi}_e(x)\gamma^{\mu}(1 -
  \gamma^5)\psi_{\nu_e}(x)]\nonumber\\
\hspace{-0.3in}&+& g_S\,[\bar{\psi}_p(x) \psi_n(x)][\bar{\psi}_e(x)(1
  - \gamma^5)\psi_{\nu_e}(x)] +
\frac{1}{2}\,g_T\,[\bar{\psi}_p(x)\sigma_{\mu\nu}\gamma^5\psi_n(x)]\,
     [\bar{\psi}_e(x) \sigma^{\mu\nu}(1 -
       \gamma^5)\psi_{\nu_e}(x)]\Big\},
\end{eqnarray}
where $g_S$ and $g_T$ are the constants of scalar and tensor weak
interactions and $\sigma_{\mu\nu} =
\frac{i}{2}(\gamma_{\mu}\gamma_{\nu} - \gamma_{\nu}\gamma_{\mu})$ is
the Dirac matrix.

In the rest frame of the neutron and in the non--relativistic
approximation for the proton the amplitude of the continuum-state
$\beta^-$--decay of the neutron takes the form \cite{SPT5}
\begin{eqnarray}\label{labelD.43}
\hspace{-0.3in}&&M(n \to p + e^- + \tilde{\nu}_e) = - 2m_n
\frac{G_F}{\sqrt{2}}\,V_{ud}\Big\{[\varphi^{\dagger}_p
  \varphi_n][\bar{u}_e\,(1 + g_S\gamma^0)\,\gamma^0\,(1 -
  \gamma^5)v_{\bar{\nu}}] + [\varphi^{\dagger}_p
  \vec{\sigma}\,\varphi_n]\cdot [\bar{u}_e (-\lambda + g_T \gamma^0
  )\,\vec{\gamma}\,(1 - \gamma^5)v_{\bar{\nu}}\Big\}.\nonumber\\
\hspace{-0.3in}&&
\end{eqnarray}
From the comparison Eq.(\ref{labelD.41}) with Eq.(\ref{labelD.43}) we
define the scalar and tensor coupling constants as 
\begin{eqnarray}\label{labelD.44}
  \hspace{-0.3in}g_S(E_e) &=& - \frac{\alpha}{2\pi}\,\frac{\sqrt{1 -
    \beta^2}}{2\beta}\,{\ell n}\Big(\frac{1 + \beta}{1 -
  \beta}\Big) = -  \frac{\alpha}{2\pi}\,g_F(E_e),\nonumber\\
\hspace{-0.3in}g_T(E_e) &=& +
\frac{\alpha}{2\pi}\,\lambda\,\frac{\sqrt{1 - \beta^2}}{2\beta}\,{\ell
  n}\Big(\frac{1 + \beta}{1 - \beta}\Big) = +
\frac{\alpha}{2\pi}\,\lambda\,g_F(E_e).
\end{eqnarray}
Using the results, obtained in \cite{SPT5}, we may define the
contribution of the electromagnetic Fierz term \cite{SPT5}
\begin{eqnarray}\label{labelD.45}
  \hspace{-0.3in}b^{(\rm em)}_F(E_e) = 2\,\frac{g_S(E_e) - 3 \lambda g_T(E_e)}{1
    + 3 \lambda^2} = - \frac{\alpha}{\pi}\,g_F(E_e),
\end{eqnarray}
induced by one--virtual photon exchanges. Summing up the contributions
of $\delta^{(j)} {\cal M}_{pe}$ with $j = 1,2,3$ we obtain the
following expression for $\delta {\cal M}^{(\gamma)}_{pe}$
\begin{eqnarray}\label{labelD.46}
\hspace{-0.3in}&&\delta {\cal M}^{(\gamma)}_{pe} = \delta^{(1)} {\cal
  M}^{(\gamma)}_{pe} + \delta ^{(2)} {\cal M}^{(\gamma)}_{pe} + \delta^{(3)}  {\cal
  M}^{(\gamma)}_{pe} = 2 m_n\,\frac{\alpha}{2\pi}\,\Big\{(\lambda +
1)\Big[- \frac{3}{2}{\ell n}\Big(\frac{\Lambda}{m_p}\Big) -
  \frac{3}{4}\Big] + \frac{1}{2\beta}\,{\ell n}\Big(\frac{1 + \beta}{1
  - \beta}\Big)\Big\}\nonumber\\
\hspace{-0.3in}&&\times\,[\varphi^{
    \dagger}_p\varphi_n][\bar{u}_e\gamma^0(1 - \gamma^5)v_{\bar{\nu}}]
+ 2m_n\,\frac{\alpha}{2\pi}\,\Big\{(\lambda + 1)\Big[+
  \frac{3}{2}\,{\ell n}\Big(\frac{\Lambda}{m_p}\Big) +
  \frac{1}{4}\Big] - \lambda\,\frac{1}{2\beta}\,{\ell n}\Big(\frac{1 +
  \beta}{1 - \beta}\Big)\Big\}[\varphi^{
    \dagger}_p\vec{\sigma}\,\varphi_n]\cdot [\bar{u}_e\vec{\gamma}\,(1
  - \gamma^5)v_{\bar{\nu}}]\nonumber\\
\hspace{-0.3in}&& - 2m_n\,\frac{\alpha}{2\pi} \,g_F(E_e)\,
\Big\{[\varphi^{\dagger}_p\varphi_n][\bar{u}_e (1 -
  \gamma^5)v_{\bar{\nu}}] - \lambda
      [\varphi^{\dagger}_p\vec{\sigma}\,\varphi_n]\cdot
      [\bar{u}_e\gamma^0 \vec{\gamma}\,(1 -
        \gamma^5)v_{\bar{\nu}}]\Big\}.
\end{eqnarray}
Summing up the contributions of Eq.(\ref{labelD.27}) and
Eq.(\ref{labelD.46}) we calculate ${\cal M}^{(\gamma)}_{pp} + {\cal
  M}^{(\gamma)}_{ee} + {\cal M}^{(\gamma)}_{pe}$. Using this
expression we get the amplitude of the continuum-state
$\beta^-$--decay of the neutron, calculated in the non--relativistic
approximation for the proton and taking into account the contributions
of one--virtual photon exchanges. We represent it in the form
\begin{eqnarray}\label{labelD.47}
\hspace{-0.3in}&&M(n \to p e^-\bar{\nu}_e) = -
2m_n\,\frac{G^{(r)}_F}{\sqrt{2}}V_{ud}\Big\{\Big(1 +
\frac{\alpha}{2\pi}\,f_{\beta^-_c}(E_e,\mu)\Big) [\varphi^{\dagger}_p
  \varphi_n][\bar{u}_e\,\gamma^0(1 -
  \gamma^5)v_{\bar{\nu}}]\nonumber\\ 
 \hspace{-0.3in}&& - \lambda^{(r)} \Big(1 +
 \frac{\alpha}{2\pi}\,f_{\beta^-_c}(E_e,\mu)\Big)[\varphi^{\dagger}_p
   \vec{\sigma}\,\varphi_n]\cdot [\bar{u}_e \vec{\gamma}\,(1 -
   \gamma^5)v_{\bar{\nu}}] -
 \frac{\alpha}{2\pi}\,g_F(E_e)\,[\varphi^{\dagger}_p
   \varphi_n][\bar{u}_e\,(1 - \gamma^5)v_{\bar{\nu}}]\nonumber\\
\hspace{-0.3in}&&+ \frac{\alpha}{2\pi}\,\lambda^{(r)}
g_F(E_e)[\varphi^{\dagger}_p \vec{\sigma}\,\varphi_n]\cdot [\bar{u}_e
  \gamma^0\vec{\gamma}\,(1 - \gamma^5)v_{\bar{\nu}}] \Big\}.
\end{eqnarray}
Here $G^{(r)}_F$ and $g^{(r)}_A$ are the renormalized Fermi and axial
coupling constants
\begin{eqnarray}\label{labelD.48}
\hspace{-0.3in}G^{(r)}_F = G_F\Big(1 +
\frac{\alpha}{2\pi}\,d_V\Big)\,,\,\lambda^{(r)} = \lambda \Big(1 +
\frac{\alpha}{2\pi}\,d_A\Big),
\end{eqnarray}
where $d_V$ and $d_A$ are ultra--violet divergent constants. In our
calculation of the radiative corrections they are equal to
\begin{eqnarray}\label{labelD.49}
\hspace{-0.3in}d_V &=& 3{\ell n}\Big(\frac{\Lambda}{m_p}\Big) -
\frac{1}{4} - c_S - (\lambda + 1)\Big[\frac{3}{2}{\ell
    n}\Big(\frac{\Lambda}{m_p}\Big) + \frac{3}{4}\Big],\nonumber\\
\hspace{-0.3in}d_A &=& - \frac{\lambda + 1}{\lambda}\Big[
  \frac{3}{2}\,{\ell n}\Big(\frac{\Lambda}{m_p}\Big) +
  \frac{1}{4}\Big] + (\lambda + 1)\Big[ \frac{3}{2}{\ell
    n}\Big(\frac{\Lambda}{m_p}\Big) + \frac{3}{4}\Big].
\end{eqnarray}
The function $f_{\beta^-_c}(E_e,\mu)$, given by
\begin{eqnarray}\label{labelD.50}
\hspace{-0.3in}&&f_{\beta^-_c}(E_e,\mu) = \frac{3}{2}{\ell
  n}\Big(\frac{m_p}{m_e}\Big) + c_S + 2 {\ell n}\Big(\frac{\mu}{
  m_e}\Big)\Big[\frac{1}{2\beta}\,{\ell n}\Big(\frac{1 +
    \beta}{1 - \beta}\Big) - 1 \Big] +
\frac{1}{\beta}L\Big(\frac{2\beta}{1 + \beta}\Big) - \frac{1}{4\beta}\,{\ell
  n}^2\Big(\frac{1 + \beta}{1 - \beta}\Big)
+ \frac{1}{2\beta}\,{\ell n}\Big(\frac{1 + \beta}{1 - \beta}\Big),\nonumber\\
\hspace{-0.3in}&&
\end{eqnarray}
contains two terms $(3/2){\ell n}(m_p/m_e)$ and $c_S$ independent of
the electron energy $E_e$.  As we show in Appendix E the term
$(3/2){\ell n}(m_p/m_e)$ is fixed by the KLN theorem.  In our
calculation the constant $c_S$ reflects an ambiguous decomposition of
the contribution of one--virtual photon exchanges to the amplitude of
the neutron $\beta^-$--decay into the renormalisation constant $d_V$
of the Fermi coupling constant and the radiative corrections to the
lifetime of the neutron. Nevertheless, following Sirlin \cite{RC8} and
Abers {\it et al.}  \cite{RC9} one can show that due to a requirement
of gauge invariance of the observable radiative corrections the value
of the constant $c_S$ is fixed and is equal to $c_S = - 11/8$. We show
this in Appendix F.

The contribution of the $W$--boson and $Z$--boson exchanges and the
QCD corrections \cite{RC1} we describe by the constant $C_{WZ}$
\cite{Gudkov2}. The numerical value $C_{WZ} = 10.249$ we discuss below
Eq.(\ref{labelD.57}). As a result the function
$f_{\beta^-_c}(E_e,\mu)$ is
\begin{eqnarray}\label{labelD.51}
\hspace{-0.3in}&&f_{\beta^-_c}(E_e,\mu) = \frac{3}{2}{\ell
  n}\Big(\frac{m_p}{m_e}\Big) - \frac{11}{8} + 2 {\ell
  n}\Big(\frac{\mu}{ m_e}\Big)\,\Big[\frac{1}{2\beta}\,{\ell n}\Big(\frac{1 +
    \beta}{1 - \beta}\Big) - 1 \Big] +
\frac{1}{\beta}L\Big(\frac{2\beta}{1 + \beta}\Big) - \frac{1}{4\beta}\,{\ell
  n}^2\Big(\frac{1 + \beta}{1 - \beta}\Big)
+ \frac{1}{2\beta}\,{\ell n}\Big(\frac{1 + \beta}{1 - \beta}\Big)\nonumber\\
\hspace{-0.3in}&& + C_{WZ}.
\end{eqnarray}
Making a replacement $G^{(r)}_F \to G_F$ and $\lambda^{(r)} \to
\lambda$ and taking into account the contributions of the ``weak
magnetism'' and the proton recoil the amplitude of the continuum-state
$\beta^-$--decay of the neutron takes the form
\begin{eqnarray}\label{labelD.52}
\hspace{-0.3in}&&M(n \to p\,e^- \,\bar{\nu}_e) = -
2m_n\,\frac{G_F}{\sqrt{2}}V_{ud} \Big\{\Big(1 +
\frac{\alpha}{2\pi}\,f_{\beta^-_c}(E_e,\mu)\Big)[\varphi^{\dagger}_p
  \varphi_n][\bar{u}_e\,\gamma^0(1 -
  \gamma^5)v_{\bar{\nu}}]\nonumber\\  
\hspace{-0.3in}&& - \tilde{\lambda} \Big(1 +
\frac{\alpha}{2\pi}\,f_{\beta^-_c}(E_e,\mu)\Big)[\varphi^{\dagger}_p
  \vec{\sigma}\,\varphi_n]\cdot [\bar{u}_e \vec{\gamma}\,(1 -
  \gamma^5)v_{\bar{\nu}}] -
\frac{\alpha}{2\pi}\,g_F(E_e)\,[\varphi^{\dagger}_p
  \varphi_n][\bar{u}_e\,(1 -
  \gamma^5)v_{\bar{\nu}}]\nonumber\\\hspace{-0.3in}&&+
\frac{\alpha}{2\pi}\,\tilde{\lambda} g_F(E_e) [\varphi^{\dagger}_p
  \vec{\sigma}\,\varphi_n]\cdot [\bar{u}_e \gamma^0\vec{\gamma}\,(1 -
  \gamma^5)v_{\bar{\nu}}] - \frac{m_e}{2
M}\,[\varphi^{\dagger}_p\varphi_n][\bar{u}_e\,(1 -
\gamma^5)v_{\bar{\nu}}]\nonumber\\
\hspace{-0.3in}&& + \frac{\tilde{\lambda}}{2
  M}[\varphi^{\dagger}_p(\vec{\sigma}\cdot \vec{k}_p) \varphi_n
]\,[\bar{u}_e\,\gamma^0 (1 - \gamma^5)v_{\bar{\nu}}] - i\,
\frac{\kappa + 1}{2 M} [\varphi^{\dagger}_p (\vec{\sigma}\times
  \vec{k}_p) \varphi_n] \cdot [\bar{u}_e\,\vec{\gamma}\,(1 -
  \gamma^5)v_{\bar{\nu}}] \Big\},
\end{eqnarray}
where $\tilde{\lambda} = \lambda (1 - E_0/2M)$ and $\vec{k}_p = -
\vec{k}_e - \vec{k}$ is the proton 3--momentum in the rest frame of
the neutron. The amplitude Eq.(\ref{labelD.52}) has been used for the
calculation of the electron--energy and angular distribution
Eq.(\ref{label6}).

The radiative corrections to the
rate of the continuum-state $\beta^-$--decay of the neutron, which we
denote as $g_{\beta^-_c}(E_e, \mu)$, acquire an additional
contribution of the electromagnetic Fierz term
\begin{eqnarray}\label{labelD.53}
\hspace{-0.3in}&&g_{\beta^-_c}(E_e, \mu) = f_{\beta^-_c}(E_e, \mu) +
\frac{\pi}{\alpha}\,b^{(\rm em]}_F(E_e)\,\frac{m_e}{E_e} =
f_{\beta^-_c}(E_e, \mu) - g_F(E_e)\,\frac{m_e}{E_e} = \frac{3}{2}{\ell
  n}\Big(\frac{m_p}{m_e}\Big) - \frac{11}{8}\nonumber\\
\hspace{-0.3in}&& + 2 {\ell n}\Big(\frac{\mu}{
  m_e}\Big)\Big[\frac{1}{2\beta}\,{\ell n}\Big(\frac{1 + \beta}{1 -
    \beta}\Big) - 1 \Big] + \frac{1}{\beta}L\Big(\frac{2\beta}{1 +
  \beta}\Big) - \frac{1}{4\beta}\,{\ell n}^2\Big(\frac{1 + \beta}{1 -
  \beta}\Big) + \frac{\beta}{2}\,{\ell n}\Big(\frac{1 + \beta}{1 -
  \beta}\Big) + C_{WZ},
\end{eqnarray}
where the term, proportional to $\beta/2$, is defined by
\begin{eqnarray}\label{labelD.54}
\hspace{-0.3in}\frac{\beta}{2}\,{\ell n}\Big(\frac{1 + \beta}{1 -
  \beta}\Big) = \frac{1}{2\beta}\,{\ell n}\Big(\frac{1 + \beta}{1 -
  \beta}\Big) - g_F(E_e)\,\frac{m_e}{E_e}.
\end{eqnarray}
For the calculation of the contributions of the effective scalar and
tensor interactions, induced by one--virtual photon exchanges, to the
correlation coefficients of the electron--energy and angular
distribution of the neutron $\beta^-$--decay we use the results
obtained in \cite{SPT5} (see Eq.(28) of Ref.\cite{SPT5}).
The corrections to the correlation coefficients from the scalar and
tensor lepton--nucleon weak interactions with the left--handed
neutrinos take the form
\begin{eqnarray}\label{labelD.55}
\delta \zeta(E_e) &=& b_F\,\frac{m_e}{E_e} = 2 \,\frac{g_S - 3 \lambda
  g_T}{1 + 3 \lambda^2 }\,\frac{m_e}{E_e},\nonumber\\ \delta a(E_e)
&=& 0,\nonumber\\ \delta A(E_e) &=& 0,\nonumber\\ \delta B(E_e) &=&
2\,\frac{(g_T - \lambda g_S) - 2 \lambda g_T}{1 + 3 \lambda^2
}\,\frac{m_e}{E_e}.
\end{eqnarray}
where we have kept only the linear terms in the scalar and tensor
coupling constant expansions. Replacing the scalar and tensor coupling
constants by their expressions, given in Eq.(\ref{labelD.44}), we obtain 
\begin{eqnarray}\label{labelD.55a}
\delta \zeta(E_e) &=& b_F\,\frac{m_e}{E_e} = 2 \,\frac{g_S - 3 \lambda
  g_T}{1 + 3 \lambda^2 }\,\frac{m_e}{E_e} = -
\frac{\alpha}{\pi}\,g_F\,\frac{m_e}{E_e},\nonumber\\ \delta a(E_e) &=&
0,\nonumber\\ \delta A(E_e) &=& 0,\nonumber\\ \delta B(E_e) &=&
2\,\frac{(g_T - \lambda g_S) - 2 \lambda g_T}{1 + 3 \lambda^2
}\,\frac{m_e}{E_e} = - B_0\,\frac{\alpha}{\pi}\,g_F\,\frac{m_e}{E_e}.
\end{eqnarray}
This gives the following radiative corrections to the correlation
coefficients
\begin{eqnarray}\label{labelD.56}
\hspace{-0.3in}&&\zeta(E_e) = \Big(1 +
\frac{\alpha}{\pi}\,g_n(E_e)\Big) + O(1/M),\nonumber\\
\hspace{-0.3in}&&a(E_e) = a_0\Big(1 +
\frac{\alpha}{\pi}\,f_n(E_e)\Big) + O(1/M),\nonumber\\
\hspace{-0.3in}&&A(E_e) = A_0 \Big(1 +
\frac{\alpha}{\pi}\,f_n(E_e)\Big) + O(1/M), \nonumber\\
\hspace{-0.3in}&& B(E_e) = B_0 + O(1/M),
\end{eqnarray}
where the terms $- g_Fm_e/E_e$ are absorbed by the function $g_n(E_e)$
(see Eq.(\ref{labelD.53})). The terms of order of $O(1/M)$ are adduced
in Eqs.(\ref{label9}) -- (\ref{label13}).

The functions $g_n(E_e)$ and  $f_n(E_e)$ are defined by
\begin{eqnarray}\label{labelD.57}
\hspace{-0.3in}&&g_n(E_e) = \lim_{\mu \to 0}[g_{\beta^-_c}(E_e,\mu) +
  g^{(1)}_{\beta^-_c\gamma} (E_e,\mu)] =\frac{3}{2}\,{\ell n}\Big(\frac{m_p}{m_e}\Big) -
\frac{3}{8} + 2\,\Big[\frac{1}{2\beta} \,{\ell n}\Big(\frac{1 +
    \beta}{1 - \beta}\Big) - 1\Big]\Big[{\ell n}\Big(\frac{2(E_0 -
    E_e)}{m_e}\Big)\nonumber\\
\hspace{-0.3in}&& - \frac{3} {2} + \frac{1}{3}\,\frac{E_0 -
  E_e}{E_e}\Big]+ \frac{2}{\beta}L\Big(\frac{2\beta}{1 + \beta}\Big) +
\frac{1}{2\beta}{\ell n}\Big(\frac{1 + \beta}{1 -
  \beta}\Big)\,\Big[(1+\beta^2) + \frac{1}{12} \frac{(E_0 -
    E_e)^2}{E^2_e} - {\ell n}\Big(\frac{1 + \beta}{1 -
    \beta}\Big)\Big] + C_{WZ}, \nonumber\\
\hspace{-0.3in}&&f_n(E_e) = \lim_{\mu \to
  0}[g^{(2)}_{\beta^-_c\gamma}(E_e,\mu) -
  g^{(1)}_{\beta^-_c\gamma}(E_e,\mu)] + g_F(E_e)\,\frac{m_e}{E_e} =
\frac{2}{3}\,\frac{E_0 - E_e}{E_e}\Big(1 + \frac{1}{8}\frac{E_0 -
  E_e}{E_e}\Big)\,\frac{1 - \beta^2}{\beta^2}\nonumber\\
\hspace{-0.3in}&&\times\, \Big[\frac{1}{2\beta}\,{\ell n}\Big(\frac{1
    + \beta}{1 - \beta}\Big) - 1\Big]- \frac{1}{12}\frac{(E_0 -
  E_e)^2}{E^2_e} + \frac{1 - \beta^2}{2\beta}\,{\ell n}\Big(\frac{1 +
  \beta}{1 - \beta}\Big),
\end{eqnarray}
where the functions $g^{(1)}_{\beta^-_c\gamma} (E_e,\mu)$ and
$g^{(2)}_{\beta^-_c\gamma} (E_e,\mu)$ are given in
Eq.(\ref{labelB.28}). 

The $g_n(E_e)$ and $f_n(E_e)$, multiplied by $\alpha/\pi$, are in
analytical agreement with results, obtained in \cite{RC8}--\cite{RC18}
and \cite{Gudkov1,Gudkov2}, respectively.  The constant $C_{WZ}$,
defined by the contributions of the $W$--boson and $Z$--boson
exchanges and the QCD corrections \cite{RC1}, is equal to $C_{WZ} =
10.249$. This numerical value is obtained from the fit of the
radiative corrections to the lifetime of the neutron
$(\alpha/\pi)\langle g_n(E_e)\rangle = 0.03886(39)$ \cite{Abele1} and
$(\alpha/\pi)\langle g_n(E_e)\rangle = 0.0390(8)$ \cite{RC1}, averaged
over the phase volume of the neutron decay.

The radiative corrections $(\alpha/ \pi)\,g_n(E_e)$ and  $(\alpha/
\pi)\,f_n(E_e)$, weighted with the
electron energy spectrum density
\begin{eqnarray}\label{labelD.58}
\hspace{-0.3in}\rho_{\beta^-_c}(E_e) = (E_0 - E_e)^2\, \sqrt{E^2_e -
  m^2_e}\, E_e\,\zeta(E_e)\,\frac{ F(E_e, Z = 1)}{f_n(E_0, Z = 1)},
\end{eqnarray}
where the functions $\zeta(E_e)$ and $F(E_e, Z = 1)$ are given in
Eq.(\ref{label7}) and Eq.(\ref{label5}), respectively, and $f_n(E_0, Z
= 1)$ is the Fermi integral Eq.(\ref{label36}), are plotted in
Fig.\,1.

Finally we would like to note that having attributed the terms,
proportional to $(\lambda + 1)$ to the renormalisation constants of
the Fermi and axial coupling constants only, we arrive at the
radiative corrections, described by the function
\begin{eqnarray}\label{labelD.59}
\hspace{-0.3in}g(E_e) &=& g_n(E_e) + 3{\ell
  n}\Big(\frac{\Lambda}{m_p}\Big) + \frac{9}{8} - C_{WZ} = \nonumber\\
\hspace{-0.3in}&=& 3{\ell n}\Big(\frac{\Lambda}{m_p}\Big) +
\frac{3}{2}\,{\ell n}\Big(\frac{m_p}{m_e}\Big) + \frac{3}{4} +
2\,\Big[\frac{1}{2\beta} \,{\ell n}\Big(\frac{1 + \beta}{1 -
    \beta}\Big) - 1\Big]\, \Big[{\ell n}\Big(\frac{2(E_0 -
    E_e)}{m_e}\Big)- \frac{3} {2} + \frac{1}{3}\,\frac{E_0 -
  E_e}{E_e}\Big]\nonumber\\
\hspace{-0.3in}&+& \frac{2}{\beta}L\Big(\frac{2\beta}{1 + \beta}\Big)
+ \frac{1}{2\beta}{\ell n}\Big(\frac{1 + \beta}{1 -
  \beta}\Big)\Big[(1+\beta^2) + \frac{1}{12} \frac{(E_0 -
    E_e)^2}{E^2_e} - {\ell n}\Big(\frac{1 + \beta}{1 -
    \beta}\Big)\Big].
\end{eqnarray}
 The function $g(E_e)$, multiplied by $\alpha/\pi$, agrees
 analytically with the result, calculated by Kinoshita and Sirlin
 \cite{RC5}.

\section*{Appendix E: Kinoshita-Lee-Nauenberg theorem for radiative 
corrections of neutron $\beta^-$--decay}
\renewcommand{\theequation}{E-\arabic{equation}}
\setcounter{equation}{0}

As has been shown by Kinoshita and Sirlin \cite{RC5}, the radiative
corrections to the rates of the muon decay $\mu^- \to e^- + \nu_{\mu}
+ \bar{ \nu}_e$ and of the neutron $\beta^-$--decay, taken in the
limit $m_e \to 0$ and integrated over the phase volume, does not
depend on the electron mass. This is so--called the
Kinoshita-Lee-Nauenberg (KLN) theorem \cite{KLN}. Thus, the radiative
corrections to the rate of the neutron decay, described by the
function $g_n(E_e)$, should obey the KLN theorem. This means that the
result of the integration of the function $g_n(E_e)$, taken in the
limit $m_e \to 0$, over the phase volume should not depend on $m_e$.

Sirlin's function $\bar{g}(E_e) = g_n(E_e) - C_{WZ}$, defining the
radiative corrections, caused by one--virtual photon exchanges, to the
lifetime of the neutron, takes the form \cite{RC8}
\begin{eqnarray}\label{labelE.1}
\hspace{-0.3in}&&\bar{g}(E_e) = g_n(E_e) - C_{WZ} = \frac{3}{2}\,{\ell
  n}\Big(\frac{m_p}{m_e}\Big) - \frac{3}{8} + 2\,\Big[\frac{1}{2\beta}
  \,{\ell n}\Big(\frac{1 + \beta}{1 - \beta}\Big) - 1\Big]\,\Big[{\ell
    n}\Big(\frac{2(E_0 - E_e)}{m_e}\Big) - \frac{3} {2} +
  \frac{1}{3}\,\frac{E_0 - E_e}{E_e}\Big]\nonumber\\
\hspace{-0.3in}&& + \frac{2}{\beta}L\Big(\frac{2\beta}{1 + \beta}\Big)
+ \frac{1}{2\beta}{\ell n}\Big(\frac{1 + \beta}{1 -
  \beta}\Big)\Big[(1+\beta^2) + \frac{1}{12} \frac{(E_0 -
    E_e)^2}{E^2_e} - {\ell n}\Big(\frac{1 + \beta}{1 -
    \beta}\Big)\Big].
\end{eqnarray}
Taking the limit $E_e \gg m_e$, corresponding to $\beta \to 1$, and
introducing the variable $x = E_e/E_0$ we transcribe the function
$g(E_e)$ into the form
\begin{eqnarray}\label{labelE.2}
\hspace{-0.3in}&&\bar{g}(E_e) \to \bar{g}(x,m_e) = \frac{3}{2}\,{\ell
  n}\Big(\frac{m_p}{2E_0}\Big) + \frac{3}{2}\,{\ell
  n}\Big(\frac{2E_0}{m_e}\Big) - \frac{3}{8} + 2\,\Big({\ell n}\,x + {\ell
  n}\Big(\frac{2E_0}{m_e}\Big) -1\Big)\,\Big[{\ell n}\Big(\frac{1 - x}{x}\Big) -
\frac{3} {2} + \frac{1}{3}\,\frac{1 - x}{x}\nonumber\\
\hspace{-0.3in}&&+ \Big({\ell n}\,x + {\ell
  n}\Big(\frac{2E_0}{m_e}\Big)\Big)\Big] + 2L(1) + \Big({\ell n}\,x +
       {\ell n}\Big(\frac{2E_0}{m_e}\Big)\Big) \Big[2 + \frac{1}{12}
         \frac{(1 - x)^2}{x^2} - 2\Big({\ell n}\,x + {\ell
           n}\Big(\frac{2E_0}{m_e}\Big)\Big)\Big].
\end{eqnarray}
Thus, the function $\bar{g}(x, m_e)$ is equal to
 \begin{eqnarray}\label{labelE.3}
\hspace{-0.3in}&&\bar{g}(x, m_e) = \frac{3}{2}\,{\ell
  n}\Big(\frac{m_p}{2E_0}\Big)- \frac{3}{8} - \frac{\pi^2}{3} + 2({\ell n}\,x -
1)\,\Big[{\ell n}\Big(\frac{1 - x}{x}\Big) -
  \frac{3}{2} + \frac{1}{3}\frac{1 - x}{x}\Big]\nonumber\\
\hspace{-0.3in}&& + \frac{1}{12}\,\frac{(1 - x)^2}{x^2}\,{\ell n}\,x+
       {\ell n}\Big(\frac{2E_0}{m_e}\Big)\Big[2{\ell n}\Big(\frac{1 -
           x}{x}\Big) - \frac{3}{2} + \frac{2}{3}\,\frac{1 - x}{x} +
         \frac{1}{12}\,\frac{(1 - x)^2}{x^2}\Big].
\end{eqnarray}
For the derivation Eq.(\ref{labelE.3}) we have used $L(1) = - \pi^2/6$
and the approximation
\begin{eqnarray}\label{labelE.4}
\,{\ell n}\Big(\frac{1 + \beta}{1 -
  \beta}\Big) \to 2{\ell n}\,x + 2{\ell n}\Big(\frac{2E_0}{m_e}\Big).
\end{eqnarray}
According to the KLN theorem \cite{KLN} (see also \cite{RC5}), the
function $\bar{g}(x, m_e)$, integrated over the phase volume, taken in
the limit $m_e \to 0$, should not depend on $m_e$. We would like to
emphasize that the quadratic terms ${\ell n}^2(2 E_0/m_e)$ are
cancelled in the function $\bar{g}(x, m_e)$ without integration over
the phase volume.

Now let check the contribution of the linear terms ${\ell n}(2
E_0/m_e)$. For this aim we have to integrate the function $g(x, m_e)$
over the phase volume. The integration of the function $g(x,m_e)$ over
the phase volume with a dimensionless element $(1 - x)^2 x^2 dx$,
obtained at $m_e \to 0$, gives the result independent of $m_e$, since
\begin{eqnarray}\label{labelE.5}
\hspace{-0.3in}\int^1_0\Big[2{\ell n}\Big(\frac{1 - x}{x}\Big) -
  \frac{3}{2} + \frac{2}{3}\,\frac{1 - x}{x} + \frac{1}{12}\,\frac{(1
    - x)^2}{x^2}\Big]\,(1 - x)^2 x^2\,dx = 0.
\end{eqnarray}
Thus, the term ${\ell n}(2E_0/m_e)$ vanishes. This reproduces the
results, obtained in \cite{RC5}, and confirms the KLN theorem
\cite{KLN}.

We would like to note that the term $- 3/2$ in the integrand of
Eq.(\ref{labelE.5}), playing an important role for the vanishing of the
integral, is given by $- 3/2 = -3 + 3/2$, where $-3$ and $+3/2$ come
from the energy depending part of the function $\bar{g}(E_e)$ and the term
$(3/2){\ell n}(m_p/m_e)$, respectively.

\section*{Appendix F: Comparison with Sirlin's calculation of 
radiative corrections \cite{RC8}}
\renewcommand{\theequation}{F-\arabic{equation}}
\setcounter{equation}{0}

In this Appendix we compare our calculation of the radiative
corrections to the continuum-state $\beta^-$--decay of the neutron,
caused by one--virtual photon exchanges, with the calculation, carried
out by Sirlin in his well--known paper \cite{RC8}.

According to Sirlin \cite{RC8}, the radiative corrections to the
amplitude of the continuum-state $\beta^-$--decay of the neutron are
defined by Eqs.(9a), (11), (14) and (19) of Ref.\cite{RC8}. Now let us
compare our expressions with Sirlin's ones.

Sirlin's Eq.(9a) corresponds to our amplitude ${\cal M}_{pe}$. In
order to show this we rewrite the amplitude ${\cal M}_{pe}$, given by
Eq.(\ref{labelD.2}), as follows
\begin{eqnarray}\label{labelF.1}
\hspace{-0.3in}{\cal M}^{(\gamma)}_{pe} = - \frac{\alpha}{4\pi}
\int\frac{d^4q}{\pi^2i}\,D_{\alpha\beta}(q)\,[\bar{u}_e\gamma^{\alpha}\frac{1}{m_e
    - \hat{k}_e - \hat{q} -i0} O_{\mu} v_{\bar{\nu}}]\,[\bar{u}_p
  \gamma^{\beta} \frac{1}{m_p - \hat{k}_p + \hat{q} -i0}\,W^{\mu}
  u_n],
\end{eqnarray}
where $D_{\alpha\beta}(q)$ is the photon Green function 
\begin{eqnarray}\label{labelF.2}
\hspace{-0.3in}D_{\alpha\beta} (q) = \frac{1}{q^2 +
  i0}\,\Big(g_{\alpha\beta} - \xi\,\frac{q_{\alpha}q_{\beta}}{q^2 +
  i0}\Big)
\end{eqnarray}
in the arbitrary gauge with a gauge parameter $\xi$, $W^{\mu}$ and
$O_{\mu}$ are the products of the Dirac matrices defined by
\begin{eqnarray}\label{labelF.3}
W^{\mu} = \gamma^{\mu}(1 + \lambda \gamma^5) \quad,\quad  O_{\mu} =
\gamma_{\mu} (1 - \gamma^5).
\end{eqnarray}
After some algebraical transformations Eq.(\ref{labelF.1}) can be
reduced to Sirlin's form
\begin{eqnarray}\label{labelF.4}
\hspace{-0.3in}{\cal M}^{(\gamma)}_{pe} = - \frac{\alpha}{4\pi}
\int\frac{d^4q}{\pi^2i}\,D_{\alpha\beta}(q)\,\frac{[\bar{u}_e(2
    k^{\alpha}_e + \gamma^{\alpha} \hat{q})O_{\mu}
    v_{\bar{\nu}}]}{q^2 + 2 k_e \cdot q + i0}\,\bar{u}_p \Big[\frac{(2 k^{\beta}_p -
    q^{\beta})W^{\mu}}{q^2 - 2 k_p\cdot q + i0} +
T^{\beta\mu}\Big]u_n,
\end{eqnarray}
where $T^{\beta\mu}$ is given by
\begin{eqnarray}\label{labelF.5}
\hspace{-0.3in}T^{\beta\mu} = \frac{R^{\beta\mu}}{q^2 - 2 k_p\cdot q +
  i0}
\end{eqnarray}
and $R^{\beta\mu}$ is equal to $R^{\beta\mu} = i \sigma^{\beta\lambda}
q_{\lambda} W^{\mu}$. As has been pointed out by Sirlin \cite{RC8},
the tensor $T^{\beta\mu}$ is obviously transverse.
$q_{\beta}T^{\beta\mu} = 0$. Then, we propose to rewrite the amplitude
Eq.(\ref{labelF.4}) as follows
\begin{eqnarray}\label{labelF.6}
\hspace{-0.3in}&&{\cal M}^{(\gamma)}_{pe} = {\cal M}^{(\rm SC)}_{pe} +
\delta {\cal M}^{(\rm SLI)}_{pe},
\end{eqnarray}
where the indices $(\rm SC)$ and $(\rm SLI)$ mean ``Sirlin's
Corrections'' and ``Strong low--energy interactions''. The amplitudes
${\cal M}^{(\rm SC)}_{pe}$ and $\delta {\cal M}^{(\rm SLI)}_{pe}$ are
equal to
\begin{eqnarray}\label{labelF.7}
\hspace{-0.3in}&&{\cal M}^{(\rm SC)}_{pe} = - \frac{\alpha}{4\pi}
\int\frac{d^4q}{\pi^2i}\,D_{\alpha\beta}(q)\,\frac{[\bar{u}_e(2
    k^{\alpha}_e + \gamma^{\alpha} \hat{q})O_{\mu}
    v_{\bar{\nu}}]}{q^2 + 2 k_e \cdot q + i0}\,\frac{[\bar{u}_p(2 k^{\beta}_p -
    q^{\beta})W^{\mu}u_n]}{q^2 - 2 k_p\cdot q + i0},\nonumber\\
\hspace{-0.3in}&&\delta {\cal M}^{(\rm SLI)}_{pe} = -
\frac{\alpha}{4\pi}
\int\frac{d^4q}{\pi^2i}\,D_{\alpha\beta}(q)\,\frac{[\bar{u}_e(2
    k^{\alpha}_e + \gamma^{\alpha} \hat{q})O_{\mu}
    v_{\bar{\nu}}]}{q^2 + 2 k_e \cdot q + i0}\,[\bar{u}_pT^{\beta\mu}u_n].
\end{eqnarray}
The contribution of Sirlin's Eq.(11) coincides with the part of the
amplitude ${\cal M}^{(\gamma)}_{ee}$, defining renormalisation of the
wave function of the electron by electromagnetic interactions. The
amplitude ${\cal M}^{(\gamma)}_{ee}$, given by Eq.(\ref{labelD.2}), we
rewrite as
\begin{eqnarray}\label{labelF.8}
{\cal M}^{(\gamma)}_{ee} = {\cal M}^{(\rm SC)}_{ee} + \delta {\cal
  M}^{(\rm SLI)}_{ee},
\end{eqnarray}
where the amplitudes ${\cal M}^{(\rm SC)}_{ee}$ and $\delta {\cal
  M}^{(\rm SLI)}_{ee}$ are equal to
\begin{eqnarray}\label{labelF.9}
\hspace{-0.3in}&&{\cal M}^{(\rm SC)}_{ee} = -\,\frac{\alpha}{8\pi
  m_e}\, [\bar{u}_pW^{\mu}u_n]
\int\frac{d^4q}{\pi^2i}\,D_{\alpha\beta}(q)\,\frac{[\bar{u}_e(2
    k^{\alpha}_e + \gamma^{\alpha} \hat{q})\hat{k}_e(2 k^{\beta}_e +
    \hat{q}\gamma^{\beta} ) O_{\mu} v_{\bar{\nu}}]}{(q^2 + 2 k_e \cdot
  q + i0)^2}\nonumber\\
\hspace{-0.3in}&&\delta {\cal M}^{(\rm SLI)}_{ee} =
\frac{\alpha}{4\pi}\,[\bar{u}_p W^{\mu}u_n]\int\frac{d^4q}{\pi^2
  i}\,D_{\alpha\beta}(q)\,\Big[\bar{u}_e\gamma^{\alpha}\frac{1}{m_e -
    \hat{k}_e - \hat{q} -i0}\gamma^{\beta}\frac{1}{m_e - \hat{k}_e -
    i0} O_{\mu} v_{\bar{\nu}}\Big]\nonumber\\
\hspace{-0.3in}&& + [\bar{u}_p W^{\mu} u_n]\Big[\bar{u}_e\Big( - \delta
  m_e + \frac{Z^{(e)}_2 - 1}{2}(m_e - \hat{k}_e)\Big)\,\frac{1}{m_e - \hat{k}_e - i0} O_{\mu}
v_{\bar{\nu}}\Big]\nonumber\\
\hspace{-0.3in}&& + \frac{\alpha}{8\pi m_e}\,[\bar{u}_pW^{\mu}u_n]
\int\frac{d^4q}{\pi^2i}\,D_{\alpha\beta}(q)\,\frac{[\bar{u}_e(2
    k^{\alpha}_e + \gamma^{\alpha} \hat{q})\hat{k}_e(2 k^{\beta}_e +
    \hat{q}\gamma^{\beta}) O_{\mu} v_{\bar{\nu}}]}{(q^2 + 2 k_e \cdot
  q + i0)^2 },
\end{eqnarray}
respectively. For the derivation of Sirlin's term ${\cal M}^{(\rm
  SC)}_{ee}$ one has to use the definition of the renormalisation
constant $Z^{(e)}_2 - 1$ of the electron wave function \cite{RC9,IZ80}
\begin{eqnarray}\label{labelF.10}
\hspace{-0.3in}&&\frac{Z^{(e)}_2 - 1}{2} = - \frac{\alpha}{8\pi}
\frac{k^{\lambda}_e}{m_e}\int\frac{d^4q}{\pi^2i} D_{\alpha\beta}(q)
\frac{\partial}{\partial k^{\lambda}}\gamma^{\alpha}\frac{1}{m_e -
  \hat{k}_e - \hat{q} }\gamma^{\beta}\Big|_{\hat{k}_e = m_e}
\!\!\!\!\!= - \frac{\alpha}{8\pi}
\frac{k^{\lambda}_e}{m_e}\int\frac{d^4q}{\pi^2i}
D_{\alpha\beta}(q)\,\frac{\partial}{\partial q^{\lambda}}
\gamma^{\alpha}\frac{1}{m_e - \hat{k}_e - \hat{q}
}\gamma^{\beta}\Big|_{\hat{k}_e = m_e} \!\!\!\!\!= \nonumber\\
\hspace{-0.3in}&&= - \frac{\alpha}{8\pi m_e} \int\frac{d^4q}{\pi^2i}
D_{\alpha\beta}(q)\,\frac{(2k^{\alpha}_e +
  \gamma^{\alpha}\hat{q})\hat{k}_e(2 k^{\beta}_e +
  \hat{q}\gamma^{\beta})}{(q^2 + 2 k_e\cdot q + i
  0)^2}\Big|_{\hat{k}_e = m_e}.
\end{eqnarray}
In comparison with Sirlin's expression, we have taken away the
operator $(\hat{k}_e + m_e)/2m_e$, which is equal to unity at
$\hat{k}_e = m_e$.

As has been pointed out by Sirlin, Eq.(14) of Ref.\cite{RC8} is
related to the emission and absorption of a photon by the proton. This
means that Eq.(14) of Ref.\cite{RC8} should be a part of the amplitude
${\cal M}^{(\gamma)}_{pp}$. This allows us to represent the amplitude
${\cal M}^{(\gamma)}_{pp}$ in the form
\begin{eqnarray}\label{labelF.11}
{\cal M}^{(\gamma)}_{pp} = {\cal M}^{(\rm SC)}_{pp} + \delta {\cal
  M}^{(\rm SLI)}_{pp}
\end{eqnarray}
with ${\cal M}^{(\rm SC)}_{pp}$ and $\delta {\cal M}^{(\rm SLI)}_{pp}$,
given by
\begin{eqnarray}\label{labelF.12}
\hspace{-0.3in}&&{\cal M}^{(\rm SC)}_{pp} =
-\,\frac{\alpha}{8\pi}\,[\bar{u}_pW^{\mu}u_n][\bar{u}_e O_{\mu}
  v_{\bar{\nu}}]
\int\frac{d^4q}{\pi^2i}\,D_{\alpha\beta}(q)\frac{(2k_p -
  q)^{\alpha}(2k_p - q)^{\beta}}{(q^2 - 2 k_p \cdot q +
  i0)^2}\nonumber\\
\hspace{-0.3in}&&\delta {\cal M}^{(\rm SLI)}_{pp} =
\frac{\alpha}{4\pi}\int\frac{d^4q}{\pi^2
  i}\,D_{\alpha\beta}(q)\,[\bar{u}_p\gamma^{\alpha}\frac{1}{m_p -
    \hat{k}_p + \hat{q} -i0}\gamma^{\beta}\,\frac{1}{m_p - \hat{k}_p - i0} W^{\mu}
u_n][\bar{u}_e O_{\mu}v_{\bar{\nu}}]\nonumber\\
\hspace{-0.3in}&&+ [\bar{u}_p\Big( - \delta m_p  + \frac{Z^{(p)}_2 - 1}{2}(m_p -
\hat{k}_p)\Big)\,\frac{1}{m_p - \hat{k}_p - i0} W^{\mu}u_n][\bar{u}_e
  O_{\mu}v_{\bar{\nu}}]\nonumber\\
\hspace{-0.3in}&& +
\frac{\alpha}{8\pi}\,[\bar{u}_pW^{\mu}u_n][\bar{u}_e O_{\mu}
  v_{\bar{\nu}}] \int\frac{d^4q}{\pi^2i}\,D_{\alpha\beta}(q)
\frac{(2k_p - q)^{\alpha}(2k_p - q)^{\beta}}{(q^2 - 2 k_p \cdot q +
  i0)^2},
\end{eqnarray}
respectively. As a result, according to Sirlin \cite{RC8}, the
observable radiative corrections to the amplitude of the
continuum-state $\beta^-$--decay of the neutron, caused by
one--virtual photon exchange, are defined by the amplitude
\begin{eqnarray}\label{labelF.13}
{\cal M}^{(\rm SC)}_{\rm RC} = {\cal M}^{(\rm SC)}_{pe} + {\cal
  M}^{(\rm SC)}_{ee} + {\cal M}^{(\rm SC)}_{pp},
\end{eqnarray}
which is gauge invariant, i.e.  invariant under the gauge
transformation of the photon Green function $D_{\alpha\beta}(q) \to
D_{\alpha\beta}(q) + c(q^2)\,q_{\alpha}q_{\beta}$ with an arbitrary
function $c(q^2)$, and suffers from the infrared divergences only
\cite{RC8}.  The subscript ${\rm RC}$ means ``Radiative
Corrections''. As we show below the additional contribution, described
by the amplitude
\begin{eqnarray}\label{labelF.14}
\hspace{-0.3in}\delta {\cal M}^{(\rm SLI)}_{\rm RC} = \delta {\cal
  M}^{(\rm SLI)}_{pe} + \delta {\cal M}^{(\rm SLI)}_{ee} + \delta
       {\cal M}^{(\rm SLI)}_{pp},
\end{eqnarray}
does not depend on the electron energy and should be absorbed by
renormalisation of the Fermi coupling constant $G_F$ and axial
coupling constant $\lambda$.

Since the amplitude ${\cal M}^{(\rm SC)}_{\rm RC}$ is invariant under
gauge transformations of the photon Green function $D_{\alpha\beta}(q)
\to D_{\alpha\beta}(q) + c(q^2)\,q_{\alpha}q_{\beta}$, we may
calculate it by using the Feynman gauge for the photon Green
function. The calculation of the amplitude $\delta {\cal
  M}^{(\rm SLI)}_{\rm RC}$, describing the contributions to
renormalisation constants of the Fermi and axial coupling constants,
can be also carried out in the Feynman gauge.

The amplitude ${\cal M}^{(\rm SC)}_{pe}$ can be rewritten as follows 
\begin{eqnarray}\label{labelF.15}
\hspace{-0.3in}&& {\cal M}^{(\rm SC)}_{pe} =
\frac{\alpha}{4\pi}\,[\bar{u}_p W^{\mu} u_n]\Big\{[\bar{u}_e O_{\mu}
  v_{\bar{\nu}}]\Big[\int \frac{d^4q}{\pi^2i}\frac{1}{q^2 +
    i0}\frac{1}{q^2 - 2k_p\cdot q + i0} + \int \frac{d^4q}{\pi^2i}\frac{1}{q^2 +
  i0}\frac{1}{q^2 + 2k_e\cdot q + i0}\nonumber\\
\hspace{-0.3in}&&- \int \frac{d^4q}{\pi^2i}\frac{1}{q^2 - 2k_p\cdot q
  + i0}\frac{1}{q^2 + 2k_e\cdot q + i0} - 4(k_e\cdot k_p)\int
\frac{d^4q}{\pi^2i}\frac{1}{q^2 + i0}\frac{1}{q^2 - 2k_p\cdot q +
  i0}\frac{1}{q^2 + 2k_e\cdot q + i0}\Big]\nonumber\\
\hspace{-0.3in}&& - 2i\int \frac{d^4q}{\pi^2i}\frac{1}{q^2 +
  i0}\frac{1}{q^2 - 2k_p\cdot q + i0}\frac{1}{q^2 + 2k_e\cdot q +
  i0}\,[\bar{u}_e  \sigma_{\alpha\beta}q^{\alpha}k^{\beta}_p
  O_{\mu} v_{\bar{\nu}}]\Big\}.
\end{eqnarray}
All momentum integrals in Eq.(\ref{labelF.15}) are calculated in
Appendix D by using the Pauli--Villars regularization of the
ultra--violet divergent integrals and the FPM regularization for the
infrared divergent contributions.  Using the results, obtained in
Appendix C and Appendix D (see Eqs.(\ref{labelC.15}),
(\ref{labelD.12}) and (\ref{labelD.40})), the amplitude $ {\cal
  M}^{(\rm SC)}_{pe}$ takes the form
\begin{eqnarray}\label{labelF.16}
\hspace{-0.3in}&& {\cal M}^{(\rm SC)}_{pe} =[\bar{u}_p W^{\mu} u_n]
\frac{\alpha}{2\pi}\,\Big\{[\bar{u}_e O_{\mu}
  v_{\bar{\nu}}]\,\Big[{\ell n}\Big(\frac{\Lambda}{m_e}\Big) +
  \frac{1}{2} + {\ell
  n}\Big(\frac{\mu}{m_e}\Big)\,\frac{1}{\beta}\,{\ell n}\Big(\frac{1 +
  \beta}{1 - \beta}\Big) + \frac{1}{\beta}\,L\Big(\frac{2\beta}{1 +
  \beta}\Big)\nonumber\\
\hspace{-0.3in}&& - \frac{1}{4\beta}\,{\ell n}^2\Big(\frac{1 +
  \beta}{1 - \beta}\Big) + \frac{1}{2\beta}\,{\ell n}\Big(\frac{1 +
  \beta}{1 - \beta}\Big)\Big] - [\bar{u}_e \gamma^0 O_{\mu}
  v_{\bar{\nu}}]\,\frac{\sqrt{1 - \beta^2}}{2\beta}\,{\ell
  n}\Big(\frac{1 + \beta}{1 - \beta}\Big)\Big\}.
\end{eqnarray}
The amplitude ${\cal M}^{(\rm SC)}_{ee} $ can be transcribed into the
form
\begin{eqnarray}\label{labelF.17}
\hspace{-0.3in}&&{\cal M}^{(\rm SC)}_{ee} = -\,\frac{\alpha}{8\pi
  m_e}\,[\bar{u}_p W^{\mu}u_n]
\int\frac{d^4q}{\pi^2i}\,D_{\alpha\beta}(q)\,\frac{[\bar{u}_e(2
    k^{\alpha}_e + \gamma^{\alpha} \hat{q})\hat{k}_e(2 k^{\beta}_e +
    \hat{q}\gamma^{\beta} )O_{\mu} v_{\bar{\nu}}]}{(q^2 + 2 k_e \cdot
  q + i0)^2 }\nonumber\\
\hspace{-0.3in}&&= -\,\frac{\alpha}{4\pi m_e}[\bar{u}_pW^{\mu}u_n]
\int\frac{d^4q}{\pi^2i}\,\frac{1}{q^2 + i0}\,\frac{[\bar{u}_e(2m^3_e +
    2m_e \hat{q} \hat{k}_e - \hat{q}
    \hat{k}_e\hat{q})O_{\mu}v_{\bar{\nu}}]}{(q^2 + 2 k_e \cdot q +
  i0)^2} \to -\,\frac{\alpha}{4\pi m_e}[\bar{u}_pW^{\mu}u_n]\nonumber\\
\hspace{-0.3in}&&\times \int^1_0 dx\, 2 x \int \frac{d^4q}{\pi^2
  i}\,\Big\{\frac{[\bar{u}_e(2m^3_e + 2m_e \hat{q} \hat{k}_e - \hat{q}
    \hat{k}_e\hat{q})O_{\mu}v_{\bar{\nu}}]}{((q + k_e x)^2 - m^2_e x^2
  - \mu^2 (1-x))^3} - \frac{[\bar{u}_e(2m^3_e + 2m_e \hat{q} \hat{k}_e
    - \hat{q} \hat{k}_e\hat{q})O_{\mu}v_{\bar{\nu}}]}{((q + k_e x)^2 -
  m^2_e x^2 - \Lambda^2 (1-x))^3}\Big\} =\nonumber\\
\hspace{-0.3in}&&= \frac{\alpha}{8\pi
}[\bar{u}_pW^{\mu}u_n][\bar{u}_eO_{\mu}v_{\bar{\nu}}] \int^1_0 dx\,
2 x \int \frac{d^4q}{\pi^2}\Big[\frac{4m^2_e(1 - x - x^2) - q^2 }{(q^2 +
    m^2_e x^2 + \mu^2 (1-x))^3} - \frac{m^2_e(4 - 4x - 2x^2) - q^2 }{(q^2 + m^2_e x^2
  + \Lambda^2 (1-x))^3}\Big]=\nonumber\\
\hspace{-0.3in}&& = \frac{\alpha}{8\pi
}[\bar{u}_pW^{\mu}u_n][\bar{u}_eO_{\mu}v_{\bar{\nu}}]\int^1_0 dx\, 2 x \int
\frac{d^4q}{\pi^2}\,\Big[\frac{m^2_e(4 - 4 x - x^2)}{(q^2 + m^2_e x^2 +
    \mu^2 (1-x))^3} - \frac{m^2_e(4 - 4 x - x^2) + \Lambda^2(1 -
  x)}{(q^2 + m^2_e x^2 + \Lambda^2 (1-x))^3} \nonumber\\
\hspace{-0.3in}&&\times - \frac{1}{(q^2 + m^2_e x^2 + \mu^2 (1-x))^2}
+ \frac{1}{(q^2 + m^2_e x^2 + \Lambda^2 (1-x))^2}\Big] = \frac{\alpha}{16\pi
}[\bar{u}_pW^{\mu}u_n][\bar{u}_eO_{\mu}v_{\bar{\nu}}]\nonumber\\
\hspace{-0.3in}&&\times \int^1_0 dx\, 2 x \,\Big\{\frac{m^2_e(4 - 4 x
  - x^2)}{m^2_e x^2 + \mu^2 (1-x)}- \frac{m^2_e(4 - 4 x - x^2) +
  \Lambda^2(1 - x)}{m^2_e x^2 + \Lambda^2 (1-x)} - 2\, {\ell
  n}\Big[\frac{\Lambda^2 (1 - x)}{m^2_e x^2}\Big]\Big\}.
\end{eqnarray}
Having integrated over $x$ we arrive at the following expression for
the amplitude ${\cal M}^{(\rm SC)}_{ee}$
\begin{eqnarray}\label{labelF.18}
\hspace{-0.3in}{\cal M}^{(\rm SC)}_{ee} =
       [\bar{u}_pW^{\mu}u_n][\bar{u}_eO_{\mu}v_{\bar{\nu}}]\, \frac{\alpha}{2\pi}\Big[ -
  \frac{1}{2}\,{\ell n}\Big(\frac{\Lambda}{m_e}\Big) - {\ell
    n}\Big(\frac{\mu}{m_e}\Big) - \frac{9}{8}\Big].
\end{eqnarray}
In the Feynman gauge the amplitude ${\cal M}^{(\rm SC)}_{pp}$ takes the
form
\begin{eqnarray*}
\hspace{-0.3in}&&{\cal M}^{(\rm SC)}_{pp} =
-\,\frac{\alpha}{8\pi}\,[\bar{u}_pW^{\mu}u_n][\bar{u}_e O_{\mu}
  v_{\bar{\nu}}]\int\frac{d^4q}{\pi^2i}\,\frac{1}{q^2 +
  i0}\,\frac{4m^2_p - 4 (k_p\cdot q) + q^2}{(q^2 - 2 k_p \cdot q +
  i0)^2} \to -\,
\frac{\alpha}{8\pi}\,[\bar{u}_pW^{\mu}u_n][\bar{u}_e O_{\mu}
  v_{\bar{\nu}}] \nonumber\\
\hspace{-0.3in}&&\times \int^1_0
dx\,2x\,\int\frac{d^4q}{\pi^2i}\,\Big[\frac{4m^2_p - 4 (k_p\cdot q) +
    q^2}{((q - k_px)^2 - m^2_p x^2 - \mu^2 (1 - x))^3} - \frac{4m^2_p
    - 4 (k_p\cdot q) + q^2}{((q - k_px)^2 - m^2_p x^2 - \Lambda^2 (1 -
    x))^3}\Big] =\nonumber\\
\hspace{-0.3in}&&= 
\frac{\alpha}{8\pi}\,[\bar{u}_pW^{\mu}u_n][\bar{u}_e O_{\mu}
  v_{\bar{\nu}}]\int^1_0
dx\,2x\,\int\frac{d^4q}{\pi^2}\,\Big[\frac{m^2_p(4 - 4x + x^2) -
    q^2}{(q^2 + m^2_p x^2 + \mu^2 (1 - x))^3}- \frac{m^2_p(4 - 4 x + x^2) - q^2}{(q^2 +
  m^2_p x^2 + \Lambda^2 (1 - x))^3}\Big]= \nonumber\\
\hspace{-0.3in}&&=
\frac{\alpha}{8\pi}\,[\bar{u}_pW^{\mu}u_n][\bar{u}_e O_{\mu}
  v_{\bar{\nu}}]\int^1_0
dx\,2x\,\int\frac{d^4q}{\pi^2}\,\Big[\frac{m^2_p(4 - 4x + 2x^2)}{(q^2
    + m^2_p x^2 + \mu^2 (1 - x))^3} - \frac{m^2_p(4 - 4 x + 2x^2) + \Lambda^2(1 -
  x)}{(q^2 + m^2_p x^2 + \Lambda^2 (1 - x))^3}\nonumber\\
\end{eqnarray*}
\begin{eqnarray}\label{labelF.19}
\hspace{-0.3in}&& - \frac{1}{(q^2 + m^2_p x^2 + \mu^2 (1 - x))^2} +
\frac{1}{(q^2 + m^2_p x^2 + \Lambda^2 (1 - x))^2}\Big] = 
\frac{\alpha}{8\pi}\,[\bar{u}_pW^{\mu}u_n][\bar{u}_e O_{\mu}
  v_{\bar{\nu}}]\int^1_0 dx\,2x\,\Big\{\frac{m^2_p(2 - 2x +
  x^2)}{m^2_p x^2 + \mu^2 (1 - x)}\nonumber\\
\hspace{-0.3in}&& - \frac{1}{2} - {\ell
    n}\Big[\frac{\Lambda^2 (1 - x)}{m^2_p x^2}\Big]\Big\}.
\end{eqnarray}
Using the results, obtained in Appendix D, we define ${\cal M}^{(\rm
  SC)}_{pp}$ as follows
\begin{eqnarray}\label{labelF.20}
\hspace{-0.3in}{\cal M}^{(\rm SC)}_{pp} =
       [\bar{u}_pW^{\mu}u_n][\bar{u}_e O_{\mu}
         v_{\bar{\nu}}]\,\frac{\alpha}{2\pi}\,\Big[- \frac{1}{2}\,{\ell
    n}\Big(\frac{\Lambda}{m_p}\Big) - {\ell
    n}\Big(\frac{\mu}{m_p}\Big) - \frac{3}{4}\Big].
\end{eqnarray}
After the summation of the contributions the amplitude ${\cal M}^{(\rm
  SC)}_{\rm RC}$ takes the form
\begin{eqnarray}\label{labelF.21}
\hspace{-0.3in}&&{\cal M}^{(\rm SC)}_{\rm RC} = [\bar{u}_p W^{\mu} u_n]
\frac{\alpha}{2\pi}\,\Big\{[\bar{u}_e O_{\mu}
  v_{\bar{\nu}}]\,\Big[\frac{3}{2}\,{\ell
    n}\Big(\frac{m_p}{m_e}\Big) - \frac{11}{8}  + 2\,{\ell
  n}\Big(\frac{\mu}{m_e}\Big)\,\Big[\frac{1}{2\beta}\,{\ell
    n}\Big(\frac{1 + \beta}{1 - \beta}\Big) - 1\Big]\nonumber\\
\hspace{-0.3in}&& + \frac{1}{\beta}\,L\Big(\frac{2\beta}{1 +
  \beta}\Big)- \frac{1}{4\beta}\,{\ell n}^2\Big(\frac{1 + \beta}{1 -
  \beta}\Big) + \frac{1}{2\beta}\,{\ell n}\Big(\frac{1 + \beta}{1 -
  \beta}\Big)\Big] - [\bar{u}_e \gamma^0 O_{\mu}
  v_{\bar{\nu}}]\,\frac{\sqrt{1 - \beta^2}}{2\beta}\,{\ell
  n}\Big(\frac{1 + \beta}{1 - \beta}\Big)\Big\}.
\end{eqnarray}
Thus, the value of the constant $c_S$ is $c_S = - 11/8$ and the
radiative corrections do not depend on the ultra--violet cut--off. It
is caused by the requirement of gauge invariance of the observable
part of the radiative corrections to the amplitude of the
continuum-state $\beta^-$--decay of the neutron.

Now let us proceed to calculating the contribution of $\delta {\cal
  M}^{(\rm SLI)}_{\rm RC}$.  The term ${\cal M}^{(\rm SLI)}_{pe}$ is
defined by
\begin{eqnarray}\label{labelF.22}
\hspace{-0.3in}&&\delta {\cal
  M}^{(\rm SLI)}_{pe} =  -
\frac{\alpha}{4\pi}
\int\frac{d^4q}{\pi^2i}\,D_{\alpha\beta}(q)\,\frac{[\bar{u}_e(2
    k^{\alpha}_e + \gamma^{\alpha} \hat{q})O_{\mu}
    v_{\bar{\nu}}]}{q^2 + 2 k_e \cdot q + i0}\,[\bar{u}_pT^{\beta\mu}u_n] =
\frac{\alpha}{4\pi}\,\Big\{[\bar{u}_p W^{\mu}u_n][\bar{u}_e O_{\mu}
  v_{\bar{\nu}}]\nonumber\\
\hspace{-0.3in}&&\times\,3\int \frac{d^4q}{\pi^2i}\frac{1}{q^2 - 2k_p\cdot q
  + i0}\frac{1}{q^2 + 2k_e\cdot q + i0} + 2(\lambda + 1)\, [\bar{u}_p R^{\mu} u_n][\bar{u}_e
  O_{\mu} v_{\bar{\nu}}] \int \frac{d^4q}{\pi^2i} \frac{1}{q^2 -
  2k_p\cdot q + i0}\frac{1}{q^2 + 2k_e\cdot q + i0}\nonumber\\
\hspace{-0.3in}&& - 2(\lambda + 1)\,[\bar{u}_p
  R^{\alpha}u_n][\bar{u}_e O^{\beta}v_{\bar{\nu}}]\int
\frac{d^4q}{\pi^2i}\frac{q_{\alpha} q_{\beta}}{q^2 + i0}\frac{1}{q^2 -
  2k_p\cdot q + i0}\frac{1}{q^2 + 2k_e\cdot q + i0}\nonumber\\
\hspace{-0.3in}&& - 2i\int \frac{d^4q}{\pi^2i}\frac{1}{q^2 +
  i0}\frac{1}{q^2 - 2k_p\cdot q + i0}\frac{1}{q^2 + 2k_e\cdot q +
  i0}\,[\bar{u}_p\sigma_{\alpha\beta} k^{\alpha}_e q^{\beta} W^{\mu}
  u_n][\bar{u}_e O_{\mu})v_{\bar{\nu}}]\Big\},
\end{eqnarray}
where $R^{\lambda} = \gamma^{\lambda}(1 + \gamma^5)$.  Using the
results, obtained in Appendix D (see Eq.(\ref{labelD.12}),
Eq.(\ref{labelD.28}), Eq.(\ref{labelD.33}) and Eq.(\ref{labelD.38}))
we obtain
\begin{eqnarray}\label{labelF.23}
\hspace{-0.3in}&&\delta {\cal M}^{(\rm SLI)}_{pe} =
\frac{\alpha}{2\pi}\,\Big[3\,{\ell n}\Big(\frac{\Lambda}{m_p}\Big) +
  \frac{3}{2}\Big][\bar{u}_p W^{\mu}u_n][\bar{u}_e
  O_{\mu}v_{\bar{\nu}}]\nonumber\\
\hspace{-0.3in}&& + \frac{\alpha}{2\pi}\,(\lambda + 1)\,\Big[- 2{\ell
    n}\Big(\frac{\Lambda}{m_p}\Big) - 1\Big]\, [\bar{u}_p
  R^{\mu}u_n][\bar{u}_e O_{\mu} v_{\bar{\nu}}]\nonumber\\
\hspace{-0.3in}&& + \frac{\alpha}{2\pi}\,(\lambda +
1)\,\Big\{g_{\alpha\beta}\Big[ \frac{1}{2}{\ell
    n}\Big(\frac{\Lambda}{m_p}\Big) + \frac{3}{4}\Big] - \frac{1}{2}
g_{o\alpha}g_{0\beta}\Big\}\,[\bar{u}_pR^{\alpha}u_n][\bar{u}_eO^{\beta}
   v_{\bar{\nu}}].
\end{eqnarray}
For ${\cal M}^{(\rm SLI)}_{ee}$ and ${\cal M}^{(\rm SLI)}_{pp}$ we
calculate the following expressions
\begin{eqnarray}\label{labelF.24}
 \hspace{-0.3in}&&{\cal M}^{(\rm SLI)}_{ee} = [\bar{u}_p
   W^{\mu}u_n][\bar{u}_e O_{\mu}v_{\bar{\nu}}]\Big\{\frac{Z^{(e)}_2 -
   1}{2} + \frac{\alpha}{2\pi}\Big(- {\ell
   n}\Big(\frac{\Lambda}{m_e}\Big) - 2 {\ell
   n}\Big(\frac{\mu}{m_e}\Big) - \frac{9}{4}\Big) +
 \frac{\alpha}{2\pi}\Big( \frac{1}{2}\,{\ell
   n}\Big(\frac{\Lambda}{m_e}\Big) + {\ell n}\Big(\frac{\mu}{m_e}\Big)
 + \frac{9}{8}\Big)\Big\}=\nonumber\\
\hspace{-0.3in}&&= [\bar{u}_p
   W^{\mu}u_n][\bar{u}_e O_{\mu}v_{\bar{\nu}}]\Big\{\frac{Z^{(e)}_2
   - 1}{2} - \frac{\alpha}{2\pi}\Big(\frac{1}{2}{\ell
  n}\Big(\frac{\Lambda}{m_e}\Big) + {\ell
  n}\Big(\frac{\mu}{m_e}\Big) + \frac{9}{8}\Big)\Big\}
\end{eqnarray}
and 
\begin{eqnarray}\label{labelF.25}
 \hspace{-0.3in}&&{\cal M}^{(\rm SLI)}_{pp} = [\bar{u}_p
   W^{\mu}u_n][\bar{u}_e O_{\mu}v_{\bar{\nu}}]\Big\{\frac{Z^{(p)}_2 -
   1}{2} + \frac{\alpha}{2\pi}\Big(- {\ell
   n}\Big(\frac{\Lambda}{m_p}\Big) - 2 {\ell
   n}\Big(\frac{\mu}{m_p}\Big) - \frac{9}{4}\Big) +
 \frac{\alpha}{2\pi}\Big( \frac{1}{2}\,{\ell
   n}\Big(\frac{\Lambda}{m_p}\Big) + {\ell n}\Big(\frac{\mu}{m_p}\Big)
 + \frac{3}{4}\Big)\Big\}=\nonumber\\
\hspace{-0.3in}&&= [\bar{u}_p W^{\mu}u_n][\bar{u}_e
  O_{\mu}v_{\bar{\nu}}]\Big\{\frac{Z^{(e)}_2 - 1}{2} -
\frac{\alpha}{2\pi}\Big(\frac{1}{2}{\ell
  n}\Big(\frac{\Lambda}{m_p}\Big) + {\ell n}\Big(\frac{\mu}{m_p}\Big)
+ \frac{3}{2}\Big)\Big\},
\end{eqnarray}
respectively. Using the definition of the renormalisation constants
$Z^{(e)}_2$ and $Z^{(p)}_2$ Eq.(\ref{labelD.25}) we obtain
\begin{eqnarray}\label{labelF.26}
 \hspace{-0.3in}&&{\cal M}^{(\rm SLI)}_{ee} = 0,\nonumber\\
\hspace{-0.3in}&& {\cal M}^{(\rm SLI)}_{pp} = [\bar{u}_p
   W^{\mu}u_n][\bar{u}_e O_{\mu}v_{\bar{\nu}}]
 \frac{\alpha}{2\pi}\Big(- \frac{3}{8}\Big).
\end{eqnarray}
The sum of the contributions to ${\cal M}^{(\rm SLI)}_{\rm RC}$ gives one
\begin{eqnarray}\label{labelF.27}
 \hspace{-0.3in}&&{\cal M}^{(\rm SLI)}_{\rm RC} =
 \frac{\alpha}{2\pi}\,\Big[3\,{\ell n}\Big(\frac{\Lambda}{m_p}\Big) +
   \frac{3}{8}\Big][\bar{u}_p W^{\mu}u_n][\bar{u}_e
   O_{\mu}v_{\bar{\nu}}]\nonumber\\
\hspace{-0.3in}&& + \frac{\alpha}{2\pi}\,(\lambda + 1)\,\Big[ - 2{\ell
    n}\Big(\frac{\Lambda}{m_p}\Big) - 1\Big]\, [\bar{u}_p
  R^{\mu}u_n][\bar{u}_e O_{\mu} v_{\bar{\nu}}]\nonumber\\
\hspace{-0.3in}&& + \frac{\alpha}{2\pi}\,(\lambda +
1)\,\Big\{g_{\alpha\beta}\Big[ \frac{1}{2}{\ell
    n}\Big(\frac{\Lambda}{m_p}\Big) + \frac{3}{4}\Big] - \frac{1}{2}
g_{o\alpha}g_{0\beta}\Big\}\,[\bar{u}_pR^{\alpha}u_n][\bar{u}_eO^{\beta}
   v_{\bar{\nu}}].
\end{eqnarray}
This shows that gauge non--invariant contributions do not depend on
the electron energy $E_e$ and the infrared cut--off $\mu$ and may be
fully absorbed by the renormalisation constants of the Fermi and axial
coupling constants (see Eq.(\ref{labelD.48}) and
Eq.(\ref{labelD.49})).

In order to prove the correctness of the term $-3/8$ in
Eq.(\ref{labelD.58}), giving the contribution to the observable
radiative corrections to the lifetime of the neutron, we propose to
sum up the contributions of ${\cal M}^{(\rm SC)}_{\rm RC}$ and the
terms, proportional to $[\bar{u}_p W^{\mu}u_n][\bar{u}_e
  O_{\mu}v_{\bar{\nu}}]$ from ${\cal M}^{(\rm SLI)}_{\rm RC}$, which
we denote as $\bar{{\cal M}}^{(\rm SLI)}_{\rm RC}$. This gives
\begin{eqnarray}\label{labelF.28}
 \hspace{-0.3in}{\cal M}_{\rm RC} = {\cal M}^{(\rm
   SC)}_{\rm RC} + \bar{{\cal M}}^{(\rm SLI)}_{\rm RC}= 
 \frac{\alpha}{2\pi}\,\Big\{[\bar{u}_p W^{\mu}u_n][\bar{u}_e
   O_{\mu}v_{\bar{\nu}}]\,F(E_e, \mu) - [\bar{u}_p
   W^{\mu}u_n][\bar{u}_e \gamma^0O_{\mu} v_{\bar{\nu}}]\frac{\sqrt{1
     - \beta^2}}{2\beta}{\ell n}\Big(\frac{1 + \beta}{1 -
   \beta}\Big),
\end{eqnarray}
where the function $F(E_e, \mu)$ takes the form
\begin{eqnarray}\label{labelF.29}
 \hspace{-0.3in}F(E_e, \mu) &=& 3{\ell n}\Big(\frac{\Lambda}{m_p}\Big)
 + \frac{3}{2}{\ell n}\Big(\frac{m_p}{m_e}\Big) - \frac{1}{4} + 2
 {\ell n}\Big(\frac{\mu}{ m_e}\Big)\Big[\frac{1}{2\beta}\,{\ell n}\Big(\frac{1 +
    \beta}{1 - \beta}\Big) - 1 \Big] + \frac{1}{\beta}L\Big(\frac{2\beta}{1 +
  \beta}\Big)\nonumber\\
\hspace{-0.3in}&-&\frac{1}{4\beta}\,{\ell n}^2\Big(\frac{1 + \beta}{1
  - \beta}\Big) + \frac{1}{2\beta}{\ell n}\Big(\frac{1 + \beta}{1 -
  \beta}\Big).
\end{eqnarray}
The contribution of the radiative corrections to the lifetime of the
neutron is defined by the function
\begin{eqnarray}\label{labelF.30}
 \hspace{-0.3in}g(E_e) &=& \lim_{\mu \to 0}[F(E_e, \mu) +
   g^{(1)}_{\beta^-_c}(E_e,\mu)] =\nonumber\\
\hspace{-0.3in}&=& 3{\ell n}\Big(\frac{\Lambda}{m_p}\Big) +
\frac{3}{2}\,{\ell n}\Big(\frac{m_p}{m_e}\Big) +
\frac{3}{4} + 2\,\Big[\frac{1}{2\beta} \,{\ell n}\Big(\frac{1 +
    \beta}{1 - \beta}\Big) - 1\Big]\, \Big[{\ell n}\Big(\frac{2(E_0 -
    E_e)}{m_e}\Big) - \frac{3} {2} + \frac{1}{3}\,\frac{E_0 - E_e}{E_e}\Big]\nonumber\\
\hspace{-0.3in}&+& \frac{2}{\beta}L\Big(\frac{2\beta}{1 + \beta}\Big)
+ \frac{1}{2\beta}{\ell n}\Big(\frac{1 + \beta}{1 -
  \beta}\Big)\Big[(1+\beta^2)+ \frac{1}{12} \frac{(E_0 -
    E_e)^2}{E^2_e} - {\ell n}\Big(\frac{1 + \beta}{1 -
    \beta}\Big)\Big],
\end{eqnarray}
where the function $ g^{(1)}_{\beta^-_c}(E_e,\mu)$is given by
Eq.(\ref{labelB.28}) and describes the contribution of the radiative
$\beta^-$--decay of the neutron.

The function Eq.(\ref{labelF.30}) reproduces our result, obtained in
Appendix D (see Eq.(\ref{labelD.59})). Then, being multiplied by
$(\alpha/\pi)$, it reproduces also the radiative corrections to the
lifetime of the neutron, calculated by Kinoshita and Sirlin
\cite{RC5}. This corroborates the correctness of our calculation of
${\cal M}^{(\rm SC)}_{\rm RC}$ and ${\cal M}^{(\rm SLI)}_{\rm RC}$.

Thus, the analysis of the radiative corrections to the neutron
$\beta^-$--decay, carried out in this Appendix, confirms Sirlin's
assertion that the requirement of gauge invariance of the amplitude of
the radiative corrections defines unambiguously the observable
radiative corrections to the lifetime of the neutron. They are
independent of the axial coupling constant $\lambda$, i.e. of strong
low--energy interactions.  The part of radiative corrections depending
on the axial coupling constant $\lambda$, i.e. on strong low--energy
interactions, is unobservable and absorbed by renormalisation
constants of the Fermi and axial coupling constants.

\section*{Appendix G: Contribution of  weak
 lepton--nucleon couplings beyond SM to correlation coefficients of
 neutron $\beta^-$--decay}
\renewcommand{\theequation}{G-\arabic{equation}}
\setcounter{equation}{0}

In this Appendix we take into account the contributions of the weak
lepton--nucleon interactions beyond the SM with left--handed and
right--handed neutrinos. For this aim we use the following Hamiltonian
of phenomenological weak lepton--nucleon interactions
\cite{SPT1}--\cite{SPT4} 
\begin{eqnarray}\label{labelG.1}
{\cal H}_W(x) &=&
\frac{G_F}{\sqrt{2}}\,V_{ud}\Big\{[\bar{\psi}_p(x)\gamma_{\mu}\psi_n(x)]
     [\bar{\psi}_e(x)\gamma^{\mu}(C_V + \bar{C}_V
       \gamma^5)\psi_{\nu_e}(x)] +
     [\bar{\psi}_p(x)\gamma_{\mu}\gamma^5\psi_n(x)][\bar{\psi}_e(x)\gamma^{\mu}(\bar{C}_A
       + C_A \gamma^5)\psi_{\nu_e}(x)] \nonumber\\ &+&
     [\bar{\psi}_p(x)\psi_n(x)][\bar{\psi}_e(x)(C_S + \bar{C}_S
       \gamma^5)\psi_{\nu_e}(x)] + [\bar{\psi}_p(x) \gamma ^5
       \psi_n(x)][\bar{\psi}_e(x)(C_P + \bar{C}_P
       \gamma^5)\psi_{\nu_e}(x)]\nonumber\\ &+&\frac{1}{2}
     [\bar{\psi}_p(x)\sigma^{\mu\nu}\gamma^5\psi_n(x)][\bar{\psi}_e(x)\sigma_{\mu\nu}
       (\bar{C}_T + C_T \gamma^5)\psi_{\nu_e}(x) \Big\}.
\end{eqnarray}
This is the most general form of the effective low--energy weak
interactions, where the coupling constants $C_i$ and $\bar{C}_i$ for
$i = V, A, S, P$ and $T$ can be induced by the left--handed and
right--handed hadronic and leptonic currents \cite{SPT1}--\cite{SPT4}
and supersymmetric interactions \cite{SUSY} as well. They are related
to the coupling constants, analogous to those which were introduced by
Herczeg \cite{SPT3}, as follows
\begin{eqnarray}\label{labelG.2}
\hspace{-0.3in} C_V &=&1 + a^h_{LL} + a^h_{LR} + a^h_{RR} +
a^h_{RL},\nonumber\\
\hspace{-0.3in} \bar{C}_V &=& - 1 - a^h_{LL} -
a^h_{LR} + a^h_{RR} + a^h_{RL},\nonumber\\
\hspace{-0.3in} C_A &=& -\lambda + a^h_{LL} - a^h_{LR} + a^h_{RR} -
a^h_{RL},\nonumber\\ \bar{C}_A &=&\lambda - a^h_{LL} + a^h_{LR} + a^h_{RR} -
a^h_{RL},\nonumber\\
\hspace{-0.3in}C_S &=& A^h_{LL} + A^h_{LR} +
A^h_{RR} + A^h_{RL},\nonumber\\
\hspace{-0.3in}\bar{C}_S &=& - A^h_{LL} -
A^h_{LR} + A^h_{RR} + A^h_{RL},\nonumber\\
\hspace{-0.3in}C_P &=& - A^h_{LL} + A^h_{LR} +
A^h_{RR} - A^h_{RL},\nonumber\\
\hspace{-0.3in}\bar{C}_P &=& A^h_{LL} - A^h_{LR} + A^h_{RR} -
A^h_{RL},\nonumber\\
\hspace{-0.3in} C_T &=& 2( \alpha^h_{LL} + \alpha^h_{RR}),\nonumber\\
\hspace{-0.3in} \bar{C}_T &=& 2( -
 \alpha^h_{LL} + \alpha^h_{RR}),
\end{eqnarray}
where the index $h$ means that the coupling constants are introduced
at the {\it hadronic} level but not at the {\it quark} level as it has
been done by Herczeg \cite{SPT3}. In addition in comparison with
Herczeg \cite{SPT3} we have taken away the common factor
$G_FV_{ud}/\sqrt{2}$ and defined the coupling constants $a^h_{LL}$ and
$a^h_{LR}$ as deviations from the coupling constants of the SM
\cite{SPT3a}.  Analogous to Herczeg \cite{SPT3}, the Hamiltonian of
phenomenological lepton--nucleon weak interactions beyond the SM may
be written in the following form \cite{SPT3a}
\begin{eqnarray}\label{labelG.3}
\hspace{-0.3in}&&{\cal H}_{W}(x) =
\frac{G_F}{\sqrt{2}}V_{ud}\Big\{[\bar{\psi}_e(x)\gamma_{\mu}(1 -
  \gamma^5)\psi_{\nu_e}(x)]\Big[\Big(\frac{1 - \lambda}{2} +
  a^h_{LL}\Big)[\bar{\psi}_p(x)\gamma^{\mu}(1 - \gamma^5)\psi_n(x)]\nonumber\\
\hspace{-0.3in}&& + \Big(\frac{1 + \lambda}{2} +
a^h_{LR}\Big)[\bar{\psi}_p(x)\gamma^{\mu}(1 + \gamma^5)\psi_n(x)]\Big] +
     [\bar{\psi}_e(x)\gamma_{\mu}(1 +
       \gamma^5)\psi_{\nu_e}(x)]\,\Big( a^h_{RL}[\bar{\psi}_p(x)\gamma^{\mu}(1 -
  \gamma^5)\psi_n(x)]\nonumber\\
\hspace{-0.3in}&& + a^h_{RR}[\bar{\psi}_p(x)\gamma^{\mu}(1 +
  \gamma^5)\psi_n(x)]\Big) + [\bar{\psi}_e(x)(1 -
  \gamma^5)\psi_{\nu_e}(x)]\Big( A^h_{LL}[\bar{\psi}_p(x)(1 -
  \gamma^5)\psi_n(x)] + A^h_{LR}[\bar{\psi}_p(x)(1 +
  \gamma^5)\psi_n(x)]\Big)\nonumber\\
\hspace{-0.3in}&& + [\bar{\psi}_e(x)(1 +
  \gamma^5)\psi_{\nu_e}(x)]\,\Big(A^h_{RL}[\bar{\psi}_p(x)(1 -
  \gamma^5)\psi_n(x)]\Big) + A^h_{RR}[\bar{\psi}_p(x)(1 +
  \gamma^5)\psi_n(x)]\Big)
+ \frac{1}{2}\,\alpha^h_{LL} [\bar{\psi}_e(x)\sigma_{\mu\nu}(1 -
  \gamma^5)\psi_{\nu_e}(x)]\nonumber\\
\hspace{-0.3in}&&\times\,[\bar{\psi}_p(x)\sigma^{\mu\nu}(1 -
  \gamma^5)\psi_n(x)] + \frac{1}{2}\,\alpha^h_{RR}
       [\bar{\psi}_e(x)\sigma_{\mu\nu}(1 +
         \gamma^5)\psi_{\nu_e}(x)]\,[\bar{\psi}_p(x)\sigma^{\mu\nu}(1
         + \gamma^5)\psi_n(x)]\Big\}.
\end{eqnarray}
In order to express the coupling constants $C_T$ and $\bar{C}_T$ in
terms of the coupling constants $\alpha^h_{LL}$ and $\alpha^h_{RR}$ we
have used the relation $\sigma_{\mu\nu} \gamma^5 =
\frac{i}{2}\varepsilon_{\mu\nu\alpha\beta} \sigma^{\alpha\beta}$
\cite{IZ80}.

The SM is defined by the coupling constants $C_S = \bar{C}_S = C_P =
\bar{C}_P = C_T = \bar{C}_T = 0$, $C_V = - \,\bar{C}_V = 1$ and $C_A =
- \,\bar{C}_A = - \lambda$.  The coupling constants $a^h_{ij}$,
$A^h_{ij}$ and $\alpha^h_{jj}$ for $i(j) = L$ or $R$ are induced by
interactions beyond the SM.

The contributions of the pseudoscalar weak interactions with the
coupling constants $C_P$ and $\bar{C}_P$ to the amplitude of the
neutron $\beta^-$--decay are of order $O(C_P/M)$ and $O(\bar{C}_P/M)$,
which are caused by the proton recoil. Since these corrections are of
order $10^{-6}$ or even smaller, the contributions of the pseudoscalar
weak interactions with the coupling constants $C_P$ and $\bar{C}_P$ to
the correlation coefficients of the neutron $\beta^-$--decay may be
neglected in comparison with contributions of order $10^{-4}$ of our
interest.

The amplitude of the continuum-state $\beta^-$--decay of the neutron,
calculated with the Hamiltonian of weak interactions
Eq.(\ref{labelG.1}) to leading order in the large proton mass
expansion, takes the form
\begin{eqnarray}\label{labelG.4}
\hspace{-0.3in}M(n \to p e^- \bar{\nu}_e) &=& -\,2m_n\,
\frac{G_F}{\sqrt{2}}\,V_{ud}\,\Big\{[\varphi^{\dagger}_p\varphi_n]
[\bar{u}_e \gamma^0(C_V + \bar{C}_V
\gamma^5) v_{\bar{\nu}}] - [\varphi^{\dagger}_p\vec{\sigma}\,\varphi_n]\cdot
       [\bar{u}_e \vec{\gamma}\,(\bar{C}_A + C_A \gamma^5)
         v_{\bar{\nu}}]\nonumber\\
\hspace{-0.3in}&&+ [\varphi^{\dagger}_p\varphi_n][\bar{u}_e (C_S +
  \bar{C}_S \gamma^5) v_{\bar{\nu}}] +
       [\varphi^{\dagger}_p\vec{\sigma}\,\varphi_n]\cdot [\bar{u}_e
         \gamma^0 \vec{\gamma}\,(\bar{C}_T + C_T \gamma^5)
         v_{\bar{\nu}}]\Big\}.
\end{eqnarray}
The hermitian conjugate amplitude is
\begin{eqnarray}\label{labelG.5}
\hspace{-0.3in}M^{\dagger}(n \to p e^- \bar{\nu}_e) &=& -\,2m_n\,
\frac{G_F}{\sqrt{2}}\,V^*_{ud}\Big\{[\varphi^{\dagger}_n\varphi_p]
[\bar{v}_{\bar{\nu}} \gamma^0(C^*_V + \bar{C}^*_V
\gamma^5) u_e] - [\varphi^{\dagger}_n\vec{\sigma}\,\varphi_p]\cdot
       [\bar{v}_{\bar{\nu}} \vec{\gamma}\,(\bar{C}^*_A + C^*_A
         \gamma^5) u_e] \nonumber\\
\hspace{-0.3in}&&+ [\varphi^{\dagger}_n\varphi_p][\bar{v}_{\bar{\nu}}
  (C^*_S - \bar{C}^*_S \gamma^5) u_e] -
       [\varphi^{\dagger}_n\vec{\sigma}\,\varphi_p]\cdot
       [\bar{v}_{\bar{\nu}} \gamma^0 \vec{\gamma}\,(\bar{C}^*_T -
         C^*_T \gamma^5) u_e]\Big\}.
\end{eqnarray}
The squared absolute value of the amplitude Eq.(\ref{labelG.4}) for
the polarized neutron and unpolarised proton and electron is given by
\begin{eqnarray}\label{labelG.5a}
\hspace{-0.3in}\sum_{\rm pol}|M(n \to p e^- \bar{\nu}_e)|^2 = 8m^2_n
G^2_F|V_{ud}|^2 E_e E \,\xi \Big(1 + b\,\frac{m_e}{E_e} +
a\,\frac{\vec{k}_e\cdot \vec{k}}{E_e E} + A\,\frac{\vec{\xi}_n\cdot
  \vec{k}_e}{E_e} + B\,\frac{\vec{\xi}_n\cdot \vec{k}}{E} +
D\,\frac{\vec{\xi}_n\cdot (\vec{k}_e \times \vec{k}\,)}{E_e E}\Big).
\end{eqnarray}
The correlation coefficients of the neutron $\beta^-$--decay,
expressed in terms of the phenomenological coupling constants $C_j$
and $\bar{C}_j$ for $j = V$, $A$, $S$ and $T$ and calculated to
leading order in the large $M$ expansion, are equal to
\begin{eqnarray}\label{labelG.5b}
\hspace{-0.3in}\xi &=& |C_V|^2 + |\bar{C}_V|^2 + 3 |C_A|^2 +
3|\bar{C}_A|^2 + |C_S|^2 + |\bar{C}_S|^2 + 3|C_T|^2 + 3|\bar{C}_T|^2
,\nonumber\\
\hspace{-0.3in}\xi a &=& |C_V|^2 + |\bar{C}_V|^2 - |C_A|^2 -
|\bar{C}_A|^2 + |C_T|^2 + |\bar{C}_T|^2 - |C_S|^2 -
|\bar{C}_S|^2,\nonumber\\
\hspace{-0.3in}\xi b &=& 2{\rm Re}\Big((C_V C^*_S + \bar{C}_V \bar{C}^*_S)
- 3(C_A C^*_T + \bar{C}_A \bar{C}^*_T)\Big),\nonumber\\
\hspace{-0.3in}\xi A &=& 2\,{\rm Re}\Big(2\,C_A \bar{C}^*_A - 2 C_T
\bar{C}^*_T - (C_V \bar{C}^*_A + \bar{C}_V C^*_A) - (C_S \bar{C}^*_T +
\bar{C}_S C^*_T)\Big),\nonumber\\
\hspace{-0.3in}\xi B &=& - 2\,{\rm Re}\Big(2 C_A \bar{C}^*_A + 2 C_T
\bar{C}^*_T + (C_V \bar{C}^*_A + \bar{C}_V C^*_A) - (C_S \bar{C}^*_T +
\bar{C}_S C^*_T)\Big)\nonumber\\
\hspace{-0.3in}&& + 2{\rm Re}\Big((C_V \bar{C}^*_T + \bar{C}_V
C^*_T) - (C_A \bar{C}^*_S + \bar{C}_A C^*_S) + 2\,(C_A \bar{C}^*_T + \bar{C}_A
C^*_T)\Big)\,\frac{m_e}{E_e},\nonumber\\
\hspace{-0.3in}\xi D &=& 2 {\rm Im}\Big((C_V C^*_A + \bar{C}_V
\bar{C}^*_A) + (C_S C^*_T + \bar{C}_S \bar{C}^*_T)\Big).
\end{eqnarray}
The correlation coefficients Eq.(\ref{labelG.5b}) reproduce well the
structure of the correlation coefficients, calculated in
\cite{SPT1}--\cite{SPT4}. In the linear approximation with respect to
the deviations of the coupling constants $C_j$ and $\bar{C}_j$ for $j
= V, A, S$ and $T$ from the coupling constants of the SM the
correlation coefficients $\xi$, $a$, $b$, $A$, $B$ and $D$ read
\begin{eqnarray}\label{labelG.5c}
\xi &=& 2(1 + 3\lambda^2)\Big[1 + \frac{1}{1 + 3\lambda^2}\Big({\rm
    Re}(\delta C_V - \delta \bar{C}_V) - 3\lambda\,{\rm Re}(\delta C_A
  - \delta \bar{C}_A)\Big)\Big] ,\nonumber\\ a &=& a_0 + \frac{1}{1 +
  3\lambda^2}\Big[{\rm Re}(\delta C_V - \delta \bar{C}_V) +
  \lambda\,{\rm Re}(\delta C_A - \delta \bar{C}_A) - a_0 \Big({\rm
    Re}(\delta C_V - \delta \bar{C}_V) - 3\lambda\,{\rm Re}(\delta C_A
  - \delta \bar{C}_A)\Big)\Big],\nonumber\\ b &=& \frac{1}{1 +
  3\lambda^2}\Big({\rm Re}(C_S - \bar{C}_S) + 3\lambda {\rm Re}(C_T -
\bar{C}_T)\Big),\nonumber\\ A &=& A_0 + \frac{1}{1 +
  3\lambda^2}\,\Big[ - \lambda\,{\rm Re}(\delta C_V - \delta
  \bar{C}_V) + (1 + 2\lambda)\,{\rm Re}(\delta C_A - \delta \bar{C}_A)
  - A_0 \Big({\rm Re}(\delta C_V - \delta \bar{C}_V) - 3\lambda\,{\rm
    Re}(\delta C_A - \delta \bar{C}_A)\Big)\Big] ,\nonumber\\ B &=&B_0
+ \frac{1}{1 + 3\lambda^2}\,\Big[ - \lambda\,{\rm Re}(\delta C_V -
  \delta \bar{C}_V) + (1 - 2\lambda)\,{\rm Re}(\delta C_A - \delta
  \bar{C}_A) - B_0 \Big({\rm Re}(\delta C_V - \delta \bar{C}_V) -
  3\lambda\,{\rm Re}(\delta C_A - \delta \bar{C}_A)\Big)\nonumber\\ &&
  + \Big( - \lambda\,{\rm Re}( C_S - \bar{C}_S) - (1 - 2\lambda)\,{\rm
    Re}(C_T - \bar{C}_T)\Big)\,\frac{m_e}{E_e}\Big],\nonumber\\ D &=&
\frac{1}{1 + 3\lambda^2}\,\Big( - \lambda\,{\rm Im}(\delta C_V -
\delta \bar{C}_V) - {\rm Im}(\delta C_A - \delta \bar{C}_A) \Big),
\end{eqnarray}
where the correlation coefficients $a_0$, $A_0$ and $B_0$ are defined
in Eq.(\ref{label8}) and the phenomenological coupling constants $C_J$
and $\bar{C}_j$ for $j = V,A$ are taken in the form $C_V = 1 + \delta
C_V$, $\bar{C}_V = -1 + \delta \bar{C}_V$, $C_A = - \lambda + \delta
C_A$ and $\bar{C}_A = \lambda + \delta \bar{C}_A$. The coupling
constants $\delta C_j$ and $\delta \bar{C}_j$ for $j = V, A$ and the
coupling constants $C_j$ and $\bar{C}_j$ for $j = S, T$ are caused by
interactions beyond the SM. For the analysis of contributions of order
$10^{-4}$ beyond the SM we may maintain the corrections, caused
deviations of the phenomenological coupling constants $\bar{C}_j$ for
$j = V, A, S$ and $T$ from the coupling constants of the SM, to linear
order only.

In terms of the correlation coefficients Eq.(\ref{labelG.5b}) the rate
of the neutron $\beta^-$--decay is equal to
\begin{eqnarray}\label{labelG.5d}
\lambda_n = \xi\,\frac{G^2_F |V_{ud}|^2}{4\pi^3}\,f_n(E_0, Z = 1)\,\Big(1 +
b\,\Big\langle \frac{m_e}{E_e}\Big\rangle\Big),
\end{eqnarray}
where the Fermi integral $f_n(E_0, Z = 1)$ is given by
\begin{eqnarray}\label{labelG.5f}
\hspace{-0.3in}&& f_n(E_0, Z = 1) = \int^{E_0}_{m_e} (E_0 - E_e
)^2\,\sqrt{E^2_e - m^2_e}\,E_e \, F(E_e, Z = 1)\,dE_e,
\end{eqnarray}
The correlation coefficients of the neutron $\beta^-$--decay,
calculated in this paper with the contributions of the ``weak
magnetism'', the proton recoil, radiative corrections and
contributions of interactions beyond the SM, are equal to
\begin{eqnarray}\label{labelG.6}
\hspace{-0.3in}\zeta(E_e) &=& \zeta_{\rm SM}(E_e) + \zeta_{\rm
  NP}(E_e),\nonumber\\
\hspace{-0.3in}a(E_e) &=& a_{\rm SM}(E_e) + a_{\rm NP}(E_e),\nonumber\\
\hspace{-0.3in}A(E_e) &=& A_{\rm SM}(E_e) + A_{\rm NP}(E_e),\nonumber\\
\hspace{-0.3in}B(E_e) &=& B_{\rm SM}(E_e) + B_{\rm NP}(E_e),\nonumber\\
\hspace{-0.3in}D(E_e) &=& D_{\rm SM}(E_e) +  D_{\rm NP}(E_e),
\end{eqnarray}
where the correlation coefficients $\zeta_{\rm SM}(E_e)$, $a_{\rm
  SM}(E_e)$, $A_{\rm SM}(E_e)$ and $B_{\rm SM}(E_e)$ are given by
Eqs.(\ref{label10}) -- (\ref{label13}), respectively.  They are
calculated within the SM with the contributions of the ``weak
magnetism'', the proton recoil and radiative corrections. The
correlation coefficient $D_{\rm SM}(E_e)$ has been calculated in
\cite{LRM1}--\cite{SRM}. For the electron kinetic energies $250\,{\rm
  keV} \le T_e \le 455\,{\rm keV}$ and the axial coupling constant
$\lambda = - 1,2750$ one may estimate that $D_{\rm SM}(E_e) \sim
10^{-5}$ \cite{LRM1}.  The correlations coefficients $\zeta_{\rm
  NP}(E_e)$, $a_{\rm NP}(E_e)$, $A_{\rm NP}(E_e)$, $B_{\rm NP}(E_e)$
and $D_{\rm NP}(E_e)$ are defined by the contributions of interactions
beyond the SM or a new physics (NP). They are taken to linear
approximation with respect to the deviations of the phenomenological
coupling constants $C_j$ and $\bar{C}_j$ for $j = V, A, S$ and $T$
from the coupling constants of the SM
\begin{eqnarray}\label{labelG.6a}
\hspace{-0.3in}\zeta_{\rm NP}(E_e) &=& \frac{1}{1 +
  3\lambda^2}\Big({\rm Re}(\delta C_V - \delta \bar{C}_V) -
3\lambda\,{\rm Re}(\delta C_A - \delta \bar{C}_A)\Big)+
b_F\,\frac{m_e}{E_e},\nonumber\\
\hspace{-0.3in}a_{\rm NP}(E_e) &=& \frac{1}{1 + 3
  \lambda^2}\,\Big({\rm Re}(\delta C_V - \delta \bar{C}_V) +
\lambda\,{\rm Re}(\delta C_A - \delta \bar{C}_A)\nonumber\\
\hspace{-0.3in}&-& a_0\,\Big(\frac{1}{1 + 3\lambda^2}\Big({\rm
  Re}(\delta C_V - \delta \bar{C}_V) - 3\lambda\,{\rm Re}(\delta C_A -
\delta \bar{C}_A)\Big) + b_F\,\frac{m_e}{E_e}\Big),\nonumber\\
\hspace{-0.3in}A_{\rm NP}(E_e) &=& \frac{1}{1 + 3
  \lambda^2}\,\Big( - \lambda\,{\rm Re}(\delta C_V - \delta \bar{C}_V)
+ (1 + 2\lambda)\,{\rm Re}(\delta C_A - \delta \bar{C}_A)\Big)\nonumber\\
\hspace{-0.3in}&-& A_0\,\Big(\frac{1}{1 + 3\lambda^2}\Big({\rm
  Re}(\delta C_V - \delta \bar{C}_V) - 3\lambda\,{\rm Re}(\delta C_A -
\delta \bar{C}_A)\Big) + b_F\,\frac{m_e}{E_e}\Big),\nonumber\\
\hspace{-0.3in}B_{\rm NP}(E_e) &=&  \frac{1}{1 + 3
  \lambda^2}\,{\rm Re}\Big( - \lambda\,{\rm Re}(\delta C_V - \delta
\bar{C}_V) + (1 - 2\lambda)\,{\rm Re}(\delta C_A - \delta
\bar{C}_A)\Big)\nonumber\\
\hspace{-0.3in}&+& \frac{1}{1 + 3 \lambda^2}\,\Big( - \lambda\,{\rm
  Re}( C_S - \bar{C}_S) - (1 - 2\lambda)\,{\rm Re}(C_T -
\bar{C}_T)\Big)\,\frac{m_e}{E_e}\nonumber\\
\hspace{-0.3in}&-& B_0\,\Big(\frac{1}{1 + 3\lambda^2}\Big({\rm
  Re}(\delta C_V - \delta \bar{C}_V) - 3\lambda\,{\rm Re}(\delta C_A -
\delta \bar{C}_A)\Big) + b_F\,\frac{m_e}{E_e}\Big),\nonumber\\
\hspace{-0.3in}D_{\rm NP}(E_e) &=& \frac{1}{1 + 3
  \lambda^2}\,\Big( - \lambda\,{\rm Im}(\delta C_V - \delta \bar{C}_V)
- {\rm Im}(\delta C_A - \delta \bar{C}_A)\Big),
\end{eqnarray}
where $b_F$ is the Fierz term \cite{Abele1,Nico1,SPT1,SPT4}, defined
by
\begin{eqnarray}\label{labelG.7}
\hspace{-0.3in}b_F = \frac{1}{1 + 3 \lambda^2}\,\Big({\rm Re}(C_S -
\bar{C}_S) + 3 \lambda \,{\rm Re}(C_T - \bar{C}_T)\Big).
\end{eqnarray}
 The rate of the neutron $\beta^-$--decay, corrected by the
 contributions of interactions beyond the SM, takes the form
\begin{eqnarray}\label{labelG.8}
\lambda_n = (\lambda_n)_{ \rm SM}\Big(1 + \frac{1}{1 + 3\lambda^2}\Big({\rm
  Re}(\delta C_V - \delta \bar{C}_V) - 3\lambda\,{\rm Re}(\delta C_A -
\delta \bar{C}_A)\Big) + b_F\Big\langle
\frac{m_e}{E_e}\Big\rangle_{\rm SM}\Big),
\end{eqnarray}
where $(\lambda_{n})_{ \rm SM}$ is defined by Eq.(\ref{label35}) and
$\langle m_e/E_e\rangle_{\rm SM}$ is the average value, calculated
with the electron--energy spectrum Eq.(\ref{labelD.58}). The lifetime
of the neutron is equal to
\begin{eqnarray}\label{labelG.9}
\tau_n = (\tau_n)_{ \rm SM}\Big(1 - \frac{1}{1 + 3\lambda^2}\Big({\rm
  Re}(\delta C_V - \delta \bar{C}_V) - 3\lambda\,{\rm Re}(\delta C_A -
\delta \bar{C}_A)\Big)- b_F\Big\langle \frac{m_e}{E_e}\Big\rangle_{\rm
  SM}\Big).
\end{eqnarray}
In terms of the correlation coefficients Eq.(\ref{labelG.6}) and
Eq.(\ref{labelG.6a}) the analysis of the sensitivity of the
asymmetries $A_{\exp}(E_e)$, $B_{\exp}(E_e)$ and $C_{\exp}$, the
proton-electron energy distribution $a(E_e, T_p)$, the proton--energy
spectrum $a(T_p)$ and the lifetime of the neutron $\tau_n$ is given in
section~\ref{sec:sensitivity}.

\section*{Appendix H: Contribution of  proton recoil, caused by 
electron--proton final state Coulomb interaction} 
\renewcommand{\theequation}{H-\arabic{equation}}
\setcounter{equation}{0}

In this Appendix we calculate the contribution of the proton recoil,
caused by the electron--proton Coulomb interaction in the final state
of the decay. A velocity of a relative motion of the electron--proton
pair is equal to
\begin{eqnarray}\label{labelH.1}
\vec{v} = \frac{\vec{k}_p}{m_p} - \frac{\vec{k}_e}{E_e}.
\end{eqnarray}
The Coulomb corrections, caused by a proton recoil and calculated to
order $\alpha/M$, change the Fermi function Eq.(\ref{label5}) as
follows
\begin{eqnarray}\label{labelH.2}
\hspace{-0.3in}&&F(E_e, Z = 1) \to F(E_e, Z = 1)\,\Big(1 - \frac{\pi
  \alpha}{\beta}\,\frac{E_e}{M} - \frac{\pi
  \alpha}{\beta^3}\,\frac{E_0 - E_e}{M}\,\frac{\vec{k}_e\cdot
  \vec{k}}{E_e E}\Big).
\end{eqnarray}
As a result the function $\zeta(E_e)$ acquires the following
correction
\begin{eqnarray}\label{labelH.3}
\zeta(E_e) \to \zeta(E_e)\eta(E_e) =\zeta(E_e)\Big(1 -
\frac{\pi \alpha}{\beta}\,\frac{E_e}{M}\Big(1 +
\frac{1}{3}\,a_0\,\frac{E_0 - E_e}{E_e}\Big)\Big).
\end{eqnarray}
The function $\eta(E_e)$, averaged over the electron energy spectrum
$\rho_{\beta^-_c}(E_e)$ Eq.(\ref{labelD.58}) and equal to $\langle
\eta(E_e)\rangle = 1 - 2.7\times 10^{-5}$, defines a correction to the
lifetime of the neutron of order $10^{-5}$. The correlation
coefficients acquire the following corrections
\begin{eqnarray}\label{labelH.4}
\hspace{-0.15in}\delta a(E_e) &=& \frac{1}{3}\,a^2_0\,\frac{\pi
  \alpha}{\beta}\,\frac{E_0 - E_e}{M} - \frac{\pi
  \alpha}{\beta^3}\,\frac{E_0 - E_e}{M}\;,\;\delta A(E_e) =
\frac{1}{3}\,a_0A_0\,\frac{\pi \alpha}{\beta}\,\frac{E_0 -
  E_e}{M}\;,\;\delta B(E_e) = \frac{1}{3}\,a_0B_0\,\frac{\pi
  \alpha}{\beta}\,\frac{E_0 - E_e}{M}, \nonumber\\
\hspace{-0.15in}\delta K_n(E_e) &=& -
A_0\,\frac{\pi\alpha}{\beta^3}\,\frac{E - E_e}{M} \;,\;\delta Q_n(E_e)
=- B_0\,\frac{\pi\alpha}{\beta^3}\,\frac{E - E_e}{M}.
\end{eqnarray}
In the electron energy region $250\,{\rm keV} \le T_e \le 455\,{\rm
  keV}$ and at $\lambda = - 1.2750$ the obtained corrections are of
order $\delta a(E_e)/a(E_e) \sim 10^{-4}$, $ \delta A(E_e)/A(E_e) =
\delta B(E_e)/B(E_e) \sim - 10^{-6}$, $\delta K_n(E_e)/K_n(E_e) \sim
\delta Q_n(E_e)/Q_n(E_e) \sim 10^{-2}$ or $\delta K_n(E_e) \sim
10^{-6}$ and $ \delta Q_n(E_e) \sim - 10^{-5}$. Thus, the relative
corrections to all correlation coefficients except $a(E_e)$ and
$Q_n(E_e)$, induced by the proton recoil in the Coulomb
electron--proton interaction in the final state of the neutron
$\beta^-$--decay and calculated to order $\alpha/M$, are smaller
compared with corrections of order $10^{-4}$ of interactions beyond
the SM in the experimentally used electron energy region $250\,{\rm
  keV} \le T_e \le 455\,{\rm keV}$. In turn, the correction
Eq.(\ref{labelH.4}) to correlation coefficient $a(E_e)$ should be
taken into account for the analysis of interactions beyond the SM at
the level of $10^{-4}$. It is important that the contribution of
$\delta Q_n(E_e)$ to $A^{(W)}(E_e)$ (see Eq.(\ref{label17})) is at the
level of $10^{-5}$.

\section*{Appendix I: Electron--proton energy distribution 
 and proton recoil asymmetry of neutron $\beta^-$--decay}
\renewcommand{\theequation}{I-\arabic{equation}}
\setcounter{equation}{0}

The correlation coefficient $a(E_e)$, which analytical expression is
given in Eq.(\ref{label11}), can be hardly used directly for the
experimental determination of the correlation coefficient $a_0$ due to
impossibility to determine experimentally a correlation between the
electron and antineutrino 3--momenta. For the experimental
determination of the correlation coefficient $a_0$ one needs to
measure the correlations of charged particles, i. e. the correlations
between 3--momenta of the proton and electron. For the experimental
analysis of the proton recoil asymmetry we need also to have the
electron--proton energy and angular distribution, including the
correlations between the neutron spin  and the proton 3--momentum.

After the integration over the antineutrino 3--momentum the
electron--proton energy--momentum and angular distribution of the
neutron $\beta^-$--decay takes the form
\begin{eqnarray}\label{labelI.1}
\hspace{-0.3in}&&\frac{d^6 \lambda_{\beta^-_c}(E_e, \vec{k}_e,
  \vec{k}_p, \vec{\xi}_n)}{dE_e dk_pd\Omega_pd \Omega_{ep}} = (1 +
3\lambda^2)\,\frac{G^2_F|V_{ud}|^2}{32\pi^5}|{\cal
  M}_{\beta^-_c}|^2\Big(1 + \frac{\alpha}{\pi}\,g_n(E_e)\Big)\,F(E_e,
Z = 1)\,\delta(m_n - E_p - E_e - E)\,\frac{m_n}{E_p}\,k_e E_e
k^2_p,\nonumber\\
\hspace{-0.3in}&&
\end{eqnarray}
where $E = |\vec{k}_p + \vec{k}_e|$, $d\Omega_p =
\sin\theta_pd\theta_pd\phi_p$ is the infinitesimal element of the
solid angle of the 3--momentum of the proton relative to the
polarization vector $\vec{\xi}_n$ of the neutron,
i.e. $\vec{\xi}_n\cdot \vec{k}_p = Pk_p\cos\theta_p$ with $P =
|\vec{\xi}_n| \le 1$. Then, $d\Omega_{ep} =
\sin\theta_{ep}d\theta_{ep}d\phi_{ep}$ is the infinitesimal element of
the solid angle of the correlations of the electron--proton 3--momenta
$\vec{k}_e\cdot \vec{k}_p = k_e k_p \cos\theta_{ep}$ and
$\vec{\xi}_n\cdot \vec{k}_e = Pk_e(\cos\theta_p \cos\theta_{ep} +
\sin\theta_p\sin\theta_{ep}\cos(\phi_p - \phi_{ep}))$. In the
non--relativistic approximation for the proton and to
next--to--leading order in the large $M$ expansion we replace
$m_n/E_p$ by

\begin{eqnarray}\label{labelI.2}
\hspace{-0.3in}\frac{m_n}{E_p} \to  1 + \frac{E_0}{M}.
\end{eqnarray}
Then, $|{\cal M}_{\beta^-_c}|^2$, multiplied by the factor
Eq.(\ref{labelI.2})), is (see Eq.(\ref{labelA.17}))
\begin{eqnarray}\label{labelI.3}
\hspace{-0.3in}&&\Big( 1 + \frac{E_0}{M}\Big) |{\cal M}_{\beta^-_c}|^2
= 1 + \tilde{a}(E_e)\,\frac{\vec{k}_e\cdot \vec{k}}{E_e E} +
\frac{1}{M}\,\frac{1}{1 + 3\lambda^2}\Big\{E_0 - \frac{m^2_e}{E_e} -
\Big(\lambda^2 + 2(\kappa + 1)\lambda\Big)\,\frac{\vec{k}\cdot
  \vec{k}_p}{E} - \Big(\lambda^2 - 2(\kappa + 1)\lambda\Big)\,
\frac{\vec{k}_e\cdot \vec{k}_p}{E_e}\Big\}\nonumber\\
\hspace{-0.3in}&& + \tilde{A}(E_e)\,\frac{\vec{\xi}_n\cdot
  \vec{k}_e}{E_e} + \tilde{B}(E_e)\,\frac{\vec{\xi}_n\cdot \vec{k}}{E}
+ \frac{1}{M}\,\frac{1}{1 + 3\lambda^2}\Big\{ - (2\kappa +
1)\,\lambda\,(\vec{\xi}_n\cdot \vec{k}_p) +
\lambda\,\frac{(\vec{\xi}_n\cdot \vec{k}_p)(\vec{k}_e\cdot
  \vec{k}\,)}{E_e E} + \Big(\lambda^2 + (\kappa + 1)\lambda + (\kappa +
1)\Big)\nonumber\\
\hspace{-0.3in}&& \times\,\frac{(\vec{\xi}_n\cdot
  \vec{k}_e)(\vec{k}\cdot \vec{k}_p)}{E_e E}- \Big(\lambda^2 - (\kappa
+ 1)\lambda + (\kappa + 1)\Big)\,\frac{(\vec{\xi}_n\cdot
  \vec{k}\,)(\vec{k}_e\cdot \vec{k}_p)}{E_e E}\Big\},\nonumber\\
\hspace{-0.3in}&& 
\end{eqnarray}
where $\vec{k} = - \vec{k}_p - \vec{k}_e$,  $E = |\vec{k}_p +
\vec{k}_e|$ and we have denoted
\begin{eqnarray}\label{labelI.4}
\hspace{-0.3in}&&\tilde{a}(E_e) = a_0\,\Big(1 + \frac{1}{1 -
  \lambda^2}\,\frac{E_0}{M}\Big)\,\Big(1 +
\frac{\alpha}{\pi}\,f_n(E_e)\Big),\nonumber\\
\hspace{-0.3in}&&\tilde{A}(E_e) = A_0\,\Big(1 + \frac{1}{2(1 +
  \lambda)}\,\frac{E_0}{M}\Big)\,\Big(1 +
\frac{\alpha}{\pi}\,f_n(E_e)\Big),\nonumber\\
\hspace{-0.3in}&&\tilde{B}(E_e) = B_0\,\Big(1 + \frac{1}{2(1 -
  \lambda)}\,\frac{1}{M}\Big(E_0 - \frac{m^2_e}{E_e}\Big)\Big).
\end{eqnarray}
Substituting Eq.(\ref{labelI.3}) into Eq.(\ref{labelI.1}) we obtain
the electron--proton energy--momentum and angular distribution of the
neutron $\beta^-$--decay with polarized neutron and unpolarised
electron and proton
\begin{eqnarray}\label{labelI.5}
\hspace{-0.3in}&&\frac{d^6 \lambda_{\beta^-_c}(E_e, \vec{k}_e,
  \vec{k}_p, \vec{\xi}_n)}{dE_e dk_pd\Omega_pd \Omega_{ep}} = (1 +
3\lambda^2)\,\frac{G^2_F|V_{ud}|^2}{32\pi^5}\,\Bigg\{ 1 +
\tilde{a}(E_e)\,\frac{\vec{k}_e\cdot \vec{k}}{E_e E} +
\frac{1}{M}\,\frac{1}{1 + 3\lambda^2}\Bigg(E_0 - \frac{m^2_e}{E_e} -
\Big(\lambda^2 + 2(\kappa + 1)\lambda\Big)\,\frac{\vec{k}\cdot
  \vec{k}_p}{E}\nonumber\\
\hspace{-0.3in}&&- \Big(\lambda^2 - 2(\kappa + 1)\lambda\Big)\,
\frac{\vec{k}_e\cdot \vec{k}_p}{E_e}\Bigg) +
\tilde{A}(E_e)\,\frac{\vec{\xi}_n\cdot \vec{k}_e}{E_e} +
\tilde{B}(E_e)\,\frac{\vec{\xi}_n\cdot \vec{k}}{E} +
\frac{1}{M}\,\frac{1}{1 + 3\lambda^2}\Bigg( - (2\kappa +
1)\,\lambda\,(\vec{\xi}_n\cdot \vec{k}_p) +
  \lambda\,\frac{(\vec{\xi}_n\cdot \vec{k}_p)(\vec{k}_e\cdot
    \vec{k}\,)}{E_e E}\nonumber\\
\hspace{-0.3in}&& + \Big(\lambda^2 +
(\kappa + 1)\lambda + (\kappa + 1)\Big)\,\frac{(\vec{\xi}_n\cdot
  \vec{k}_e)(\vec{k}\cdot \vec{k}_p)}{E_e E}- \Big(\lambda^2 - (\kappa
+ 1)\lambda + (\kappa + 1)\Big)\,\frac{(\vec{\xi}_n\cdot
  \vec{k}\,)(\vec{k}_e\cdot \vec{k}_p)}{E_e E}\Bigg)\Bigg\}\nonumber\\
\hspace{-0.3in}&&\times\,\Big(1 +
\frac{\alpha}{\pi}\,g_n(E_e)\Big)\,F(E_e, Z = 1)\,\delta(m_n - E_p -
E_e - E)\,k_e E_e k^2_p,
\end{eqnarray}
Now let us analyse the case of unpolarised neutrons. Integrating over
the solid angles $d\Omega_p$ and $d\Omega_{ep}$ we obtain the
electron-proton energy spectrum
\begin{eqnarray}\label{labelI.6}
\hspace{-0.3in}\frac{d^2\lambda_{\beta^-_c}(E_e,T_p)}{dE_edT_p} = M\, (1 +
3 \lambda^2) \frac{G^2_F|V_{ud}|^2}{4\pi^3}\,a(E_e,T_p)\,\Big(1 +
\frac{\alpha}{\pi}\,g_n(E_e)\Big)\,F(E_e, Z = 1)\,E_e,
\end{eqnarray}
where $T_p = k^2_p/2M$ is the kinetic energy of the proton.  The
electron-proton energy distribution $a(E_e,T_p)$ takes the form
\begin{eqnarray}\label{labelI.7}
\hspace{-0.3in}a(E_e,T_p) = \zeta_1(E_e,T_p) +
\tilde{a}(E_e)\,\zeta_2(E_e,T_p).
\end{eqnarray}
The functions $\zeta_1(E_e,T_p)$ and $\zeta_2(E_e,T_p)$ are determined
by the integrals over $\cos\theta_{ep} = \vec{k}_e\cdot \vec{k}_p/k_e
k_p$
\begin{eqnarray}\label{labelI.8}
\hspace{-0.3in}&&\zeta_1(E_e,T_p) = k_e k_p\int^{+1}_{-1}\Big\{1 +
\frac{1}{M}\,\frac{1}{1 + 3\lambda^2}\,\Big[E_0 - \frac{m^2_e}{E_e} +
  \Big(\lambda^2 + 2(\kappa + 1)\lambda \Big)\,\frac{k^2_p +
    \vec{k}_e\cdot \vec{k}_p}{|\vec{k}_p + \vec{k}_e|} -
  \Big(\lambda^2 - 2(\kappa + 1)\lambda\Big)\,\frac{\vec{k}_e\cdot
    \vec{k}_p}{E_e}\Big]\Big\}\nonumber\\
\hspace{-0.3in}&&\hspace{1in}\times\,\delta\Big(m_n - E_p - E_e -
|\vec{k}_p + \vec{k}_e|\Big)\,d\cos\theta_{ep}
\end{eqnarray}
and 
\begin{eqnarray}\label{labelI.9}
\hspace{-0.3in}\zeta_2(E_e,T_p) = - k_e k_p\int^{+1}_{-1}\frac{k^2_e +
  \vec{k}_e\cdot \vec{k}_p}{E_e |\vec{k}_p +
    \vec{k}_e|}\,
\delta\Big(m_n - E_p - E_e - |\vec{k}_p + \vec{k}_e|\Big)\,
d\cos\theta_{ep}.
\end{eqnarray}
The integration over $\cos\theta_{ep}$ we perform by making a change
of variables $E = |\vec{k}_p + \vec{k}_e|$.  This gives
\begin{eqnarray}\label{labelI.10}
\hspace{-0.3in}\zeta_1(E_e,T_p) &=& \Big\{E + \frac{1}{M}\,\frac{1}{1
  + 3\lambda^2}\,\Big[\Big(E_0 - \frac{m^2_e}{E_e}\Big)E + \frac{1}{2}
  \Big(\lambda^2 + 2(\kappa + 1)\lambda \Big)\,(E^2 + k^2_p -
  k^2_e)\nonumber\\
\hspace{-0.3in}&&- \frac{1}{2}\,\frac{E}{E_e}\,\Big(\lambda^2 -
2(\kappa + 1)\lambda\Big)\,(E^2 - k^2_p - k^2_e)\Big]\Big\},\nonumber\\
\hspace{-0.3in}\zeta_2(E_e,T_p) &=& - \frac{1}{2}\,\frac{1}{E_e}\,(E^2 -
k^2_p + k^2_e)
\end{eqnarray}
at $E = m_n - E_p - E_e$. The energy region of the definition of the
distribution $a(E_e, T_p)$ is given by \cite{Nachtmann68}
\begin{eqnarray}\label{labelI.11}
\hspace{-0.3in}0 \le &T_p& \le (T_p)_{\rm max} = \frac{(m_n - m_p)^2 -
  m^2_e}{2 m_n} = \frac{E^2_0 - m^2_e}{2 M} = 0.751\,{\rm
  keV},\nonumber\\
\hspace{-0.3in}(E_e)_{\rm min} \le &E_e& \le (E_e)_{\rm max},
\end{eqnarray}
where $(E_e)_{\rm min/max}$ are equal to
\begin{eqnarray}\label{labelI.12}
\hspace{-0.3in}(E_e)_{\rm min/max} = \frac{(m_n - E_p \mp k_p)^2 +
  m^2_e}{2 (m_n - E_p \mp k_p)} = \frac{(E_0 \mp \sqrt{2MT_p})^2 +
  m^2_e}{2(E_0 \mp \sqrt{2MT_p})} + \frac{1}{2}\Big(1 -
\frac{m^2_e}{(E_0 \mp \sqrt{2MT_p})^2}\Big)((T_p)_{\rm max} - T_p).
\end{eqnarray}
Here we have kept the next--to--leading terms in the large $M$
expansion and used the relation
\begin{eqnarray}\label{labelI.13}
\hspace{-0.3in}m_n - m_p = E_0 + \frac{E^2_0 - m^2_e}{2 M} = E_0 +
(T_p)_{\rm max}.
\end{eqnarray}
To next--to--leading order in the large $M$ expansion the variable $E$
is defined by
\begin{eqnarray}\label{labelI.14}
\hspace{-0.3in} E = m_n - E_p - E_e = E_0 - E_e + ((T_p)_{\rm max} -
T_p).
\end{eqnarray}
The functions $\zeta_1(E_e, T_p)$ and $\zeta_2(E_e, T_p)$, determined
to order $1/M$, are equal to
\begin{eqnarray}\label{labelI.15}
\hspace{-0.3in}&&\zeta_1(E_e,T_p) = (E_0 - E_e) + ((T_p)_{\rm
  max} - T_p) + \frac{1}{M}\,\frac{1}{1 + 3\lambda^2}\,\Big[\Big(E_0 -
  \frac{m^2_e}{E_e}\Big)(E_0 - E_e)\nonumber\\
\hspace{-0.3in}&& - \frac{1}{2}\frac{E_0}{E_e}\Big((E_0 - E_e)^2 + E^2_e
- m^2_e - 2MT_p\Big) + \frac{1}{2} \Big(\lambda^2 + 2(\kappa + 1)\lambda
\Big)\,\Big((E_0 - E_e)^2 - E^2_e + m^2_e - 2M T_p\Big)\nonumber\\
\hspace{-0.3in}&&- \frac{1}{2}\,\Big(\lambda^2 - 2(\kappa +
1)\lambda\Big)\,\frac{E_0 - E_e}{E_e}\,(\Big(E_0 - E_e)^2 - E^2_e +
m^2_e - 2MT_p\Big)\Big],\nonumber\\
\hspace{-0.3in}&&\zeta_2(E_e,T_p) =  -
\frac{1}{2}\,\frac{1}{E_e}\,\Big((E_0 - E_e)^2 + E^2_e - m^2_e - 2MT_p
+ 2 (E_0 - E_e)((T_p)_{\rm max} - T_p)\Big).
\end{eqnarray}
Integrating the electron--proton energy spectrum Eq.(\ref{labelI.6})
over the electron energy we obtain the proton--energy spectrum
\begin{eqnarray}\label{labelI.16}
\hspace{-0.3in}\frac{d\lambda_{\beta^-_c}(T_p)}{dT_p} = M\,
(1 + 3 \lambda^2) \frac{G^2_F|V_{ud}|^2}{4\pi^3}\,a(T_p),
\end{eqnarray}
where $a(T_p)$ is defined by
\begin{eqnarray}\label{labelI.17}
\hspace{-0.3in}a(T_p) = g_1(T_p) + a_0\Big(1 + \frac{1}{1 -
  \lambda^2}\,\frac{E_0}{M}\Big)\,g_2(T_p)
\end{eqnarray}
with the functions $g_1(T_p)$ and $g_2(T_p)$, given by the
integrals
\begin{eqnarray}\label{labelI.18}
\hspace{-0.3in}g_1(T_p) &=& \int^{(E_e)_{\rm max}}_{(E_e)_{\rm min}}
\zeta_1(E_e,T_p) \,\Big(1 + \frac{\alpha}{\pi}\,g_n(E_e)\Big)\,F(E_e,
Z = 1)\,E_e\,dE_e,\nonumber\\
\hspace{-0.3in}g_2(T_p) &=& \int^{(E_e)_{\rm max}}_{(E_e)_{\rm min}}
\zeta_2(E_e,T_p) \,\Big(1 + \frac{\alpha}{\pi}\,g_n(E_e) +
\frac{\alpha}{\pi}\,f_n(E_e)\Big)\,F(E_e, Z = 1)\,E_e\,dE_e.
\end{eqnarray}
For the calculation of the proton recoil asymmetry $C$, first, we have
to integrate over the azimuthal angles. After the integration we
arrive at the expression
\begin{eqnarray}\label{labelI.19}
\hspace{-0.3in}&&\frac{d^4 \lambda_{\beta^-_c}(E_e, k_p,
  \theta_p,\theta_{ep}, P)}{dE_e dk_pd\cos\theta_pd\cos\theta_{ep}} =
(1 + 3\lambda^2)\,\frac{G^2_F|V_{ud}|^2}{8\pi^3}\,\Bigg\{ 1 -
\tilde{a}(E_e)\,\frac{k^2_e + k_ek_p\cos\theta_{ep}}{E_e E} +
\frac{1}{M}\,\frac{1}{1 + 3\lambda^2}\nonumber\\
\hspace{-0.3in}&&\times\,\Bigg(E_0 - \frac{m^2_e}{E_e} +
\Big(\lambda^2 + 2(\kappa + 1)\lambda\Big)\frac{k^2_p + k_ek_p
  \cos\theta_{ep}}{E} - \Big(\lambda^2 - 2(\kappa + 1)\lambda\Big)\,
\frac{k_e k_p \cos\theta_{ep}}{E_e}\Bigg)\nonumber\\
\hspace{-0.3in}&& + P
\cos\theta_p\Bigg[\tilde{A}(E_e)\,\frac{k_e\cos\theta_{ep}}{E_e} -
  \tilde{B}(E_e)\,\frac{k_e\cos\theta_{ep} + k_p}{E} +
  \frac{1}{M}\,\frac{1}{1 + 3\lambda^2}\Bigg(\Big(- (2\kappa +
  1)\,\lambda- \lambda\,\frac{k^2_e + k_e k_p
  \cos\theta_{ep}}{E_e E}\Big)\,k_p \nonumber\\
\hspace{-0.3in}&& - 2(\kappa +
1)\lambda\,\frac{k_ek_p\cos\theta_{ep}}{E_e E}(k_p + k_e
\cos\theta_{ep})\Bigg)\Bigg]\Bigg\}\,\Big(1 +
\frac{\alpha}{\pi}\,g_n(E_e)\Big)\,F(E_e, Z = 1)\,\delta(m_n - E_p -
E_e - E)\,k_e E_e k^2_p.
\end{eqnarray}
Having integrated over $\cos\theta_{ep}$ we obtain the following
result
\begin{eqnarray}\label{labelI.20}
\hspace{-0.3in}&&\frac{d^3 \lambda_{\beta^-_c}(E_e, k_p, \theta_p,
  P)}{dE_e dk_pd\cos\theta_p} = (1 +
3\lambda^2)\,\frac{G^2_F|V_{ud}|^2}{16\pi^3}\,\Bigg\{2 (E_0 - E_e)E_e
k_p - \tilde{a}(E_e)\,((E_0 - E_e)^2 - k^2_p + k^2_e)k_p +
\frac{1}{M}\,\frac{1}{1 + 3\lambda^2}\nonumber\\
\hspace{-0.3in}&&\times\,\Bigg(2(E_0E_e - m^2_e) (E_0 - E_e) k_p + (1
+ 3\lambda^2) E_e (E^2_0 - m^2_e - k^2_p)k_p - (1 - \lambda^2)(E_0 -
E_e)(E^2_0 - m^2_e - k^2_p)k_p\nonumber\\
\hspace{-0.3in}&& + (\lambda^2 + 2(\kappa + 1)\lambda)\, E_e ((E_0
- E_e)^2 + k^2_p - k^2_e)k_p - (\lambda^2 - 2(\kappa +
1)\lambda)\,(E_0 - E_e) ((E_0 - E_e)^2 - k^2_p -
k^2_e)k_p\Bigg)\nonumber\\
\hspace{-0.3in}&& + P \cos\theta_p\Bigg[ \tilde{A}(E_e)\,(E_0 -
  E_e)((E_0 - E_e)^2 - k^2_p - k^2_e) - \tilde{B}(E_e)\,E_e((E_0 -
  E_e)^2 + k^2_p - k^2_e) + \frac{1}{M}\,\frac{1}{1 +
    3\lambda^2}\nonumber\\
\hspace{-0.3in}&&\times\, \Bigg(- (2\kappa + 1)\,\lambda\, 2 (E_0 -
E_e) E_e k^2_p - \lambda\,((E_0 - E_e)^2 - k^2_p + k^2_e) k^2_p -
(\kappa + 1)\lambda\,(((E_0 - E_e)^2 - k^2_e)^2 - k^4_p)\nonumber\\
\hspace{-0.3in}&& - \lambda(1 + \lambda)\,(3 (E_0 - E_e)^2 - k^2_p -
k^2_e)(E^2_0 - m^2_e - k^2_p) + \lambda(1 - \lambda) 2 (E_0 - E_e) E_e
(E^2_0 - m^2_e - k^2_p)\Bigg)\Bigg]\Bigg\}\nonumber\\
\hspace{-0.3in}&&\times\,\Big(1 +
\frac{\alpha}{\pi}\,g_n(E_e)\Big)\,F(E_e, Z = 1),
\end{eqnarray}
where we have kept the terms of order $1/M$ and the radiative
corrections of order $\alpha/\pi$. The integration over the electron
energy $E_e$ and the proton momentum $k_p$ gives one
\begin{eqnarray}\label{labelI.21}
\hspace{-0.3in}&&\frac{d \lambda_{\beta^-_c}(\theta_p,
  P)}{d\cos\theta_p} = (1 +
3\lambda^2)\,\frac{G^2_F|V_{ud}|^2}{16\pi^3}\,\Big\{X_1 +
\frac{\alpha}{\pi}\,X_2 +  \frac{1}{M}\,\frac{1}{1 +
  3\lambda^2}\,\Big[X_3 + (1 + 3\lambda^2)\, (X_4 + Y_1) - (1 -
\lambda^2)\, (X_5 + Y_2)\nonumber\\
\hspace{-0.3in}&& + \Big(\lambda^2 + 2(\kappa + 1)\lambda\Big)\, X_6 -
\Big(\lambda^2 - 2(\kappa + 1)\lambda\Big)\,X_7\Big] + P
\cos\theta_p\Big[ -(A_0 + B_0)\,X_8 + A_0\,X_9 +  A_0\,\frac{\alpha}{\pi}\,X_{10} -
B_0\,\frac{\alpha}{\pi}\,X_{11}\nonumber\\
\hspace{-0.3in}&& + \frac{1}{M}\,\frac{1}{1 + 3\lambda^2}\,\Big(
\lambda\,X_{12} - (\kappa + 1)\lambda\,X_{13} - (2\kappa +
1)\,\lambda\,X_{14}- \lambda\,(1 + \lambda)\,(X_{15} + Y_3) +
\lambda\,(1 - \lambda)\,(X_{16} + Y_4)\Big)\Big]\Big\}.
\end{eqnarray}
The factors $X_j$ for $j = 1, \ldots, 16$ are calculated for the
limits of the integration over $E_e$ and $k_p$ equal to
\begin{eqnarray}\label{labelI.23}
\hspace{-0.3in} 0 &\le& k_p \le (k_p)_{\rm max} = \sqrt{E^2_0 - m^2_e},
\nonumber\\
 \hspace{-0.3in} (E_e)_{\rm min} = \frac{(E_0 - k_p)^2 +
   m^2_e}{2(E_0 - k_p)} &\le& E_e \le (E_e)_{\rm max} = \frac{(E_0 +
   k_p)^2 + m^2_e}{2(E_0 + k_p)}.
\end{eqnarray}
The factors $Y_j$ for $j = 1,2,3,4$ are defined by the contributions of
the $1/M$ corrections to the limits of the integration over $E_e$ and
$k_p$.  The $1/M$ corrections to the limits of the integration over
$E_e$ are given in Eq.(\ref{labelI.12}). The limits of the integration
over $k_p$ have no $1/M$ corrections
\begin{eqnarray}\label{labelI.24}
\hspace{-0.3in}(k_p)_{\rm max} = \sqrt{\frac{m_p}{m_n}((m_n - m_p)^2 -
  m^2_e)} = \sqrt{E^2_0 - m^2_e}.
\end{eqnarray}
The account for the $1/M$ corrections to the limits of the integration
over $E_e$ can be carried out by the formula
\begin{eqnarray}\label{labelI.25}
\hspace{-0.3in}&&\int^{(k_p)_{\rm max}}_0\int^{(E_e)_{\rm max} +
    \Delta E_+(k_p)}_{(E_e)_{\rm min} + \Delta
    E_-(k_p)}\Phi(E_e,k_p)dE_e dk_p = \int^{(k_p)_{\rm
      max}}_0\int^{(E_e)_{\rm max}(k_p)}_{(E_e)_{\rm
        min}(k_p)}\Phi(E_e,k_p)dE_e dk_p\nonumber\\
\hspace{-0.3in}&& + \int^{(k_p)_{\rm max}}_0\Big[\Phi((E_e)_{\rm
    max},k_p)\,\Delta E_+(k_p) - \Phi((E_e)_{\rm
    min},k_p)\,\Delta E_-(k_p)\Big]dk_p,
\end{eqnarray}
where $\Delta E_{\pm}(k_p)$ are equal to
\begin{eqnarray}\label{labelI.25a}
\hspace{-0.3in} \Delta E_{\pm}(k_p) = \frac{1}{2}\Big(1 -
\frac{m^2_e}{(E_0 \pm \sqrt{2MT_p})^2}\Big)((T_p)_{\rm max} - T_p).
\end{eqnarray}
The results of the integration are
\begin{eqnarray*}
\hspace{-0.3in}&&X_1 = \int^{(k_p)_{\rm max}}_0\int^{(E_e)_{\rm
    max}}_{(E_e)_{\rm min}}2 (E_0 - E_e)E_e k_p F(E_e, Z = 1)\,dE_edk_p =
    0.235040\,{\rm MeV^5},\nonumber\\
\hspace{-0.3in}&&X_2 = \int^{(k_p)_{\rm max}}_0\int^{(E_e)_{\rm
    max}}_{(E_e)_{\rm min}}2 (E_0 - E_e)E_e k_p g_n(E_e) F(E_e, Z = 1)\,dE_edk_p
= 3.932201 \,{\rm MeV^5},\nonumber\\
\hspace{-0.3in}&&X_3 = \int^{(k_p)_{\rm max}}_0\int^{(E_e)_{\rm
    max}}_{(E_e)_{\rm min}} 2 (E_0E_e - m^2_e) (E_0 - E_e) k_p F(E_e,
Z = 1)\, dE_edk_p = 0.225097\,{\rm MeV^6},\nonumber\\
\hspace{-0.3in}&&X_4 = \int^{(k_p)_{\rm max}}_0\int^{(E_e)_{\rm
    max}}_{(E_e)_{\rm min}} E_e (E^2_0 - m^2_e - k^2_p)k_p F(E_e, Z =
1)\,dE_edk_p = 0.190889\,{\rm MeV^6},\nonumber\\
\hspace{-0.3in}&&X_5 = \int^{(k_p)_{\rm max}}_0\int^{(E_e)_{\rm
    max}}_{(E_e)_{\rm min}} (E_0 - E_e)(E^2_0 - m^2_e - k^2_p)k_p
F(E_e, Z = 1)\,dE_edk_p = 0.112935\,{\rm MeV^6},\nonumber\\
\hspace{-0.3in}&&X_6 = \int^{(k_p)_{\rm max}}_0\int^{(E_e)_{\rm
    max}}_{(E_e)_{\rm min}} E_e((E_0 - E_e)^2 + k^2_p - k^2_e)k_p
F(E_e, Z = 1)\,dE_edk_p = 0.112935\,{\rm MeV^6},\nonumber\\
\hspace{-0.3in}&&X_7 = \int^{(k_p)_{\rm max}}_0\int^{(E_e)_{\rm
    max}}_{(E_e)_{\rm min}} (E_0 - E_e)((E_0 - E_e)^2 - k^2_p - k^2_e)k_p
F(E_e, Z = 1)\,dE_edk_p = - 0.112162\,{\rm MeV^6},\nonumber\\
\hspace{-0.3in}&&X_8 = \int^{(k_p)_{\rm max}}_0\int^{(E_e)_{\rm
    max}}_{(E_e)_{\rm min}} E_e((E_0 - E_e)^2 + k^2_p - k^2_e)\,
F(E_e, Z = 1)\,dE_edk_p = 0.129699\,{\rm MeV^5},\nonumber\\
\hspace{-0.3in}&&X_9 = \int^{(k_p)_{\rm max}}_0\int^{(E_e)_{\rm
    max}}_{(E_e)_{\rm min}}\Big[(E_0 - E_e)((E_0 - E_e)^2-k^2_p -
  k^2_e) + E_e((E_0 - E_e)^2 + k^2_p - k^2_e)\Big]\, F(E_e, Z =
1)\,dE_edk_p =\nonumber\\
\hspace{-0.3in}&&= 0.001059\,{\rm MeV^5},\nonumber\\
\hspace{-0.3in}&&X_{10} = \int^{(k_p)_{\rm max}}_0\int^{(E_e)_{\rm
    max}}_{(E_e)_{\rm min}}(E_0 - E_e)((E_0 - E_e)^2-k^2_p -
k^2_e)(g_n(E_e) + f_n(E_e))\, F(E_e, Z = 1) \,dE_edk_p=\nonumber\\
\hspace{-0.3in}&&= - 2.157085\,{\rm
  MeV^5},\nonumber\\
\hspace{-0.3in}&&X_{11} = \int^{(k_p)_{\rm max}}_0\int^{(E_e)_{\rm
    max}}_{(E_e)_{\rm min}}E_e((E_0 - E_e)^2 + k^2_p - k^2_e)\, g_n(E_e)
\, F(E_e, Z = 1)\,dE_edk_p = 2.244146\,{\rm MeV^5},\nonumber\\
\hspace{-0.3in}&&X_{12} = \int^{(k_p)_{\rm max}}_0\int^{(E_e)_{\rm
    max}}_{(E_e)_{\rm min}}[ E_0 E_e ((E_0 - E_e)^2 + k^2_p - k^2_e) -
  E_0 (E_0 - E_e)((E_0 - E_e)^2 - k^2_p - k^2_e)\nonumber\\
\hspace{-0.3in}&& - ((E_0 - E_e)^2 - k^2_p + k^2_e)k^2_p]\, F(E_e, Z =
1)\,dE_edk_p = 0.353804\,{\rm MeV^6},\nonumber\\
\hspace{-0.3in}&&X_{13} = \int^{(k_p)_{\rm max}}_0\int^{(E_e)_{\rm
    max}}_{(E_e)_{\rm min}}(((E_0 - E_e)^2 - k^2_e)^2 - k^4_p)\,
F(E_e, Z = 1)\,dE_edk_p = - 0.119181\,{\rm MeV^6},\nonumber\\
\hspace{-0.3in}&&X_{14} = \int^{(k_p)_{\rm max}}_0\int^{(E_e)_{\rm
    max}}_{(E_e)_{\rm min}}2 (E_0 - E_e) E_e k^2_p\, F(E_e, Z =
1)\,dE_edk_p = 0.186815\,{\rm MeV^6},\nonumber\\
\hspace{-0.3in}&&X_{15} = \int^{(k_p)_{\rm max}}_0\int^{(E_e)_{\rm
    max}}_{(E_e)_{\rm min}}(3 (E_0 - E_e)^2 - k^2_p - k^2_e)(E^2_0 -
m^2_e - k^2_p)\, F(E_e, Z = 1)\,dE_edk_p = - 0.002707\,{\rm
  MeV^6},\nonumber\\
\end{eqnarray*}
\begin{eqnarray}\label{labelI.22}
\hspace{-0.3in}&&X_{16} = \int^{(k_p)_{\rm max}}_0\int^{(E_e)_{\rm
    max}}_{(E_e)_{\rm min}}2 (E_0 - E_e) E_e (E^2_0 - m^2_e - k^2_p)\, F(E_e, Z =
1)\,dE_edk_p = 0.291289\,{\rm MeV^6},\nonumber\\
\hspace{-0.3in}&&X_{17} = \int^{(k_p)_{\rm max}}_0\int^{(E_e)_{\rm
    max}}_{(E_e)_{\rm min}}\frac{m_e}{E_e}\,(E_0 - E_e) ((E_0 - E_e)^2
- k^2_p - k^2_e)\, F(E_e, Z = 1)\,dE_edk_p = - 0.076306\,{\rm
  MeV^5},\nonumber\\
\hspace{-0.3in}&&X_{18} = \int^{(k_p)_{\rm max}}_0\int^{(E_e)_{\rm
    max}}_{(E_e)_{\rm min}} m_e\,((E_0 - E_e)^2 + k^2_p - k^2_e)\, F(E_e, Z =
1)\,dE_edk_p = 0.093560\,{\rm MeV^5}.
\end{eqnarray}
The factors $X_{17}$ and $X_{18}$ are related to contributions of
interactions beyond the SM (see section~\ref{sec:sensitivity}).  The
factors $Y_j$ for $j = 1,2,3,4$ are given by
\begin{eqnarray}\label{labelI.26}
\hspace{-0.3in}&&Y_1 = \frac{1}{2}\int^{(k_p)_{\rm max}}_0\Big[\Big(E_0 -
  (E_e)_{\rm max}\Big)(E_e)_{\rm max}\Big(1 - \frac{m^2_e}{(E_0 +
    k_p)^2}\Big)\,F((E_e)_{\rm max}, Z = 1) \nonumber\\
\hspace{-0.3in}&&- \Big(E_0 - (E_e)_{\rm min}\Big)(E_e)_{\rm
  min}\Big(1 - \frac{m^2_e}{(E_0 - k_p)^2}\Big)\,F((E_e)_{\rm min}, Z
= 1)\Big]\,(E^2_0 - m^2_e - k^2_p)\, k_p\, dk_p = 0.077953\,{\rm
  MeV^6},\nonumber\\
\hspace{-0.3in}&&Y_2 = \frac{1}{4}\int^{(k_p)_{\rm max}}_0\Big[\Big(\Big(E_0 -
  (E_e)_{\rm max}\Big)^2 - k^2_p + (E_e)^2_{\rm max} -
  m^2_e\Big)\Big(1 - \frac{m^2_e}{(E_0 + k_p)^2}\Big)\nonumber\\
\hspace{-0.3in}&&\times\,F((E_e)_{\rm max}, Z = 1) - \Big(\Big(E_0 -
(E_e)_{\rm min}\Big)^2 - k^2_p + (E_e)^2_{\rm min} - m^2_e\Big)\Big(1
- \frac{m^2_e}{(E_0 - k_p)^2}\Big)\nonumber\\
\hspace{-0.3in}&&\times\,F((E_e)_{\rm min}, Z = 1)\Big]\,(E^2_0 -
m^2_e - k^2_p)\,k_p\, dk_p = - 0.000772\,{\rm MeV^6},\nonumber\\
\hspace{-0.3in}&&Y_3 = \frac{1}{2}\int^{(k_p)_{\rm
    max}}_0\Big[\Big(E_0 - (E_e)_{\rm max}\Big)\Big(\Big(E_0 -
  (E_e)_{\rm max}\Big)^2 - k^2_p - (E_e)^2_{\rm max} +
  m^2_e\Big)\Big(1 - \frac{m^2_e}{(E_0 + k_p)^2}\Big)\nonumber\\
\hspace{-0.3in}&&\times\,F((E_e)_{\rm max}, Z = 1) - \Big(E_0 -
(E_e)_{\rm min}\Big)\Big(\Big(E_0 - (E_e)_{\rm min}\Big)^2 - k^2_p -
(E_e)^2_{\rm min} + m^2_e\Big)\Big(1 - \frac{m^2_e}{(E_0 -
  k_p)^2}\Big)\nonumber\\
\hspace{-0.3in}&&\times\,F((E_e)_{\rm min}, Z = 1)\Big]\,(E^2_0 - m^2_e -
k^2_p)\, dk_p= - 0.204003\,{\rm MeV^6},\nonumber\\
\hspace{-0.3in}&&Y_4 = \frac{1}{2}\int^{(k_p)_{\rm max}}_0\Big[
  (E_e)_{\rm max}\Big(\Big(E_0 - (E_e)_{\rm max}\Big)^2 + k^2_p -
  (E_e)^2_{\rm max} + m^2_e\Big)\Big(1 - \frac{m^2_e}{(E_0 +
    k_p)^2}\Big)\nonumber\\
\hspace{-0.3in}&&\times\,F((E_e)_{\rm max}, Z = 1) - (E_e)_{\rm min}
\Big(\Big(E_0 - (E_e)_{\rm min}\Big)^2 + k^2_p - (E_e)^2_{\rm min} +
m^2_e\Big)\Big(1 - \frac{m^2_e}{(E_0 - k_p)^2}\Big)\nonumber\\
\hspace{-0.3in}&&\times\,F((E_e)_{\rm min}, Z = 1)\Big]\,(E^2_0 -
m^2_e - k^2_p)\, dk_p= - 0.082343\,{\rm MeV^6}.
\end{eqnarray}
The correctness of our calculation of the electron--proton energy
distribution $a(E_e, T_p)$ we may verify by calculating the lifetime
of the neutron. Having integrated Eq.(\ref{labelI.21}) over
$\cos\theta_p$ we obtain the rate of the neutron $\beta^-$--decay
\begin{eqnarray}\label{labelI.27}
\hspace{-0.3in} (\lambda_{\beta^-_c})_{\rm SM} &=& (1 +
3\lambda^2)\,\frac{G^2_F|V_{ud}|^2}{8\pi^3}\,\Big\{X_1 +
\frac{\alpha}{\pi}\,X_2 + \frac{1}{M}\,\frac{1}{1 +
  3\lambda^2}\,\Big[X_3 + (1 + 3\lambda^2)\, (X_4 + Y_1) - (1 -
  \lambda^2)\, (X_5 + Y_2)\nonumber\\
\hspace{-0.3in}&+& \Big(\lambda^2 + 2(\kappa + 1)\lambda\Big)\, X_6 -
\Big(\lambda^2 - 2(\kappa + 1)\lambda\Big)\,X_7\Big]\Big\},
\end{eqnarray}
related to the lifetime as $ (\lambda_{\beta^-_c})^{-1}_{\rm SM} =
(\tau_n)_{\rm SM}$. The numerical factors $X_j$ and $Y_k$ for $j = 1,
\ldots, 7$ and $k = 1,2$ define the lifetime of the neutron with the
account for the radiative ($X_2 \neq 0$) and the $1/M$ ($X_j \neq 0$
for $j = 3, \ldots, 7$ and $Y_k \neq 0$ for $k =1,2$) corrections.

Let us consider three possibilities: 1) $X_j = 0$ for
$j = 2,\ldots, 7$ and $Y_k = 0$ for $k = 1,2$ (i.e. without radiative
and $1/M$ corrections), 2) $X_k \neq 0$ for $k = 1,2$ and $X_j = 0 $
for $j = 3, \ldots, 7$ and $Y_k = 0$ for $k = 1,2$ (i.e. with
radiative but without $1/M$ corrections) and 3) $X_k \neq 0$ for $k =
1, \dots , 7$ and $Y_k \neq 0$ for $k = 1,2$ (with radiative and $1/M$
corrections).  For these cases we obtain 1) $(\tau_n)_{\rm SM} =
915.3\,{\rm s}$, 2) $(\tau_n)_{\rm SM} = 881.0\,{\rm s}$ and 3)
$(\tau_n)_{\rm SM} = 879.6\,{\rm s}$, respectively.  These results
agree well with the lifetimes of the neutron, adduced in Table I (see
section~\ref{sec:conclusion}).

The correlation of the neutron spin and the proton 3--momentum is
described by the angular distribution \cite{PRA1}
\begin{eqnarray}\label{labelI.28}
\hspace{-0.3in} 4\pi\frac{dW(\theta_p)}{d\Omega_p} = 1 + 2 P C
\cos\theta_p,
\end{eqnarray}
where $P$ is a neutron polarization. The correlation coefficient $C$
is defined by
\begin{eqnarray}\label{labelI.29}
\hspace{-0.3in}&&C = - \frac{1}{2}\frac{X_8}{X_1}\,(A_0 + B_0) +
\frac{1}{2}\frac{X_9}{X_1}\,A_0 +
\frac{\alpha}{\pi}\,\frac{1}{2}\frac{X_{10}}{X_1}\,A_0 -
\frac{\alpha}{\pi}\,\frac{1}{2}\frac{X_{11}}{X_1}\,B_0 +
\frac{1}{M}\,\frac{1}{1 +
  3\lambda^2}\,\Big(\lambda\,\frac{1}{2}\frac{X_{12}}{X_1} - (\kappa +
1)\,\lambda\,\frac{1}{2}\frac{X_{13}}{X_1}\nonumber\\
\hspace{-0.3in}&& - (2\kappa +
1)\,\lambda\,\frac{1}{2}\frac{X_{14}}{X_1} -\lambda(1 +
\lambda)\,\frac{1}{2}\frac{X_{15} + Y_3}{X_1} + \lambda(1 -
\lambda)\,\frac{1}{2}\frac{X_{16} + Y_4}{X_1}\Big) + (A_0 +
B_0)\,\frac{X_8}{X_1}\,\Big\{
\frac{\alpha}{\pi}\,\frac{1}{2}\frac{X_2}{X_1} +
\frac{1}{M}\,\frac{1}{1 + 3\lambda^2}\nonumber\\
\hspace{-0.3in}&&\times\,\Big(\frac{1}{2}\frac{X_3}{X_1} + (1 +
3\lambda^2)\,\frac{1}{2}\frac{X_4 + Y_1}{X_1} - (1 - \lambda^2)\,
\frac{1}{2}\frac{X_5 + Y_2}{X_1}+ (\lambda^2 + 2(\kappa +
1)\lambda)\,\frac{1}{2}\frac{X_6}{X_1} - (\lambda^2 - 2(\kappa +
1)\lambda)\,\frac{1}{2}\frac{X_7}{X_1}\Big)\Big\},
\end{eqnarray}
where the factor $x_C = X_8/2X_1 = 0.27591$ agrees well with the
factor $x_C = 0.27594$, calculated by Gl\"uck \cite{PRA2}. The
apperance of the term $A_0 X_9/2X_1$ is related to the deviation of
the Fermi function $F(E_e, Z = 1)$ from unity.

The contributions of interactions beyond the SM (see Appendix G)
changes the electron--proton energy distribution $a(E_e, T_p)$ as
follows 
\begin{eqnarray}\label{labelI.30}
\hspace{-0.3in}&&a_{\rm eff}(E_e, T_p) = \Big(1 + \frac{1}{1 +
  3\lambda^2}\Big({\rm Re}(\delta C_V - \delta \bar{C}_V) -
3\lambda\,{\rm Re}(\delta C_A - \delta \bar{C}_A)\Big) +
b_F\,\frac{m_e}{E_e}\Big)\zeta_1(E_e, T_p)\nonumber\\
\hspace{-0.3in}&&+ \Big[a_0 + \frac{1}{1 + 3
    \lambda^2} \Big({\rm Re}(\delta C_V - \delta \bar{C}_V) +
  \lambda\,{\rm Re}(\delta C_A - \delta \bar{C}_A)\Big)\Big]\,\Big(1 +
\frac{1}{1 - \lambda^2}\,\frac{E_0}{M}\Big)\,\Big(1 +
\frac{\alpha}{\pi}\,f_n(E_e)\Big)\,\zeta_2(E_e, T_p).
\end{eqnarray}
Multiplying $a(E_e, T_p)$ by the lifetime of the neutron $\tau_n$ (see
Eq.(\ref{labelG.9})) we obtain 
\begin{eqnarray}\label{labelI.31}
\hspace{-0.3in}&&\tau_n a_{\rm eff}(E_e, T_p) = (\tau_n)_{\rm SM} \Big\{\Big(1 +
b_F\,\frac{m_e}{E_e} - b_F\,\Big\langle
\frac{m_e}{E_e}\Big\rangle_{\rm SM}\Big)\zeta_1(E_e, T_p) + \Big[a_0 +
 \frac{1}{1 + 3 \lambda^2} \Big({\rm Re}(\delta C_V - \delta
  \bar{C}_V) + \lambda\nonumber\\
\hspace{-0.3in}&&\times\,{\rm Re}(\delta C_A - \delta
\bar{C}_A)\Big) - a_0\Big(\frac{1}{1 + 3\lambda^2}\Big({\rm
  Re}(\delta C_V - \delta \bar{C}_V) - 3\lambda\,{\rm Re}(\delta C_A -
\delta \bar{C}_A)\Big) + b_F\,\Big\langle
\frac{m_e}{E_e}\Big\rangle_{\rm SM}\Big)\Big]\,\Big(1 + \frac{1}{1 -
  \lambda^2}\,\frac{E_0}{M}\Big)\nonumber\\
\hspace{-0.3in}&&\times\,\Big(1 +
\frac{\alpha}{\pi}\,f_n(E_e)\Big)\,\zeta_2(E_e, T_p)\Big\}.
\end{eqnarray}
This allows to reduce the analysis of contributions of vector and
axial--vector interactions beyond the SM to the electron--proton
energy distribution to the analysis of these contributions to the
axial coupling constant $\lambda$ (see section~\ref{sec:a0exp} and
section~\ref{sec:sensitivity}). Contributions of interactions beyond
the SM to the proton recoil asymmetry $C$ are given in
section~\ref{sec:sensitivity}.

The radiative corrections to the electron--proton energy distribution
$a(E_e, T_p)$ Eq.(\ref{labelI.7}) or the proton--energy spectrum
$a(T_p)$ Eq.(\ref{labelI.17}) and the proton recoil asymmetry $C$
Eq.(\ref{labelI.29}) have been calculated at the neglect the
contributions of the proton--photon correlations in the proton recoil
energy and angular distribution of the radiative $\beta^-$--decay of
the neutron. As has been pointed out by Gl\"uck \cite{Gluck1997}, the
contributions of the radiative $\beta^-$--decay of the neutron to the
proton recoil energy and angular distribution, caused by the
proton--photon correlations, demand a detailed analysis.  For the aim
of a consistent calculation of the contributions of the
nucleus--photon and hadron--photon correlations in the radiative
nuclear and hadronic $\beta$--decays Gl\"uck has used the Monte Carlo
simulation method. The calculation of the proton--photon correlations
has been recently performed in \cite{Ivanov2013}. There has been shown
that the contributions of the proton--photon correlations to the
lifetime of the neutron $\tau_n$ and the electron--proton energy
distribution $a(E_e, T_p)$ (or the proton--energy spectrum $a(T_p)$)
are of order $10^{-5}$ and can be neglected with respect to the
contributions of order $10^{-4}$ of interactions beyond the SM. Such a
neglect of the contributions of the proton--photon correlations
confirms the correctness of the use the functions
$(\alpha/\pi)\,g_n(E_e)$ and $(\alpha/\pi)\,f_n(E_e)$ for the
description of the radiative corrections in the lifetime of the
neutron and the electron--proton energy distribution (or the
proton--energy spectrum), respectively. In turn, the contributions of
the proton--photon correlations to the proton recoil angular
distribution and the proton recoil asymmetry $C$ are of order
$10^{-4}$. Hence they should be taken into account for the correct
determination of contributions of order $10^{-4}$ beyond the SM. The
contributions of these corrections changes the proton recoil asymmetry
$C$ as follows \cite{Ivanov2013}
\begin{eqnarray}\label{labelI.32}
\hspace{-0.3in} &&C = - \Big(x_C + \frac{\alpha}{\pi}\,x_{\rm
  eff}\Big)\,(A_0 + B_0) + \frac{1}{2}\frac{X_9}{X_1}\,A_0 +
\frac{1}{M}\,\frac{1}{1 +
  3\lambda^2}\,\Big(\lambda\,\frac{1}{2}\frac{X_{12}}{X_1} - (\kappa +
1)\,\lambda\,\frac{1}{2}\frac{X_{13}}{X_1} - (2\kappa +
1)\,\lambda\,\frac{1}{2}\frac{X_{14}}{X_1}\nonumber\\
\hspace{-0.3in}&& -\lambda(1 + \lambda)\,\frac{1}{2}\frac{X_{15} +
  Y_3}{X_1} + \lambda(1 - \lambda)\,\frac{1}{2}\frac{X_{16} +
  Y_4}{X_1}\Big) + (A_0 +
B_0)\,\frac{X_8}{X_1}\,\Big\{\frac{\alpha}{\pi}\,\frac{1}{2}\frac{X_2}{X_1}
+ \frac{1}{M}\,\frac{1}{1 +
  3\lambda^2}\,\Big(\frac{1}{2}\frac{X_3}{X_1} + (1 +
3\lambda^2)\nonumber\\
\hspace{-0.3in}&&\times\,\frac{1}{2}\frac{X_4 + Y_1}{X_1} - (1 -
\lambda^2)\, \frac{1}{2}\frac{X_5 + Y_2}{X_1}+ (\lambda^2 + 2(\kappa +
1)\lambda)\,\frac{1}{2}\frac{X_6}{X_1} - (\lambda^2 - 2(\kappa +
1)\lambda)\,\frac{1}{2}\frac{X_7}{X_1}\Big)\Big\},
\end{eqnarray}
where $x_{\rm eff} = X_{\rm eff}/2X_1 = 4.712120$
\cite{Ivanov2013}. One may see that the contributions of the
proton--photon correlations make the radiative corrections to the
proton recoil asymmetry $C$ symmetric with respect to a change $A_0
\longleftrightarrow B_0$ as well as the main term $C_0 = - x_C (A_0 +
B_0)$.


\begin{thebibliography}{9}
\bibitem{SPT1a} 
There is also the pseudoscalar lepton--baryon weak
  coupling constant, the contribution of which can be neglected with
  respect to the contributions of order $10^{-4}$.
\bibitem{Wilkinson1} 
In comparison with \cite{Wilkinson}, we have
  replaced $\lambda \to - \lambda$ and included the factor $2 (\lambda
  - \kappa - 1)$ into the definition of the coefficients $A^{(\rm
    W)}_j$ for $j = 1,2,3$.
\bibitem{Cooper1} 
In Eq.(\ref{label3}) of Ref.\cite{Cooper} the sum of
the momenta $(\vec{p}_e + 2 \vec{k}\,)$ should be replaced by $(2
\vec{p}_e + \vec{k}\,)$. After such a replacement Eq.(\ref{labelB.10})
for the unpolarised neutron coincides fully with Eq.(\ref {label3}) of
Ref.\cite{Cooper}.
\bibitem{SPT3a} 
With respect to the definition of the coupling
  constants in the original paper \cite{SPT3} we have extracted the
  common factor $G_FV_{ud}/\sqrt{2}$ and replaced $a^h_{LL}$ and
  $a^h_{LR}$ by $a^h_{LL} \to a^h_{LL} + (1 - \lambda)/2$ and $a^h_{LR} \to
  a^h_{LR} + (1 + \lambda)/2$, respectively, where the coupling
  constants $a^h_{LL}$ and $a^h_{LR}$ describe deviations from the
  coupling constants of the SM.
\end{thebibliography}
\end{document}